\documentclass[11pt]{article}
\usepackage[top=2cm,left=2cm,right=2cm]{geometry}

\usepackage[inline]{showlabels}
\usepackage[numbers,square]{natbib}

\usepackage{axodraw}
\usepackage{epsfig}
\usepackage{cancel}
\usepackage{caption}
\usepackage{feynmf} 
\usepackage{amssymb}
\usepackage{amsfonts}
\usepackage{epsf}
\usepackage{rotating}
\usepackage{graphicx}
\usepackage{amsmath}
\usepackage{fancyhdr}
\usepackage{subfigure}
\usepackage{graphics}
\usepackage{pstricks}
\usepackage{color}
\usepackage{frontespizio}
\usepackage{hyperref}
\hypersetup{
    colorlinks,
    citecolor=green,
    filecolor=black,
    linkcolor=blue,
    urlcolor=black
}
\usepackage{type1ec}
\usepackage[T1]{fontenc}
\usepackage{lettrine}
\usepackage{bbold}
\usepackage{calligra}
\usepackage{tikz}
\usepackage{subfigure}
\usepackage{mathrsfs}
\usepackage{curve2e}
\usepackage{setspace}
\usepackage{indentfirst}
\usepackage{emptypage}
\usepackage[babel]{csquotes} 
\usepackage[font=small,labelfont=bf,labelsep=quad]{caption} 
\usepackage{graphicx} 
\usepackage{listings} 
\usetikzlibrary{patterns}
\usepackage{relsize}
\usetikzlibrary{intersections,positioning}
\usetikzlibrary{decorations.pathmorphing,decorations.markings,arrows,positioning}
\usepackage{braket}
\usepackage{mathrsfs}
\usepackage{stackengine}
\usepackage{calc}
\newlength\shlength
\newcommand\xshlongvec[2][0]{\setlength\shlength{#1pt}%
  \stackengine{-5.6pt}{$#2$}{\smash{$\kern\shlength%
    \stackengine{7.55pt}{$\mathchar"017E$}%
      {\rule{\widthof{$#2$}}{.57pt}\kern.4pt}{O}{r}{F}{F}{L}\kern-\shlength$}}%
      {O}{c}{F}{T}{S}}

\newcommand{\mO}{\mathcal{O}}

\newcommand{\sdfrac}[2]{\mbox{\small$\displaystyle\frac{#1}{#2}$}}

\IfFileExists{dsfont.sty}
{\usepackage{dsfont}
	\let\mathbb=\mathds
	\newcommand{\id}{\mathds{1}}}
{\typeout{Package dsfont.sty was not found, using alternative macros.}
	\let\mathds=\mathbb
	\newcommand{\id}{\mbox{1 \kern-.59em {\rm l}}}}

\usepackage{slashed}
\usepackage{units}
\usepackage{setspace}
\newcommand{\eeqa}{\end{eqnarray}}
\newcommand{\beqa}{\begin{eqnarray}}

\newcommand{\nn}{\nonumber}
\let\a=\alpha   \let\b=\beta   \let\g=\gamma   \let\d=\delta
    \let\h=\eta     
    \let\k=\kappa  \let\l=\lambda  \let\m=\mu
\let\n=\nu           \let\p=\pi      
\let\s=\sigma        
\let\c=\chi         
\let\D=\Delta
\let\d=\delta
\let\s=\sigma

\let\G=\Gamma
\newcommand{\figref}[1]{Fig.~\ref{#1}}			
\renewcommand{\a}{\alpha}

\newcommand{\Tr}{\text{Tr}}

\newcommand{\ul}{\underline}
\def\nbox#1#2{\vcenter{\hrule \hbox{\vrule height#2in
			\kern#1in \vrule} \hrule}}
\def\sq{\,\raise.5pt\hbox{$\nbox{.09}{.09}$}\,}
\def\sqb{\,\raise.5pt\hbox{$\overline{\nbox{.09}{.09}}$}\,}

\newcommand{\bea}{\begin{eqnarray}}
\newcommand{\eea}{\end{eqnarray}}
\newcommand{\be}{\begin{equation}}
\newcommand{\ee}{\end{equation}}

\newcommand{\bes}{\begin{subequations}}
	\newcommand{\ees}{\end{subequations}}

\def\nn{\nonumber\\}

\numberwithin{equation}{section}

\usepackage{accents}

\begin{document}
\begin{center}
\vspace{1.5cm}
{\Large\bfseries
Exact Correlators from Conformal Ward Identities in Momentum Space}
\vspace{0.2cm}
{\Large\bfseries and the Perturbative $TJJ$ Vertex \\}
\vspace{0.3cm}
{\Large\bfseries}

\vspace{0.3cm}

\vspace{2 cm}

{\bf  Claudio Corian\`o and Matteo Maria Maglio \\}

\vspace{0.5cm}

{\em Dipartimento di Matematica e Fisica "Ennio De Giorgi"\\ Universit\`a del Salento and INFN Lecce, \\
Via Arnesano, 73100 Lecce, Italy}

\vspace{0.5cm}

\end{center}

\begin{abstract}
We present a general study of 3-point functions of conformal field theory in momentum space, following a reconstruction method for tensor correlators, based on the solution of the conformal Ward identities (CWI' s), introduced in recent works by Bzowski, McFadden and Skenderis (BMS). We investigate and detail the structure of the CWI's, their non-perturbative solutions and the transition to momentum space, comparing them to perturbation theory by taking QED as an example. We then proceed with an analysis of the $TJJ$ correlator, presenting independent and detailed re-derivations of the conformal equations in the reconstruction method of BMS, originally formulated using a minimal tensor basis in the transverse traceless sector. A careful comparison with a second basis introduced in previous studies shows that this correlator is affected by one anomaly pole in the graviton (T) line, induced by renormalization. The result shows that the origin of the anomaly, in this correlator, should be necessarily attributed to the exchange of a massless effective degree of freedom.  
Our results are then exemplified in massless QED at one-loop in $d$-dimensions, expressed in terms of perturbative master integrals. 
An independent analysis of the Fuchsian character of the solutions, which bypasses the 3K integrals, is also presented. We show that the combination of field theories at one-loop - with a specific field content of degenerate massless scalar and fermions - is sufficient to generate the complete non-perturbative solution, in agreement with a previous study in coordinate space.
The result shows that free conformal field theories, in specific dimensions, arrested at one-loop, reproduce the general result for the $TJJ$. Analytical checks of this correspondence are presented in $d=3,4$ and $5$ spacetime dimensions. This implies that the generalized 3K integrals of the BMS solution can be expressed in terms of the two single master integrals $B_0$ and $C_0$  of 2- and 3-point functions, with significant simplifications. 

\end{abstract}

\newpage
\section{Introduction} 
The analysis of multi-point correlation functions in conformal field theory (CFT) is of outmost importance 
in high energy physics and in string theory, where exact results for lower (2- and 3-)point functions are combined with the operator product expansion (OPE) in order to characterize the structure of correlators of higher orders. This is the key motivation for a bootstrap program in $d=4$ spacetime dimensions. \\
The enlarged $SO(2,4)$ symmetry of CFT's - respect to Poincar\'{e} invariance - has been essential for establishing the form of some of their correlation functions. For 3-point functions, the solution of the conformal constraints in coordinate space allows to determine such correlators only up to few constants \cite{Osborn:1993cr, Erdmenger:1996yc}, which can then be fixed within a specific realization of a theory.  In the case of a Lagrangian realization of a given CFT, such constants are expressed in terms of its (massless) field content (number of scalars, vectors, fermions), according to rather simple algebraic relations.  

Except for perturbative studies performed at Lagrangian level, such as in the case of the $N=4$ super Yang-Mills theory, which reach considerably high orders in the gauge coupling expansion, most of these analysis are performed in coordinate space, with no reference to any specific Lagrangian. 

 There are obvious reasons for this. The first is that the inclusion of the conformal constraints is more straightforward to obtain in coordinate space, compared to momentum space. The second is that the operator product expansion (OPE) in momentum space is  difficult to perform, especially for correlators of higher orders ($\geq 3$ ), in the Minkowski region. 
 However, there are also some advantages which are typical of a momentum space analysis, and these are related to 
 the availability of dimensional regularization (DR), at least at perturbative level, and to the technology of master integrals, which has allowed to compute large classes of multiloop amplitudes. \\ 
 Another advantage has to do with the identification of the conformal anomaly \cite{Capper:1975ig}, which can be automatically extracted  in DR (in $d$ spacetime dimensions), being proportional to the $1/(d-4)$ singularity of the corresponding correlators. In coordinate space, instead, the anomaly contributions has to be added by hand by the inclusion of an inhomogeneous local term 
 (i.e by pinching all of its external coordinates), whose structure has to be inferred indirectly  \cite{Osborn:1993cr}. \\  
Finally, a crucial issue concerns the physical character of the anomaly, which does not find any simple particle interpretation in position space, while it is clearly associated to the appearance of an anomaly pole in momentum 
space \cite{Giannotti:2008cv,2009PhLB..682..322A,Armillis:2009sm} in an uncontracted anomaly vertex. One finds, by a perturbative one-loop analysis of any anomalous correlator, that the anomaly is always associated with such massless exchanges in the corresponding diagram. It is therefore possible to identify them as effective degrees of freedom induced by the anomaly, present in the 1PI (one-particle irreducible) effective action. The physical significance of such contributions has been stressed in several previous works \cite{Giannotti:2008cv,Armillis:2009pq, Armillis:2009sm} along the years. They have recently discussed in condensed matter theory in the context of topological insulators and of Weyl semimetals \cite{Chernodub:2017jcp,Rinkel:2016dxo}.\\
One of the goals of our work is to compare and extend previous perturbative analysis of the $TJJ$ correlator with more 
recent ones based on the solution of the conformal Ward identities (CWI's) in momentum space \cite{Bzowski:2013sza,Bzowski:2015yxv,Bzowski:2017poo,2014JHEP...03..111B}. This correlator is the simplest one describing the coupling of gravity to ordinary matter in QED and it has been 
investigated in perturbation theory from several directions \cite{Bastianelli:2012es,Bastianelli:2012bz,Bonora:2017gzz,Bonora:2014qla, Bastianelli:2016nuf}. 
  \subsection{Direct Fourier transform and the reconstruction program}
In principle, one can move from coordinate space to position space in a CFT by a Fourier transform. This was the approach of \cite{Coriano:2012wp} for 3-point functions, which can be explicitly worked out by introducing a regulator ($\omega$) for the transform very much alike DR. The regulator serves as an intermediate step since some of the components of the correlators 
in position space are apparently non-transformable. It  has been shown that $1/\omega$ poles generated by the transform {\em cancel} in all the correlators analysed, giving a complete expression for these in momentum space. The result is expressed in terms of ordinary and logarithmic master integrals of Feynman type, for which, in the latter case, it is possible to derive recursion relations as for ordinary ones \cite{Coriano:2012wp}. The advantage of such approach is of being straightforward and algorithmic. It may be essential and probably the only manageable way to re-express the bootstrap program of CFT's in momentum space beyond 3- and 4- point functions, from the original coordinate space analysis. 
Consistency with the analysis presented in \cite{Coriano:2012wp} implied rather directly that such logarithmic integrals had to be re-expressed in terms of ordinary Feynman integrals. In fact, it was shown in the same study that the TJJ correlator  was entirely reproduced by a free field theory in coordinate space. Our analysis in momentum space is in complete agreement with this former result.

\subsection{Reconstruction}
An alternative method has been developed more recently,  based on the direct solution of the conformal Ward identities in momentum space. The method has been proposed in \cite{2014JHEP...03..111B} and \cite{Coriano:2013jba} for scalar 3-point functions and extensively generalized to tensor correlators in \cite{2014JHEP...03..111B}.

 Several issues related to the renormalization of the solutions of the conformal Ward identities have been investigated in \cite{Bzowski:2015yxv, Bzowski:2017poo}, adopting the formalism of the 3K integrals (i.e. parametric integrals of 3 Bessel functions). 
Several analysis in momentum  space, for specific applications, have been worked out \cite{Coriano:2012hd, Bzowski:2011ab}, but the generality of the approach is clearly a significant feature of \cite{2014JHEP...03..111B}, which reconstructs a tensor correlator starting from its transverse/traceless components and using the conservation/trace Ward identities (local terms). The latter are reconstructed from lower point functions.

 The result is expressed in terms of two sets of primary and secondary conformal Ward identities (CWI' s), the first involving the form factors of the transverse/traceless contributions, which are parametrized on a symmetry basis, the second emerging from CWI's of lower point functions. For 3-point functions, the secondary CWI's involve conservation, trace and special WI's. In all the cases, the reconstructed solutions for 3-point functions can be given in terms of generalized hypergeometrics of type $F_4$, \cite{Coriano:2012wp}, also known as Appell's hypergeometric function of two variables ($F_4$), related to 3-K integrals \cite{2014JHEP...03..111B}. 
\subsubsection{ The anomaly pole of the TJJ}
One of the results of our analysis will be to show how such contributions originate from the process of renormalization, taking as an example the case of the $TJJ$, filling in the intermediate steps of the discussion presented in \cite{Coriano:2018zdo}. We follow the general (BMS) approach introduced in \cite{Bzowski:2013sza} for the solution of the conformal constraints, which we detail in several of its parts, not offered in \cite{Bzowski:2013sza}. It has been compelling to proceed with an independent re-derivation of all the lengthy equations. The method can be directly generalized to higher point functions and implemented algorithmically, as we are going to show in a separate work.
When coming to discuss the momentum space approach in CFT, there are several gaps in the literature, which are of methodological nature and need to be addressed. These concern the correct form of the differential equations, the treatment of the derivatives of the 
Dirac $\delta$'s induced by momentum conservation, violations of the Leibnitz rule for the special conformal transformations, or the choice of the Lorentz (spin) singlet operator in the action of the conformal group on a specific correlator. These are points that we will address systematically.
We will illustrate how to merge the results of the BMS approach on the structure of the minimal set of (4) form factors 
(the $A-$basis), solutions of the CWI's  for the $TJJ$ correlator, with a basis of 13 ones (the $F$-basis) defined in previous perturbative studies. We will show how to extract from the $F-$basis 4 combinations of the 13 and we will verify 
that they respect the scalar equations identified within the BMS approach. \\ 
The use of this second basis is essential in order to prove that the WI's and the renormalization procedure for this correlator, imply that the anomaly can be attributed to the appearance of an anomaly pole in a single tensor structure of nonzero trace.

\subsection{Our work} 
As we have just mentioned, one of the goals of this work, in a first part, is to present a systematic approach to the analysis of the CWI's in momentum space, closing a gap in the literature. The transition to momentum space raises the issue of how to include momentum conservation (i.e. translational invariance in coordinate space) in the presence of the dilatation and the special conformal generators. We will be dealing, in particular, with a rigorous treatment of such contributions which show up after a Fourier transform of the conformal generators to momentum space. \\
We are going to investigate in detail the role of these contributions relying on the theory of tempered distributions. In particular, the discussion of these terms will be performed using a Gaussian basis which converges - in a distributional sense -  to a covariant 
$\delta$ function in $D=4$ and allows to define a formal calculus for such distributions. \\
Such contributions do not cancel, but lead to specific forms of the conformal generators in momentum space which are, however, in agreement with those presented in \cite{Coriano:2013jba, 2014JHEP...03..111B, Maldacena:2011nz}. We define operational methods for the treatment of the covariant derivatives of $\delta$ functions in a consistent way, which may find application also beyond the scope of the current treatment, being quite general. \\
In a second part we move to discuss scalar and tensor correlators and the solutions of the CWI's. We elaborate, in particular, 
one the apparent violation of the Leibnitz rule for the 
special conformal (SC) generator ($K^\kappa$), which emerges whenever we impose momentum conservation and eliminate one of the momenta, and the symmetric action of this operator, at an intermediate stage, is not evident. In position space this corresponds to 
choosing one coordinate to be zero, and treating the corresponding operator in a given correlation function, as spin singlet.
The derivation of the constraints on the form factors is performed, in our case, by using Lorentz Ward identities, on which we elaborate in detail, confirming the results of \cite{2014JHEP...03..111B}. 
 We show how different choices for the singlet operator leads to an equivalent set of conformal equations. We then illustrate how to derive the solutions of the various form factors using some properties of the hypergeometric equations, bypassing the 3K integrals, showing that the Fuchsian indices of all the equations remain the same for all the solutions.  

Such analysis is followed by a perturbative study of $TJJ$ correlator in the transverse traceless basis both in QED and in scalar QED, deriving the associated anomalous conformal Ward identities from this perspective. \\
In the final part of our work we show how the perturbative solutions for the $A_i$, which are given in an appendix, reproduce the exact BMS result in a simplified way. We use the cases of $d=3$ and $d=5$ to show the exact correspondence between the two. This correspondence is studied by fixing an appropriate normalization of the photon two-point functions, on which we elaborate. This shows that the choice of different perturbative sectors (scalar, fermion) in both cases are sufficient to reproduce the entire nonperturbative result. This implies that only arbitrary constant in the nonperturbative solution, expressed in terms of the 3K integrals, has to simplify and be expressible in terms of simple 
integrals $B_0$ and $C_0$, the scalar 2- and 3-point functions. In our conclusions we briefly comment on the possible origin of such simplifications.  


\section{Special Conformal Ward identities in the operatorial approach}
 In this section, to make our treatment self-contained, we briefly illustrate the operatorial derivation of the CWI's for correlators involving 3-point functions of stress energy tensors.
 
 An infinitesimal transformation
\be
x^\mu(x)\to x'^\mu(x)=x^\mu + v^\mu(x)
\ee
is classified as an isometry if it leaves the metric $g_{\mu \nu}(x)$ invariant in form. If we denote with 
$g'_{\mu\nu}(x')$ the new metric in the coordinate system $x'$, then an isometry is such that
\be
g^\prime_{\mu\nu}(x')=g_{\mu\nu}(x').
\label{met1}
\ee
This condition can be inserted into the ordinary covariant transformation rule for $g_{\mu\nu}(x)$ to give 
\be
g^\prime_{\mu\nu}(x')=\frac{\partial x^\rho  }{\partial x'^\mu }\frac{\partial x^\sigma}{\partial x'^\nu} g_{\rho\sigma}(x)= g_{\mu\nu}(x')
\ee
from which one derives the Killing equation for the metric
\be
v^\alpha\partial_\alpha g_{\mu\nu} + g_{\mu\sigma} \partial_\nu v^\sigma + 
g_{\sigma \nu} \partial_\mu v^\sigma =0.
 \ee
For a conformal transformation the metric condition (\ref{met1}) is replaced by the condition 
\be
g'_{\mu\nu}(x')=\Omega^{-2} g_{\mu\nu}(x')
\ee
generating the conformal Killing equation (with $\Omega(x)= 1-\sigma(x)$)
\be
v^\alpha\partial_\alpha g_{\mu\nu} + g_{\mu\sigma} \partial_\nu v^\sigma + 
g_{\sigma \nu} \partial_\mu v^\sigma=2 \sigma g_{\mu\nu}.
\ee
In the flat spacetime limit this becomes 
\be
\label{sigma}
 \partial_\mu v_\nu + 
\partial_\nu v_\mu=2 \sigma\, \eta_{\mu\nu},\qquad  \sigma=\frac{1}{d} \partial \cdot v.
\ee
From now on we switch to the Euclidean case, neglecting the index positions. Using the fact that every conformal transformation can be written as a local rotation matrix of the form 
\be
\label{rot1}
R^\mu_\alpha=\Omega \frac{\partial x'^\mu}{\partial x^\alpha} 
\ee
we can first expand genericaly $R$ around the identity as 
\be
R=\mathbf{ 1  } + \left[\mathbf{\epsilon}\right] +\ldots
\ee
with an antisymmetric matrix $\left[\epsilon\right]$, which we can re-express in terms of antisymmetric parameters 
($\tau_{\rho\sigma}$) and $1/2 \,d\, (d-1)$ generators $\Sigma_{\rho\sigma}$ of $SO(d)$ as 
\bea
\left[\epsilon\right]_{\mu\alpha}&=&\frac{1}{2} \tau_{\rho\sigma}\left(\Sigma_{\rho\sigma}\right)_{\mu\alpha}\nonumber \\
\left(\Sigma_{\rho\sigma}\right)_{\mu\alpha}&=&\delta_{\rho\mu}\delta_{\sigma\alpha}-\delta_{\rho\alpha}\delta_{\sigma\mu}
\eea
from which, using also (\ref{rot1}) we derive a constraint between the parameters of the conformal transformation $(v)$ and the parameters $\tau_{\mu\alpha}$ of $R$
\be
R_{\mu\alpha}= \delta_{\mu\alpha} + \tau_{\mu\alpha}=\delta_{\mu\alpha} + \frac{1}{2}\partial_{[\alpha }v_{\mu]}
\ee
with $ \partial_{[\alpha }v_{\mu]}\equiv
\partial_{\alpha }v_{\mu}-\partial_{\mu }v_{\alpha}$.\\
Denoting with $\Delta_A$ the scaling dimensions of a vector field $A_\mu(x)'$,  its variation under a conformal transformation can be expressed via $R$ in the form 
\bea
\label{trans1}
A'^\mu(x')&=&\Omega^{\Delta_A} R_{\mu \alpha} A^\alpha(x)\nonumber \\
&=&(1-\sigma+\ldots)^{\Delta_A}(\delta_{\mu\alpha}+\frac{1}{2}\partial_{[\alpha }v_{\mu]} +\ldots )A^\alpha(x)
\eea
from which one can easily deduce that 
\be
\label{trans2}
\delta A^\mu(x)\equiv A'^\mu(x)-A^\mu(x)=-(v\cdot \partial +\Delta_A \sigma)A^\mu(x) +\frac{1}{2} \partial_{[\alpha }v_{\mu]}A^\alpha(x), 
\ee
which is defined to be the Lie derivative of $A^\mu$ in the $v$ direction, modulo a sign
\be
L_v A^\mu(x) \equiv -\delta A^\mu(x).
\ee
As an example, in the case of a generic rank-2 tensor field ($\phi^{I \, K}$) of scaling dimension $\Delta_\phi$, transforming according to a representation $D^I_J(R)$ of the rotation group $SO(d)$, (\ref{trans1}) takes the form 
\be
\phi'^{I\, K}(x')=\Omega^{\Delta_{\phi}} D^I_{I'}(R) D^K_{K'}(R) \phi^{I' \,K'}(x).
\ee
In the case of the stress energy tensor ($D(R)=R$), with scaling (mass) dimension $\Delta_T$  $(\Delta_T=d)$ the analogue of (\ref{trans1}) is 
\bea
T'^{\mu\nu}(x')&=&\Omega^{\Delta_T} R^\mu_\alpha R^\nu_\beta T^{\alpha\beta}(x)\nonumber \\
 &=&(1- \Delta_T \sigma +\ldots)(\delta_{\mu\alpha}+\frac{1}{2}\partial_{[\alpha }v_{\mu]}+\ldots)
(\delta_{\mu\alpha}+\frac{1}{2}\partial_{[\alpha }v_{\mu]}+\ldots)\,T^{\a\b}(x)
\eea
where $\partial_{[\alpha }v_{\mu]}\equiv\partial_\a v_\mu -\partial_\mu v_\a$. One gets 
\be
\delta T^{\mu\nu}(x)=-\Delta_T\, \sigma\, T^{\mu\nu} -v\cdot \partial \,T^{\mu\nu}(x) +
\frac{1}{2}\partial_{[\alpha }v_{\mu]}\,T^{\alpha\nu} +\frac{1}{2}\partial_{[\nu }v_{\alpha]}T^{\mu\alpha}.
\ee

For a special conformal transformation (SCT) one chooses 
\be
v_{\mu}(x)=b_\mu x^2 -2 x_\mu b\cdot x
\ee
with a generic parameter $b_\mu$ and  $\sigma=-2 b\cdot x$ (from \ref{sigma}) to obtain
\be
\delta T^{\mu\nu}(x)=-(b^\alpha x^2 -2 x^\alpha b\cdot x )\, \partial_\alpha  T^{\mu\nu}(x)   - \Delta_T \sigma T^{\mu\nu}(x)+
2(b_\mu x_\alpha- b_\alpha x_\mu)T^{\alpha\nu} + 2 (b_\nu x_\alpha -b_\alpha x_\nu)\, T^{\mu\alpha}(x).
\ee

It is sufficient to differentiate this expression respect to $b_\kappa$ in order to derive the form of the 
SCT  $K^\kappa$ on $T$ in its finite form 
\bea
\mathcal{K}^\kappa T^{\mu\nu}(x)&\equiv &\delta_\kappa T^{\mu\nu}(x) =\frac{\partial}{\partial b^\kappa} (\delta T^{\mu\nu})\nonumber \\
&=& -(x^2 \partial_\kappa - 2 x_\kappa x\cdot \partial) T^{\mu\nu}(x) + 2\Delta_T x_\kappa T^{\mu\nu}(x) +
2(\delta_{\mu\kappa}x_\alpha -\delta_{\alpha \kappa}x_\mu) T^{\alpha\nu}(x) \nonumber \\ 
&& + 2 (\delta_{\kappa\nu} x_{\alpha} -\delta_{\alpha \kappa} x_\nu )T^{\mu\alpha}. 
\label{ith}
\eea
The approach can be generalized to correlators built out of several operators. 
In the case of a $TJJ$ correlator, 

\be
\Gamma^{\mu\nu\alpha\beta}(x_1,x_2,x_3)=\langle T^{\mu\nu}(x_1) J^\alpha(x_2)J^\beta(x_3)\rangle
\ee

with a vector current of dimension $\Delta_J$, the CWI's  take the explicit form 
\bea
\mathcal{K}^\kappa \Gamma^{\mu\nu\a\b}(x_1,x_2,x_3) 
&=& \sum_{i=1}^{3} {K_i}^{ \kappa}_{scalar}(x_i) \Gamma^{\mu\nu\a\b}(x_1,x_2,x_3) \nn
&& + 2 \left(  \delta^{\mu\kappa} x_{1\rho} - \delta_{\rho}^{\kappa }x_1^\mu  \right)\Gamma^{\rho \nu\alpha\beta}
 + 2 \left(  \delta^{\nu\kappa} x_{1\rho} - \delta_{\rho}^{\kappa }x_1^\nu  \right)\Gamma^{\mu\rho \alpha\beta}\nonumber
\\ &&  2 \left(  \delta^{\a\kappa} x_{2\rho} - \delta_{\rho}^{\kappa }x_2^\a  \right)\Gamma^{\mu\nu \rho\beta}
 +  2 \left(  \delta^{\beta\kappa} x_{3\rho} - \delta_{\rho}^{\kappa }x_3^\beta  \right)\Gamma^{\mu\nu \alpha\rho}=0,
 \eea
where 
\be
\label{ki}
{\mathcal{K}_i}^{\kappa}_{scalar}=-x_i^2 \frac{\partial }{\partial x_\kappa} + 2 x_i^\kappa x_i^\tau\frac{\partial}{\partial x_i^\tau} + 2 \Delta_i x_i^\kappa
\ee
is the scalar part of the special conformal operator acting on the $i_{\textrm{th}}$ coordinate and $\Delta_i\equiv(\Delta_T,\Delta_J,\Delta_J)$ are the scaling dimensions of the operators in the correlation function.

\subsection{Constraints from translational symmetry and the Leibniz rule}
One of the main issues, when moving to momentum space, is to include the constraint from translational symmetry 
on a tensor correlator. The inclusion of this constraint at the beginning, for a tensor 3-point function of the form 
\begin{equation} 
\langle H_1^{\mu_1\nu_1}(x_1) H_2^{\mu_2\nu_2}(x_2) H_3^{\mu_3\nu_3}(x_3)\rangle ,
\label{v1}
\end{equation}
with each of the $H_i$'s of scaling dimensions $\Delta^H_i$, reduces it to the form 
\begin{equation} 
\langle H_1^{\mu_1\nu_1}(x_{13}) H_2^{\mu_2\nu_2}(x_{23}) H_3^{\mu_3\nu_3}(0)\rangle ,
\label{v2}
\end{equation}
and the action of $K^\kappa$, the special conformal generator, on (\ref{v1}) and (\ref{v2}) will obviously change. The transition to momentum space in the two cases above takes to two different forms of the special CWI. The first form will be symmetric in momentum space, but at the cost of generating derivatives of the delta-function, which enforce conservation of the total momentum, while the second one will be asymmetric, treating one of the momenta as dependent from the other two. The final result for the scalar equations of the corresponding form factors will obviously be symmetric respect the three momenta. \\
In particular, the Lorentz (spin) generator, in the case of (\ref{v2}), will act only the indices of $H_1$ and $H_2$, but not on those of $H_3$, 
although the differentiation respect to the 4-momentum $p_3$ will be performed implicitly by a the chain rule  in this second case, once we move to momentum space.\\
We are going to illustrate this point in detail.\\ 
Identifying $K^\kappa_i H_i^{\mu\nu}$ with the expression (\ref{ith}) (with $\Delta_T\to \Delta^H_i$), 
then the action of the special conformal transformation on (\ref{v1}) will take the forms 
\begin{eqnarray} 
 K^\kappa \langle H_1^{\mu_1\nu_1}(x_1) H_2^{\mu_2\nu_2}(x_2) H_3^{\mu_3\nu_3}(x_3)\rangle 
&=& \sum_{i=1}^3 K_i^\kappa \langle H_1^{\mu_1\nu_1}(x_1) H_2^{\mu_2\nu_2}(x_2) H_3^{\mu_3\nu_3}(x_3)\rangle
\nonumber \\
&=&e^{-iPx_3}	K^{\kappa\prime} \langle H_1^{\mu_1\nu_1}(x_{13}) H_2^{\mu_2\nu_2}(x_{23}) H_3^{\mu_3\nu_3}(0)\rangle
\end{eqnarray}
with $P$ the total translation operator  ($P=P_1+P_2+P_3$) and 
\begin{equation}
	{K}^{\kappa\prime}=e^{iPx_3}\,K^\kappa e^{-iPx_3}.
\label{trans}
	\end{equation}
	Using the relations of the conformal algebra 
	\begin{equation}
	\begin{split}
	[{K}^\kappa,P^\nu]=2i(\eta^{\kappa\nu}D+M^{\kappa\nu})&,\qquad[D,P^\mu]=-iP^\mu\notag\\[1.5ex]
	[M^{\kappa\nu},P^\mu]=-i(\eta^{\kappa\mu}P^\nu-\eta^{\mu\nu}P^\kappa)&,
	\end{split}
	\end{equation}
 and expanding (\ref{trans}) we obtain the relation
	\begin{equation}
	\begin{split}
	{K}^{\kappa\prime}&={K}^\kappa+i x_{3\mu}[P^\mu,{K}^\kappa]+\frac{i^2}{2}x_{3\mu}x_{3\nu}[P^\mu,[P^\nu,{K}^\kappa]]+\dots\\[1.5ex]
	&={K}^\kappa+2x_{3}^\kappa D+2x_{3\mu}M^{\kappa\mu}-2 x_{3}^\kappa x_{3\mu}P^\mu+x_3^2P^\kappa,
	\end{split}
	\end{equation}
	since the commutator of higher order vanish.\\
	The explicit form of the operators (dilatation, Lorentz and special conformal)  $D,M^{\mu\nu},{K}^\kappa$ is
	$ \,K^\kappa={K}^\kappa_1+{K}^\kappa_2+{K}^\kappa_3$, 
	$D=D_1+D_2+D_3$, $M^{\mu\nu}=M_1^{\mu\nu} + M_2^{\mu\nu} + M_3^{\mu\nu}$ and 
	\begin{equation}
	M_i^{\mu\nu}=L_i^{\mu\nu} + \Sigma_i^{\mu\nu}, \, L_l=i (x_l^{\mu}\frac{\partial}{\partial x_l^{\nu}}
	-x_l^{\nu}\frac{\partial}{\partial x_l^{\mu}}),\qquad\end{equation}
	split into angular momentum ($L$) and spin ($\Sigma$). We illustrate this crucial point in some detail, 
	since it shows how the action of the Lorenz generators on the field at $x_3$ vanishes. We get  (using $\hat{p}_l^\kappa\equiv i\partial/\partial x_l^\kappa$)

	\begin{equation}
	\begin{split}
	\,K^{\kappa\prime}&=\ \ x_1^2\hat{p}_1^\kappa-2x_1^\kappa x_{1\nu} \hat{p}_1^\nu-2ix_1^\kappa \Delta_1-2x_{1\nu}\Sigma^{\kappa\nu}_1\\[0.6ex]
	&\ \ +x_2^2\hat{p}_2^\kappa-2x_2^\kappa x_{2\nu}\hat{ p}_2^\nu-2ix_2^\kappa \Delta_2-2x_{2\nu}\Sigma^{\kappa\nu}_2\\[0.6ex]
	&\ \ +x_3^2 \hat{p}_3^\kappa-2x_3^\kappa x_{3\nu}\hat{ p}_3^\nu-2ix_3^\kappa \Delta_3-2x_{3\nu}\Sigma^{\kappa\nu}_3\\[0.6ex]
	&\ \ +2ix_3^\kappa(\Delta_1+\Delta_2+\Delta_3)+2x_3^\kappa(x_{1\nu}\hat{p}_1^\nu+x_{2\nu}\hat{p}_2^\nu+x_{3\nu}\hat{p}_3^\nu)\\[0.6ex]
	&\ \ +2x_{3\nu}(\Sigma^{\kappa\nu}_1+\Sigma^{\kappa\nu}_2+\Sigma^{\kappa\nu}_3)+2x_{3\nu}(x_1^\kappa \hat{p}_1^\nu-x_1^\nu\hat{ p}_1^\kappa+x_2^\kappa\hat{p}_2^\nu-x_2^\nu\hat{p}_2^\kappa+x_3^\kappa\hat{p}_3^\nu-x_3^\nu\hat{p}_3^\kappa)\\[0.6ex]
	&\ \ -2x_3^\kappa x_{3\mu}(\hat{p}_1^\mu+\hat{p}_2^\mu+\hat{p}_3^\mu)+x_3^2(\hat{p}_1^\kappa+\hat{p}_2^\kappa+\hat{p}_3^\kappa)
	\end{split}
	\end{equation}
which shows the cancellation of the contribution from the generator $M_3$ since

	\begin{equation}
\label{spins}
	\begin{split}
	{K}^{\kappa\prime}&=\ \ (x_1-x_3)^2 \hat{p}_1^\kappa-2(x_1^\kappa-x_3^\kappa) (x_{1\nu} -x_{3\nu})\hat{p_1}^\nu-2i(x_1^\kappa-x_3^\kappa) \Delta_1-2(x_{1\nu}-x_{3\nu})\Sigma^{\kappa\nu}_1\\[0.6ex]
	&\ \ +(x_2-x_3)^2 \hat{p}_2^\kappa-2(x_2^\kappa-x_3^\kappa) (x_{2\nu} -x_{3\nu})\hat{p}_2^\nu-2i(x_2^\kappa-x_3^\kappa) \Delta_2-2(x_{2\nu}-x_{3\nu})\Sigma^{\kappa\nu}_2.\\[0.6ex]
	\end{split}
	\end{equation}
Notice that both $\Sigma_1$ and $\Sigma_2$ denote the two spin matrices, which act only on $H_1$ and $H_2$. In other words, in coordinate space the choice $x_3=0$ implies that
$H_3$ behaves as a Lorentz singlet respect to the spin part. The result can be rewritten in the compact form 
\begin{equation}
	{K}^{\kappa\prime}={K}^\kappa_{13}+{K}^\kappa_{23},
	\end{equation}
	where the action of ${K}^\kappa_{13}$ on a rank-2 tensor $H^{\mu\nu}$, for instance, is given by 
	\bea
{K}^\kappa  H^{\mu\nu}(x_{13})
&=&  {K}_{scalar}^\kappa(x_{13})H^{\mu\nu}  + 2 \left(  \delta^{\mu\kappa} x_{13\rho} - \delta_{\rho}^{\kappa }x_{13}^\mu  \right)H^{\rho \nu}
 + 2 \left(  \delta^{\nu\kappa} x_{13\rho} - \delta_{\rho}^{\kappa }x_{13}^\nu  \right)H^{\mu\rho},\nn
 \eea
 with ${K}_{scalar}(x_{13})$ being given as in (\ref{ki}) with $x_i\to x_{13}$ and $\Delta_i\to \Delta_H$,
and we obtain the equation
	\begin{equation}
	0={K}^\kappa  \langle H_1^{\mu_1\nu_1}(x_1) H_2^{\mu_2\nu_2}(x_2) H_3^{\mu_3\nu_3}(x_3)\rangle  =e^{-iPx_3}	\left({K}^\kappa_{13}+{K}^\kappa_{23}\right) \langle H_1^{\mu_1\nu_1}(x_{13}) H_2^{\mu_2\nu_2}(x_{23}) H_3^{\mu_3\nu_3}(0)\rangle.
	\label{8}
	\end{equation}
Notice that the solution of this equation can be obtained by solving the reduced equation 
\be
({K}^\kappa_{13}+{K}^\kappa_{23}) \langle H_1^{\mu_1\nu_1}(x_{13}) H_2^{\mu_2\nu_2}(x_{23}) H_3^{\mu_3\nu_3}(0)\rangle =0
\ee
which is equivalent to finding the solution of (\ref{8}) with $x_3=0$ and acting afterwards with the translation operator $e^{-i P x_3}$ to restore the full dependence on the third coordinate. Setting
\be
\chi^{\mu_1\nu_1\mu_2\nu_2\mu_3\nu_3}(x_{12},x_{13})\equiv \langle H_1^{\mu_1\nu_1}(x_{13}) H_2^{\mu_2\nu_2}(x_{23}) H_3^{\mu_3\nu_3}(0)\rangle,
\ee
anticipating the discussion that will be presented for these equations in momentum space, the special CWI then can be cast into the form
	\begin{equation}
\int d^4p_1\,d^4p_2 e^{-i(p_1x_{13}+p_2x_{23})}({K}^\kappa_{p_1}+
{K}^\kappa_{p_2})\chi(p_1,p_2)=0
\end{equation}
giving 
\be
({K}^\kappa_{p_1}+
{K}^\kappa_{p_2})\chi(p_1,p_2)=0.
\ee
Notice that the previous form of $\chi(p_1,p_2)$, which is a function of the independent momenta $p_1$ and $p_2$, conjugate  to $x_{12}$ and $x_{13}$, is the final form of the function, having re-expressed $p_3$ in terms of $p_1$ and $p_2$. In a direct explicit computation, one has to act with the transforms of ${K}^\kappa(x_{12})$ and 
${K}^\kappa(x_{13})$, that we denote as ${K}^\kappa(p_1)$ and ${K}^\kappa(p_2)$, 
on the transform of the initial correlator 
\begin{align}
&({K}^\kappa(p_1)+{K}^\kappa(p_2)) \langle H_1^{\mu_1\nu_1}(p_1) H_2^{\mu_2\nu_2}(p_2) H_3({\bar{p}_3)}\rangle=\nonumber \\
&= \int d^d x_{13}d^d x_{23}\,e^{-i p_1\cdot x_{12} - i p_2\cdot x_{23}} ({K}^\kappa(x_{12})+{K}^\kappa(x_{13}))\times\langle H_1^{\mu_1\nu_1}(x_{13}) H_2^{\mu_2\nu_2}(x_{23}) H_3^{\mu_3\nu_3}(0)\rangle
\end{align}
with $p_3 \to \bar{p}_3=-p_1-p_2$ and the Leibniz rule is violated. The symmetry respect to the external invariants $(p_1^2,p_2^2,p_3^2)$ of the conformal generator is only reobtained at the end, after applying the chain rule for the differentiation of $p_3$ respect to the two independent momenta. \\
The final result is that in momentum space we can treat $H_3( \bar{p}_3)$ as a single particle operator, in the sense that the differentials in $p_1$ and $p_2$ will act separately on $H_1$ and $H_2$, but also on $H_3$ implicitly, via a chain rule.\\
 At the same time, as clear from (\ref{spins}), the spin rotation matrix $\Sigma^{\mu\nu} $ contained in the Lorentz generator $M^{\mu\nu}$ will act only on $H_1$ and $H_2$, treating $H_3$ as a Lorentz (spin) singlet. We will present on the sections below complete worked out examples of this action. Since we are free to set any of the 3 coordinates to zero, the intermediate steps of the computations of the CWI's will be completely different, and the choice of the Lorentz singlet operator can be dictated by convenience. \\
The choice of the point $x$ which will be set to zero (e.g. $x_3=0$) is obviously arbitrary, but preferably should be suggested by the symmetry of the correlator. For instance, for correlators such as $\langle T(x_1)T(x_2)T(x_3)\rangle$  and $\langle T(x_1)T(x_2)O(x_3)\rangle$ setting $x_3=0$ and removing momentum $p_3$ in terms of $p_1$ and $p_2$ is the natural choice. In the $\langle T(x_1)J(x_2)J(x_3) \rangle$ case it is convenient to set $x_1=0$ and re-express the momentum $p_1$ in terms of $p_2$ and $p_3$.   

\section{The conformal generators in momentum space } 
In this section we discuss two formulations of the dilatation and SCT's, with the goal of clarifying the treatment of the constraints coming from the conservation of the total momentum in a generic correlator. We will be using some condensed notations in order to shorten the expressions 
of the transforms in momentum space. We will try to avoid the proliferation of indices, whenever necessary, with the conventions
\bea
\label{conv}
&& \Phi(\underline{x})\equiv \langle \phi_1(x_1)\phi_2(x_2)\ldots \phi_n(x_n)\rangle \qquad e^{i \underline{p x}}\equiv e^{i(p_1 x_1 + p_2 x_2 + \ldots p_n x_n)} \qquad \nn
&& \ul{d p}\equiv dp_1 dp_2 \ldots d p_n \qquad 
\Phi(\ul{p})\equiv \Phi(p_1,p_2,\ldots, p_n).\qquad 
\eea
It will also be useful to introduce the total momentum $P=\sum_{j=1}^{n} p_j$.\\
The momentum constraint is enforced via a delta function $\delta(P)$ under integration. For instance, translational invariance of $\Phi(\ul{x})$ gives 
\be
\label{ft1}
\Phi(\ul{x})=\int \ul{dp}\,\delta(P) \,e^{i\ul{p x}} \,\Phi(p_1,p_2,p_3).
\ee
In general, for an n-point function $\Phi(x_1,x_2,\ldots, x_n)=\langle \phi_1(x_1)\phi_2(x_2)...\phi_n(x_n)\rangle $, the condition of translation invariance  
\bea
\langle 
 \phi_1(x_1)\phi_2(x_2),\ldots, \phi_n(x_n)\rangle = \langle\phi_1(x_1+a )\phi_2(x_2+a)\ldots \phi_n(x_n+a)\rangle
 \eea 
 generates the expression in momentum space of the form (\ref{ft1}), from which we can remove one of the momenta, conventionally the last one, $p_n$, which is replaced by its "on shell" version 
 $\overline{p}_n=-(p_1+p_2 +\ldots p_{n-1})$
 \be
 \Phi(x_1,x_2,\ldots,x_n)=\int dp_1 dp_2... dp_{n-1}e^{i(p_1 x_1 + p_2 x_2 +...p_{n-1} x_{n-1} + 
 \overline{p}_n x_n)}\Phi(p_1,p_2,\ldots,\overline{p}_n).
 \ee
We start by considering the dilatation WI. \\
 The condition of scale covariance for the fields $\phi_i$ of scale dimensions $\Delta_i$ (in mass units)
 \be
 \label{scale1}
 \Phi(\lambda x_1,\lambda x_2,\ldots,\lambda x_n)=\lambda^{-\Delta} \Phi(x_1,x_2,\ldots, x_n), \qquad 
 \Delta=\Delta_1 +\Delta_2 +\ldots \Delta_n
 \ee
 after setting $\lambda=1 +\epsilon$ and Taylor expanding up to $O(\epsilon)$ gives the 
 scaling relation 
 \be
 \label{ft2}
(D_n + \Delta)\Phi\equiv \sum_{j=1}^n \left(x_j^\alpha \frac{\partial}{\partial x_j^\alpha} +\Delta_j\right) \Phi(x_1,x_2,\ldots,x_n) =0.
 \ee 
 with 
 \be
 D_n=\sum_{j=1}^n x_j^\alpha\frac{\partial}{\partial x_j^\alpha}. 
 \ee
The corresponding equation in momentum space can be obtained either by a Fourier transform 
of (\ref{ft2}), which can give either symmetric or asymmetric expressions of the equations in the respective momenta $p_i$ or, more simply, exploiting directly (\ref{scale1}). In the latter case, using the translational invariance of the correlator under the integral, by removing the $\delta$-function constraint, one obtains
 \begin{align}
 \Phi(\lambda x_1,\lambda x_2,\ldots,\lambda x_n)&= \int d^d p_1 d^d p_2\ldots d^d p_{n-1} e^{i\lambda (p_1 x_1 + p_2 x_2 +...p_{n-1} x_{n-1} + 
 \overline{p}_n x_n)}\Phi(p_1,p_2,\ldots,\overline{p}_n)\nonumber \\
&=\lambda^{-\Delta}  \int d^d p_1 d^d p_2\ldots d^d p_{n-1} e^{i(p_1 x_1 + p_2 x_2 +...p_{n-1} x_{n-1} + 
 \overline{p}_n x_n)}\Phi(p_1,p_2,\ldots,\overline{p}_n).
 \end{align}
 It is simply a matter of performing the change of variables $ p_i=p'_i/\lambda$ on the rhs of the equation above (first line) with $d p_1...d p_{n-1}=(1/\lambda)^{d(n-1)} d^d p'_1 \ldots d^d p'_{n-1}$ to derive the relation 
 \be
 \frac{1}{{\lambda}^{d(n-1)}}\Phi(\frac{p_1}{\lambda},\frac{p_2}{\lambda},\ldots,\frac{\overline{p}_n}{\lambda})=\lambda^{-\Delta} \Phi(p_1,p_2,\ldots,\overline{p}_n).
\ee
Setting $\lambda=1/s$ this generates the condition
\be
 s^{(n-1) d-\Delta}\Phi(s p_1,s p_2,\ldots, s \overline{p}_n)=\Phi(p_1,p_2,\ldots,\overline{p}_n)
\ee
and with $s\sim 1 +\epsilon$, expanding at $O(\epsilon)$ we generate the equation 
\be
\label{sc1}
\left[\sum_{j=1}^n \Delta_j  -(n-1) d -\sum_{j=1}^{n-1}p_j^\alpha \frac{\partial}{\partial p_j^\alpha}\right]
\Phi(p_1,p_2,\ldots,\overline{p}_n)=0.
\ee
It is straightforward to reobtain the same equation from the direct Fourier transform of (\ref{ft2}) if we use translational invariance since
\begin{align}
&(D_n +\Delta)\Phi(x_1,\ldots, x_n)=\notag\\
&\quad=(D_n +\Delta)\int d^d p_1\ldots d^d p_n\delta(P) e^{i p_1 x_1 +\ldots p_n x_n}\Phi(p_1,\ldots, p_n) \nn
&\quad=\left(\sum_{j=1}^{n}\Delta_j + \sum_{i=1}^{n-1} x_{i\, n}\frac{\partial}{\partial x_{i n}}\right) \int d^d p_1\ldots d^d p_{n -1}e^{i p_1 x_{1 n} +\ldots p_{n-1} x_{n-1 \, n}}\Phi(p_1,\ldots, p_{n-1},\bar{p}_n)\nn
&\quad=\int d^d p_1\ldots d^d p_{n -1}\left(\sum_{j=1}^n \Delta_j + \sum_{j=1}^{n-1}  p_j
\frac{\partial}{\partial p_j}\right) e^{i p_1 x_{1 n} +\ldots p_{n-1} x_{n-1 \, n}}\Phi(p_1,\ldots, p_{n-1},\bar{p}_n).
\end{align}
At this point we perform a partial integration $n-1$ times, moving the derivatives from the exponential to the correlator $\Phi$ to reobtain (\ref{sc1})
\begin{align}
0&=(D_n +\Delta)\Phi(x_1,\ldots, x_n)\notag\\
&=\int d^d p_1\ldots d^d p_{n -1}\left(\sum_{j=1}^n \Delta_j -(n-1)d - \sum_{j=1}^{n-1}  p_j^\alpha
\frac{\partial}{\partial p_j^\alpha}\right) \Phi(p_1,\ldots, p_{n-1},\bar{p}_n)\,e^{i p_1 x_{1 n} +\ldots i p_{n-1} x_{n-1 \, n}}.
\end{align}
A rigorous way to reobtain this result is to consider directly the conformal algebra for the dilatation operator 
$D=(iD_n+\Delta)$ and use the commutation relations. 
For our purpose we can consider a realization of $\Phi(x_1,\ldots x_n)$ via some operators $O_j(x_j)$
\begin{align}
D\braket{\mO_1(x_1)\dots\mO_n(x_n)}&\equiv\sum_{j=1}^{n}D(x_j)\braket{\mO_1(x_1)\dots\mO_n(x_n)}\notag\\
&=\left[\sum_{j=1}^n\,\D_j+\sum_{j=1}^n\,x_j^\a\sdfrac{\partial}{\partial x_j^\a}\right]\braket{\mO_1(x_1)\dots\mO_n(x_n)}.
\end{align}
 
This n-point function is translationally invariant, so that we can shift the fields using the translation operator $\exp(iP\cdot x_n)$, with $P$ the total translation operator $P=\sum_{j=1}^nP_j$
\begin{equation}
D\braket{\mO_1(x_1)\dots\mO_n(x_n)}=e^{-i x_n\cdot P}D'\braket{\,\mO_1(x_1-x_n)\dots\mO_n(x_{n-1}-x_n)\,\mO_n(0)}=0\label{Rel}
\end{equation}
where
\begin{equation}
D'=e^{ix_n^\m\cdot P_\m }\,D\,e^{-i x_n^\m\cdot P_\m}=\sum_{k=0}^\infty\sdfrac{i^k}{k!}\,x_n^{\n_1}\dots x_n^{\n_k}\,[P_{\n_1},[\dots[P_{\n_k},D]\dots]].\label{hauss}
\end{equation}
Using the commutation relations of the conformal algebra, there are at most two non-vanishing terms in this sum. Evaluating the finite multiple commutators we get
\begin{equation}
D'=D+ix_n^\n\,[P_\n,D]=D-x_n^\n P_\n
\end{equation}
and explicitly 
\begin{align}
D' &=\sum_{j=1}^{n}\left[\D_j+ x_j^\n (P_j)_\n\right]-x_n^\n \sum_{j=1}^n(P_j)_\n\notag\\
&=\sum_{j=1}^{n-1} (x_j-x_n)^\n (P_j)_\n+\sum_{j=1}^n\D_j\ =\ \sum_{j=1}^{n-1} (x_j-x_n)^\n \frac{\partial}{\partial x_{j}}_\n+\sum_{j=1}^n\D_j.
\end{align}
Notice that the solution of \eqref{Rel} can be obtained by solving the reduced equation 
\begin{equation}
D'\braket{\,\mO_1(x_1-x_n)\dots\mO_n(x_{n-1}-x_n)\,\mO_n(0)}=0
\end{equation}
and then acting with a total translation. In momentum space this relation generates the Ward identity
\begin{align}
\int\prod_{j=1}^n\,d^dx_j \ e^{i x_1\cdot p_1+\ldots i p_n\cdot x_n}\ D'\braket{\,\mO_1(x_1-x_n)\dots\mO_n(x_{n-1}-x_n)\,\mO_n(0)}=0,
\end{align}
that is 
\begin{equation}
\int d^d p_1\ldots d^d p_n \d^{(d)}\left(\sum_{j=1}^n p_j\right)\ \left[\sum_{j=1}^n\D_j-(n-1)d-\sum_{j=1}^{n-1}p_j^\a\sdfrac{\partial}{\partial p_j^\a}\right]\braket{{\,\mO_1(p_1)\dots\mO_n(p_{n-1})\,\mO_n(\bar p_n)}}=0,
\end{equation}
where $\bar p_n=-\sum_{j=1}^{n-1}p_j$. \\
If we decided to work with symmetric expressions of the transform, the approach would be more cumbersome
since it would involve $\delta'$ (derivative) terms in the integrand. We are going to discuss this second approach both for the dilatation and for the special conformal transformations. It requires a brief digression on the use of some relations for the covariant $\delta$-functions which we are going to formulate below and that will be essential in order to clarify the correct way to treat such contributions.

\subsection{Delta calculus and symmetric Gaussians} 

It is possible to derive a formal calculus for the derivatives of $\delta^d(P)$ using as defining conditions that, for a generic function $f(p)$ which is regular at $P^\mu=0$ the rules
\be
\label{oneder}
\int d^d P\, \partial_\alpha \delta^d(P) f(P)=-\partial_\alpha f(0)
\ee
and 
\be
\label{twoder}
\int d^d P\,\partial_\beta \partial_\alpha \delta^d(P) f(P)=\partial_\beta\partial_\alpha f(0)
\ee
hold.
For instance one easily obtains formally
\be
\label{delta11}
\partial_\alpha \delta^d(P)\equiv \frac{\partial}{\partial  P^\alpha}\delta^d( P) =-\frac{d}{P^2} \delta^d( P) P^\alpha, 
\ee
which can be checked using the following rule for symmetric integration in (\ref{oneder})
\be
\int d^d P \frac{\delta^d(P)}{P^2} P^\alpha P^\beta f(0) =\frac{1}{d} \eta^{\alpha\beta}.
\label{rule0}
\ee
where $f(0)$ is a constant.
Similarly, one can show that 
\be
\int d^d P\, P^\alpha \frac{\delta^d(P)}{P^2}f(0) =0,
\label{rule11}
\ee
which is consistent with the fact that the integral 
\be
\int d^d P\,\partial_\alpha\delta^d(P)=0,
\ee
 has a zero boundary value.
Notice that (\ref{rule11}) can be extended to the more general form 
\be
\int d^d P\, P^\alpha \frac{\delta^d(P)}{P^2}f(P) =0,
\label{rule12}
\ee
if the function $f(P)$ is regular at $P=0.$\\
To derive such relations on a rigorous basis, we need to introduce a suitable family of functions converging to the $\delta^d(P)$ in the distributional limit.\\
 We will be needing the relations
\be
\delta^d (P)=\frac{\delta(P)\delta(\Omega)}{P^{d-1}C(\theta_1,...,\theta_{d-1})},\qquad \delta(\Omega)\equiv 
\prod_{l=1}^{D-1} \delta(\theta_l)\qquad  C(\theta_1,...\theta_{d-1})=\prod_{l=1}^{d-2}  \sin\theta_l^{d-l-1}
\ee
between the cartesian and the polar coordinates versions of the delta function.
 Then clearly 
\be
\int d^d P \delta^d(P)=\int_0^\infty d P\int_{0}^\pi d\theta_1 \int_0^\pi d\theta_2\ldots \int_0^{2 \pi} d\theta_{d-1} \delta(P)\delta(\Omega_d)=1.
\label{bound}
\ee
It is easily shown that the integral of a vector $n$ of unit norm, expressed in the same variables
\bea
\label{bd}
n^\alpha(\theta_1,\ldots,\theta_{d-1})&=&(\cos\theta_1,\cos\theta_2 \sin\theta_1,\ldots ,\sin\theta_{d-1}\ldots \sin\theta_1), \nn 
&& \qquad 0\leq \theta_i \leq \pi,   i=1, 2,\ldots d-2,\qquad    0\leq \theta_{d-1} \leq 2 \pi, 
\eea
 vanishes
\be
\label{vano}
\int d\Omega\, n^\alpha(\theta_1,\ldots,\theta_{d-1})=0, \qquad d\Omega\equiv d\theta_1 d\theta_2\ldots d\theta_{d-1}.
\ee
These relations will be used to parametrize the tensor integrals over the (total) momentum $P^\mu$ of the correlators 
as $n^\mu |P|$, with $|P|$ being the magnitude of P.
Notice that respect to the $d$-dimensional angular integration measure $d\Omega_d$, $d\Omega$ is stripped of the angular factors 
\be
d\Omega_d\equiv d \Omega \prod_{l=1}^{d-2} \sin\theta_l^{d-l-1}.
\ee

The vanishing of (\ref{vano}) is simply due to the symmetry of the angular integrations. This may not be obvious if we separate the angular from the radial parts of the $\delta$ function and performs the angular integration first, since the $d$-dimensional rotational symmetry is broken 

\be
\label{vano1}
\int d\Omega\, n^\alpha(\theta_1,\ldots,\theta_{d-1})\delta(\Omega)=\delta^{\alpha 0},
\ee
where the nonzero component surviving in (\ref{vano1}) depends on the directions chosen for the polar axis, and  the integration measure has been stripped off of the angular factors. Notice that only one component $(n^0)$, proportional to $\cos\theta_{d-1}$, is nonvanishing after the integration with $\delta(\Omega)$, the remaining ones being zero. Therefore, the vanishing of (\ref{rule11}) has to be shown by using a {\em rotationally symmetric} sequence of functions which avoid the formal manipulation in 
(\ref{vano}). For this purpose we use a sequence of normalized Gaussians 
\be
G_k(P)=\frac{1}{(2 \pi)^{d/2} k^d}e^{-\frac{P^2}{2 k^2}}, \qquad \int d^d P G_k(P)=1
\ee

converging to $\delta^d(P)$ as $k\to 0$. We need to consider the distributional limit of
\bea
 \int d^d P \frac{P^\alpha}{P^2} G_k(P) 
 &=& \int_0^\infty  P^{d-2} dP\,  G_k(P))\int n^\alpha(\theta_1,\ldots, \theta_{d-1}) d \Omega_d\nn
 &=&\frac{1}{2 \sqrt{2} \pi^{d/2} k^{d/2 +1/2}}\Gamma(\frac{d-1}{2})\int n^\alpha(\theta_1,\ldots, \theta_{d-1}) d \Omega_d
 \eea
 which vanishes after angular integration since
 \be
Y^\alpha(n)\equiv \int n^\alpha(\theta_1,\ldots, \theta_{d-1}) 
d \Omega_d=0
\ee
with the boundaries given in (\ref{bd}). Therefore, the correct angular average should be taken before the distributional limit of $k\to 0$, giving a vanishing result, thereby proving (\ref{rule11}).
The result for the rank-2 integral in ({\ref{rule0})
which has been justified above by covariance and symmetric integration, can also be obtained by a similar method. In this case we get 
\begin{equation}
 \int d^d P \frac{P^\alpha P^\beta}{P^2} G_k(P)=\int_0^{\infty} P^{d-1}G_k(P) dP\, Y^{\alpha\beta}(n),
\label{fact3}
\end{equation}
where we have defined the angular part
\be
\label{Y2}
Y^{\alpha\beta}(n)\equiv \int n^\alpha n^\beta(\theta_1,\ldots, \theta_{d-1}) d\Omega_d= \frac{1}{d}V_d \delta^{\alpha \beta}, \qquad V_d=\frac{2 \pi^{d/2}}{\Gamma[d/2]}.
\ee

Also in this case the integral (\ref{fact3}) is factorized with 
\be
\label{P2}
\int_0^{\infty} d P P^{d-1}G_k(P)=\frac{1}{V_d}, 
\ee
which is independent of the $k$ parameter of the distributional limit. Therefore 
\be
 \int d^d P \frac{P^\alpha P^\beta}{P^2} G_k(P)=\int_0^{\infty} P^{d-1}G_k(P) dP\, Y^{\alpha\beta}(n)=\frac{1}{d}\delta^{\alpha\beta}
\label{fact2}
\ee
for any value of the parameter $k$. This takes directly to (\ref{rule0}) if the parameter of the Gaussian family approaches 
$k=0$ in order to extract the value of a test function $f(P)$ at $P=0$. \\
One can expand on this result using the rules of the ordinary calculus formally, by taking multiple derivatives of $\delta(P)$ using (\ref{oneder}), which imply that 
\bea
\partial_\beta\partial_\alpha \delta^d(P)&=&\partial_\beta\left(-\frac{d}{P^2}\delta^d(P) P^\alpha\right)\nonumber \\
&=&\frac{d(d+2)}{(P^2)^2}\delta^d(P)P^\alpha P^\beta 
-\frac{d}{P^2}\delta^d(P)\delta^{\alpha\beta}.
\label{derp0}
\eea
These correctly generate (\ref{twoder}) using symmetric integration 
\bea
\label{delta2}
\int d^d P \frac{\delta^d(P)}{(P^2)^2} P^\alpha P^\beta P^\rho P^\sigma f(0)&=&
\frac{1}{d (d+2)}\left( \delta^{\alpha \beta}\delta^{\rho \sigma} +\delta^{\alpha \rho}\delta^{\beta \sigma} +
\delta^{\alpha \sigma} \delta^{\beta\rho}\right).\nn
\eea

Another useful relation is 
\bea
\label{derp}
\sum_{j=1}^n p_j^\alpha \frac{\partial}{\partial p_j^\alpha} \delta^d(P)&=&\sum_{j=1}^n p_j^\alpha \frac{\partial}{\partial P^\alpha} \delta^d(P)\nn
&=&P^\alpha \partial_\alpha\delta^d(P)
=-d\, \delta^d(P),
\eea
using (\ref{delta11}), which is of immediate derivation. We will be using the relations above to illustrate the elimination of one of the momenta from the differential equations which characterize the CWI's in momentum space.

\subsection{The dilatation Ward identity with one less momentum}
We can reobtain the results of the previous sections for the dilatation WI by using the calculus derived above.
The dilatation Ward identity of (\ref{ft2}) can indeed be written in a $(p_1,p_2,\ldots p_n)$ symmetric form 
using 
\begin{align}
 \label{ft3}
 \sum_{j=1}^n \left(x_j^\alpha \frac{\partial}{\partial x_j^\alpha} +\Delta_j\right) \Phi(x_1,x_2,\ldots,x_n)
 & = \sum_{j=1}^3 \left(x_j^\alpha \frac{\partial}{\partial x_j^\alpha} +\Delta_j\right) \int \ul{d^d p}\,e^{i\ul{px}}\,\delta(P)\,\Phi(\ul{p})\nonumber \\
 &=\int \ul{d^d p} \,\delta^d(P)\,e^{i\ul{p x}}\,\left( 
 \sum_{j=1}^n \Delta_j - n d -\sum_{j=1}^n p_j^\alpha \frac{\partial}{\partial p_j^\alpha}\right)\Phi(\ul{p})
 +\delta'_\textrm{term}, \nonumber \\
 \end{align}
with 
\begin{equation}
\label{delta1}
\delta'_\textrm{term}=-\int \ul{d^d p}\, P^\alpha\partial_\alpha \delta^d(P) e^{i \ul{p x}}\Phi(\ul{p})
= d \int \ul{d^d p}\,\delta^d(P) e^{i\ul{ p x}}\Phi(\ul{p}),
\end{equation}
where we have used (\ref{derp}). Inserting (\ref{delta1}) into (\ref{ft3}) we obtain the {\em symmetric} expression of the scaling relation in momentum space 
\begin{equation}
 \label{ft4}
 \sum_{j=1}^n \left(x_j^\alpha \frac{\partial}{\partial x_j^\alpha} +\Delta_j\right) \Phi(x_1,x_2,\ldots,x_n)
 =\int \ul{d^d p} \,\delta^d(P)e^{i\ul{p x}}\left( 
 \sum_{j=1}^n \Delta_j - (n-1) d -\sum_{j=1}^n p_j^\alpha \frac{\partial}{\partial p_j^\alpha}\right)\Phi(\ul{p}).
 \end{equation}
The expression given above depends only on $n-1$ momenta, since one of them can be eliminated.
If we choose as independent ones $p_1,\ldots p_{n-1}$ with  $(p_1,\ldots,p_n)\to (p_1,\ldots,p_{n-1},P)$, and define 
$q=p_1+p_2 \ldots + p_{n-1}$, then the dependence on $p_n$ in (\ref{ft4}) can be re-expressed  in terms of the total momentum $P$ and of the sum of the independent momenta $q$ as 
\bea
\Phi(\ul{p})&=&\Phi(p_1,\ldots ,p_{n-1},P) \qquad p_n^\alpha= P^\alpha - q^\alpha \nn
\sum_{j=1}^n p_j^\alpha \frac{\partial}{\partial p_j^\alpha}\Phi(\ul{p})&=&\sum_{j=1}^{n-1} p_j^\alpha \frac{\partial}{\partial p_j^\alpha}\Phi(p_1,\ldots p_{n-1},P) + (P^\alpha -q^\alpha)\frac{\partial}{\partial P^\alpha}\Phi(p_1,\ldots p_{n-1},P). 
\eea
The last ($n_{th}$) term in (\ref{ft4}) is given by 
\be
\label{ex1}
\sigma_n\equiv\int \ul{d^d p} \,\delta^d(P)e^{i\ul{p x}}p_n^\alpha \frac{\partial}{ p_n^\alpha}\Phi(\ul{p}).
 \ee
Rewriting the exponential as
 $e^{i \ul{p\cdot x}}\to  e^{i( p_1\cdot x_{1 n} +\ldots p_{n-1}\cdot x_{n-1 \, n} + i P\cdot x_n )}$ and using the $\delta^d(P)$ to remove the $P\cdot x_n$ term, $\sigma_n$ takes the form
\bea
\label{ex2}
\sigma_n&=&\int {d^d p_1d^d p_2 \ldots d^d p_{n-1} d^d P} \,\delta^d(P)e^{i{p_1 x_{1 n} + i p_2 x_{2 n} +\ldots i p_{n-1} x_{n-1 n} }}\times \nn
&& \qquad \times (P-q)^\alpha \frac{\partial}{ P^\alpha}\Phi({p_1},\ldots p_{n-1},P)\nn
&=&-\int {d^d p_1d^d p_2 \ldots d^d p_{n-1}} e^{i{p_1 x_{1 n} + i p_2 x_{2 n} +\ldots i p_{n-1} x_{n-1 n} }}q^\alpha \int d^d P \,  \frac{\partial}{ \partial P^\alpha}\Phi({p_1},\ldots p_{n-1},P) \delta^d(P).\nn
 \eea
 Notice that in the expression above we have removed the $\delta^d(P) P^\alpha\frac{\partial}{\partial P^\alpha}\Phi(\ul{p})$ term, which after a partial integration becomes
 \bea
 \int d^d P\delta^d(P) P^\alpha \frac{\partial}{\partial P^\alpha}\Phi(p_1,\ldots p_{n-1},P)e^{i{p_1 x_{1 n} + i p_2 x_{2 n} +\ldots i p_{n-1} x_{n-1 n} }}&& \nn
 =
 \int d^d Pe^{i{p_1 x_{1 n} + i p_2 x_{2 n} +\ldots i p_{n-1} x_{n-1 n} }}
  \left(-d  +P^\alpha  \frac{\partial}{\partial P^\alpha}\delta^d(P)\right)\Phi(p_1,\ldots p_{n-1},P) && \nn
 \eea
and vanishes by (\ref{delta11}). Similarly, we derive the vanishing relation
 \bea
  \int d^d P \, P^\alpha \frac{\partial}{ \partial P^\alpha}\Phi({p_1},\ldots p_{n-1},P) \delta^d(P)&=&
 d \int d^d P \delta^d(P)\frac{P^\alpha}{P^2} \Phi(p_1,\ldots p_{n-1},P)\nn\
 &=& d \int d^d P \delta^d(P)\frac{P^\alpha}{P^2} \Phi(p_1,\ldots p_{n-1},0)\\
 &=0&
 \eea
which has been obtained as a result of (\ref{rule11}) and (\ref{rule12}). Therefore we find that 
$\sigma_n=0$, reobtaining the expected scaling equation in momentum space

\be
\left(\sum_{j=1}^n\D_j-(n-1)d-\sum_{j=1}^{n-1}p_j^\a\frac{\partial}{\partial p_j^\a} \right)\Phi(p_1,\ldots p_{n-1},\bar{p}_n)=0.
\ee

 \subsection{Special Conformal WI's for scalar correlators} 
 We now turn to the analysis of the special conformal transformations in momentum space. 
In this second case the  $\delta'$ terms cancel identically. Also in this case we discuss both the symmetric and the asymmetric forms of the equations, focusing our attention first on the scalar case. 
The Ward identity in the scalar case is given by
\be
\sum_{j=1}^{n} \left(- x_j^2\frac{\partial}{\partial x_j^\kappa}+ 2 x_j^\kappa x_j^\alpha \frac{\partial}
{\partial x_j^\alpha} +2 \Delta_j x_j^\kappa\right)\Phi(x_1,x_2,\ldots,x_n) =0
\ee
which in momentum space, using 
\be
x_j^\alpha\to -i \frac{\partial}{\partial p_j^\alpha} \qquad \frac{\partial}{\partial x_j^\kappa}\to i p_j^\kappa
\ee
becomes 
\be
\sum_{j=1}^n \int \ul {d^d p}\left(p_j^\kappa \frac{\partial^2}{\partial p_j^\alpha \partial p_j^\kappa} -
2 p_j^\alpha \frac{\partial^2}{\partial p_j^\alpha \partial p_j^\kappa}  -2 \Delta_j\frac{\partial}{\partial p_j^\kappa}\right) 
e^{i\ul{p\cdot x}} \delta^d(P)\phi(\ul{p})=0,
\ee
where the action of the operator is only on the exponential. At this stage we integrate by parts, bringing the derivatives from the exponential to the correlator and on the Dirac $\delta$ function obtaining

\be
\int \ul{d^d p}e^{i \ul{p x}}  \,K_s^k\Phi(\ul{p})\delta^d(P) +\delta'_\textrm{term}  =0
\ee
in the notations of Eq. (\ref{conv}), where we have introduced the differential operator acting on a scalar correlator in a symmetric form
\be
 \,K_s^k=\sum_{j=1}^n\left(p_j^\kappa \frac{\partial^2}{\partial p_j^\alpha\partial p_j^\alpha} + 2(\Delta_j- d)\frac{\partial}{\partial p_j^\kappa}-2 p_j^\alpha\frac{\partial^2}{\partial p_j^\kappa \partial p_j^\alpha}\right).
\ee

Some of the terms containing first and second derivatives of the Dirac delta function can be rearranged using also the intermediate relation
 \begin{align}
 \label{uno}
\,K_s^\k \delta^d(P)&=\left(P^k\frac{\partial^2}{\partial P^\alpha \partial P_\alpha}  -2 P^\alpha \frac{\partial^2}{\partial P^\alpha\partial P^k} 
+2 (\Delta- n\, d)\frac{\partial }{\partial P^k}\right)\delta^d(P)\nn
&=2 d( d \,n -  d -\Delta)P^k\frac{\delta^d(P)}{P^2}\nn
&=-2 ( d \,n -  d -\Delta)\frac{\partial}{\partial P_k} {\delta^d(P)},
\qquad \qquad \Delta=\sum_{j=1}^n\Delta_j,
\end{align}
where we have repeatedly used (\ref{delta11}) together with (\ref{derp0}) and (\ref{derp}). Combining all the derivative terms, on the other hand, we obtain
\begin{align}
\delta'_\textrm{term}&= \int {\ul{d^d p}}\,e^{i\ul{p\cdot x}}
\left[ \frac{\partial}{\partial P^\alpha} \delta^d(P)\sum_{j=1}^n\left(p_j^\alpha\frac{\partial}{\partial p_j^\kappa}- p_j^\kappa\frac{\partial}{\partial p_j^\alpha}\right)\Phi(\ul{p}) \right.\notag\\
&\hspace{5cm}\left..+ 2 \frac{\partial}{\partial P^\kappa} \delta^d(P)\left(\sum_{j=1}^n (\Delta_j -  p_j^\alpha 
\frac{\partial}{\partial p_j^\alpha} )- (n-1) d \right)\Phi(\ul{p})\right].
\end{align}
Notice that such terms vanish by using rotational invariance of the scalar correlator 
\be
\sum_{j=1}^3\left(p_j^\alpha\frac{\partial}{\partial p_j^\kappa}- p_j^\kappa\frac{\partial}{\partial p_j^\alpha}\right)\Phi(\ul{p})=0,
\ee
as a consequence of the SO(4) symmetry 
 \be
\sum_{j=1}^3 L_{\mu\nu}(x_j) \langle \phi(x_1)\phi(x_2)\phi(x_3)\rangle =0, 
\ee
with 
\be
L_{\mu\nu}(x)=i\left(x_\mu\partial_\nu - x_\nu\partial_\mu \right),
\ee
and the symmetric scaling relation,
\be
\left(\sum_{j=1}^n \Delta_j - \sum_j^{n-1} p_j^\alpha 
\frac{\partial}{\partial p_j^\alpha} - (n-1) d \right)\Phi(\ul{p})=0,
\ee
 where we have used (\ref{delta11}).  Using (\ref{uno}) and the vanishing of the $\delta'_\textrm{term}$ terms, the structure of the CWI on the correlator $\Phi(p)$ then takes the symmetric form 
 \be
 \label{sm1}
  \sum_{j=1}^n \int \ul {d^d p}e^{i\ul{p\cdot x}}\left(p_j^\kappa \frac{\partial^2}{\partial p_{j\alpha} \partial p_j^\alpha} -
2 p_j^\alpha \frac{\partial^2}{\partial p_j^\alpha \partial p_j^\kappa}  + 2( \Delta_j-d)\frac{\partial}{\partial p_j^\kappa}\right) 
 \phi(\ul{p})\delta^d(P)=0.
\ee
This {\em symmetric} expression is the starting point in order to proceed with the elimination of one of the momenta, say $p_n$. 
Also in this case, one can proceed by following the same procedure used in the derivation of the dilatation identity, dropping the contribution coming from the dependent momentum $p_n$, thereby obtaining the final form of the equation

\be
\sum_{j=1}^{n-1}\left(p_j^\kappa \frac{\partial^2}{\partial p_j^\alpha\partial p_j^\alpha} + 2(\Delta_j- d)\frac{\partial}{\partial p_j^\kappa}-2 p_j^\alpha\frac{\partial^2}{\partial p_j^\kappa \partial p_j^\alpha}\right)\Phi(p_1,\ldots p_{n-1},\bar{p}_n)=0.
\ee 
Also in this case the differentiation respect to $p_n$ requires the chain rule. For a certain sequence of scalar single particle operators 
\be
 \Phi(p_1,\ldots p_{n-1},\bar{p}_n)=\langle \phi(p_1)\ldots \phi (\bar{p}_n)\rangle 
\ee
the Leibnitz rule is therefore violated. As we have already mentioned, the complete symmetry of the solution respect to the three  momenta is however respected. We will now move to a discussion of the general structure of the method, focusing first on scalars and then on tensor correlators.

\section{Reduction of the action of $K^\kappa_{scalar}$ {}}

In the case of a scalar correlator all the anomalous conformal WI's can 
be re-expressed in scalar form by taking as independent momenta the magnitude $ {p}_i=\sqrt{p_i^2}$ as the 
three independent variables. Defining $\mathcal{F}(p_1,p_2)=\Phi(p_1,p_2,\bar{p_3}) $
and using the relation 

\be
p_1^\alpha \frac{\partial \mathcal{F}}{{p_1}^\alpha}+ p_2^\alpha \frac{\partial \mathcal{F}}{{p_2}^\alpha}=
 {p}_1\frac{ \partial \Phi}{\partial   p_1} +   p_2\frac{ \partial \Phi}{\partial   p_2} +   p_3\frac{ \partial \Phi}{\partial   p_3}
\ee
the anomalous scale equation becomes 
\be
\label{scale}
\left(\Delta -2 d - \sum_{i=1}^3    p_i \frac{ \partial}{\partial   p_i}\right)\Phi(p_1,p_2,\bar{p}_3)=0.\ee
The relation above is derived using the chain rule
\be
\label{chainr}
\frac{\partial \Phi}{\partial p_i^\mu}=\frac{p_i^\mu}{  p_i}\frac{\partial\Phi}{\partial  p_i} 
-\frac{\bar{p}_3^\mu}{  p_3}\frac{\partial\Phi}{\partial   p_3}.
\ee

It is a straightforward but lengthy computation to show that the special (non anomalous) conformal transformation in 
$d$ dimension takes the form, for  the scalar component  

\be
{{K}_{scalar}}^{\kappa}\Phi=0
\ee
with
\be
{{K}}^{\kappa}_{scalar}=\sum_{i=1}^3 p_i^\kappa \,{K}_i 
\label{kappa2}
\ee
\be
{ K}_i\equiv \frac{\partial^2}{\partial    p_i \partial    p_i} 
+\frac{d + 1 - 2 \Delta_i}{   p_i}\frac{\partial}{\partial   p_i}
\ee
with the expression (\ref{kappa2}) which can be split into the two independent equations
\be
\frac{\partial^2\Phi}{\partial   p_i\partial   p_i}+
\frac{1}{  p_i}\frac{\partial\Phi}{\partial  p_i}(d+1-2 \Delta_1)-
\frac{\partial^2\Phi}{\partial   p_3\partial   p_3} -
\frac{1}{  p_3}\frac{\partial\Phi}{\partial  p_3}(d +1 -2 \Delta_3)=0\qquad i=1,2.
\label{3k1}
\ee
Notice that in the derivation of (\ref{kappa2})
one needs at an intermediate step the derivative of the scaling WI 
\be
  p_1\frac{\partial^2\Phi}{\partial   p_3\partial  p_1} 
+   p_2\frac{\partial^2\Phi}{\partial   p_3\partial  p_2}=(\Delta -2 d -1)
\frac{\partial\Phi}{\partial  p_3} -  p_3\frac{\partial^2\Phi}{\partial   p_3\partial   p_3}.
\ee
Defining 
\be
\label{kij}
K_{ij}\equiv {K}_i-{K}_j
\ee
Eqs. (\ref{3k1}) take the form 
\be
\label{3k2}
K^\kappa_{13}\Phi=0 \qquad \textrm{and} \qquad K^\kappa_{23}\Phi=0.
\ee

\section{Transverse Ward identities} 
To fix the form of the correlator we need to impose the transverse WI on the vector lines and the conservation WI for $T^{\mu\nu}$. 
In this section we briefly discuss their derivation and their explicit expressions.
We consider the functional 
\be
W[g,A] =\int {D\bar\psi}D\psi e^{-(S_0[g,\psi] + S_1[A,\psi ])}
\label{GF}
\ee
integrated over the fermions $\psi$, in the background of the metric $g_{\mu\nu}$ and of the gauge field $A_\mu^a$. In the case of a nonabelian gauge theory the action is given by
\bea
S_0[g,A,\psi]&=&-\frac{1}{4}\int d^4 x \sqrt{-g_x}F_{\mu\nu}^{{a}} F^{\mu\nu a} +\int d^4 x \sqrt{-g_x} i \bar{\psi}\gamma^\mu D_\mu \psi \nn
 S_1[g,A]&=&\int  \sqrt{-g_x}J_\mu^{a} A^{\mu a}
\eea 
with
\bea
F_{\mu\nu}^{a} &=&\nabla_\mu A_{\nu}^{a} - \nabla_{\nu} A_{\mu}^{ a} + g_c f^{a  b  c} A_\mu^{ b} A_\nu^c\nn 
&=&\partial_\mu A_{\nu}^{a} - \partial_{\nu} A_{\mu}^{ a} + g_c f^{a  b  c} A_\mu^{ b} A_\nu^c, \nn
\nabla_\mu A^{\nu a} &=&\partial_\mu A^{\nu a} +\Gamma_{\mu\nu}^\lambda A^{\lambda a}
\eea
with $J^{\mu  a}=g_c\bar{\psi}\gamma^\mu T^{a}\psi$ denoting the fermionic current, with $T^a$ the generators of the theory and $\nabla_\mu$ denoting the covariant derivative in the curved background on a vector field. The local Lorentz and gauge covariant derivative $(D)$ on the fermions acts via the spin connection 
\be
D_\mu \psi=\left( \partial_\mu \psi  + A_\mu^a T^a +\frac{1}{4}\omega_{\mu}^{\ul a \ul b}\gamma_{\ul a \ul b}\right)\psi
\ee
having denoted with $\ul{a}\ul{b}$ the local Lorentz indices.  A local Lorentz covariant derivative $(D)$ can be similarly defined for a vector field, say $V^{\ul a}$, via the Vielbein $e^{\ul a}_{\, \,\mu}$ and its inverse  $e^{\,\,\mu}_{\ul a}$
\be
D_\mu V^{\ul a}= \partial_\mu V^{\ul a} +\omega^{\ul a}_{\mu \ul b} V^{\ul b}
\ee
with 
\be
\nabla_\mu V^\rho = e_{\ul a}^{\,\,\rho} D_\mu V^{\ul a}
\ee
with the Christoffel and the spin connection related via the holonomic relation 
 \be
 \Gamma^\rho_{\mu\nu}=e^{\,\,\rho}_{\ul a}\left( \partial_\mu e^{\ul a}_{\,\,\nu} +\omega^{\ul a}_{\mu \ul b} e^{\ul b}_{\,\,\nu}\right).
\ee
Diffeomorphism invariance of the generating functional (\ref{GF}) gives 
 \be
 \int d^d x \left(\frac{\delta W}{\delta g_{\mu\nu}}\delta g_{\mu\nu}(x) +  \frac{\delta W}{\delta A_{\mu}^a}\delta A_{\mu}^a(x)\right)=0
\label{aver}
\ee
where the variation of the metric and the gauge fields are the corresponding Lie derivatives, for a change of variables 
$x^\mu\to x^\mu + \epsilon^\mu(x)$
\bea
\label{lie}
\delta A_\mu^a(x) &=& -\nabla_\alpha A_\mu^a \epsilon^\alpha - A^a_\alpha \nabla_\mu \epsilon^\alpha\nn
\delta g_{\mu\nu}&=&-\nabla_\mu \epsilon_\nu  - \nabla_\nu \epsilon_\mu 
\eea
while for a gauge transformation with a parameter $\theta^a(x)$
\be
\delta A_\mu^a=\ul{D}_\mu \theta^a \equiv\partial_\mu \theta^a + g_c f^{a b c} A_\mu^b \theta^c.
\ee
Using (\ref{lie}), Eq. (\ref{aver}) becomes 
\begin{align}
0=&\,\left\langle\, \int d^4 x \left( \frac{\delta (S_0 + S_1)}{\delta g_{\mu\nu}} \delta g_{\mu\nu} +\frac{\delta S_1}{\delta A_\mu^a} \delta A_{\mu}^a\right)
\right\rangle\nn
=&\,\left\langle \int d^4 x \sqrt{-g_x}\left[\nabla_\mu T^{\mu\nu} +(\nabla_\mu A_\nu^a -\nabla_\nu A_\mu^a) J^{\mu a} +
\nabla_\mu J^{\mu a} A_\nu^a\right]\epsilon^\nu(x)\right\rangle
\label{interm}
\end{align}
while the condition of gauge invariance gives 
\be
 \int d^d x \frac{\delta W}{\delta A_{\mu}^a}\delta A_{\mu}^a=\left\langle \int d^4 x \sqrt{-g_x}J^\mu_a\ul{D}_\mu \theta^a \right\rangle=0
\ee
which, in turn, after an integration by parts, generates the gauge WI
\be
\langle \nabla_\mu J^{\mu a} \rangle = g_c f^{a b c} \langle J_\mu^b\rangle A^{\mu c}. 
\ee
Inserting this relation into (\ref{interm}) we obtain the conservation WI
\be
\langle \nabla^\mu T_{\mu\nu}\rangle +F_{\mu\nu}^a \langle J^{\mu a}\rangle =0.
\ee
In the abelian case, diffeomorphism and gauge invariance  then give the relations
\begin{equation}
\begin{split}
0&=\nabla_\n \braket{T^{\m \n}}+F^{\m\n}\braket{J_\n}\\
0&=\nabla_\n\braket{J^\nu}
\end{split}\label{transverse}
\end{equation}
with naive scale invariance gives the traceless condition
\begin{equation}
g_{\mu\nu}\braket{T^{\m\n}}=0.\label{trace}
\end{equation}
The functional differentiation of \eqref{transverse} and \eqref{trace} allows to derive ordinary Ward identities for the various correlators. In the $TJJ$ case we obtain, after a Fourier transformation, the conservation equation 
\begin{align}
\label{tr}
p_{1\n_1}\braket{T^{\mu_1\nu_1}(p_1)\,J^{\m_2}(p_2)\,J^{\m_3}(p_3)}&=4\,\big[\d^{\m_1\m_2}p_{2\l}\braket{J^\l(p_1+p_2)\,J^{\m_3}(p_3)}-p_2^{\m_1}\braket{J^{\m_2}(p_1+p_2)\,J^{\m_3}(p_3)}\big]\notag\\
&+4\,\big[\d^{\m_1\m_3}p_{3\l}\braket{J^\l(p_1+p_3)\,J^{\m_2}(p_2)}-p_3^{\m_1}\braket{J^{\m_3}(p_1+p_3)\,J^{\m_2}(p_2)}\big]
\end{align}
and vector current Ward identities
\begin{align}
\label{x1}
p_{2\m_2}\braket{T^{\mu_1\nu_1}(p_1)\,J^{\m_2}(p_2)\,J^{\m_3}(p_3)}&=0\\
p_{3\m_3}\braket{T^{\mu_1\nu_1}(p_1)\,J^{\m_2}(p_2)\,J^{\m_3}(p_3)}&=0,
\end{align}
while the naive identity \eqref{trace} gives the non-anomalous condition
\begin{equation}
\label{x2}
\d_{\m_1\n_1}\braket{T^{\mu_1\nu_1}(p_1)\,J^{\m_2}(p_2)\,J^{\m_3}(p_3)}=0,
\end{equation}
valid in the $d\ne4$ case. We recall that the 2-point function of two conserved vector currents $J_i$ $(i=2,3)$ \cite{Coriano:2013jba} in any conformal field theory in $d$ dimension is given by 
\beqa
\langle J_2^\alpha(p)J_3^\beta(-p) \rangle =\delta_{\Delta_2\, \Delta_3}\left(c_{123} \Gamma_J \right)\pi^{\alpha\beta}(p) (p^2)^{\Delta_2-d/2},
\qquad \Gamma_J=\frac{\pi^{d/2}}{ 4^{\Delta_2 -d/2}}\frac{\Gamma(d/2-\Delta_2)}{\Gamma(\Delta_2)},
\eeqa
with $c_{123}$ an overall constant and $\Delta_2=d-1$. In our case $\D_2=\D_3=d-1$ and Eq. (\ref{tr}) then takes the form 
\begin{align}
\label{2point}
p_{1\mu_1}\braket{T^{\mu_1\nu_1}(p_1)\,J^{\m_2}(p_2)\,J^{\m_3}(p_3)}&=4 c_{123} \Gamma_J
\left( \delta^{\nu_1\mu_2}\frac{p_{2\lambda}}{(p_3^2)^{d/2 -\Delta_2}} {\pi^{\lambda \mu_3}}(p_3) -
\frac{p_2^{\nu_1}}{(p_3^2)^{d/2 -\Delta_2}}\pi^{\mu_2\mu_3}(p_3) \right.\nn
& \left. + \delta^{\nu_1\mu_3}\frac{p_{3\lambda}}{(p_2^2)^{d/2-\Delta_2}}\pi^{\lambda \mu_2}(p_2) -\frac{p_3^{\nu_1}}{(p_2^2)^{d/2-\Delta_2}}\pi^{\mu_3\mu_2}(p_2)\right).
\end{align}
Explicit expressions of the secondary CWI's are determined using (\ref{x1}) and (\ref{x2}) and the explicit form (\ref{2point}).

\section{TJJ reconstruction the BMS way}\label{Skend}
We are now going investigate the BMS approach, which is technically quite involved, highlighting several steps which are crucial in order to clarify the basic structure of the method. The method is exemplified in the case of the $TJJ$. Several intermediate steps, which we believe are necessary in order to characterize the approach, have been worked out independently and are based on the use of the Lorentz Ward identities.

 Given the partial symmetry of the $TJJ$ correlator, for instance, respect to the $TTT$ case, one can choose as independent momenta either $p_1$ and $p_2$ or, more conveniently, $p_2$ and $p_3$, given the symmetry of the two $J$ currents.\\
With the first choice, outlined below, the current $J(p_3)$ is singlet under the (spin) Lorentz generators. With the second choice, the two currents are treated symmetrically and the stress energy tensor is treated as a singlet under the same generators. The derivation of the CWI's in this second case will be outlined in section (\ref{p1section}). The equations obtained in the two cases are obviously the same. 

First of all we discuss the canonical Ward identities for the $\braket{TJJ}$ correlation function in momentum space. From the general definition of the global Ward identities in position space
\begin{equation}
\sum_{j=1}^3G_g(x_j)\braket{T^{\m_1\n_1}(x_1)\,J^{\mu_2}(x_2)\,J^{\mu_3}(x_3)}=0,
\end{equation}
where $G_g$ is the generator of the infinitesimal symmetry transformation. The dilatation Ward identities take the form
\begin{equation}
0=\left[\sum_{j=1}^3\Delta_j-(n-1)d-\sum_{j=1}^2\,p_j^\a\sdfrac{\partial}{\partial p_j^\alpha}\right]\braket{{T^{\mu_1\nu_1}(p_1)\,J^{\m_2}(p_2)\,J^{\mu_3}(\bar p_3)}}.\label{DilatationTJJ}
\end{equation}
To proceed towards the analysis of the constraints, it is essential to introduce the Lorentz covariant Ward identities
\begin{align}
0&=\sum_{j=1}^{2}\left[p_j^\nu\sdfrac{\partial}{\partial p_{j\mu}}-p_j^{\mu}\sdfrac{\partial}{\partial p_{j\nu}}\right]\braket{{T^{\mu_1\nu_1}(p_1)\,J^{\m_2}(p_2)\,J^{\mu_3}(\bar p_3)}}\notag\\
&\qquad+2\left(\delta^{\nu}_{\a_1}\d^{\mu(\mu_1}-\delta^{\mu}_{\alpha_1}\delta^{\nu(\mu_1}\right)\braket{{T^{\nu_1)\alpha_1}(p_1)\,J^{\m_2}(p_2)\,J^{\mu_3}(\bar p_3)}}\notag\\
&\qquad+\left(\delta^{\nu}_{\a_2}\d^{\mu\mu_2}-\delta^{\mu}_{\alpha_2}\delta^{\nu\mu_2}\right)\braket{{T^{\mu_1\nu_1}(p_1)\,J^{\alpha_2}(p_2)\,J^{\mu_3}(\bar p_3)}}\notag\\
&\qquad+\left(\delta^{\nu}_{\a_3}\d^{\mu\mu_3}-\delta^{\mu}_{\alpha_3}\delta^{\nu\mu_3}\right)\braket{{T^{\mu_1\nu_1}(p_1)\,J^{\mu_2}(p_2)\,J^{\alpha_3}(\bar p_3)}}\label{RotationTJJ},
\end{align}
where 
\begin{equation}
\d^{\n(\m_1}\,\braket{{T^{\nu_1)\alpha_1}\,J^{\m_2}\,J^{\mu_3}}}\equiv\sdfrac{1}{2}\big(\d^{\n\m_1}\,\braket{{T^{\nu_1\alpha_1}\,J^{\m_2}\,J^{\mu_3}}}+\d^{\n\n_1}\,\braket{{T^{\mu_1\alpha_1}\,J^{\m_2}\,J^{\mu_3}}}\big),
\end{equation}
and finally the special conformal Ward identities
\begin{align}
0&=\sum_{j=1}^{2}\left[2(\Delta_j-d)\sdfrac{\partial}{\partial p_j^\k}-2p_j^\a\sdfrac{\partial}{\partial p_j^\a}\sdfrac{\partial}{\partial p_j^\k}+(p_j)_\k\sdfrac{\partial}{\partial p_j^\a}\sdfrac{\partial}{\partial p_{j\a}}\right]\braket{{T^{\mu_1\nu_1}(p_1)\,J^{\m_2}(p_2)\,J^{\mu_3}(\bar p_3)}}\notag\\
&\qquad+4\left(\d^{\k(\mu_1}\sdfrac{\partial}{\partial p_1^{\a_1}}-\delta^{\k}_{\alpha_1}\delta^{\l(\mu_1}\sdfrac{\partial}{\partial p_1^\l}\right)\braket{{T^{\nu_1)\alpha_1}(p_1)\,J^{\m_2}(p_2)\,J^{\mu_3}(\bar p_3)}}\notag\\
&\qquad+2\left(\d^{\k\mu_2}\sdfrac{\partial}{\partial p_2^{\a_2}}-\delta^{\k}_{\alpha_2}\delta^{\l\mu_2}\sdfrac{\partial}{\partial p_2^\l}\right)\braket{T^{\mu_1\nu_1}(p_1)\,J^{\alpha_2}(p_2)\,J^{\mu_3}(\bar p_3)}.\label{SCWTJJ}
\end{align} 
which we will use in the next sections in order to determine the tensor structure of this correlator. 
\subsection{Projectors}
The basic observation in the BMS approach is that the action of  the special conformal, trace and conservation (longitudinal) WI' s take a simpler form if we enforce a decomposition of the tensor correlators  in terms of transverse traceless, longitudinal  and trace parts and project the Ward identities on the same subspaces. Recall that for a symmetric tensor 
such as the EMT, this decomposition is performed using the following projectors 

\begin{align}
\pi^{\mu}_{\alpha} & = \delta^{\mu}_{\alpha} - \frac{p^{\mu} p_{\alpha}}{p^2},  \qquad \tilde{\pi}^{\mu}_{\alpha} =\frac{1}{d-1}\pi^{\mu}_{\alpha} \\\
\Pi^{\mu \nu}_{\alpha \beta} & = \frac{1}{2} \left( \pi^{\mu}_{\alpha} \pi^{\nu}_{\beta} + \pi^{\mu}_{\beta} \pi^{\nu}_{\alpha} \right) - \frac{1}{d - 1} \pi^{\mu \nu}\pi_{\alpha \beta}, \\
\mathcal{I}^{\mu\nu}_{\alpha\beta}&=\frac{1}{p^2}p^{\beta}\left( p^{\mu}\delta^{\nu}_{\alpha} +p^{\nu}\delta^{\mu}_{\alpha} -
\frac{p_{\alpha}p_{\beta}}{p^2}( \delta^{\mu\nu} +(d-2)\frac{p^\mu p^\nu}{p^2})    \right)\\
\mathcal{L}^{\mu\nu}_{\alpha\beta}&=\frac{1}{2}\left(\mathcal{I}^{\mu\nu}_{\alpha\beta} +\mathcal{I}^{\mu\nu}_{\beta\alpha}\right) \qquad \tau^{\mu\nu}_{\alpha\beta} =\tilde{\pi}^{\mu \nu}\delta_{\alpha \beta}
\end{align}
with 
\begin{align}
\delta^{\mu\nu}_{\alpha\beta}&=\Pi^{\mu \nu}_{\alpha \beta} +\Sigma^{\mu\nu}_{\alpha\beta} \\
\Sigma^{\mu\nu}_{\alpha\beta}&\equiv\mathcal{L}^{\mu\nu}_{\alpha\beta} +\tau^{\mu\nu}_{\alpha\beta}
\end{align}
The previous identities allow to decompose a symmetric tensor into its transverse traceless (via $\Pi$), longitudinal (via $\mathcal{L}$) and trace parts (via $\tau$), or on the sum of the combined longitudinal and trace contributions (via $\Sigma$). We are now going to illustrate the approach in the case of the $TTT$ correlation function. The transverse traceless projections will be denoted as $(t)$ and the trace parts with $(s)$.
 We will be denoting such correlator as ${\phi}$ and try to resort to a condensed notation in order to characterize the algebraic structure of the procedure. For a rank-6 correlator of 3 $T$'s, $\phi$ will denote the tensor 
\be
\phi\equiv \langle T^{\mu_1\nu_1}(p_1)T^{\mu_2\nu_2}(p_2)T^{\mu_3\nu_3}(p_3)\rangle .
\ee
We can act on this correlator with the two projectors $\Pi$ and $\Sigma$ on each (combined) pair of indices and momenta. For instance, the transverse traceless part of $\phi$ is obtained by acting with 3 $\Pi$ projectors on the indices of the EMT's 
\begin{align}
\phi_{t t t}&\equiv \Pi_1\Pi_2\Pi_3 \phi\\
&\equiv {\Pi_1}^{\mu_1 \nu_1}_{\alpha_1 \beta_1} {\Pi_2}^{\mu_2 \nu_2}_{\alpha_2 \beta_2} {\Pi_3}^{\mu_3 \nu_3}_{\alpha_3 \beta_3}\langle T^{\alpha_1\beta_1}(p_1)T^{\alpha_2\beta_2}(p_2)T^{\alpha_3\beta_3}(p_3)\rangle,
 \end{align}
 where ${\small {\Pi_1}^{\mu_1 \nu_1}_{\alpha_1 \beta_1} \equiv{\Pi}^{\mu_1 \nu_1}_{\alpha_1 \beta_1}(p_1)}$. Similarly, the remaining 7 components of the $\phi$ correlator can be obtained by acting with all the other combinations of projectors $(\Sigma, \Pi)$, 
 to obtain 
 \be
 \phi=\phi_{ttt}+ \phi_{tst}+ \phi_{tss}+\phi_{sss}+\phi_{stt}+\phi_{sst}+\phi_{sts}.
\ee
Acting with the special conformal transformation $K^\kappa$ (both with the scalar and the spin parts) on $\phi$ we can again project the result onto the orthogonal subspaces $ttt, tss,$ etc., and try to solve the equations separately in each of these 8 sectors. In our condensed notation the equation for the special special conformal transformation takes the form  
\begin{align}
\label{ecco}
\phi'&=K^\kappa \phi=0,
\end{align} 
and its projection into the 8 independent sectors, such as, for instance 
\begin{align}
\label{eccoti}\phi'_{ttt}&\equiv\Pi_1 \Pi_2 \Pi_3 K^\kappa \phi=0,\\
& \phi'_{stt}\equiv\Sigma_1 \Pi_2 \Pi_3 K^\kappa \phi=0,\qquad  \phi'_{sst}\equiv\Sigma_1 \Sigma_2 \Pi_3 K^\kappa \phi=0, \ldots
 \end{align}
and so on, can be obtained by the action of the $\Pi$'s and $\Sigma$'s on (\ref{ecco}). It is important to realize that only the equation $\phi'_{t t t}=0$ involves 3-point functions beside 2 point functions, and needs to be solved. The remaining sectors do not give any new equation, since they involve only 2-point functions, being related to the conservation and trace WI's. However they define consistency conditions for lower point functions that will introduce some constraint on the arbitrary constants appearing in the solutions of the primary WI's. \\
To illustrate these points we will treat the correlators $\phi, \phi'$ as vectors in a functional space on which the  
$K^{\kappa}$ operator will act both in differential form and algebraically via its spin rotation matrices. For instance we define 
\be
\phi_{ttt}\equiv P_{t t t}\phi,\,\, \textrm{where}\,\, P_{t t t}\equiv \Pi_1 \Pi_2 \Pi_3, \,\, \phi_{tst}\equiv P_{t s t}\phi \,\, \textrm{with}\,\, P_{t s t}\equiv \Pi_1 \Sigma_2 \Pi_3, \ldots
\ee
and similarly in the other cases. 
In general, is is convenient to characterize the action of $K^\kappa$ on each subspace via  a projection, such as 
\be
\phi'^{ (t t t)}_{t t t}\equiv  P_{t t t} K^\kappa P_{t t t}\phi \qquad \phi'^{ (t t t)}_{s t t}= P_{s t t} K^\kappa P_{t t t}\phi,\ldots 
 \ee
 and so on, for a total of $64=8\times 8$ sectors. In the first expressions above, for example, the original transverse traceless projection $(ttt)$ is acted upon by $K^\kappa$ and then it is re-projected onto the $ttt$ sector. 
 There are several simplifications among these matrix elements. For example, a direct computation, that we will prove below, gives 
 \be
  P_{stt}K^\kappa P_{ttt}\phi=0, \qquad P_{tst}K^\kappa P_{ttt}\phi=0, \qquad P_{tts}K^\kappa P_{ttt}\phi=0, \ldots
 \label{inter1}
 \ee
showing that the $\phi_{t t t}$ amplitude is mapped only into another amplitude in the same $t t t$ subspace by the action of $K^\kappa$. 
\subsubsection{Endomorphic action of \texorpdfstring{$K^\kappa$}{} on the transverse-traceless sector}
$K^\kappa$ acts as endomorphism on the transverse traceless sector of a tensor correlator. 
To illustrate this point we consider the case of the $TTT$, though the approach is generic.
Define 
\be
Y^{\mu_1\nu_1\mu_2\nu_2\mu_3\nu_3}={\Pi_1}^{\mu_1\nu_1}_{\alpha_1\beta_1}{\Pi_2}^{\mu_2\nu_2}_{\alpha_2\beta_2}{\Pi_3}^{\mu_3\nu_3}_{\alpha_3\beta_3}\langle {T}^{\alpha_1\beta_1}{T}^{\alpha_2\beta_2}{T}^{\alpha_3\beta_3} \rangle
\ee
to be the transverse traceless projection of the $TTT$.
One can check the transversality of the action of $K^\kappa_{1 scalar}$ (the other contribution of $K^\kappa$ being similar) by contracting $Y$ with $p_1$  
\begin{align}
p_1^{\mu_1}K^\kappa_{1 scalar} Y^{\mu_1\nu_1\mu_2\nu_2\mu_3\nu_3}&=
\left(-2 p_1^\alpha p_1^{\mu_1}\frac{\partial^2}{\partial p_1^\alpha \partial p_1^\kappa} 
+ p_1^\kappa p_1^{\mu_1}\frac{\partial^2}{\partial p_1^\alpha \partial p_1^\alpha}  \right) Y^{\mu_1\nu_1\mu_2\nu_2\mu_3\nu_3}\nn
&= -2 p_1^\alpha\frac{\partial}{\partial p_1^\alpha}\left(  p_{1}^{\mu_1}\frac{\partial}{\partial p_1^\kappa} Y^{\mu_1\nu_1\mu_2\nu_2\mu_3\nu_3}\right) + 2 p_1^{\mu_1}\frac{\partial}{ p_1^\kappa}Y^{\mu_1\nu_1\mu_2\nu_2\mu_3\nu_3} \nn
& + p_1^\kappa\frac{\partial}{\partial p_1^\alpha}\left(p_1^{\mu_1}\frac{\partial}{p_1^\alpha}Y^{\mu_1\nu_1\mu_2\nu_2\mu_3\nu_3}\right) -
p_1^\kappa \frac{\partial}{\partial p_1^{\mu_1}}Y^{\mu_1\nu_1\mu_2\nu_2\mu_3\nu_3}\nn
&=2 p_1^\alpha \frac{\partial}{\partial p_1^\kappa}Y^{\kappa\nu_1\mu_2\nu_2\mu_3\nu_3} -
2\, Y^{\kappa\nu_1\mu_2\nu_2\mu_3\nu_3} -2 p_1^\kappa \frac{\partial}{\partial p_1^{\mu_1}}Y^{\mu_1\nu_1\mu_2\nu_2\mu_3\nu_3}
\label{onec}
\end{align}
where we have rearranged the partial derivatives. For the spin part we obtain 
\begin{align}
\label{twoc}
p_1^{\mu_1}K^\kappa_{1 spin}Y^{\mu_1\nu_1\mu_2\nu_2\mu_3\nu_3}&=
2 p_1^{\mu_1}\left(\delta^{\kappa \mu_1} \frac{\partial}{\partial p_1^\alpha}- 
\delta^{\kappa \alpha}\frac{\partial}{p_1^{\mu_1}}\right)Y^{\alpha\nu_1\mu_2\nu_2\mu_3\nu_3} \nn
&\quad+
2 p_1^{\mu_1}\left(\delta^{\kappa \nu_1} \frac{\partial}{\partial p_1^\alpha}- 
\delta^{\kappa \alpha}\frac{\partial}{p_1^{\nu_1}}\right)Y^{\mu_1\alpha\mu_2\nu_2\mu_3\nu_3}\nn
&=2 p_1^\kappa \frac{\partial}{\partial p_1^\alpha}Y^{\alpha\nu_1\mu_2\nu_2\mu_3\nu_3}-
2 p_1^{\mu_1}\frac{\partial}{\partial p_1^{\mu_1}}Y^{\kappa\nu_1\mu_2\nu_2\mu_3\nu_3} + 2 Y^{\kappa\nu_1\mu_2\nu_2\mu_3\nu_3}.
\end{align}
Adding (\ref{onec}) and (\ref{twoc}) it is shown that 
\be
p_1^{\mu_1}K^\kappa_1 Y^{\mu_1\nu_1\mu_2\nu_2\mu_3\nu_3}=0,
\ee
which clearly holds for the entire $K$ operator since $\Pi_1$ filters to the left of $K_2$, obtaining
\be
p_1^{\mu_1}K^\kappa Y^{\mu_1\nu_1\mu_2\nu_2\mu_3\nu_3}=0.
\ee
Notice that in the derivation of this result the nonlinear character of the action of $K^\kappa_{scalar}$, which induces 
mixed derivative terms  terms does not play any role.  Due to the trace-free property of the projectors, then we obtain in our condensed notation
\be
\Sigma_1 K^\kappa \phi_{t t t}=0. 
\ee
The solution of the CWI's are then constructed, in this method, by acting on the entire correlator having parametrized its transverse traceless parts in therms of a minimal set of form factors plus trace/longitudinal terms (the semilocal or pinched terms). Semilocal terms are those containing a single delta function which will pinch two of the three external coordinates. The term ultralocal 
(or local) refers to the contribution of the anomaly itself, which is obtained when all the 3 point of the correlator coalesce.

\subsection{Application to the \texorpdfstring{$TJJ$}{}}
Turning to the $TJJ$ case, we can divide the 3-point function into two parts: the \emph{transverse-traceless} part and the \emph{semi-local} part (indicated by subscript $loc$) expressible through the transverse and trace Ward Identities. These parts are obtained by using the projectors $\Pi$ and $\Sigma$, previously defined. 
We can then decompose the full 3-point function as follows
\begin{align}
\braket{T^{\mu_1\nu_1}\,J^{\mu_2}\,J^{\mu_3}}&=\braket{t^{\mu_1\nu_1}\,j^{\mu_2}\,j^{\mu_3}}+\braket{T^{\mu_1\nu_1}\,J^{\mu_2}\,j_{loc}^{\mu_3}}+\braket{T^{\mu_1\nu_1}\,j_{loc}^{\mu_2}\,J^{\mu_3}}+\braket{t_{loc}^{\mu_1\nu_1}\,J^{\mu_2}\,J^{\mu_3}}\notag\\
&\quad-\braket{T^{\mu_1\nu_1}\,j_{loc}^{\mu_2}\,j_{loc}^{\mu_3}}-\braket{t_{loc}^{\mu_1\nu_1}\,j_{loc}^{\mu_2}\,J^{\mu_3}}-\braket{t_{loc}^{\mu_1\nu_1}\,J^{\mu_2}\,j_{loc}^{\mu_3}}+\braket{t_{loc}^{\mu_1\nu_1}\,j_{loc}^{\mu_2}\,j_{loc}^{\mu_3}}.
\end{align}
All the terms on the right-hand side, apart from the first one, may be computed by means of transverse and trace Ward Identities. The exact form of the Ward identities depends on the exact definition of the operators involved, but more importantly, all these terms depend on 2-point function only. The main goal now is to write the general form of the transverse-traceless part of the correlator and to give the solution using the Conformal Ward identities. 

Using the projectors $\Pi$ and $\pi$ one can write the most general form of the transverse-traceless part as
\begin{equation}
{\braket{t^{\m_1\n_1}(p_1)\,j^{\mu_2}(p_2)\,j^{\mu_3}(p_3)}}=\Pi^{\mu_1\nu_1}_{\alpha_1\beta_1}(p_1)\pi^{\mu_2}_{\alpha_2}(p_2)\pi^{\mu_3}_{\alpha_3}(p_3)\,\,X^{\alpha_1\beta_1\,\alpha_3\alpha_3},
\end{equation}
where $X^{\alpha_1\beta_1\,\alpha_3\alpha_3}$ is a general tensor of rank four built from the metric and momenta. We can enumerate all possible tensor that can appear in $X^{\alpha_1\beta_1\,\alpha_3\alpha_3}$ preserving the symmetry of the 
correlator, as illustrated in \cite{2014JHEP...03..111B}
\begin{align}
\langle t^{\mu_1\nu_1}(p_1)j^{\mu_2}(p_2)j^{\mu_3}(p_3)\rangle& =
{\Pi_1}^{\mu_1\nu_1}_{\alpha_1\beta_1}{\pi_2}^{\mu_2}_{\alpha_2}{\pi_3}^{\mu_3}_{\alpha_3}
\left( A_1\ p_2^{\alpha_1}p_2^{\beta_1}p_3^{\alpha_2}p_1^{\alpha_3} + 
A_2\ \delta^{\alpha_2\alpha_3} p_2^{\alpha_1}p_2^{\beta_1} + 
A_3\ \delta^{\alpha_1\alpha_2}p_2^{\beta_1}p_1^{\alpha_3}\right. \notag\\
& \left. + 
A_3(p_2\leftrightarrow p_3)\delta^{\alpha_1\alpha_3}p_2^{\beta_1}p_3^{\alpha_2}
+ A_4\  \delta^{\alpha_1\alpha_3}\delta^{\alpha_2\beta_1}\right).\label{DecompTJJ}
\end{align}
where we have used the symmetry properties of the projectors, and the coefficients $A_i\  i=1,\dots,4$ are the form factors, functions of $p_1^2, p_2^2$ and $p_3^2$. This ansatz introduces a minimal set of form factors which will be later determined by the solutions of the CWI's. For future discussion, we will refer to this basis as to the $A$-basis. \\
We can now consider the dilatation Ward identities for the transverse-traceless part obtained by the decomposition of \eqref{DilatationTJJ}. We are then free to apply the projectors $\Pi$ and $\pi$ to this decomposition in order to obtain the final result
\begin{align}
0&=\Pi^{\mu_1\nu_1}_{\alpha_1\beta_1}(p_1)\pi^{\mu_2}_{\alpha_2}(p_2)\pi^{\mu_3}_{\alpha_3}(p_3)\left[\sum_{j=1}^3\,\Delta_j-2d-\sum_{j=1}^{2}\,p_j^\alpha\sdfrac{\partial}{\partial p_j^\alpha}\right]\bigg[ A_1 p_2^{\alpha_1}p_2^{\beta_1}p_3^{\alpha_2}p_1^{\alpha_3} \notag\\[-1ex]
&\qquad+ 
A_2\delta^{\alpha_2\alpha_3} p_2^{\alpha_1}p_2^{\beta_1} + 
A_3 \delta^{\alpha_1\alpha_2}p_2^{\beta_1}p_1^{\alpha_3} + 
A_3(p_2\leftrightarrow p_3)\delta^{\alpha_1\alpha_3}p_2^{\beta_1}p_3^{\alpha_2}
+ A_4 \delta^{\alpha_1\alpha_3}\delta^{\alpha_2\beta_1}\bigg].
\end{align}
It is possible to obtain from this projection a set of differential equations for all the form factors. These equations are expressed as
\begin{equation}
\left[2d+N_n-\sum_{j=1}^{3}\Delta_j+\sum_{j=1}^2\,p_j^\alpha\sdfrac{\partial}{\partial p_j^\alpha}\right]\,A_n(p_1,p_2,p_3)=0,\label{DilatationFactor}
\end{equation}
where $N_n$ is the tensorial dimension of $A_n$, i.e. the number of momenta multiplying the form factor $A_n$ and the projectors $\Pi$ and $\pi$. \\
Turning to the special CWI's, ${\braket{TJJ}}$ in  \eqref{SCWTJJ}, we can write the same equation in the form 
\begin{equation}
{K}^\k\,{\braket{T^{\mu_1\nu_1}(p_1)\,J^{\m_2}(p_2)\,J^{\mu_3}( p_3)}}=0,
\end{equation}
where $K^\k$ is the special conformal generator.
 As before, we introduce the decomposition of the 3-point function to obtain
\begin{align}
0&={K}^\kappa\bigg[ \braket{{t^{\mu_1\nu_1}\,j^{\mu_2}\,j^{\mu_3}}}+{\braket{t_{loc}^{\mu_1\nu_1}\,j^{\mu_2}\,j^{\mu_3}}}+{\braket{t^{\mu_1\nu_1}\,j_{loc}^{\mu_2}\,j^{\mu_3}}}+{\braket{t^{\mu_1\nu_1}\,j^{\mu_2}\,j_{loc}^{\mu_3}}}\notag\\
&\hspace{2cm} +{\braket{t^{\mu_1\nu_1}_{loc}\,j_{loc}^{\mu_2}\,j^{\mu_3}}}+{\braket{t_{loc}^{\mu_1\nu_1}\,j_{loc}^{\mu_2}\,j^{\mu_3}}}+{\braket{t^{\mu_1\nu_1}\,j_{loc}^{\mu_2}\,j_{loc}^{\mu_3}}}+{\braket{t_{loc}^{\mu_1\nu_1}\,j_{loc}^{\mu_2}\,j_{loc}^{\mu_3}}}\bigg].
\end{align}
In order to isolate the equations for the form factors appearing in the decomposition, we are free to apply the projectors $\Pi$ and $\pi$ defined previously. Through a lengthy calculation we find
\begin{align}
\Pi^{\rho_1\sigma_1}_{\mu_1\nu_1}(p_1)\pi^{\rho_2}_{\mu_2}(p_2)\pi^{\rho_3}_{\mu_3}(p_3)\ \,K^\k
{\braket{t_{loc}^{\mu_1\nu_1}\,j^{\mu_2}\,j^{\mu_3}}}&=\Pi^{\rho_1\sigma_1}_{\mu_1\nu_1}\,\pi^{\rho_2}_{\mu_2}\,\pi^{\rho_3}_{\mu_3}\,\left[\sdfrac{4d}{p_1^2}\,\delta^{\kappa\mu_1}\,p_{1\alpha_1}\,{\braket{T^{\alpha_1\nu_1}J^{\mu_2}J^{\mu_3}}}\right]\notag\\
\Pi^{\rho_1\sigma_1}_{\mu_1\nu_1}(p_1)\pi^{\rho_2}_{\mu_2}(p_2)\pi^{\rho_3}_{\mu_3}(p_3)\ \,K^\k
\braket{t^{\mu_1\nu_1}\,j_{loc}^{\mu_2}\,j^{\mu_3}}&=\Pi^{\rho_1\sigma_1}_{\mu_1\nu_1}\,\pi^{\rho_2}_{\mu_2}\,\pi^{\rho_3}_{\mu_3}\,\left[\sdfrac{2(d-2)}{p_2^2}\delta^{\kappa\mu_2}\,p_{2\alpha_2}\,{\braket{T^{\alpha_1\nu_1}J^{\alpha_2}J^{\mu_3}}}\right]\notag\\
\Pi^{\rho_1\sigma_1}_{\mu_1\nu_1}(p_1)\pi^{\rho_2}_{\mu_2}(p_2)\pi^{\rho_3}_{\mu_3}(p_3)\ \,K^\k
\braket{t^{\mu_1\nu_1}\,j^{\mu_2}\,j_{loc}^{\mu_3}}&=\Pi^{\rho_1\sigma_1}_{\mu_1\nu_1}\,\pi^{\rho_2}_{\mu_2}\,\pi^{\rho_3}_{\mu_3}\,\left[\sdfrac{2(d-2)}{p_3^2}\,\delta^{\kappa\mu_3}p_{3\alpha_3}\,{\braket{T^{\alpha_1\nu_1}J^{\mu_2}J^{\alpha_3}}}\right]
\label{sec}
\end{align}
and all the terms with at least two insertion of local terms are zero. T
We have verified, as expected, that the equations above remain invariant if we choose as independent momenta $p_2$ and $p_3$ while acting on $p_1$ indirectly 
by the derivative chain rule. More details on this analysis will be given in a section below. In this way we may rewrite \eqref{SCWTJJ} in the form

\begin{align}
0&=\Pi^{\rho_1\sigma_1}_{\mu_1\nu_1}(p_1)\pi^{\rho_2}_{\mu_2}(p_2)\pi^{\rho_3}_{\mu_3}(p_3)\  \bigg(\,K^\k\,{\braket{T^{\mu_1\nu_1}(p_1)\,J^{\m_2}(p_2)\,J^{\mu_3}( p_3)}}\bigg)\notag\\
&=\Pi^{\rho_1\sigma_1}_{\mu_1\nu_1}(p_1)\pi^{\rho_2}_{\mu_2}(p_2)\pi^{\rho_3}_{\mu_3}(p_3)\ \bigg\{\,K^\kappa\,{\braket{t^{\mu_1\nu_1}(p_1)\,j^{\mu_2}(p_2)\,j^{\mu_3}(p_3)}}+\sdfrac{4d}{p_1^2}\,\delta^{\kappa\mu_1}\,p_{1\alpha_1}\,{\braket{T^{\alpha_1\nu_1}(p_1)J^{\mu_2}(p_2)J^{\mu_3}(p_3)}}\notag\\
&\hspace{2cm}+\sdfrac{2(d-2)}{p_2^2}\,\delta^{\kappa\mu_2}p_{2\alpha_2}\,\braket{{T^{\alpha_1\nu_1}J^{\alpha_2}J^{\mu_3}}}+\sdfrac{2(d-2)}{p_3^2}\,\delta^{\kappa\mu_3}p_{3\alpha_3}\,{\braket{T^{\alpha_1\nu_1}J^{\mu_2}J^{\alpha_3}}}\bigg\}.\label{StrucGenSWIS}
\end{align}
The equation above is an independent derivation of the corresponding BMS result, which is not offered in \cite{2014JHEP...03..111B}. Notice that our derivation, which details the various contributions coming from the local terms in the $TJJ$, has been derived using heavily the Lorentz Ward identities.

The last three terms may be re-expressed in terms of 2-point functions via the transverse Ward identities. After other rather lengthy computations, we find that the first term in the previous expression, corresponding to the transverse traceless contributions, can be written in the form
\begin{align}
&\Pi^{\rho_1\sigma_1}_{\mu_1\nu_1}(p_1)\pi^{\rho_2}_{\mu_2}(p_2)\pi^{\rho_3}_{\mu_3}(p_3)\ \bigg[\,K^\kappa\,\braket{{t^{\mu_1\nu_1}(p_1)\,j^{\mu_2}(p_2)\,j^{\mu_3}(p_3)}}\bigg]\notag\\
&=\Pi^{\rho_1\sigma_1}_{\mu_1\nu_1}(p_1)\pi^{\rho_2}_{\mu_2}(p_2)\pi^{\rho_3}_{\mu_3}(p_3)\times\notag\\
&\times\bigg[p_1^\kappa\left(C_{11}\,p_1^{\mu_3}p_2^{\mu_1}p_2^{\nu_1}p_3^{\mu_2}+C_{12}\,\delta^{\mu_2\mu_3}p_2^{\mu_1}p_2^{\nu_1}+C_{13}\delta^{\mu_1\mu_2}p_2^{\nu_1}p_1^{\mu_3}+C_{14}\delta^{\mu_1\mu_3}p_2^{\nu_1}p_3^{\mu_2}+C_{15}\delta^{\mu_1\mu_2}\delta^{\nu_1\mu_3}\right)\notag\\
&\quad+p_2^\kappa\left(C_{21}\,p_1^{\mu_3}p_2^{\mu_1}p_2^{\nu_1}p_3^{\mu_2}+C_{22}\,\delta^{\mu_2\mu_3}p_2^{\mu_1}p_2^{\nu_1}+C_{23}\delta^{\mu_1\mu_2}p_2^{\nu_1}p_1^{\mu_3}+C_{24}\delta^{\mu_1\mu_3}p_2^{\nu_1}p_3^{\mu_2}+C_{25}\delta^{\mu_1\mu_2}\delta^{\nu_1\mu_3}\right)\notag\\	
&\quad+\delta^{\mu_1\k}\left(C_{31}\,p_1^{\mu_3}p_2^{\nu_1}p_3^{\mu_2}+C_{32}\,\delta^{\mu_2\mu_3}p_2^{\nu_1}+C_{33}\,\delta^{\mu_2\nu_1}p_1^{\mu_3}+C_{34}\,\delta^{\mu_3\nu_1}p_3^{\mu_2}\right)\notag\\
&\qquad+\delta^{\mu_2\k}\left(C_{41}\,p_1^{\mu_3}p_2^{\mu_1}p_2^{\nu_1}+C_{42}\,\delta^{\mu_1\mu_3}p_2^{\nu_1}\right)+\delta^{\mu_3\k}\left(C_{51}\,p_3^{\mu_2}p_2^{\nu_1}p_3^{\mu_2}+C_{52}\,\delta^{\mu_1\mu_2}p_2^{\nu_1}\right)\bigg]\label{StrucSWIS}
\end{align}
where now $C_{ij}$ are differential equations involving the form factors $A_1,\ A_2,\ A_3,\ A_4$ of the representation of the $\langle{tjj}\rangle$ in \eqref{DecompTJJ}. For any 3-point function, the resulting equations can be divided into two groups, the \emph{primary} and the \emph{secondary} conformal Ward identities. The primary are second-order differential equations and appear as the coefficients of transverse or transverse-traceless tensor containing $p_1^\kappa$ and $p_2^\kappa$, where $\kappa$ is the special index related to the conformal operator ${K}^\kappa$. The remaining equations, following from all other transverse or transverse-traceless terms, are then secondary conformal Ward identities and are first-order differential equations. 

\subsection{Primary CWI's}
\label{primsection}
From \eqref{StrucGenSWIS} and \eqref{StrucSWIS} one finds that the primary CWI's are equivalent to the vanishing of the coefficients $C_{1j}$ and $C_{2j}$ for $j=1,\dots, 5$.  The CWI's can be rewritten in terms of the operators defined in Eq.  (\ref{kij}) as
\begin{equation}
\begin{split}
0&=C_{11}=K_{13}A_1\\
0&=C_{12}=K_{13}A_2+2A_1\\
0&=C_{13}=K_{13}A_3-4A_1\\
0&=C_{14}=K_{13}A_3(p_2\leftrightarrow p_3)\\
0&=C_{15}=K_{13}A_4-2A_3(p_2\leftrightarrow p_3)
\end{split}
\hspace{1.5cm}
\begin{split}
0&=C_{21}=K_{23}A_1\\
0&=C_{22}=K_{23}A_2\\
0&=C_{23}=K_{23}A_3-4A_1\\
0&=C_{24}=K_{23}A_3(p_2\leftrightarrow p_3)+4A_1\\
0&=C_{25}=K_{23}A_4+2A_3-2A_3(p_2\leftrightarrow p_3)
\end{split}\label{Primary}
\end{equation}

\subsection{Secondary CWI's}
\label{secsection}
The secondary conformal Ward identities are first-order partial differential equations and in principle involve the semi-local information contained in $j_{loc}^\m$ and $t^{\m\n}_{loc}$. In order to write them compactly, one defines the two differential operators
\begin{align}
L_N&= p_1(p_1^2 + p_2^2 - p_3^2) \frac{\partial}{\partial p_1} + 2 p_1^2\, p_2 \frac{\partial}{\partial p_2} + \big[ (2d - \Delta_1 - 2\Delta_2 +N)p_1^2 + (2\Delta_1-d)(p_3^2-p_2^2)  \big] \label{Ldef} \\
R &= p_1 \frac{\partial}{\partial p_1} - (2\Delta_1-d) \label{Rdef}\,. 
\end{align}
The reason for introducing such operators comes from (\ref{StrucSWIS}), once the action of 
$K^\kappa$ is made explicit. The separation between the two sets of constraints comes from the same equation, and in particular from the terms trilinear in the momenta within the square bracket. One needs also the symmetric versions of such operators
\begin{align}
&L'_N=L_N,\quad\text{with}\ p_1\leftrightarrow p_2\ \text{and}\ \D_1\leftrightarrow\D_2,\\
&R'=R,\qquad\text{with}\ p_1\mapsto p_2\ \text{and}\ \D_1\mapsto\D_2.
\end{align}
These operators depend on the conformal dimensions of the operators involved in the 3-point function under consideration, and additionally on a single parameter $N$ determined by the Ward identity in question. In the $\braket{TJJ}$ case one finds considering the structure of Eqs. \eqref{StrucGenSWIS} and \eqref{StrucSWIS} 
\begin{equation}
\begin{split}
C_{31}&=-\sdfrac{2}{p_1^2}\left[L_4 A_1+R A_3-R A_3(p_2\leftrightarrow p_3)\right]\\
C_{32}&=-\sdfrac{2}{p_1^2}\left[L_2\,A_2-p_1^2(A_3-A_3(p_2\leftrightarrow p_3))\right]\\
C_{33}&=-\sdfrac{1}{p_1^2}\left[L_4\,A_3-2R\,A_4\right]\\
C_{34}&=-\sdfrac{1}{p_1^2}\left[L_4\,A_3(p_2\leftrightarrow p_3)+2R\,A_4-4p_1^2A_3(p_2\leftrightarrow p_3)\right]\\
\end{split}
\end{equation}
\begin{equation}
\begin{split}
C_{41}&=\sdfrac{1}{p_2^2}\left[L'_3\,A_1-2R'A_2+2R'A_3\right]\\
C_{42}&=\sdfrac{1}{p_2^2}\left[L'_1\,A_3(p_2\leftrightarrow p_3)+p_2^2(4A_2-2A_3)+2R'A_4\right]\\
C_{51}&=\sdfrac{1}{p_3}\left[(L_4-L'_3)A_1-2(2d+R+R')A_2+2(2d+R+R')A_3(p_2\leftrightarrow p_3)\right]\\
C_{52}&=\sdfrac{1}{p_3^2}\left[(L_2-L'_1)A_3-4p_3^2A_2+2p_3^2A_3(p_2\leftrightarrow p_3)+2(2d-2+R+R')A_4\right]
\end{split}
\end{equation}
From (\ref{StrucGenSWIS}) and (\ref{StrucSWIS}) using (\ref{2point}) the secondary CWI's take the explicit form 
\begin{equation}
\begin{split}
&C_{31}=C_{41}=C_{42}=C_{51}=C_{52}=0, \qquad C_{32}=\frac{16\,d\, c_{123}\,\Gamma_J}{p_1^2}\left[ \frac{1}{(p_3^2)^{\sigma_0}} - \frac{1}{(p_2^2)^{\sigma_0}} \right],\\[2ex]
& \hspace{1cm}C_{33}=\frac{16\,d\,c_{123}\,\Gamma_J}{p_1^2 (p_3^2)^{\sigma_0}}, \hspace{2.5cm} C_{34}=-\frac{16\,d\,c_{123}\,\Gamma_J }{p_1^2\,(p_2^2)^{\sigma_0}},
\end{split}
\end{equation}
where in our $\sigma_0=d/2-\Delta_2$. 
Expressed in this form all the scalar equations for the $A_i$ are not apparently symmetric in the exchange of $p_2$ and $p_3$, and it may not be immediately evident that they can be recast in such a way that the symmetry is respected. 
\section{Symmetric treatment of the \texorpdfstring{$J$}{} currents }
\label{p1section}
Let's now consider $p_1$ as dependent momentum, showing the equivalence of the CWI's with this second choice. As we have just mentioned above, this choice is the preferred one in the search for the solutions of the $TJJ$. In this case, the action of the spin (Lorentz) part of the transformation will leave the stress energy tensor 
as a singlet, acting implicitly on $p_1$ via the chain rule. As we are going to show, the resulting equations will be linear combinations of the original part. This extends the analysis presented by BMS. 

The structure of the decomposition in \eqref{DecompTJJ} of the $\braket{TJJ}$ correlator is still valid but now the explicit form of the special conformal operator ${K}^\k$ has to be modified as
\begin{align}
{K}^\k{\braket{T^{\m_1\n_1}\,J^{\m_2}\,J^{\m_3}}}&=\sum_{j=2}^{3}\left[2(\Delta_j-d)\sdfrac{\partial}{\partial p_j^\k}-2p_j^\a\sdfrac{\partial}{\partial p_j^\a}\sdfrac{\partial}{\partial p_j^\k}+(p_j)_\k\sdfrac{\partial}{\partial p_j^\a}\sdfrac{\partial}{\partial p_{j\a}}\right]{\braket{T^{\mu_1\nu_1}(\bar p_1)\,J^{\m_2}(p_2)\,J^{\mu_3}(p_3)}}\notag\\
&\qquad+2\left(\d^{\k\mu_2}\sdfrac{\partial}{\partial p_2^{\a_2}}-\delta^{\k}_{\alpha_2}\delta^{\l\mu_2}\sdfrac{\partial}{\partial p_2^\l}\right){\braket{T^{\mu_1\nu_1}(\bar p_1)\,J^{\alpha_2}(p_2)\,J^{\mu_3}(p_3)}}\notag\\
&\qquad+2\left(\d^{\k\mu_3}\sdfrac{\partial}{\partial p_3^{\a_3}}-\delta^{\k}_{\alpha_3}\delta^{\l\mu_3}\sdfrac{\partial}{\partial p_3^\l}\right){\braket{T^{\mu_1\nu_1}(\bar p_1)\,J^{\mu_2}(p_2)\,J^{\a_3}( p_3)}}
\end{align}
where $\bar p_1^\m=-p_2^\m-p_3^\m$. Considering the SCWI's for the 3-point function we can write
\[{K}^\k(p_2,p_3){\braket{T^{\m_1\n_1}(\bar p_1)\,J^{\m_2}(p_2)\,J^{\m_3}(p_3)}}=0,\]
in which we have stress the $p_2$ and $p_3$ dependence of the special conformal operator. Then one has to take the decomposition of the 3-point function as in \eqref{DecompTJJ} and using the relations \eqref{sec}, that are still valid in this case, one derives \eqref{StrucGenSWIS}, in which now the $K$ operator is defined in terms of $p_2$ and $p_3$ only. As in the previous case, one finds CWI's which are similar to those given in \eqref{StrucSWIS} 
\begin{align}
&\Pi^{\rho_1\sigma_1}_{\mu_1\nu_1}(p_1)\pi^{\rho_2}_{\mu_2}(p_2)\pi^{\rho_3}_{\mu_3}(p_3)\ \bigg[{K}^\kappa\,{\braket{t^{\mu_1\nu_1}(p_1)\,j^{\mu_2}(p_2)\,j^{\mu_3}(p_3)}}\bigg]\notag\\
&=\Pi^{\rho_1\sigma_1}_{\mu_1\nu_1}(p_1)\pi^{\rho_2}_{\mu_2}(p_2)\pi^{\rho_3}_{\mu_3}(p_3)\times\notag\\
&\times\bigg[p_2^\kappa\left(\tilde{C}_{11}\,p_1^{\mu_3}p_2^{\mu_1}p_2^{\nu_1}p_3^{\mu_2}+\tilde{C}_{12}\,\delta^{\mu_2\mu_3}p_2^{\mu_1}p_2^{\nu_1}+\tilde{C}_{13}\delta^{\mu_1\mu_2}p_2^{\nu_1}p_1^{\mu_3}+\tilde{C}_{14}\delta^{\mu_1\mu_3}p_2^{\nu_1}p_3^{\mu_2}+\tilde{C}_{15}\delta^{\mu_1\mu_2}\delta^{\nu_1\mu_3}\right)\notag\\
&\quad+p_3^\kappa\left(\tilde{C}_{21}\,p_1^{\mu_3}p_2^{\mu_1}p_2^{\nu_1}p_3^{\mu_2}+\tilde{C}_{22}\,\delta^{\mu_2\mu_3}p_2^{\mu_1}p_2^{\nu_1}+\tilde{C}_{23}\delta^{\mu_1\mu_2}p_2^{\nu_1}p_1^{\mu_3}+\tilde{C}_{24}\delta^{\mu_1\mu_3}p_2^{\nu_1}p_3^{\mu_2}+\tilde{C}_{25}\delta^{\mu_1\mu_2}\delta^{\nu_1\mu_3}\right)\notag\\	
&\quad+\delta^{\mu_1\k}\left(\tilde{C}_{31}\,p_1^{\mu_3}p_2^{\nu_1}p_3^{\mu_2}+\tilde{C}_{32}\,\delta^{\mu_2\mu_3}p_2^{\nu_1}+\tilde{C}_{33}\,\delta^{\mu_2\nu_1}p_1^{\mu_3}+\tilde{C}_{34}\,\delta^{\mu_3\nu_1}p_3^{\mu_2}\right)\notag\\
&\qquad+\delta^{\mu_2\k}\left(\tilde{C}_{41}\,p_1^{\mu_3}p_2^{\mu_1}p_2^{\nu_1}+\tilde{C}_{42}\,\delta^{\mu_1\mu_3}p_2^{\nu_1}\right)+\delta^{\mu_3\k}\left(\tilde{C}_{51}\,p_3^{\mu_2}p_2^{\nu_1}p_3^{\mu_2}+\tilde{C}_{52}\,\delta^{\mu_1\mu_2}p_2^{\nu_1}\right)\bigg].
\end{align}
In this case we obtain the primary WI's by imposing the vanishing of the coefficients $\tilde{C}_{ij}$, for $i=1,2$ and $j=1,\dots, 5$. In this way we get
\begin{equation}
\label{listeq}
\begin{split}
0&=\tilde{C}_{11}=K_{21}A_1\\
0&=\tilde{C}_{12}=K_{21}A_2-2A_1\\
0&=\tilde{C}_{13}=K_{21}A_3\\
0&=\tilde{C}_{14}=K_{21}A_3(p_2\leftrightarrow p_3)+4A_1\\
0&=\tilde{C}_{15}=K_{21}A_4+2A_3
\end{split}
\hspace{1.5cm}
\begin{split}
0&=\tilde{C}_{21}=K_{31}A_1\\
0&=\tilde{C}_{22}=K_{31}A_2-2A_1\\
0&=\tilde{C}_{23}=K_{31}A_3 +4A_1\\
0&=\tilde{C}_{24}=K_{31}A_3(p_2\leftrightarrow p_3)\\
0&=\tilde{C}_{25}=K_{31}A_4+2A_3(p_2\leftrightarrow p_3)
\end{split}
\end{equation}
and it is simple to verify that these equations are equivalent to those given in \eqref{Primary}. 
In the case of the secondary WI's we have to consider some further properties of the form factors. For instance the coefficient $\tilde{C}_{31}$ has the explicit form
\begin{align}
\tilde{C}_{31}&=\sdfrac{2}{p_1^2}\bigg[p_2(p_1^2-p_2^2+p_3^2)\sdfrac{\partial}{\partial p_2}A_1-p_1^2p_3\sdfrac{\partial}{\partial p_3}A_1-p_2^2p_3\sdfrac{\partial}{\partial p_3}A_1+p_3^3\sdfrac{\partial}{\partial p_3}A_1-p_2\sdfrac{\partial}{\partial p_2}A_3-p_3\sdfrac{\partial}{\partial p_3}A_3\notag\\
&\hspace{1cm}+p_2\sdfrac{\partial}{\partial p_2}A_3(p_2\leftrightarrow p_3)+p_3\sdfrac{\partial}{\partial p_3}A_3(p_2\leftrightarrow p_3)-6 (p_2^2-p_3^2)A_1-4(A_3-A_3(p_2\leftrightarrow p_3))
\bigg]\label{C31new}
\end{align}
in which it is possible to substitute the derivative with respect to $p_3$ in terms of derivatives with respect to $p_2$ and $p_1$ using the dilatation Ward identities
\begin{equation}
\sdfrac{\partial}{\partial p_3}\,A_n=\sdfrac{1}{p_3}\bigg[(d-2-N_n)A_n-\sum_{j=1}^2p_j\sdfrac{\partial}{\partial p_j}A_n\bigg]\label{rel}.
\end{equation}
Using the identity given above in \eqref{C31new}, one derives the relation
\begin{equation}
\tilde{C}_{31}=\sdfrac{2}{p_1^2}\bigg[L_4\,A_1+R\,A_3-R\,A_3(p_2\leftrightarrow p_3)\bigg]
\end{equation}
with the identification of the differential operators $L$ and $R$ defined in \eqref{Ldef} and \eqref{Rdef}. In this way it is possible to show that all the coefficients related to the secondary Ward identities are the same of those obtained with $p_3$ as the dependent momentum. This argument proves that in spite of the choice of the dependent momentum, the scalar equations for the form factors related to the CWI's remain identical. 

\section{The Fuchsian approach to the solutions of the primary CWI's and universality}
\label{fuchs}
In this section we are going to investigate the Fuchsian structure of the equations. The goal of the section is 
to present a new method of solution which differs from the one based on 3K integrals presented in \cite{2014JHEP...03..111B}.
We should mention that the number of integration constants introduced by the primary CWI's, using this method, may not necessarily coincide with those presented in \cite{2014JHEP...03..111B}, and the constraints imposed by the secondary CWIs, that we will not discuss, will obviously be different. \\
The goal of this section is twofold. We want to show first of all that the Fuchsian exponents (defined as $(a_i,b_j)$ below), are 
universal and characterize the entire system of equations. In the scalar case, as well as for all the 3-point functions that we have investigated, we have verified that always the same set of exponents $(a_i,b_j)$ are generated. \\
The second important feature is that the method allows to characterize particular solutions of 4 and higher point functions in some restricted kinematics, allowing a significant generalization of the analysis presented here, with new special functions appearing in the solutions. Details of this study will be presented in a separate work. \\ 
 Being the CWI's a system of equations, we will first solve for each of the form factors, starting from the equations for $A_1$, which are homogeneous, and then proceed towards the inhomogenous ones, from $A_2$ to  $A_4$. For each form factor we identify the general solution and a particular solution, which are added together. Then we impose the symmetry constraints on the two vector lines, due to Bose symmetry. For example, the solution for $A_2$ will constraint the constants appearing in the general solution of $A_1$, and so on for $A_3$ and $A_4$.
The independent constants of integration are identified only at the end, once all the constraints from $A_1$ to $A_4$ are put together.  We have included a small section where we summarize the final expressions of the form factors by this method.

\subsection{Scalar 3-point functions}
To illustrate our approach we start reviewing the case of the scalar correlator 
$\Phi(p_1,p_2, p_3)$, which is simpler, defined by the two homogeneous conformal equations
\begin{equation}
K_{31}\Phi=0  \qquad K_{21}\Phi=0
\end{equation}
combined with the scaling equation 
\begin{equation}
\label{scale}
\sum_{i=1}^3 p_i\frac{\partial}{\partial p_i} \Phi=(\Delta-2 d) \Phi. 
\end{equation}
Following the approach presented in \cite{Coriano:2013jba}, the ansatz for the solution can be taken of the form 
\begin{equation}
\label{ans}
\Phi(p_1,p_2,p_3)=p_1^{\Delta - 2 d} x^{a}y^{b} F(x,y)
\end{equation}

with $x=\frac{p_2^2}{p_1^2}$ and $y=\frac{p_3^2}{p_1^2}$. Here we are taking $p_1$ as "pivot" in the expansion, but we could equivalently choose any of the 3  momentum invariants. $\Phi$ is required to be homogenous of degree $\Delta-2 d$ under a scale transformation, according to (\ref{scale}), and in (\ref{ans}) this is taken into account by the factor $p_1^{\Delta - 2 d}$.
The use of the scale invariant variables $x$ and $y$ takes to the hypergeometric form of the solution. One obtains 
\begin{align}
K_{21}\phi &= 4 p_1^{\Delta -2d -2} x^a y^b
\left(  x(1-x)\frac{\partial }{\partial x \partial x}  + (A x + \gamma)\frac{\partial }{\partial x} -
2 x y \frac{\partial^2 }{\partial x \partial y}- y^2\frac{\partial^2 }{\partial y \partial y} + 
D y\frac{\partial }{\partial y} + (E +\frac{G}{x})\right) \notag\\
& \hspace{3cm}\times F(x,y)=0
\label{red}
\end{align}
with
\begin{align}
&A=D=\Delta_2 +\Delta_3 - 1 -2 a -2 b -\frac{3 d}{2} \qquad \gamma(a)=2 a +\frac{d}{2} -\Delta_2 + 1
\notag\\
& G=\frac{a}{2}(d +2 a - 2 \Delta_2)
\notag\\
&E=-\frac{1}{4}(2 a + 2 b +2 d -\Delta_1 -\Delta_2 -\Delta_3)(2 a +2 b + d -\Delta_3 -\Delta_2 +\Delta_1).
\end{align}
Similar constraints are obtained from the equation $K_{31}\Phi=0$, with the obvious exchanges $(a,b,x,y)\to (b,a,y,x)$
\begin{align}
K_{31}\phi &= 4 p_1^{\Delta -2 d -2} x^a y^b
\left(  y(1-y)\frac{\partial }{\partial y \partial y}  + (A' y + \gamma')\frac{\partial }{\partial y} -
2 x y \frac{\partial^2 }{\partial x \partial y}- x^2\frac{\partial^2 }{\partial x \partial x} + 
D' x\frac{\partial }{\partial x} + (E' +\frac{G'}{y})\right) \notag\\
& \hspace{3cm}\times F(x,y)=0
\label{red}
\end{align}
with
\begin{align}
&A'=D'= A   \qquad \qquad \gamma'(b)=2 b +\frac{d}{2} -\Delta_3 + 1
\notag\\
& G'=\frac{b}{2}(d +2 b - 2 \Delta_3)
\notag\\
&E'= E.
\end{align}
Notice that in (\ref{red}) we need to set $G/x=0$ in order to perform the reduction to the hypergeometric form of the equations, which implies that
\begin{equation}
\label{cond1}
a=0\equiv a_0 \qquad \textrm{or} \qquad a=\Delta_2 -\frac{d}{2}\equiv a_1.
\end{equation}
From the equation $K_{31}\Phi=0$ we obtain a similar condition for $b$ by setting $G'/y=0$, thereby fixing the two remaining indices
\begin{equation}
\label{cond2}
b=0\equiv b_0 \qquad \textrm{or} \qquad b=\Delta_3 -\frac{d}{2}\equiv b_1.
\end{equation}
The four independent solutions of the CWI's will all be characterised by the same 4 pairs of indices $(a_i,b_j)$ $(i,j=1,2)$.
Setting 
\begin{equation}
\alpha(a,b)= a + b + \frac{d}{2} -\frac{1}{2}(\Delta_2 +\Delta_3 -\Delta_1) \qquad \beta (a,b)=a +  b + d -\frac{1}{2}(\Delta_1 +\Delta_2 +\Delta_3) \qquad 
\label{alphas}
\end{equation}
then
\begin{equation}
E=E'=-\alpha(a,b)\beta(a,b) \qquad A=D=A'=D'=-\left(\alpha(a,b) +\beta(a,b) +1\right).
\end{equation}
the solutions take the form 
\begin{align}
\label{F4def}
F_4(\alpha(a,b), \beta(a,b); \gamma(a), \gamma'(b); x, y) = \sum_{i = 0}^{\infty}\sum_{j = 0}^{\infty} \frac{(\alpha(a,b), {i+j}) \, 
	(\beta(a,b),{i+j})}{(\gamma(a),i) \, (\gamma'(b),j)} \frac{x^i}{i!} \frac{y^j}{j!} 
\end{align}
where $(\alpha,i)=\Gamma(\alpha + i)/ \Gamma(\alpha)$ is the Pochammer symbol. We will refer to $\alpha\ldots \gamma'$ as to the first,$\ldots$, fourth parameters of $F_4$.\\ 
The 4 independent solutions are then all of the form $x^a y^b F_4$, where the 
hypergeometric functions will take some specific values for its parameters, with
$a$ and $b$ fixed by (\ref{cond1}) and (\ref{cond2}). Specifically we have
\begin{equation}
\Phi(p_1,p_2,p_3)=p_1^{\Delta-2 d} \sum_{a,b} c(a,b,\vec{\Delta})\,x^a y^b \,F_4(\alpha(a,b), \beta(a,b); \gamma(a), \gamma'(b); x, y) 
\label{compact}
\end{equation}
where the sum runs over the four values $a_i, b_i$ $i=0,1$ with arbitrary constants $c(a,b,\vec{\Delta})$, with $\vec{\Delta}=(\Delta_1,\Delta_2,\Delta_3)$. Notice that \eqref{compact} is a very compact way to write down the solution. However, once this type of solutions of a homogeneous hypergeometric system are inserted into an inhomogenous system of equations, the sum over $a$ and $b$ 
needs to be made explicit. For this reason it is convenient to define 
\begin{align}
&\alpha_0\equiv \alpha(a_0,b_0)=\frac{d}{2}-\frac{\Delta_2 + \Delta_3 -\Delta_1}{2},\, && \beta_0\equiv \beta(b_0)=d-\frac{\Delta_1 + \Delta_2 +\Delta_3}{2},  \nn
&\gamma_0 \equiv \gamma(a_0) =\frac{d}{2} +1 -\Delta_2,\, &&\gamma'_0\equiv \gamma(b_0) =\frac{d}{2} +1 -\Delta_3.
\end{align}
to be the 4 basic (fixed) hypergeometric parameters, and define all the remaining ones by shifts respect to these. The 4 independent solutions can be re-expressed in terms of the parameters above as 

\bea
\label{F4def}
S_1(\alpha_0, \beta_0; \gamma_0, \gamma'_0; x, y)\equiv F_4(\alpha_0, \beta_0; \gamma_0, \gamma'_0; x, y) = \sum_{i = 0}^{\infty}\sum_{j = 0}^{\infty} \frac{(\alpha_0,i+j) \, 
(\beta_0,i+j)}{(\gamma_0,i )\, (\gamma'_0,j)} \frac{x^i}{i!} \frac{y^j}{j!} 
\eea
and
\bea
\label{solutions}
S_2(\alpha_0, \beta_0; \gamma_0, \gamma'_0; x, y) &=& x^{1-\gamma_0} \, F_4(\alpha_0-\gamma_0+1, \beta_0-\gamma_0+1; 2-\gamma_0, \gamma'_0; x,y) \,, \nn \\
S_3(\alpha_0, \beta_0; \gamma_0, \gamma'_0; x, y) &=& y^{1-\gamma'_0} \, F_4(\alpha_0-\gamma'_0+1,\beta_0-\gamma'_0+1;\gamma_0,2-\gamma'_0 ; x,y) \,, \nn \\
S_4(\alpha_0, \beta_0; \gamma_0, \gamma'_0; x, y) &=& x^{1-\gamma_0} \, y^{1-\gamma'_0} \, 
F_4(\alpha_0-\gamma_0-\gamma'_0+2,\beta_0-\gamma_0-\gamma'_0+2;2-\gamma_0,2-\gamma'_0 ; x,y) \, . \nn
\eea
Notice that in the scalar case, one is allowed to impose the complete symmetry of the correlator under the exchange of the 3 external momenta and scaling dimensions, as discussed in \cite{Coriano:2013jba}. This reduces the four
constants to just one. We are going first to extend this analysis to the case of the $A_1-A_4$ form factors of the $TJJ$. 

\subsection{Form factors: the solution for $A_1$}

The solutions for the form factors $A_1-A_4$ can be derived using a similar, but modified approach, being the equations also inhomogenous. As previously we take as a pivot $p_1^2$, and assume a symmetry under the $(P_{23})$ exchange of $(p_2,\Delta_2)$ with $(p_3,\Delta_3)$ in the correlator. In the case of two photons 
$\Delta_2=\Delta_3=d-1$.

We start from $A_1$ by solving the two equations from (\ref{Primary}) 
\begin{equation}
K_{21}A_1=0   \qquad K_{31}A_1=0.
\end{equation}
In this case we introduce the ansatz 
\begin{equation}
A_1=p_1^{\Delta-2 d - 4}x^a y^b  F(x,y)
\end{equation}
and derive two hypergeometric equations, which are characterised by the same indices 
$(a_i, b_j)$ as before in (\ref{cond1}) and (\ref{cond2}), but new values of the 4 defining parameters. 
We obtain 
\begin{equation}
\label{A1}
A_1(p_1,p_2,p_3)=p_1^{\Delta-2 d - 4}\sum_{a,b} c^{(1)}(a,b,\vec{\Delta})\,x^a y^b \,F_4(\alpha(a,b) +2, \beta(a,b)+2; \gamma(a), \gamma'(b); x, y) 
\end{equation}
with the expression of $\alpha(a,b),\beta(a,b), \gamma(a), \gamma'(b)$ as given before, with the obvious switching of the 
$\Delta_i$ in order to comply with the new choice of the pivot ($p_1^2$)
\begin{align}
\label{cons1}
\alpha(a,b)&= a + b + \frac{d}{2} -\frac{1}{2}(\Delta_2 + \Delta_3 -\Delta_1) \notag\\
\beta(a,b)&= a + b + d -\frac{1}{2}(\Delta_1 + \Delta_2 +\Delta_3) 
\end{align}
which are $P_{23}$ symmetric and 
\begin{align} 
\label{cons2}
\gamma(a)& =2 a +\frac{d}{2} -\Delta_2 + 1 \notag\\
\gamma'(b)&=2 b +\frac{d}{2} -\Delta_3 + 1 
\end{align}
with $P_{23}\gamma(a)=\gamma'(b)$.
If we require that $\Delta_2=\Delta_3$, as in the $TJJ$ case, the symmetry constraints are easily implemented. 
Given that the 4 indices, if we choose $p_1$ as a pivot, are given by 
\begin{equation}
a_0=0, b_0=0, a_1=\Delta_2- \frac{d}{2}, b_1=\Delta_3-\frac{d}{2} 
\end{equation}
clearly in this case $a=b$ and $\gamma(a)=\gamma(b)$. 
$F_4$ has the symmetry 
\begin{align}
F_4(\a,\b; \g, \g' ; x,y)=F_4(\a,\b; \g', \g ; y,x),
\end{align}
and this reflects in the Bose symmetry of $A_1$ if we impose the constraint
\be
c^{(1)}(a_1,b_0)=c^{(1)}(a_0,b_1).
\ee

\subsection{The solution for $A_2$}

The equations for $A_2$ are inhomogeneous. In this case the solution can be identified using some properties of the hypergeometric differential operators $K_i$, appropriately splitted. We recall that in this case they are 
\begin{align}
K_{21}A_2 &= 2 A_1\label{inhom}\\
K_{31}A_2& = 2A_1.\label{inhom1}
\end{align}
We take an ansatz of the form 
\begin{equation}
A_2(p_1,p_2,p_3)=p_1^{\Delta-2 d - 2}F(x,y)
\end{equation}
which provides the correct scaling dimensions for $A_2$. 
Observe that the action of $K_{21}$ and $K_{3}$ on $A_2$ can be rearranged as follows
\begin{align}
K_{21} A_2&=4 x^a y^b p_1^{\Delta-2 d -4}\bigg( \bar{K}_{21}F(x,y) +\frac{\partial}{\partial x} F(x,y)\bigg)\\[1.5ex]
K_{31} A_2&=4 x^a y^b p_1^{\Delta-2 d -4}\bigg( \bar{K}_{31}F(x,y) +\frac{\partial}{\partial y} F(x,y)\bigg)
\end{align}
where
\begin{align}
\label{k1bar}
\bar{K}_{21}F(x,y)&=\bigg\{x(1-x) \frac{\partial^2}{\partial x^2} - y^2 \frac{\partial^2}{\partial y^2} - 2 \, x \, y \frac{\partial^2}{\partial x \partial y} +\big[  (\gamma(a)-1) - (\alpha(a,b) + \beta(a,b) + 3) x \big] \frac{\partial}{\partial x}\notag\\
&\hspace{3cm}
+ \frac{a (a-a_1)}{x} - (\alpha(a,b) + \beta(a,b) + 3) y \frac{\partial}{\partial y}  - (\alpha +1)(\beta +1) \bigg\} F(x,y),
\end{align}
and 
\begin{align}
\label{k2bar}
\bar{K}_{31} A_2&=\bigg\{ y(1-y) \frac{\partial^2}{\partial y^2} - x^2 \frac{\partial^2}{\partial x^2} - 2 \, x \, y \frac{\partial^2}{\partial x \partial y} +  \big[ (\gamma'(b)-1)- (\alpha(a,b) + \beta(a,b) + 3) y \big]\frac{\partial}{\partial y}\notag\\
&\hspace{2cm} +  \frac{b(b- b_1)}{y} - (\alpha(a,b) + \beta(a,b) + 3) x \frac{\partial}{\partial x}  -(\alpha(a,b) +1)(\beta(a,b) +1)\bigg\} F(x,y).
\end{align}
At this point observe that the hypergeometric function solution of the equation
\begin{equation}
\label{ffirst1}
\bar{K}_{21}F(x,y)=0
\end{equation}
can be taken of the form
\begin{equation}
\label{rep1}
\Phi_1^{(2)}(x,y)=p_1^{\Delta-2 d - 2}\sum_{a,b} c^{(2)}_1(a,b,\vec{\Delta})\,x^a y^b \,F_4(\alpha(a,b) +1, \beta(a,b)+1; \gamma(a)-1, \gamma'(b) ; x, y) 
\end{equation}
with $c^{(2)}_1$ a constant and the parameters $a,b$ fixed at the ordinary values $(a_i,b_j)$ as in the previous cases (\ref{cond1}) and (\ref{cond2}), in order to get rid of the $1/x$ and $1/y$ poles in the coefficients of the differential operators. The sequence of parameters in (\ref{rep1}) will obviously solve the related equation 
\begin{equation}
\label{sec}
{K}_{31}\Phi_1^{(2)}(x,y)=0. 
\end{equation}
Eq. (\ref{ffirst1}) can be verified by observing that the sequence of  parameters 
$(\alpha(a,b)+1,\beta(a,b)+1 \gamma(a)-1)$ allows to define a solution of (\ref{k2bar}) set to zero, for an arbitrary $\gamma'(b)$, since this parameter does not play any role in the solution of the corresponding equation. 
The sequence $(\alpha(a,b)+1,\beta(a,b)+1, \gamma'(b))$, on the other hand, solves the homogeneous equations associated to $K_{31}$ (i.e. Eq. (\ref{sec})) for any value of the third parameter of $F_4$, which in this case takes the value $\gamma(a)-1$.
A similar result holds for the mirror solution
\begin{equation} 
\label{rep}
\Phi_2^{(2)}(x,y)= p_1^{\Delta-2 d - 2}\sum_{a,b} c_2^{(2)}(a,b,\vec{\Delta})\,x^a y^b \,F_4(\alpha(a,b) +1, \beta(a,b)+1; \gamma(a), \gamma'(b)-1 ; x, y) 
\end{equation}
which satisfies 
\begin{equation}
\label{first}
\bar{K}_{31}\Phi_2^{(2)}(x,y)=0 \qquad  {K}_{21}\Phi_2^{(2)}(x,y)=0. 
\end{equation}
As previously remarked, the values of the exponents $a$ and $b$ remain the same for any equation involving either a $K_{i,j}$ or a $\bar{K}_{i j}$, as can be explicitly verified. This implies that the fundamental solutions of the conformal equations are essentially the 4 functions of the type $S_1,\ldots S_4$, for appropriate values of their parameters.  \\
At this point, to show that $F_1$ and $F_2$ is a solution of  Eqs. (\ref{inhom}) we use the property 
\begin{equation}
\frac{\partial^{p+q} F_4(\alpha,\beta;\gamma_1,\gamma_2;x,y)}{\partial x^p\partial y^q} =\frac{(\alpha,p+q)(\beta,p+q)}{(\gamma_1,p)(\gamma_2,q)}
F_4(\alpha + p + q,\beta + p + q; \gamma_1 + p ; \gamma_2 + q;x,y)
\end{equation}
 which gives (for generic parameters $\alpha,\beta,\gamma_1,\gamma_2$)
\begin{align}
& \frac{\partial F_4(\alpha,\beta;\gamma_1,\gamma_2;x,y)}{\partial x} =\frac{\alpha \beta}{\gamma_1}F_4(\alpha+1,\beta+1,\gamma_1+1,\gamma_2,x,y) \notag\\
&  \frac{\partial F_4(\alpha,\beta;\gamma_1,\gamma_2;x,y)}{\partial y} =\frac{\alpha \beta}{\gamma_2}F_4(\alpha+1,\beta+1,\gamma_1,\gamma_2 +1,x,y).
\end{align} 
Obviously, such relations are valid whatever dependence the four parameters $\alpha,\beta,\gamma_1,\gamma_2$ may have on 
the Fuchsian exponents $(a_i,b_j)$. 
The actions of $K_{21}$ and $K_{31}$ on the the $\Phi_2^{(i)}$'s (i=1,2)  in (\ref{rep}) are then given by 
\begin{align}
&K_{21}\Phi_1^{(2)}(x,y) = 4p_1^{\Delta-2 d -4} \sum_{a,b} c^{(2)}_1(a,b,\vec{\Delta})\,x^a y^b \frac{\partial}{\partial x} \,F_4(\alpha(a.b) +1, \beta(a,b)+1; \gamma(a)-1, \gamma'(b) ; x, y)   \notag\\
&= 4 p_1^{\Delta-2 d -4} \sum_{a,b} c^{(2)}_1(a,b,\vec{\Delta})\,x^a y^b \frac{(\alpha(a,b)+1)(\beta(a,b)+1)}{(\gamma(a)-1)}F_4(\alpha(a,b) +2, \beta(a,b)+2; \gamma(a), \gamma'(b); x, y) \notag\\
&K_{31}\Phi_1^{(2)}(x,y)= 0\\[3ex]
&K_{31}\Phi_2^{(2)}(x,y) = 4 p_1^{\Delta-2 d -4} \sum_{a,b} c^{(2)}_2(a,b,\vec{\Delta})\, x^a y^b \frac{\partial}{\partial y} \,F_4(\alpha(a,b) +1, \beta(a,b)+1; \gamma(a), \gamma'(b)-1 ; x, y) \notag\\
&=4 p_1^{\Delta-2 d -4} \sum_{a,b} c^{(2)}_2(a,b,\vec{\Delta})\,  x^a y^b  \frac{(\alpha(a,b)+1)(\beta(a,b)+1)}{(\gamma'(b)-1)} F_4(\alpha(a,b) +2, \beta(a,b)+2; \gamma(a), \gamma'(b); x, y)
\notag\\
&K_{21} \Phi_2^{(2)}(x,y)=0,
\end{align}
where it is clear that the non-zero right-hand-side of both equations are proportional to the form factor $A_1$ given in (\ref{A1}). Once this particular solution is determined, Eq. \eqref{A1}, by comparison, gives the conditions on $c_1^{(2)}$ and $c_1^{(2)}$ as
\begin{align}
c_1^{(2)}(a,b,\vec{\Delta})&=\frac{\gamma(a)-1}{2(\alpha(a,b)+1)(\beta(a,b)+1)}\ c^{(1)}(a,b,\vec \Delta)\,,\\
c_2^{(2)}(a,b,\vec{\Delta})&=\frac{\gamma'(b)-1}{2(\alpha(a,b)+1)(\beta(a,b)+1)}\ c^{(1)}(a,b,\vec \Delta)\,.
\label{condc2}
\end{align}
Therefore, the general solution for $A_2$ in the $TJJ$ case (in which $\g(a)=\g'(b)$ ) is given by superposing the solution of the homogeneous form of \eqref{A1} and the particular one \eqref{rep1} and \eqref{rep}, by choosing the constants appropriately using \eqref{condc2}. Its explicit form is written as
\begin{align}
A_2&= p_1^{\Delta-2 d - 2}\sum_{a b} x^a y^b\Bigg[c^{(2)}(a,b,\vec{\Delta})\,F_4(\alpha(a,b)+1, \beta(a,b)+1; \gamma(a), \gamma'(b); x, y)\notag\\
&\hspace{1cm}+ \frac{(\gamma(a)-1)\,c^{(1)}(a,b,\vec \Delta)}{2(\alpha(a,b)+1)(\beta(a,b)+1)}\bigg(F_4(\alpha(a,b) +1, \beta(a,b)+1; \gamma(a)-1, \gamma'(b); x, y)\notag\\
&\hspace{7cm}+ F_4(\alpha(a,b) +1, \beta(a,b)+1; \gamma(a), \gamma'(b)-1; x, y)\bigg)\Bigg],
\end{align}
since $\g(a)=\g'(b)$. 
\subsection{The solution for $A_3$}
Using a similar strategy, the particular solution for the form factor $ A_3$ of the equations 
\begin{equation}
K_{21} A_3 =0 \qquad K_{3 1} A_3=-4 A_1 \label{A3eq}
\end{equation}
can be found in the form 
\begin{equation}
\Phi^{(3)}(x,y)= p_1^{\Delta-2 d -2} \sum_{a b} c_1^{(3)}(a,b,\vec\Delta) x^a y^b   F_4(\alpha(a,b)+1,\beta(a,b)+1; \gamma(a),\gamma'(b)-1;x,y). 
\end{equation}

Also in this case the inhomogeneous equation in \eqref{A3eq} fixes the integration constants to be those appearing in $A_1$ 
\begin{equation}
c_1^{(3)}(a,b,\vec \Delta)=-\frac{\gamma'(b)-1}{(\alpha(a,b)+1)(\beta(a,b)+1)}\, c^{(1)}(a,b,\vec\Delta).
\end{equation}
Therefore the general solution of the equations \eqref{A3eq} can be written as
\begin{align}
A_3=&p_1^{\Delta-2 d -2} \sum_{a b} x^a y^b\,\bigg[c^{(3)}(a,b,\vec\Delta)\,F_4(\a(a,b)+1,\b(a,b)+1;\g(a),\g'(b);x,y)\notag\\
&\hspace{3cm}-\frac{(\gamma(a)-1)\,c^{(1)}(a,b,\vec\Delta) }{(\alpha(a,b)+1)(\beta(a,b)+1)}F_4(\alpha(a,b)+1,\beta(a,b)+1; \gamma(a),\gamma'(b)-1;x,y)\bigg]
\end{align} 
since $\g(a)=\g'(b)$ in the $TJJ$ case. 

\subsection{The $A_4$ solution}
The last pair of equations 
\begin{equation}
K_{21} A_4 =-2 A_3 \qquad \qquad K_{31} A_4=-2 A_3 (p_2\leftrightarrow p_3)\label{eqA4}
\end{equation}
admit three particular solutions
\begin{align}
\Phi_1^{(4)}&= p_1^{\Delta-2d}\sum_{a b} x^a y^b\, c_1^{(4)}(a,b,\vec{\Delta})\,F_4(\alpha(a,b),\beta(a,b),\gamma(a)-1,\gamma'(b),x,y)\\
\Phi_2^{(4)}&= p_1^{\Delta-2d}\sum_{a b} x^a y^b\, c_2^{(4)}(a,b,\vec{\Delta})\,F_4(\alpha(a,b),\beta(a,b),\gamma(a),\gamma'(b)-1,x,y)\\
\Phi_3^{(4)}&= p_1^{\Delta-2d}\sum_{a b} x^a y^b\,c_3^{(4)}\,(a,b,\vec{\Delta})\,F_4(\a(a,b),\b(a,b),\g(a)-1,\g'(b)-1;x,y)
\end{align}
with the action of $K_{21}$ and $K_{31}$ on them as
\begin{align}
K_{21}\Phi^{(4)}_1&=4p_1^{\Delta-2d-2}\sum_{a b} x^a y^b\, c_1^{(4)}(a,b,\vec{\Delta})\,\frac{\a(a,b)\b(a,b)}{(\g(a)-1)}\,F_4(\alpha(a,b)+1,\beta(a,b)+1,\gamma(a),\gamma'(b),x,y)\\
K_{31}\Phi^{(4)}_1&=0\\[2.2ex]
K_{21}\Phi^{(4)}_2&=0\\
K_{31}\Phi^{(4)}_2&=4p_1^{\Delta-2d-2}\sum_{a b} x^a y^b\, c_2^{(4)}(a,b,\vec{\Delta})\,\frac{\a(a,b)\b(a,b)}{(\g'(b)-1)}\,F_4(\alpha(a,b)+1,\beta(a,b)+1,\gamma(a),\gamma'(b),x,y)\\[2.2ex]
K_{21}\Phi^{(4)}_3&=4p_1^{\Delta-2d-2}\sum_{a b} x^a y^b\,c_3^{(4)}(a,b,\vec{\Delta})\,\frac{\a(a,b)\b(a,b)}{(\g(a)-1)}\,F_4(\a(a,b)+1,\b(a,b)+1,\g(a),\g'(b)-1;x,y)\\
K_{31}\Phi^{(4)}_3&=4p_1^{\Delta-2d-2}\sum_{a b} x^a y^b\,c_3^{(4)}(a,b,\vec{\Delta})\,\frac{\a(a,b)\b(a,b)}{(\g'(b)-1)}\,F_4(\a(a,b)+1,\b(a,b)+1,\g(a)-1,\g'(b);x,y)
\end{align}
The inhomogeneous equations \eqref{eqA4} fix the integration constants to be those appearing in $A_3$ and $A_3(p_2\leftrightarrow p_3)$ as
\begin{align}
c_1^{(4)}&=-\frac{(\g(a)-1)}{2\,\a(a,b)\b(a,b)}\,c^{(3)}(a,b,\vec{\Delta})\\[1.2ex]
c_2^{(4)}&=-\frac{(\g'(b)-1)}{2\,\a(a,b)\b(a,b)}\,c^{(3)}(a,b,\vec{\Delta})\\[1.2ex]
c_3^{(4)}&=\frac{(\a(a,b)+1)(\b(a,b)+1)}{2\,\a(a,b)\b(a,b)}\,c^{(1)}(a,b,\vec{\Delta}).
\end{align}
Finally, using the properties $\g(a)=\g'(b)$, we give the general solution for the $A_4$ as
\begin{align}
A_4&=p_1^{\D-2d}\,\sum_{ab}\,x^a\,y^b\bigg[c^{(4)}(a,b,\vec{\Delta})\,F_4(\a(a,b),\b(a,b),\g(a),\g'(b);x,y)\notag\\
&\hspace{0.5cm}+\frac{(\a(a,b)+1)(\b(a,b)+1)}{2\,\a(a,b)\b(a,b)}\,c^{(1)}(a,b,\vec{\Delta})\,F_4(\a(a,b),\b(a,b),\g(a)-1,\g'(b)-1;x,y)\notag\\
&\hspace{0.5cm}-\frac{(\g(a)-1)}{2\,\a(a,b)\b(a,b)}\,c^{(3)}(a,b,\vec{\Delta})\,\bigg(F_4(\alpha(a,b),\beta(a,b),\gamma(a)-1,\gamma'(b),x,y)\notag\\
&\hspace{7.5cm}+\,F_4(\alpha(a,b),\beta(a,b),\gamma(a),\gamma'(b)-1,x,y)\bigg)
\bigg]
\end{align}
Notice that, differently from this case, number of free constants can be significantly reduced in the case of a fully symmetric correlator, such as the $TTT$, where the  number of constants reduces to 4, as in the BMS case.  
\subsection{Summary}
\label{finfin}
To summarize, the solutions of the primary WI' s in the $TJJ$ case are expressed as sums of 4 hypergeometrics of universal indicial points 
\begin{equation}
a_0 =0,\quad b_0=0,\quad a_1=\Delta_2- \frac{d}{2},\quad b_1=\Delta_3-\frac{d}{2}
\end{equation}
and parameters 
\begin{align}
&\alpha(a,b)= a + b + \frac{d}{2} -\frac{1}{2}(\Delta_2 +\Delta_3 -\Delta_1)\,,  &&\beta (a,b)=a +  b + d -\frac{1}{2}(\Delta_1 +\Delta_2 +\Delta_3) \\
&\gamma(a) =2 a +\frac{d}{2} -\Delta_2 + 1\,, &&\gamma'(b)=2 b +\frac{d}{2} -\Delta_3 + 1. \label{cons2}
\end{align}
where $\D_2=\D_3=d-1$ and $\D_1=d$. In particular they are given by
\begin{equation}
\begin{split}
A_1&=p_1^{\Delta-2 d - 4}\sum_{a,b} c^{(1)}(a,b,\vec{\Delta})\,x^a y^b \,F_4(\alpha(a,b) +2, \beta(a,b)+2; \gamma(a), \gamma'(b); x, y)
\end{split}
\end{equation}
\begin{equation}
\begin{split}
A_2&= p_1^{\Delta-2 d - 2}\sum_{a b} x^a y^b\Bigg[c^{(2)}(a,b,\vec{\Delta})\,F_4(\alpha(a,b)+1, \beta(a,b)+1; \gamma(a), \gamma'(b); x, y)\\
&\hspace{1cm}+ \frac{(\gamma(a)-1)\,c^{(1)}(a,b,\vec \Delta)}{2(\alpha(a,b)+1)(\beta(a,b)+1)}\bigg(F_4(\alpha(a,b) +1, \beta(a,b)+1; \gamma(a)-1, \gamma'(b); x, y)\\
&\hspace{7cm}+ F_4(\alpha(a,b) +1, \beta(a,b)+1; \gamma(a), \gamma'(b)-1; x, y)\bigg)\Bigg]
\end{split}
\end{equation}
\begin{align}
A_3=&p_1^{\Delta-2 d -2} \sum_{a b} x^a y^b\,\bigg[c^{(3)}(a,b,\vec\Delta)\,F_4(\a(a,b)+1,\b(a,b)+1;\g(a),\g'(b);x,y)\notag\\
&\hspace{3cm}-\frac{(\gamma(a)-1)\,c^{(1)}(a,b,\vec\Delta) }{(\alpha(a,b)+1)(\beta(a,b)+1)}F_4(\alpha(a,b)+1,\beta(a,b)+1; \gamma(a),\gamma'(b)-1;x,y)\bigg]
\end{align}
\begin{align}
A_4&=p_1^{\D-2d}\,\sum_{ab}\,x^a\,y^b\bigg[c^{(4)}(a,b,\vec{\Delta})\,F_4(\a(a,b),\b(a,b),\g(a),\g'(b);x,y)\notag\\
&\hspace{0.5cm}+\frac{(\a(a,b)+1)(\b(a,b)+1)}{2\,\a(a,b)\b(a,b)}\,c^{(1)}(a,b,\vec{\Delta})\,F_4(\a(a,b),\b(a,b),\g(a)-1,\g'(b)-1;x,y)\notag\\
&\hspace{0.5cm}-\frac{(\g(a)-1)}{2\,\a(a,b)\b(a,b)}\,c^{(3)}(a,b,\vec{\Delta})\,\bigg(F_4(\alpha(a,b),\beta(a,b),\gamma(a)-1,\gamma'(b),x,y)\notag\\
&\hspace{7.5cm}+\,F_4(\alpha(a,b),\beta(a,b),\gamma(a),\gamma'(b)-1,x,y)\bigg)
\bigg]
\end{align}
in terms of the constants $c^{(i)}(a,b)$ given above. The method has the advantage of being generalizable to higher point functions, 
in the search of specific solutions of the corresponding correlation functions.
\section{Perturbative analysis in the Conformal case: QED and scalar QED}
\label{pchecks}
In this section we turn to discuss the connection between the solutions of the CWI's presented by BMS and the perturbative $TJJ$ vertex. The QED case has been previously studied in \cite{Giannotti:2008cv,Armillis:2009pq}, where more details can be found. The expressions fo the form factors  had been given in the F-basis of 13 form factors, which will be reviewed in the next section. We will have to recompute them in order to present them expressed in terms of the two basic fundamental master integrals $B_0$ and $C_0$ of the tensor reduction rather then in their final form, given in 
\cite{Armillis:2009pq}.\\ 
Here we are also going to introduce the diagrammatic expansion for the $TJJ$ in scalar QED, since it will be needed in the last part of the work when we are going to compare the general BMS solution against the perturbative one in $d=3$ and $d=5$. 

\begin{figure}[t]
	\centering
	\vspace{-1.6cm}
	\subfigure{\includegraphics[scale=0.18]{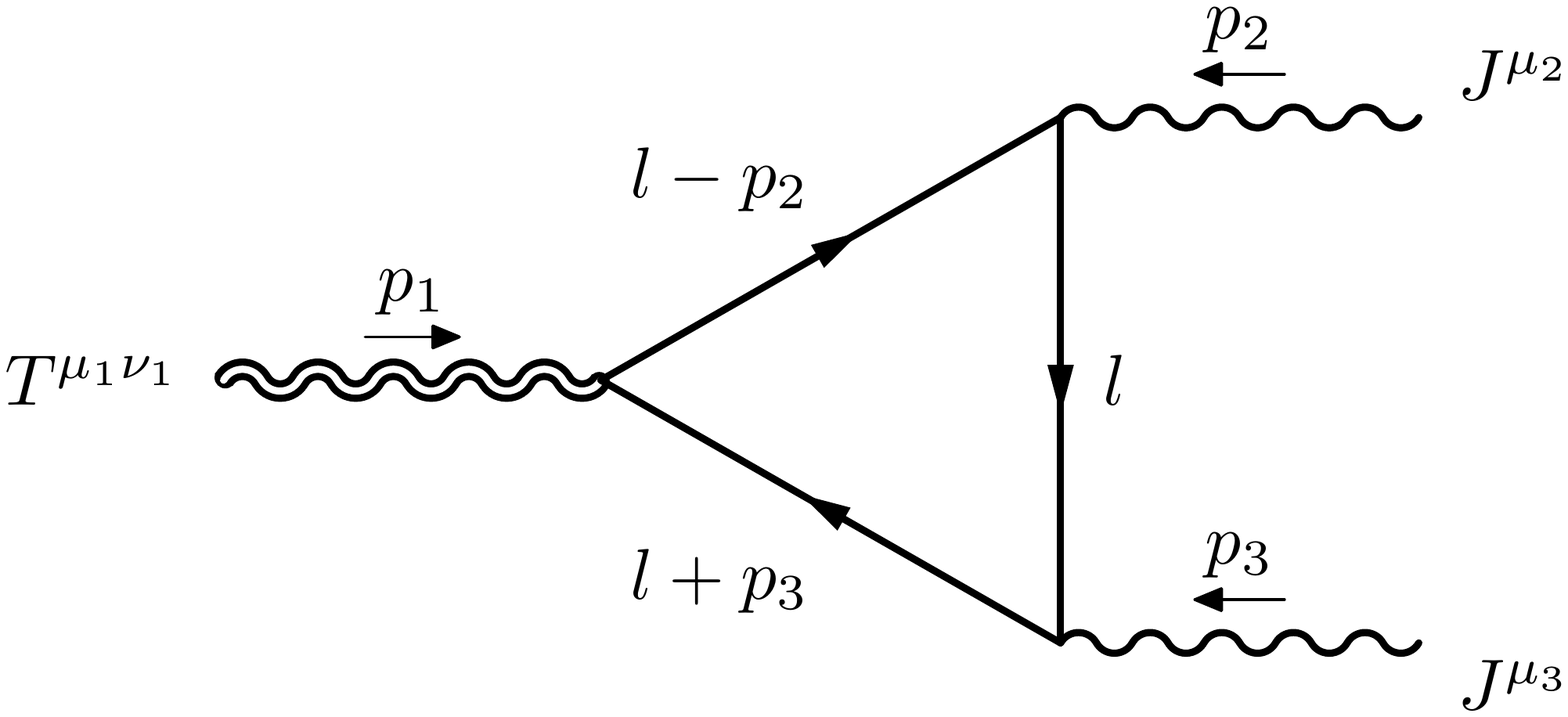}} \hspace{.3cm}
	\subfigure{\includegraphics[scale=0.18]{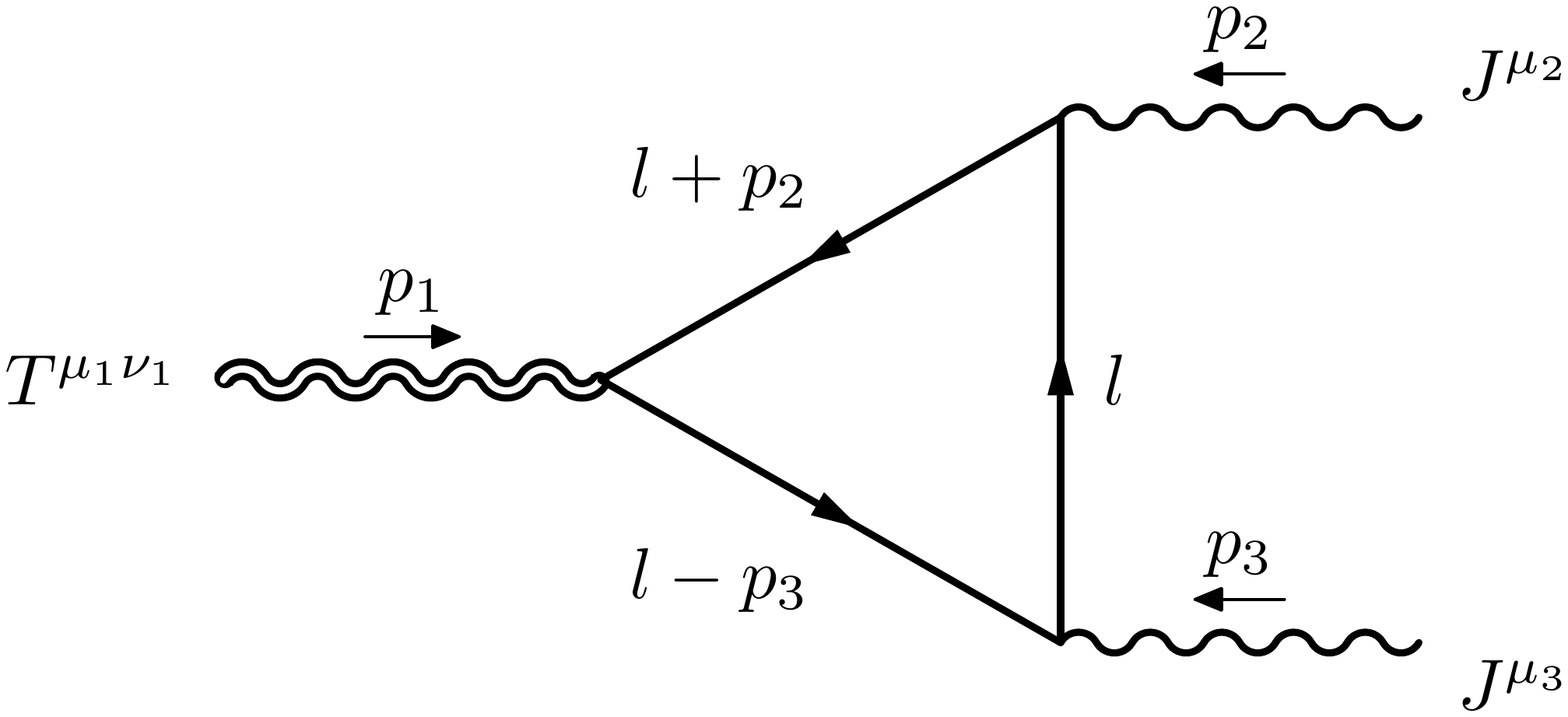}} \hspace{.3cm}
	\raisebox{.11\height}{\subfigure{\includegraphics[scale=0.16]{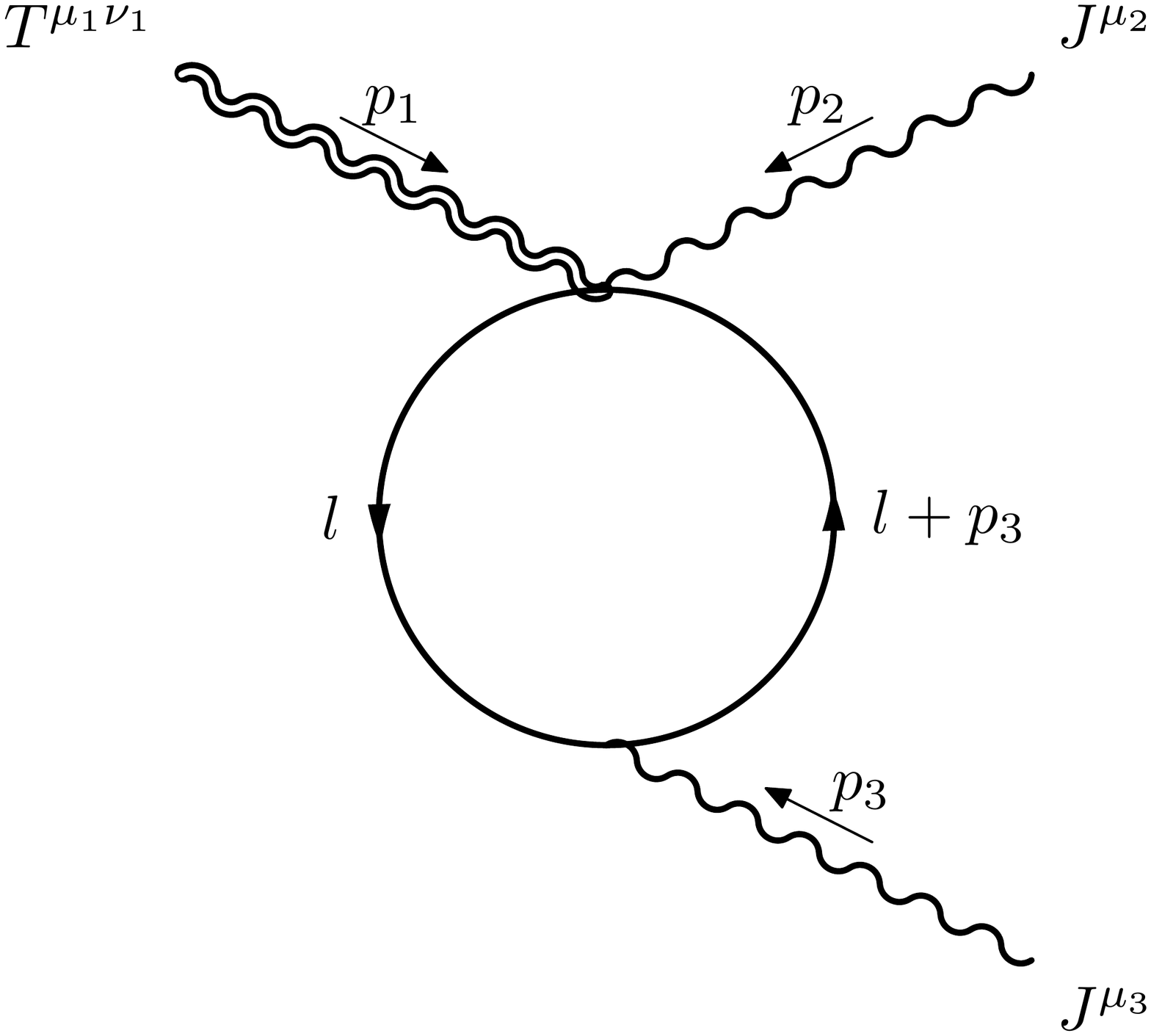}}}\hspace{.3cm}	\raisebox{.11\height}{\subfigure{\includegraphics[scale=0.16]{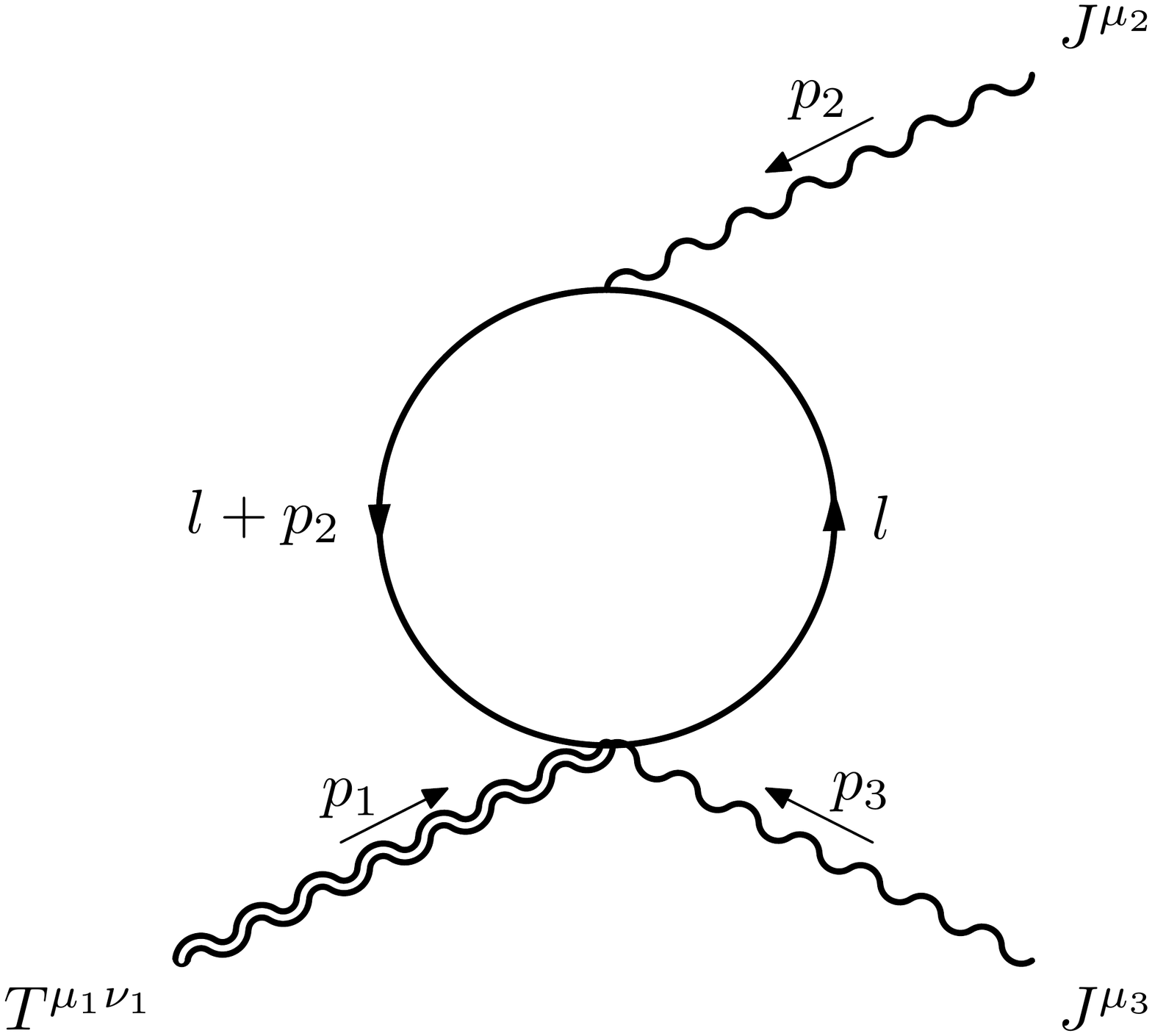}}}\\
	\vspace{-1.9cm}
	\subfigure{\includegraphics[scale=0.16]{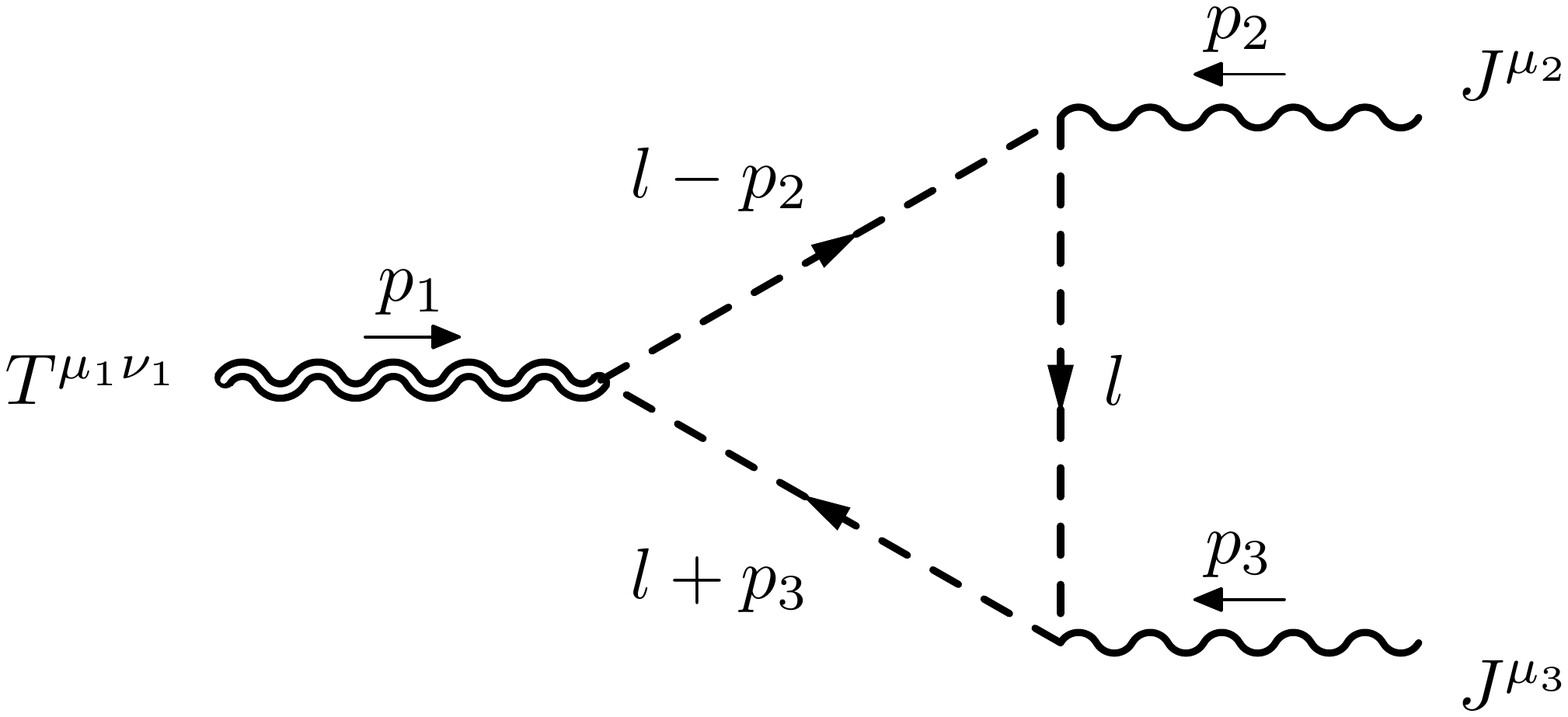}}
	\subfigure{\includegraphics[scale=0.16]{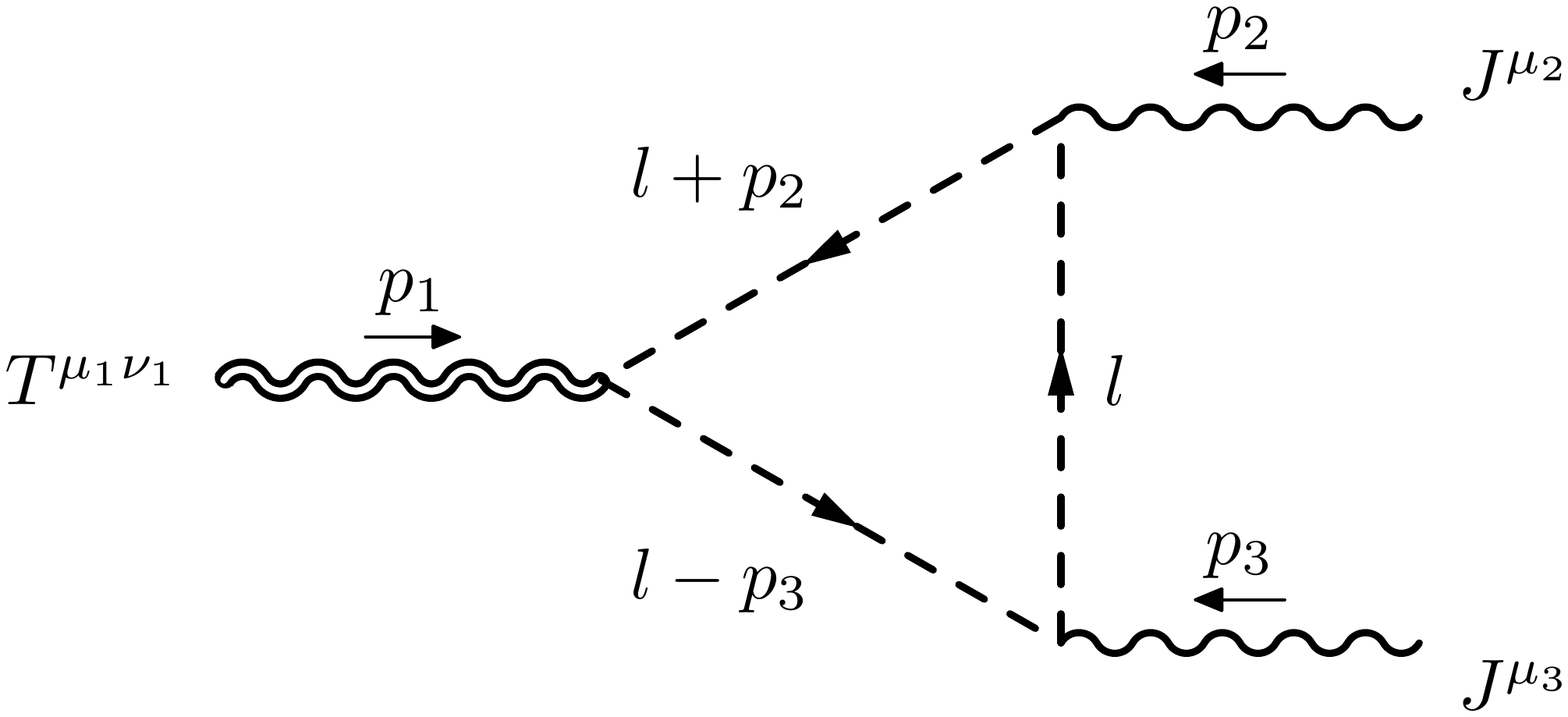}} 
	\raisebox{-.05\height}{\subfigure{\includegraphics[scale=0.15]{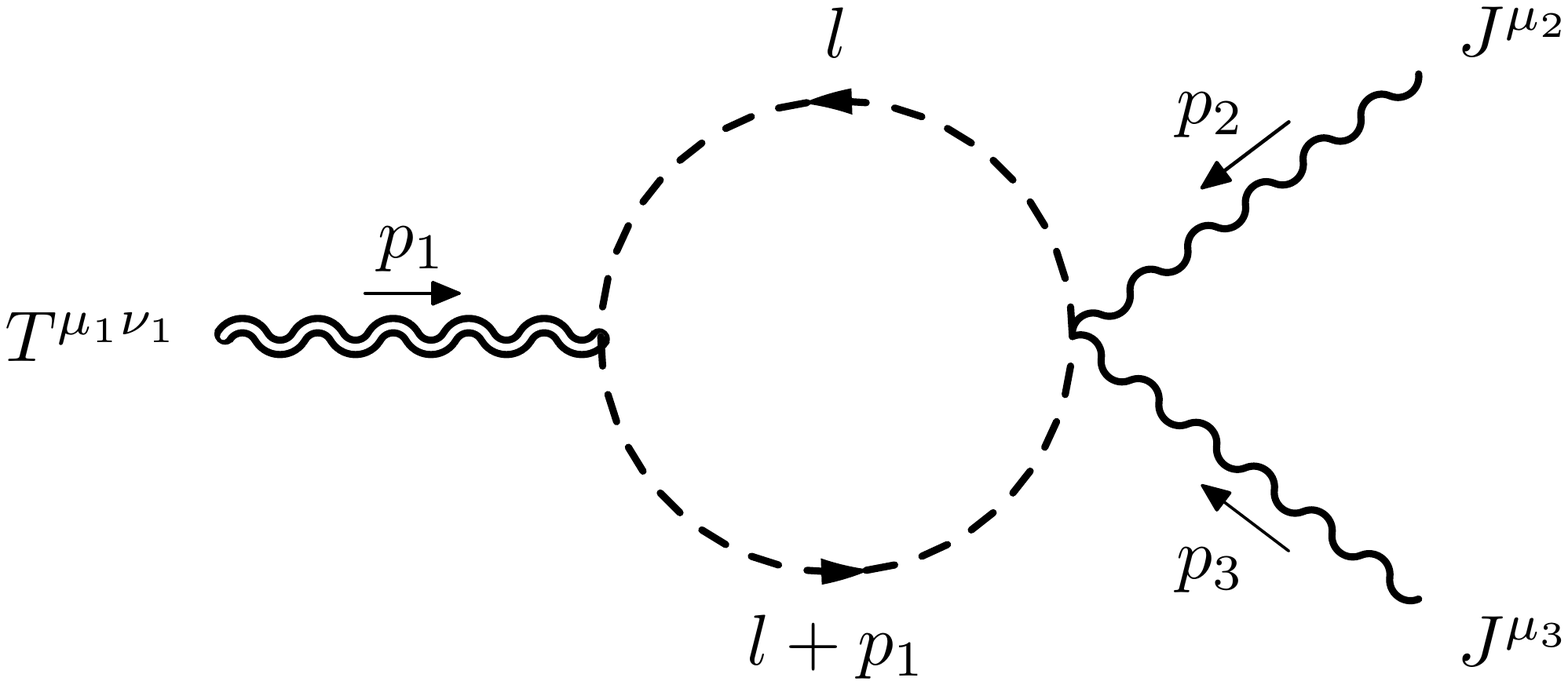}}}
	\raisebox{.11\height}{\subfigure{\includegraphics[scale=0.15]{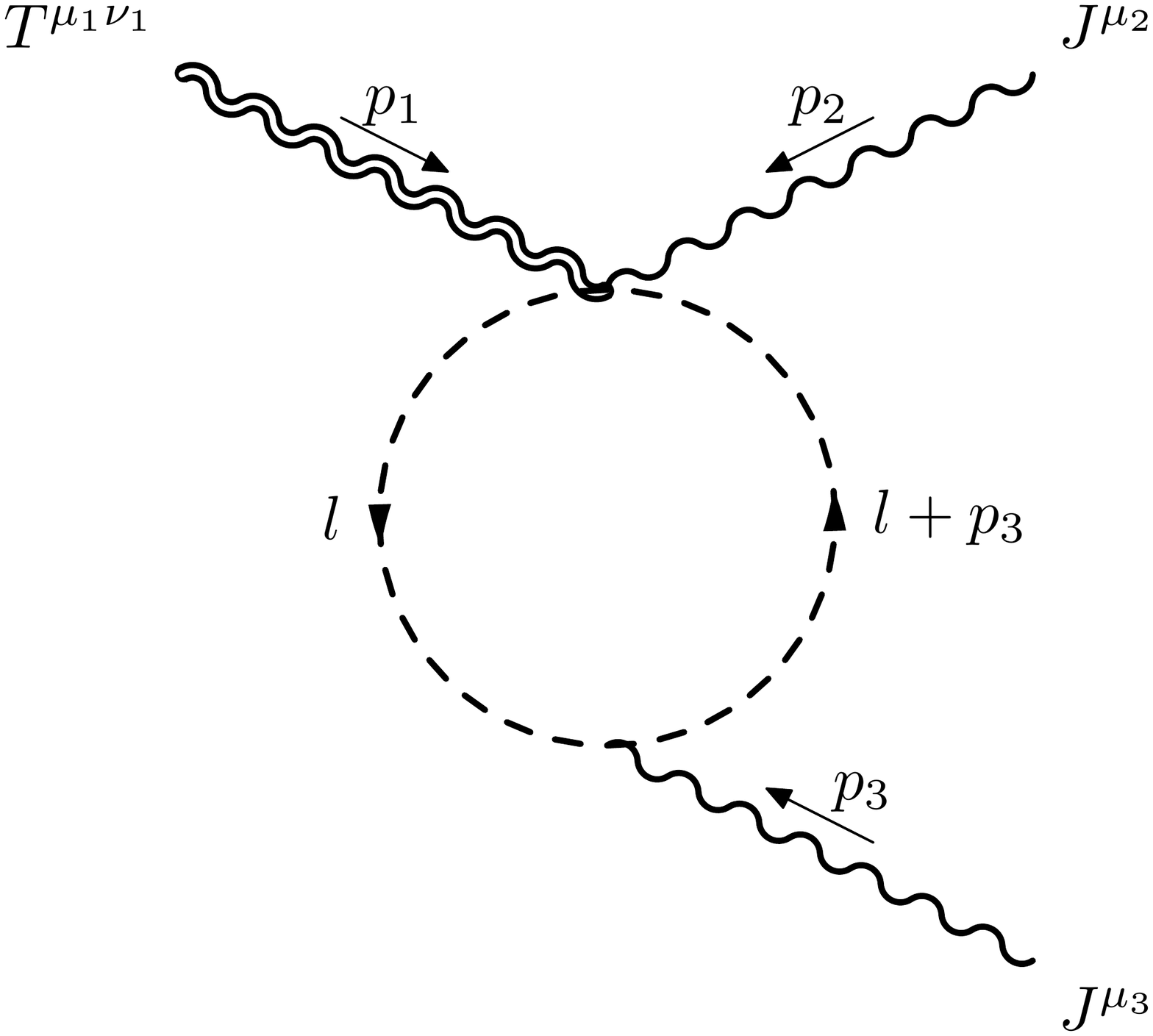}}}
	\raisebox{.11\height}{\subfigure{\includegraphics[scale=0.15]{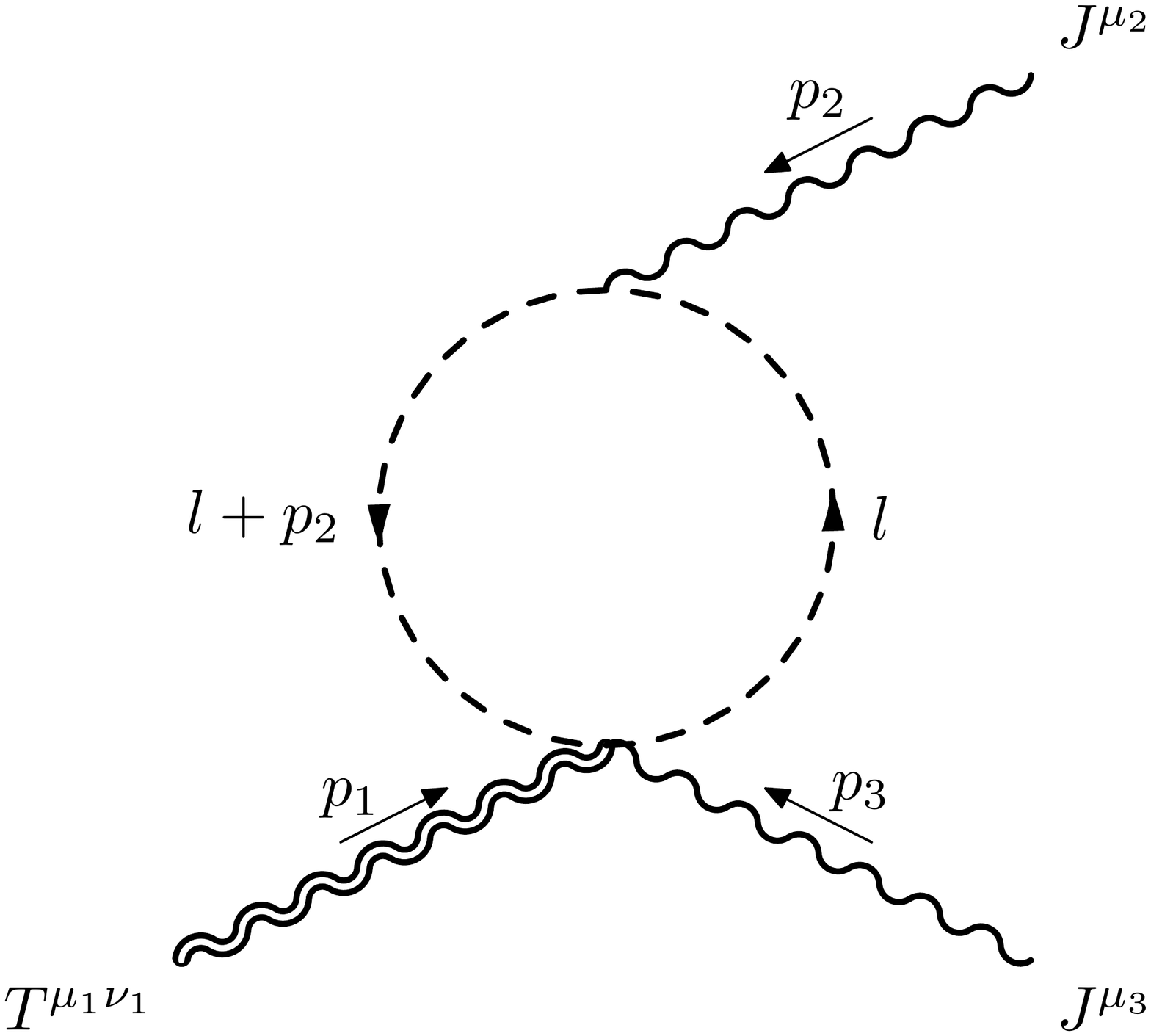}}}
	\vspace{-0.8cm}\caption{One-loop diagrams for the TJJ in QED (continuous fermion lines). A similar set of diagrams describes the scalar QED case (dashed lines), by replacing the internal fermion with a charged scalar. \label{Figura1}}
	\end{figure}

 The quantum actions for the fermion field is
\begin{align}
S_{fermion}&=\sdfrac{i}{2}\int\, d^dx\ e\ e^{\m}_a\left[\bar{\psi}\g^a(D_\m\psi)-(D_\m\bar{\psi})\g^a\psi\right],
\end{align}
$e^\m_a$ is the vielbein and $e$ its determinant, with its covariant derivative $D_\m$ as
\begin{equation}
D_\m=\partial_\m+ie\,A_\mu+\G_\m=\partial_\m+ie\,A_\mu+\sdfrac{1}{2}\Sigma^{ab}\,e^\s_a\nabla_\m\,e_{b\,\s}.
\end{equation}
The $\Sigma^{ab}$ are the generators of the Lorentz group in the case of a spin $1/2$-field. The gravitational field is expanded, as usual, in the form $g_{\mu\nu}=\eta_{\mu\nu} + h_{\mu\nu}$ around the flat background metric with fluctuations $h_{\mu\nu}$.As usual, the Latin anf Greek indices are related to the locally flat and curved backgrounds respectively.
We take the external momenta as incoming. In order to simplify the notation, we introduce the tensor components
\begin{align}
A^{\m_1\n_1\m\n}&\equiv\h^{\mu_1\nu_1}\h^{\m\n}-2\h^{\m(\m_1}\h^{\n_1)\n}
\end{align}
where we indicate with the round brackets the symmetrization of the indices and the square brackets their anti-symmetrization 
\begin{equation}
\h^{\m(\m_1}\h^{\n_1)\n}\equiv\sdfrac{1}{2}\bigg(\h^{\m\m_1}\h^{\n_1\n}+\h^{\m\n_1}\h^{\m_1\n}\bigg)
\end{equation}
and the vertices in the fermion sector are
\begin{align}
V^\m_{J\psi\bar\psi}(k_1,k_2) &=-ie\,\g^\m\\[2ex]
V^{\m_1\n_1}_{T\psi\bar\psi}(p_1k_1,k_2) &=-\sdfrac{i}{4}\,A^{\m_1\n_1\m\n}\,\g_\n\,(k_1+k_2)_\m\\[2ex]
V^{\m_1\n_1\m_2}_{TJ\psi\bar\psi}(k_1,k_2) &=\frac{i\,e}{2}A^{\m_1\n_1\m_2\n}\,\g_\m.
\end{align}

In the one-loop approximation the contribution to the correlation functions are given by the diagrams in \figref{Figura1}, with vertices shown in \figref{Figura2}. We calculate all the diagram contributions in momentum space for he fermion sector as
\begin{eqnarray}
\Gamma^{\m_1\n_1\m_2\m_3}(p_2,p_3)&\equiv&\braket{T^{\m_1\n_1}(p_1)\,J^{\m_2}(p_2)\,J^{\m_3}(p_3)}_F\nn
&&=2\,\bigg(\sum_{i=1}^2V_{F,i}^{\m_1\n_1\m_2\m_3}(p_1,p_2,p_3)+\sum_{i=1}^{2}W_{F,i}^{\m_1\n_1\m_2\m_3}(p_1,p_2,p_3)\bigg)
\end{eqnarray}
where the $V_{F,i}$ terms are related to the triangle topology contributions, while the $W_{F,i}$ terms denote the two bubble contributions in \figref{Figura1}. All these terms are explicitly given as
\begin{align}
V_{F,1}^{\m_1\n_1\m_2\m_3}&=-i^3\,\int\frac{d^d\ell}{(2\p)^d}\frac{\Tr\left[V^{\m_1\n_1}_{T\psi\bar\psi}(\ell-p_2,\ell+p_3)\left(\slashed{\ell}+\slashed{p}_3\right)V^{\m_2}_{J\psi\bar\psi}(\ell,\ell-p_2)\,\slashed{\ell}\,V^{\m_3}_{J\psi\bar\psi}(\ell,\ell+p_3)\,\left(\slashed{\ell}-\slashed{p}_2\right)\right]}{\ell^2\,(\ell-p_2)^2(\ell+p_3)^2}\\
V_{F,2}^{\m_1\n_1\m_2\m_3}&=-i^3\,\int\frac{d^d\ell}{(2\p)^d}\frac{\Tr\left[V^{\m_1\n_1}_{T\psi\bar\psi}(\ell-p_3,\ell+p_2)\left(\slashed{\ell}+\slashed{p}_2\right)V^{\m_2}_{J\psi\bar\psi}(\ell,\ell-p_3)\,\slashed{\ell}\,V^{\m_3}_{J\psi\bar\psi}(\ell,\ell+p_2)\,\left(\slashed{\ell}-\slashed{p}_3\right)	\right]}{\ell^2\,(\ell-p_3)^2(\ell+p_2)^2}\\
W_{F,3}^{\m_1\n_1\m_2\m_3}&=-i^2\,\int\frac{d^d\ell}{(2\p)^d}\frac{\Tr\left[V^{\m_1\n_1\m_2}_{TJ\psi\bar\psi}(\ell+p_3,\ell)\left(\slashed{\ell}+\slashed{p}_3\right)V^{\m_3}_{J\psi\bar\psi}(\ell,\ell+p_3)\,\slashed{\ell}\right]}{\ell^2\,(\ell+p_3)^2}\\[1.5ex]
W_{F,2}^{\m_1\n_1\m_2\m_3}&=-i^2\,\int\frac{d^d\ell}{(2\p)^d}\frac{\Tr\left[V^{\m_1\n_1\m_3}_{TJ\psi\bar\psi}(\ell+p_2,\ell)\left(\slashed{\ell}+\slashed{p}_2\right)V^{\m_2}_{J\psi\bar\psi}(\ell,\ell+p_2)\,\slashed{\ell}\right]}{\ell^2\,(\ell+p_2)^2}
\end{align}

\subsection{The $TJJ$ in scalar QED}
Now we turn to consider scalar QED. The action, in this case, can be written as
\begin{equation}
S_{scalar}=\int\,d^dx\,\sqrt{-g}\,\left(\,\left|D_\m\,\phi\right|^2+\frac{(d-2)}{8(d-1)}\,R\,|\phi|^2\,\right)
\end{equation}
where $R$ is the scalar curvature and $\phi$ denotes a complex scalar. We have explicitly reported the coefficient of the term of improvement, and with $D_\m\phi=\partial_\m\phi+ie\,A_\m$ being the covariant derivative for the coupling to the gauge field $A_\mu$. 
At one-loop the contribution to the $TJJ$ is given by the diagram in \figref{Figura1}, with the obvious replacement of a fermion by a scalar in the internal loop corrections. In this case they are given by 
\begin{equation}
\braket{T^{\m_1\n_1}(p_1)\,J^{\m_2}(p_2)\,J^{\m_3}(p_3)}_S=2\,\bigg(V_{S}^{\m_1\n_1\m_2\m_3}(p_1,p_2,p_3)+\sum_{i=1}^{3}W_{S,i}^{\m_1\n_1\m_2\m_3}(p_1,p_2,p_3)\bigg)
\end{equation}
where the $V_S$ terms are related to the triangle topology contribution and the $W_{S,i}$'s  are the three bubble contributions in \figref{Figura1}. All these are explicitly given as
\begin{align}
V_{S}^{\m_1\n_1\m_2\m_3}(p_1,p_2,p_3)&=i^3\,\int\frac{d^d\ell}{(2\p)^d}\frac{V^{\m_1\n_1}_{T\phi\phi^*}(\ell-p_2,\ell+p_3)\,V^{\m_2}_{J\phi\phi^*}(\ell,\ell-p_2)\,V^{\m_3}_{J\phi\phi^*}(\ell,\ell+p_3)}{\ell^2\,(\ell-p_2)^2(\ell+p_3)^2}\\[1.5ex]
W_{S,1}^{\m_1\n_1\m_2\m_3}(p_1,p_2,p_3)&=\frac{i^2}{2}\,\int\frac{d^d\ell}{(2\p)^d}\frac{V^{\m_1\n_1}_{T\phi\phi^*}(\ell+p_1,\ell)\,V^{\m_2\m_3}_{JJ\phi\phi^*}(\ell,\ell+p_1)}{\ell^2\,(\ell+p_1)^2}\\[1.5ex]
W_{S,2}^{\m_1\n_1\m_2\m_3}(p_1,p_2,p_3)&=\frac{i^2}{2}\,\int\frac{d^d\ell}{(2\p)^d}\frac{V^{\m_1\n_1\m_2}_{TJ\phi\phi^*}(\ell+p_3,\ell)\,V^{\m_3}_{J\phi\phi^*}(\ell,\ell+p_3)}{\ell^2\,(\ell+p_3)^2}\\[1.5ex]
W_{S,3}^{\m_1\n_1\m_2\m_3}(p_1,p_2,p_3)&=\frac{i^2}{2}\,\int\frac{d^d\ell}{(2\p)^d}\frac{V^{\m_1\n_1\m_3}_{TJ\phi\phi^*}(\ell+p_2,\ell)\,V^{\m_2}_{J\phi\phi^*}(\ell,\ell+p_2)}{\ell^2\,(\ell+p_2)^2}
\end{align}
where we have included the symmetry factors and the vertices are given by
\begin{align}
&V^\m_{J\phi\phi^*}(k_1,k_2) =ie\,(k_1^\m+k_2^\m)\\[2ex]
&V^{\m_1\n_1}_{T\phi\phi^*}(p_1k_1,k_2) =\frac{i}{2}A^{\m\n\m_1\n_1}\,k_{1\m}k_{2\n}+i\,\c\,\left(p_1^{\m_1}p_1^{\n_1}-\h^{\m_1\n_1}p_1^2\right)\\[2ex]
&V^{\m_1\n_1\m_2}_{TJ\phi\phi^*}(k_1,k_2) =\frac{i\,e}{2}\,A^{\m_1\n_1\m\m_2}\,(k_{1\m}+k_{2\m})\\[2ex]
&V^{\m_1\m_2}_{JJ\phi\phi^*}(k_1,k_2) =2\,i\,e^2\,\h^{\m_1\m_2}
\end{align}
where $\c=(d-2)/[8(d-1)]$ is the coefficient for the term of improvement. They are shown in \figref{Figura2}.
\begin{figure}[t]
	\centering
	\vspace{-.8cm}
	\raisebox{.14\height}{\subfigure{\includegraphics[scale=0.12]{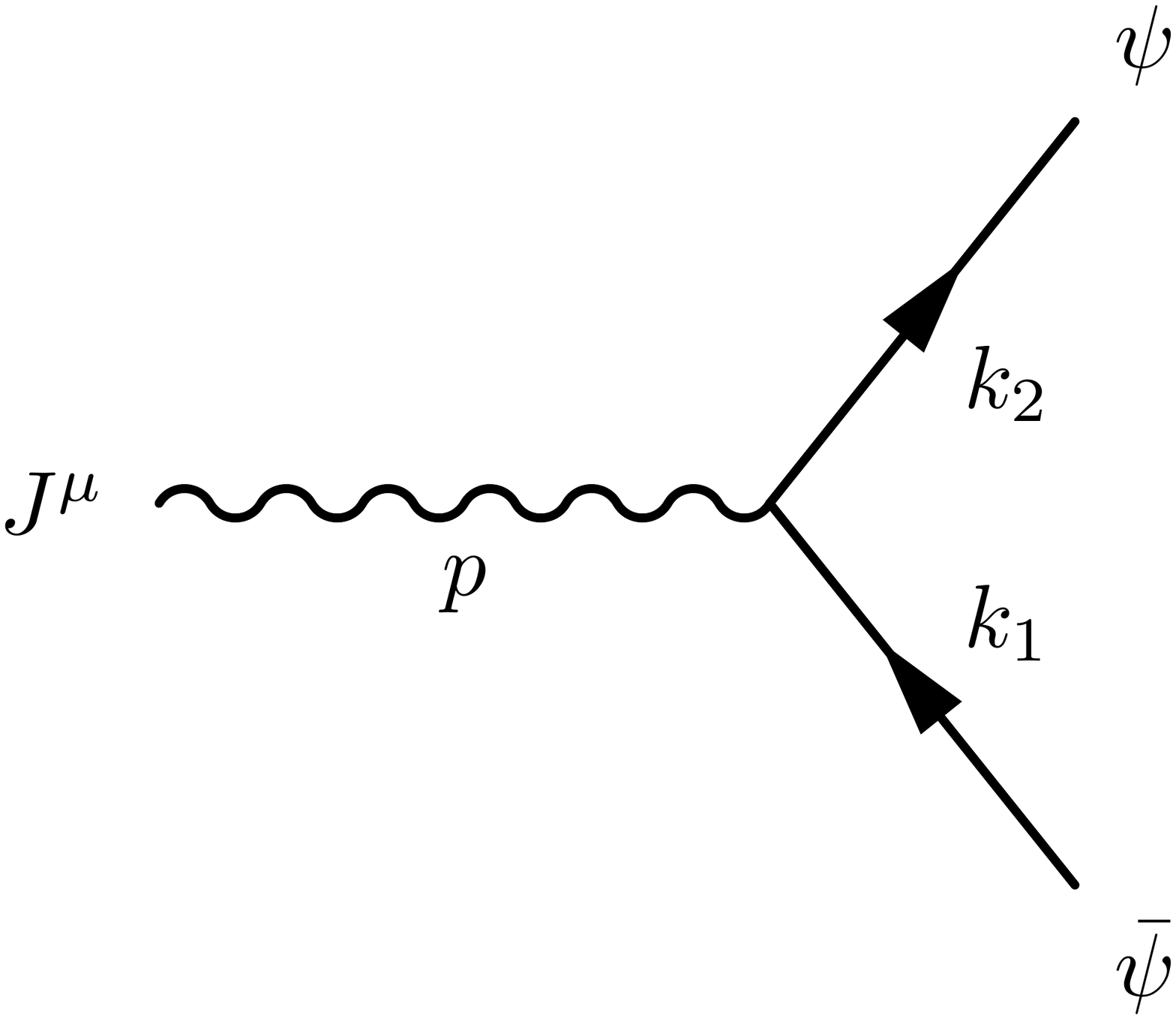}}} \hspace{.3cm}
	\raisebox{.05\height}{\subfigure{\includegraphics[scale=0.12]{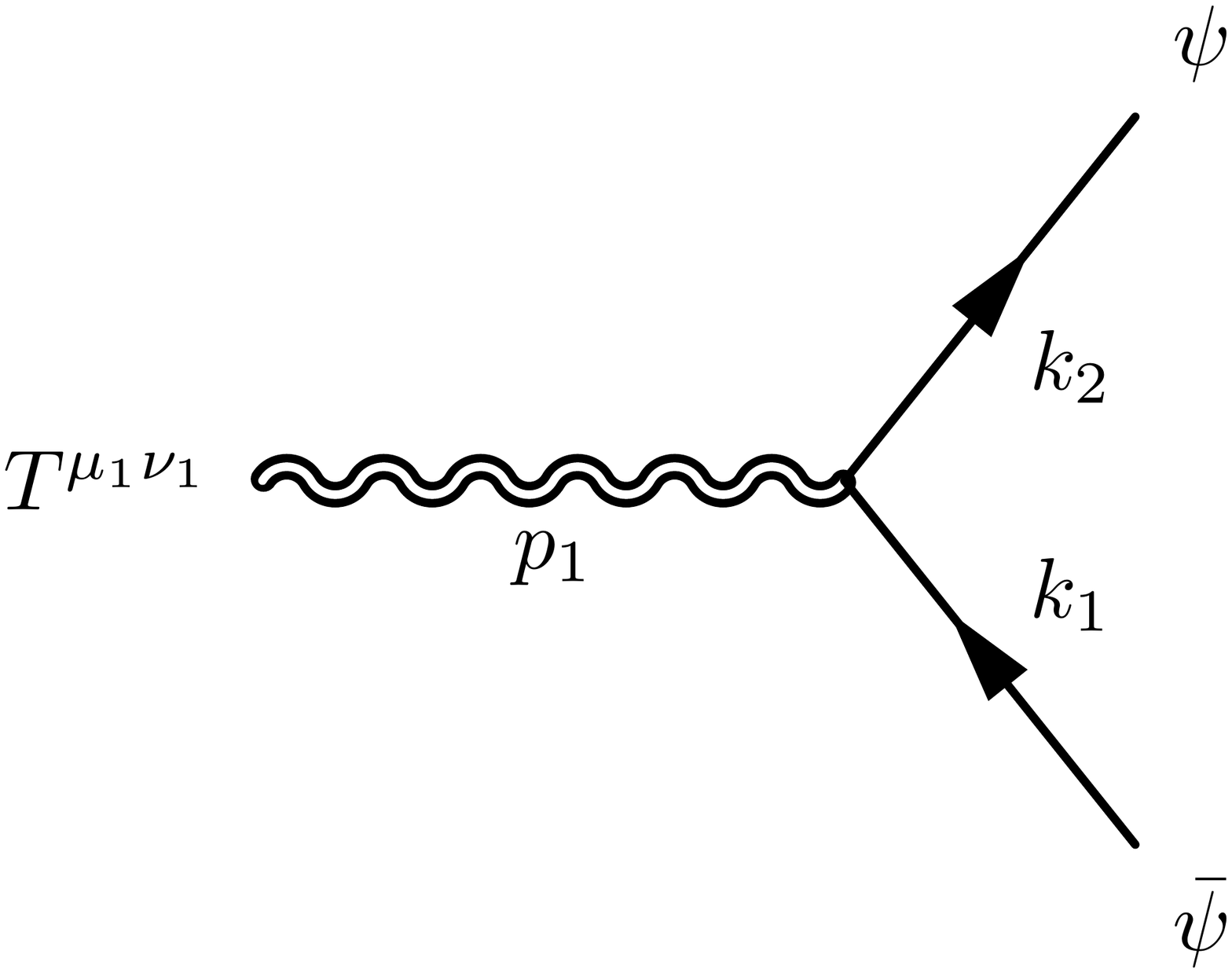}}} \hspace{.3cm}
	\raisebox{.12\height}{\subfigure{\includegraphics[scale=0.12]{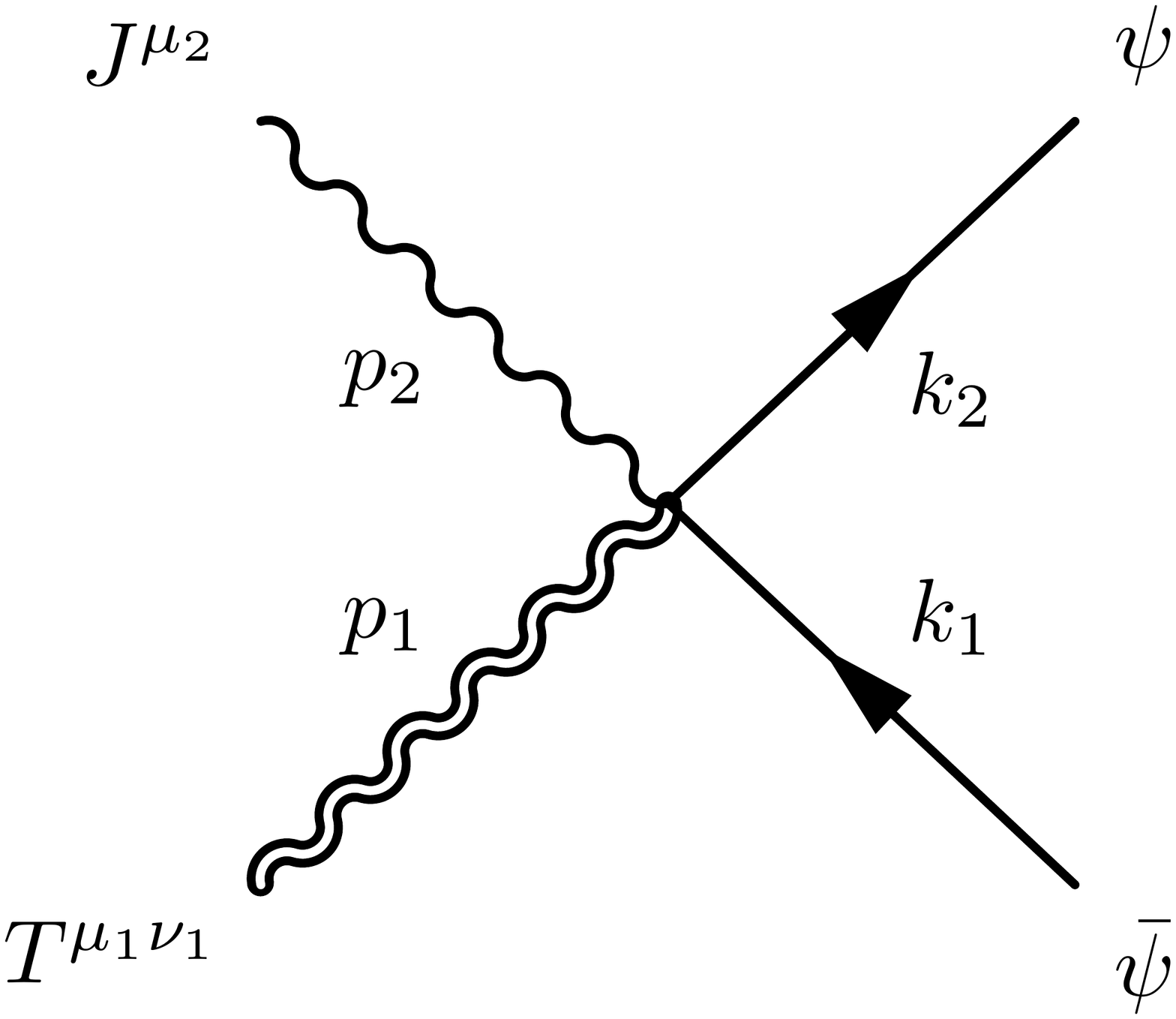}}}\hspace{.3cm}\vspace{-1.6cm}
	\vspace{-.6cm}
	\raisebox{.14\height}{\subfigure{\includegraphics[scale=0.12]{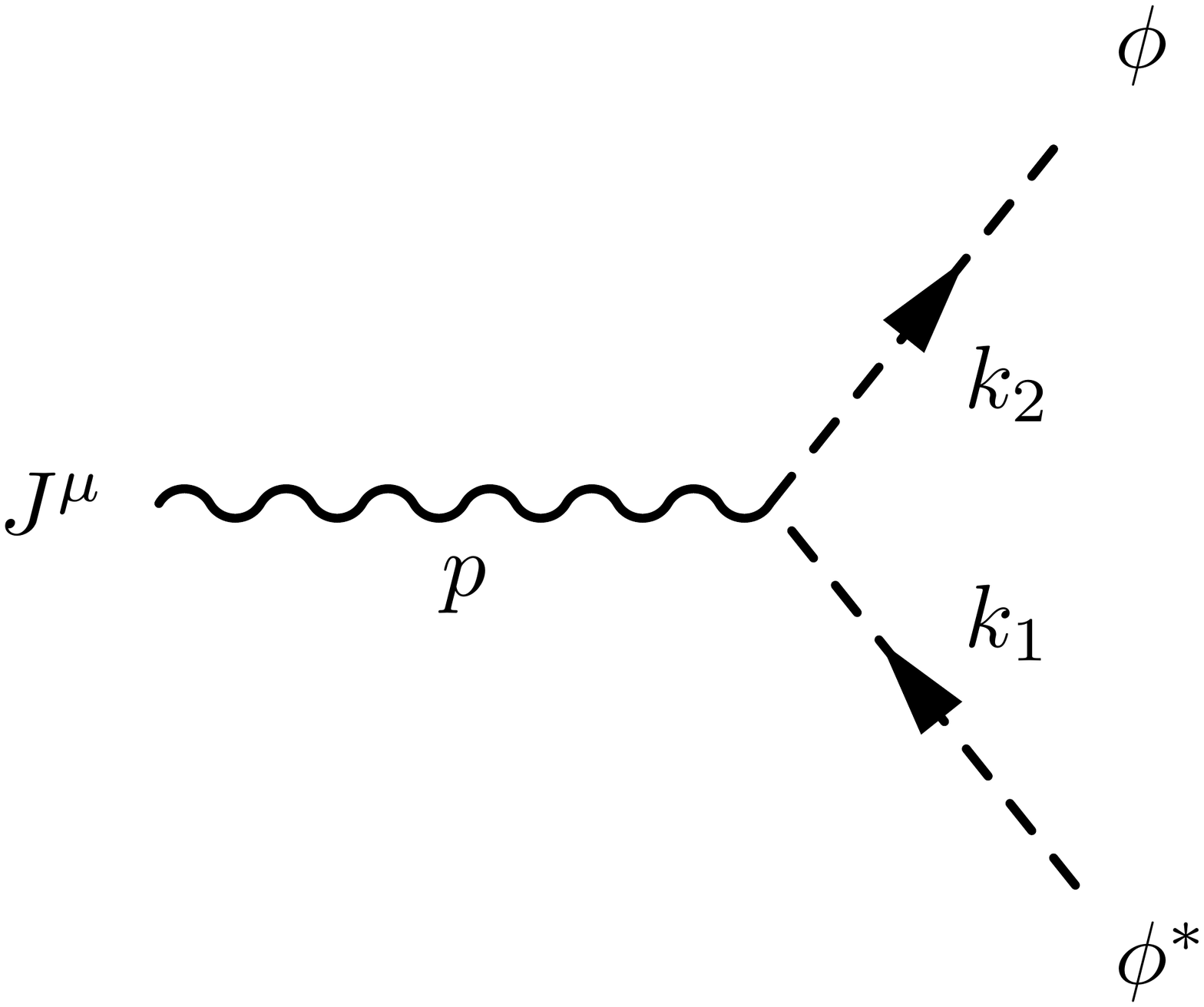}}} \hspace{.3cm}
	\raisebox{.05\height}{\subfigure{\includegraphics[scale=0.12]{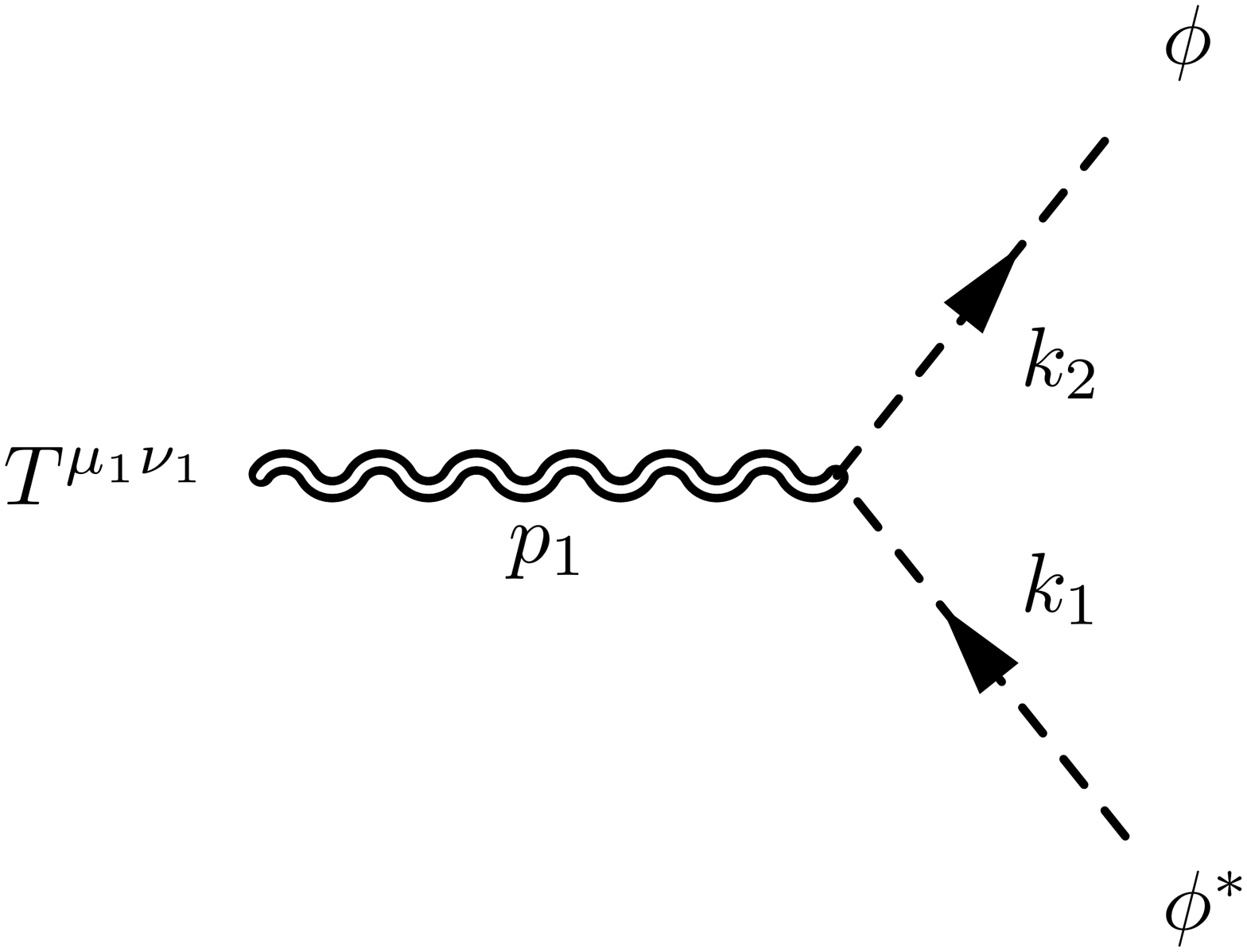}}} \hspace{.3cm}
	\raisebox{.12\height}{\subfigure{\includegraphics[scale=0.12]{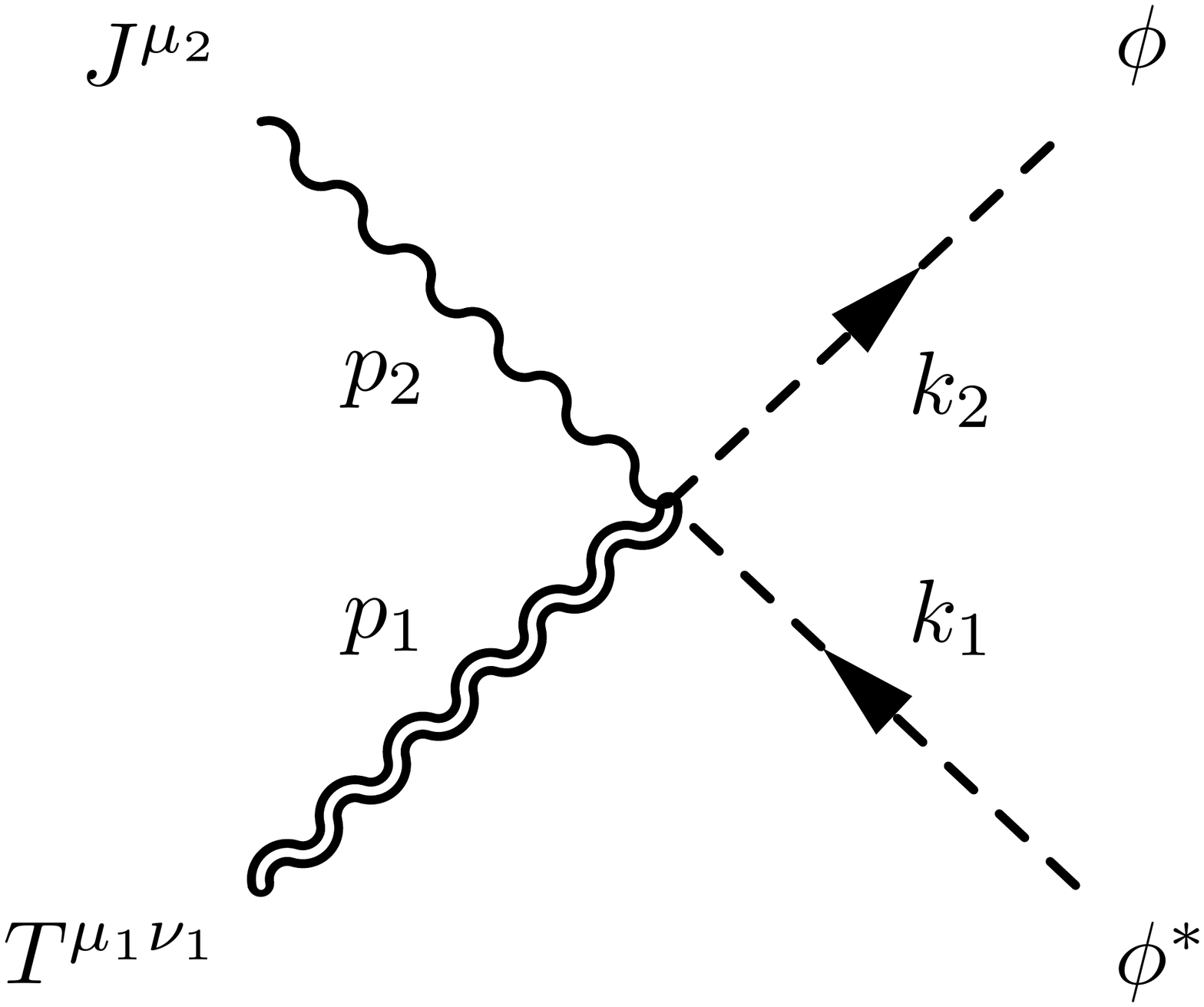}}}\hspace{.3cm}
	\raisebox{.12\height}{\subfigure{\includegraphics[scale=0.12]{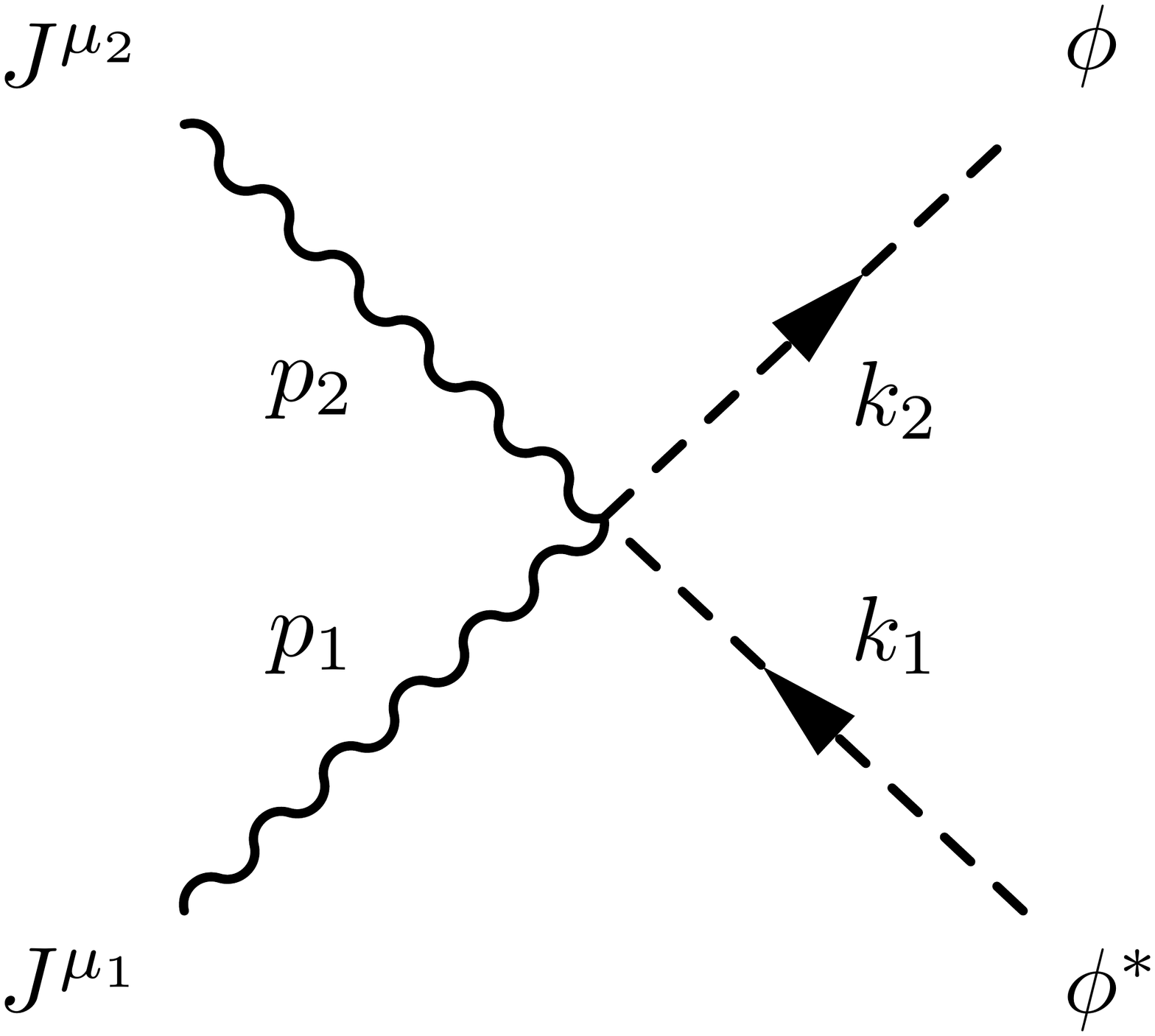}}}
	\vspace{-1cm}
\caption{The vertices in QED and scalar QED. }\label{Figura2}
\end{figure}

\section{ The F-basis of the expansion for the $TJJ$ in QED}
The $TJJ$ amplitude  can be expanded on the basis proposed by Giannotti and Mottola \cite{Giannotti:2008cv}, in terms of 13 independent tensors structures given in table (\ref{genbasis}). In this scheme, the amplitude can be written as
\begin{equation}
\Gamma^{\m_1\n_1\m_2\m_3}(p_2,p_3)=\sum_{i=1}^{13}\,F_i(s;s_1,s_2,0)\,t_i^{\m_1\n_1\m_2\m_3}(p_2,p_3),
\end{equation} 
where the invariant amplitudes $F_i$ are functions of the kinematic invariants $s=p_1^2=(p_2+p_3)^2$, $s_1=p_2^2$, $s_2=p_3^2$, and the $t_i^{\m_1\n_1\m_2\m_3}$ form the basis of independent tensor structures. 

This set of $13$ tensors is linearly independent in $d$ dimensions, for generic $k^2, p^2, q^2$
different from zero. Five of the $13$ are Bose symmetric, 
\be
t_i^{\mu\nu\alpha\beta}(p,q) = t_i^{\mu\nu\beta\alpha}(q,p)\,,\qquad i=1,2,7,8,13\,,
\ee
while the remaining eight tensors are Bose symmetric pairwise
\bea
\label{pair}
&&t_3^{\mu\nu\alpha\beta}(p,q) = t_5^{\mu\nu\beta\alpha}(q,p)\,,\\
&&t_4^{\mu\nu\alpha\beta}(p,q) = t_6^{\mu\nu\beta\alpha}(q,p)\,,\\
&&t_9^{\mu\nu\alpha\beta}(p,q) = t_{10}^{\mu\nu\beta\alpha}(q,p)\,,\\
&&t_{11}^{\mu\nu\alpha\beta}(p,q) = t_{12}^{\mu\nu\beta\alpha}(q,p)\,.
\eea

\begin{table}
	$$
	\begin{array}{|c|c|}\hline
	i & t_i^{\mu\nu\alpha\beta}(p,q)\\ \hline\hline
	1 &
	\left(k^2 g^{\mu\nu} - k^{\mu } k^{\nu}\right) u^{\alpha\beta}(p.q)\\ \hline
	2 &
	\left(k^2g^{\mu\nu} - k^{\mu} k^{\nu}\right) w^{\alpha\beta}(p.q)  \\ \hline
	3 & \left(p^2 g^{\mu\nu} - 4 p^{\mu}  p^{\nu}\right)
	u^{\alpha\beta}(p.q)\\ \hline
	4 & \left(p^2 g^{\mu\nu} - 4 p^{\mu} p^{\nu}\right)
	w^{\alpha\beta}(p.q)\\ \hline
	5 & \left(q^2 g^{\mu\nu} - 4 q^{\mu} q^{\nu}\right)
	u^{\alpha\beta}(p.q)\\ \hline
	6 & \left(q^2 g^{\mu\nu} - 4 q^{\mu} q^{\nu}\right)
	w^{\alpha\beta}(p.q) \\ \hline
	7 & \left[p\cdot q\, g^{\mu\nu}
	-2 (q^{\mu} p^{\nu} + p^{\mu} q^{\nu})\right] u^{\alpha\beta}(p.q) \\ \hline
	8 & \left[p\cdot q\, g^{\mu\nu}
	-2 (q^{\mu} p^{\nu} + p^{\mu} q^{\nu})\right] w^{\alpha\beta}(p.q)\\ \hline
	9 & \left(p\cdot q \,p^{\alpha}  - p^2 q^{\alpha}\right)
	\big[p^{\beta} \left(q^{\mu} p^{\nu} + p^{\mu} q^{\nu} \right) - p\cdot q\,
	(g^{\beta\nu} p^{\mu} + g^{\beta\mu} p^{\nu})\big]  \\ \hline
	10 & \big(p\cdot q \,q^{\beta} - q^2 p^{\beta}\big)\,
	\big[q^{\alpha} \left(q^{\mu} p^{\nu} + p^{\mu} q^{\nu} \right) - p\cdot q\,
	(g^{\alpha\nu} q^{\mu} + g^{\alpha\mu} q^{\nu})\big]  \\ \hline
	11 & \left(p\cdot q \,p^{\alpha} - p^2 q^{\alpha}\right)
	\big[2\, q^{\beta} q^{\mu} q^{\nu} - q^2 (g^{\beta\nu} q^ {\mu}
	+ g^{\beta\mu} q^{\nu})\big]  \\ \hline
	12 & \big(p\cdot q \,q^{\beta} - q^2 p^{\beta}\big)\,
	\big[2 \, p^{\alpha} p^{\mu} p^{\nu} - p^2 (g^{\alpha\nu} p^ {\mu}
	+ g^{\alpha\mu} p^{\nu})\big] \\ \hline
	13 & \big(p^{\mu} q^{\nu} + p^{\nu} q^{\mu}\big)g^{\alpha\beta}
	+ p\cdot q\, \big(g^{\alpha\nu} g^{\beta\mu}
	+ g^{\alpha\mu} g^{\beta\nu}\big) - g^{\mu\nu} u^{\alpha\beta} \\
	& -\big(g^{\beta\nu} p^{\mu}
	+ g^{\beta\mu} p^{\nu}\big)q^{\alpha}
	- \big (g^{\alpha\nu} q^{\mu}
	+ g^{\alpha\mu} q^{\nu }\big)p^{\beta}  \\ \hline
	\end{array}
	$$
	\caption{Basis of 13 fourth rank tensors satisfying the vector current conservation on the external lines with momenta $p$ and $q$. \label{genbasis}}
\end{table}

In the set are present two tensor structures
\bes\bea
&&u^{\alpha\beta}(p,q) \equiv (p\cdot q) g^{\alpha\beta} - q^{\alpha}p^{\beta}\,,\\
&&w^{\alpha\beta}(p,q) \equiv p^2 q^2 g^{\alpha\beta} + (p\cdot q) p^{\alpha}q^{\beta}
- q^2 p^{\alpha}p^{\beta} - p^2 q^{\alpha}q^{\beta}\,,
\eea \label{uwdef}\ees
which appear in $t_1$ and $t_2$ respectively.
Each of them satisfies the Bose symmetry  requirement,
\bes\bea
&&u^{\alpha\beta}(p,q) = u^{\beta\alpha}(q,p)\,,\\
&&w^{\alpha\beta}(p,q) = w^{\beta\alpha}(q,p)\,,
\eea\ees
and vector current conservation,
\bes\bea
&&p_{\alpha} u^{\alpha\beta}(p,q) = 0 = q_{\beta}u^{\alpha\beta}(p,q)\,,\\
&&p_{\alpha} w^{\alpha\beta}(p,q) = 0 = q_{\beta}w^{\alpha\beta}(p,q)\,,
\eea\ees
obtained from the variation of gauge invariant quantities
$F_{\mu\nu}F^{\mu\nu}$ and $(\partial_{\mu} F^{\mu}_{\ \,\lambda})(\partial_{\nu}F^{\nu\lambda})$

\bea
&&u^{\alpha\beta}(p,q) = -\frac{1}{4}\int\,d^4x\,\int\,d^4y\ e^{ip\cdot x + i q\cdot y}\ 
\frac{\delta^2 \{F_{\mu\nu}F^{\mu\nu}(0)\}} {\delta A_{\alpha}(x) A_{\beta}(y)} \,,
\label{one}\\
&&w^{\alpha\beta}(p,q) = \frac{1}{2} \int\,d^4x\,\int\,d^4y\ e^{ip\cdot x + i q\cdot y}\
\frac{\delta^2 \{\partial_{\mu} F^{\mu}_{\ \,\lambda}\partial_{\nu}F^{\nu\lambda}(0)\}} 
{\delta A_{\alpha}(x) A_{\beta}(y)}\,.\label{two}
\eea\label{three}
 All the $t_i$'s are transverse in their photon indices
\bea
q^\alpha t_i^{\mu\nu\alpha\beta}=0  \qquad p^\beta t_i^{\mu\nu\alpha\beta}=0.
\eea
$t_2\ldots t_{13}$ are traceless, $t_1$ and $t_2$ are tracefull. With this decomposition, the two vector Ward identities are automatically satisfied by all the amplitudes, as well as the Bose symmetry. \\ 
Coming to the conservation WI for the graviton line, this is automatically satisfied by the two tensor structures $t_1$ and $t_2$, which are completely transverse, while it has to be imposed on the second set ($t_3\ldots t_{13}$) giving 3 constraints  
 
\bea
&& - p^2 F_3 + (3 q^2 + 4 p\cdot q) F_5 + (2 p^2 + p\cdot q) F_7 - p^2 q^2 F_{10}
- p^2 (p^2 + p\cdot q) F_9 + p^2 q^2 F_{11} = 0\,, \label{WIcons1a}\nn 
&& p^2 F_4 - (3 q^2 + 4 p\cdot q) F_6 - (2 p^2 + p\cdot q) F_8 - p\cdot q F_{10}
+ (q^2 + 2 p\cdot q) F_{11} = 0\,,\label{WIcons1b}\nn
&& -p\cdot q \,(p^2 + p\cdot q) F_9 - q^2 (q^2 + p\cdot q) F_{11} + F_{13} + \Pi(p^2)  =0 \,,
\label{cwi}\eea
plus 3 symmetric additional ones, obtained by the exchange of the two photon momenta and the symmetries of the form factors corresponding to \eqref{pair}
\bea
\label{pair1}
&&F_3(p,q) = F_5(q,p)\,,\\
&&F_4(p,q) = F_6(q,p)\,,\\
&&F_9(p,q) = F_{10}(q,p)\,,\\
&&F_{11}(p,q) = F_{12}(q,p)\,.
\eea
In other words, if we decided to identify from the $F_3\ldots F_{13}$ components a complete transverse traceless sector, using \eqref{pair1} and \eqref{cwi} we would identify only 4 components in this sector. Such 4 components, obviously, would be related to the transverse and traceless form factors $(A_1\ldots A_4)$ introduced in the BMS parametrization. Their explicit expressions will be given below. An important aspect of the $F-$basis is that {\em only 1-form factor has to be renormalized} for dimensional reasons, the others being finite. Such form factor, $F_{13}$, plays an important role in the description of the behaviour of the trace parts 
of the same expansion, which involve $t_1$ and $t_2$ (i.e. $F_1$ and $F_2$).
 
 \subsection{Dilatation Ward Identities in the \texorpdfstring{$F$}{}-basis}
The identification of the combination of form factors $F_i$ which span the transverse traceless sector of the correlator can proceed in several ways. In this and in the next section we will proceed by starting from its general expansion in the $F$-basis, and perform a transverse traceless projection, after acting on it with the dilatation and the special conformal transformations. 
This allows to gather the result of the action of the dilatation in terms of coefficients $D_i$ in the form 

\begin{subequations}
\begin{align}
0&=\Pi^{\m_1\n_1}_{\a_1\b_1}(p_1)\pi^{\m_2}_{\a_2}(p_2)\pi^{\m_3}_{\a_3}(p_3)\big\{D_1\,p_2^{\a_1}p_2^{\b_1}p_3^{\a_2}p_1^{\a_3} \notag\\
&\hspace{2cm}+ D_2\,\d^{\a_2\a_3} p_2^{\a_1}p_2^{\b_1}+ D_3\d^{\a_1\a_2}p_2^{\b_1}+p_1^{\a_3}D_4\,\d^{\a_1\a_3}p_2^{\b_1}p_3^{\a_2}+D_5\,\d^{\a_1\a_3}\d^{\a_2\b_1}\big\}
\end{align}
\end{subequations}
where $D_i$, $i=1,\dots,5$ are differential operator acting on the $F_i$ form factors. In order to verify the previous relation, the coefficients $D_i$ multiplying the independent tensor structures have to vanish, giving a set of differential equation for particular combination of the $F_i$'s. The first equation for $D_1$ will be of the form
\begin{align}
D_1&=4\left[\sum_{i=1}^3p_i\sdfrac{\partial}{\partial p_i}-(d-6)\right](F_7-F_3-F_5)\notag\\
&\hspace{2cm}-2p_2^2\left[\sum_{i=1}^3p_i\sdfrac{\partial}{\partial p_i}-(d-8)\right] F_9-2p_3^2\left[\sum_{i=1}^3p_i\sdfrac{\partial}{\partial p_i}-(d-8)\right]F_{10}=0
\end{align}
which can be rearranged as
\begin{align}
&D_1=\left(\sum_{i=1}^3p_i\sdfrac{\partial}{\partial p_i}-(d-6)\right)\big[ 4(F_7-F_5-F_3)-2p_2^2F_9-2p_3^2F_{10}\big]=0,
\end{align}
and similarly for the other $D_i$'s, which correspond to
\begin{align}
&D_2\left(\sum_{i=1}^3p_i\sdfrac{\partial}{\partial p_i}-(d-4)\right)\big[2(p_1^2-p_2^2-p_3^2)(F_7-F_5-F_3)-4p_2^2p_3^2(F_6-F_8+F_4)-2F_{13}\big]=0\\
&D_3=\left(\sum_{i=1}^3p_i\sdfrac{\partial}{\partial p_i}-(d-4)\right)\big[p_3^2(p_1^2-p_2^2-p_3^2)F_{10}-2p_2^2\,p_3^2 F_{12}-2F_{13}\big]=0\\
&D_4=\left(\sum_{i=1}^3p_i\sdfrac{\partial}{\partial p_i}-(d-4)\right)\big[p_2^2(p_1^2-p_2^2-p_3^2)F_9-2p_2^2p_3^2F_{11}-2F_{13}\big]=0\\
&D_5=\left(\sum_{i=1}^3p_i\sdfrac{\partial}{\partial p_i}-(d-2)\right)\big[(p_1^2-p_2^2-p_3^2)F_{13}\big]=0.
\end{align}
This allows us to identify specific combinations of the $F$'s which will span the transverse traceless sector of the $TJJ$.

\subsection{Special Conformal Ward identities in the \texorpdfstring{$F$}{}-basis}
A similar approach can be followed in the case of the primary and secondary CWI's.
Also in this case we project the special CWI's onto the transverse traceless sector, obtaining
\begin{align}
0&=\Pi^{\rho_1\sigma_1}_{\mu_1\nu_1}(p_1)\pi^{\rho_2}_{\mu_2}(p_2)\pi^{\rho_3}_{\mu_3}(p_3)\ \bigg\{\,K^\kappa\,\braket{{t^{\mu_1\nu_1}(p_1)\,j^{\mu_2}(p_2)\,j^{\mu_3}(p_3)}}\notag\\
&\hspace{5.5cm}+\sdfrac{4d}{p_1^2}\,\delta^{\kappa\mu_1}\,p_{1\alpha_1}\,\braket{{T^{\alpha_1\nu_1}(p_1)J^{\mu_2}(p_2)J^{\mu_3}(p_3)}}\bigg\}\label{Constr}
\end{align}
where we have used the conservation Ward Identities
\begin{subequations}
\begin{align}
p_{2\m_2}\braket{{T^{\m_1\n_1}(p_1)\,J^{\m_2}(p_2)\,J^{\m_3}(p_3)}}&=0 \nn
p_{3\m_3}\braket{{T^{\m_1\n_1}(p_1)\,J^{\m_2}(p_2)\,J^{\m_3}(p_3)}}&=0.
\end{align}
\end{subequations}

Also in this case one can express the first term in \eqref{Constr} in the form of \eqref{StrucSWIS} in order to isolate the primary and secondary WI's for the form factors. 
\subsection{Primary WI's}
A first set of primary conformal WI's is given by
\begin{subequations}
\begin{align}
0=&K_{13}\big[ 4(F_7-F_5-F_3)-2p_2^2F_9-2p_3^2F_{10}\big]\\[2ex]
0=&K_{13}\big[2(p_1^2-p_2^2-p_3^2)(F_7-F_5-F_3)-4p_2^2p_3^2(F_6-F_8+F_4)-2F_{13}\big]\notag\\
&\hspace{5cm}+2\big[ 4(F_7-F_5-F_3)-2p_2^2F_9-2p_3^2F_{10}\big]\\[2ex]
0=&K_{13}\big[p_3^2(p_1^2-p_2^2-p_3^2)F_{10}-2p_2^2\,p_3^2 F_{12}-2F_{13}\big]\notag\\
&\hspace{5cm}-4\big[ 4(F_7-F_5-F_3)-2p_2^2F_9-2p_3^2F_{10}\big]\\[2ex]
0=&K_{13}\big[p_2^2(p_1^2-p_2^2-p_3^2)F_9-2p_2^2p_3^2F_{11}-2F_{13}]\\[2ex]
0=&K_{13}\big[(p_1^2-p_2^2-p_3^2)F_{13}\big]-2\big[p_2^2(p_1^2-p_2^2-p_3^2)F_9-2p_2^2p_3^2F_{11}-2F_{13}\big]
\end{align}
\end{subequations}
and a second set as
\begin{subequations}
	\begin{align}
	0=&K_{23}\big[ 4(F_7-F_5-F_3)-2p_2^2F_9-2p_3^2F_{10}\big]\\[2ex]
	0=&K_{23}\big[2(p_1^2-p_2^2-p_3^2)(F_7-F_5-F_3)-4p_2^2p_3^2(F_6-F_8+F_4)-2F_{13}\big]\\[2ex]
	0=&K_{23}\big[p_3^2(p_1^2-p_2^2-p_3^2)F_{10}-2p_2^2\,p_3^2 F_{12}-2F_{13}\big]\notag\\
	&\hspace{5cm}-4\big[ 4(F_7-F_5-F_3)-2p_2^2F_9-2p_3^2F_{10}\big]\\[2ex]
	0=&K_{23}\big[p_2^2(p_1^2-p_2^2-p_3^2)F_9-2p_2^2p_3^2F_{11}-2F_{13}]\notag\\
	&\hspace{5cm}+4\big[ 4(F_7-F_5-F_3)-2p_2^2F_9-2p_3^2F_{10}\big]\\[2ex]
	0=&K_{23}\big[(p_1^2-p_2^2-p_3^2)F_{13}\big]+2\big[p_3^2(p_1^2-p_2^2-p_3^2)F_{10}-2p_2^2\,p_3^2 F_{12}-2F_{13}\big]\notag\\
	&\hspace{5cm}-2\big[p_2^2(p_1^2-p_2^2-p_3^2)F_9-2p_2^2p_3^2F_{11}-2F_{13}\big].
	\end{align}
\end{subequations}
It is clear from the way in which we have organized the contributions in square brackets ($[\,\,\,]$) 
that they correspond to the same structures identified in the projections of the dilatation WI's. 
\subsection{Secondary WI's}
For completeness we list the secondary Ward Identities obtained in a similar way, which are given by
\begin{align}
0&=L'_3\big[4(F_7-F_3-F_5)-2p_2^2F_9-2p_3^2F_{10}\big]+2R'\big[p_3^2(p_1^2-p_2^2-p_3^2)F_{10}-2p_2^2\,p_3^2 F_{12}-2F_{13}\big]\notag\\
&\hspace{1cm}-2R'\big[2(p_1^2-p_2^2-p_3^2)(F_7-F_5-F_3)-4p_2^2p_3^2(F_6-F_8+F_4)-2F_{13}\big]\\[2.5ex]
0&=L'_1\big[p_2^2(p_1^2-p_2^2-p_3^2)F_9-2p_2^2p_3^2F_{11}-2F_{13}\big]-2p_2^2\big[p_3^2(p_1^2-p_2^2-p_3^2)F_{10}-2p_2^2\,p_3^2 F_{12}-2F_{13}\big]\notag\\
&\hspace{1cm}+4p_2^2\big[2(p_1^2-p_2^2-p_3^2)(F_7-F_5-F_3)-4p_2^2p_3^2(F_6-F_8+F_4)-2F_{13}\big]+2R'\big[(p_1^2-p_2^2-p_3^2)F_{13}\big]\notag\\[2.5ex]
\end{align}
\begin{align}
0&=-\sdfrac{2}{p_1^2}\bigg\{L_4\big[4(F_7-F_3-F5)-2p_2^2F_9-2p_3^2F_{10}\big]+R\big[p_3^2(p_1^2-p_2^2-p_3^2)F_{10}-2p_2^2\,p_3^2 F_{12}-2F_{13}\big]\notag\\
&\hspace{0.5cm}-R\big[p_2^2(p_1^2-p_2^2-p_3^2)F_9-2p_2^2p_3^2F_{11}-2F_{13}\big]\bigg\}+\sdfrac{4d}{p_1^2}\bigg[p_3^2(p_1\cdot p_3-p_1\cdot p_2-p_2\cdot p_3)F_{10}-4F_3p_1\cdot p_2\notag\\
&\hspace{2cm}-p_2^2p_3^2(F_{11}-F_{12})+4F_5p_1\cdot p_3+2(p_1\cdot p_2-p_1\cdot p_3)F_7+p_2^2(p_1\cdot p_3-p_1\cdot p_2+p_2\cdot p_3)F_9\bigg]\notag\\[2.5ex]
\end{align}
\begin{align}
0&=-\sdfrac{2}{p_2^2}\Big\{\,L_2\big[2(p_1^2-p_2^2-p_3^2)(F_7-F_5-F_3)-4p_2^2p_3^2(F_6-F_8+F_4)-2F_{13}\big]\notag\\
&\hspace{1cm}-p_1^2(p_1^2-p_2^2-p_3^2)(p_3^2F_{10}-p_2^2F_9)+2p_3^2p_2^2p_1^2(F_{12}-F_{11})\Big\}+\sdfrac{4d}{p_1^2}\big[(p_2^2-p_3^2)F_{13}\notag\\
&\hspace{1cm}+(p_3^4+p_1^4-p_2^4-2p_1^2p_3^2)F_3+2p_2^2p_3^2(p_1^2+p_2^2-p_3^2)F_4+(2p_1^2p_2^2-p_1^4-p_2^4+p_3^4)F_5\notag\\
&\hspace{1cm}+2p_2^2p_3^2(p_2^2-p_3^2-p_1^2)F_6+2p_2^2p_3^2(p_3^2-p_2^2)F_8+(p_2^4-p_3^4+p_1^2p_3^2-p_1^2p_2^2)F_7\big]\\[2.5ex]
0&=-\sdfrac{1}{p_1^2}\Big\{L_4\big[p_3^2(p_1^2-p_2^2-p_3^2)F_{10}-2p_2^2\,p_3^2 F_{12}-2F_{13}\big]-2R\big[(p_1^2-p_2^2-p_3^2)F_{13}\big]\Big\}\notag\\
&\hspace{1cm}+\sdfrac{4d}{p_1^2}\big[p_1\cdot p_3p_2\cdot p_3p_3^2F_{10}+p_\cdot p_2p_2^2p_3^2F_{12}-p_2^2F_{13}\big]\\[2.5ex]
0&=-\sdfrac{1}{p_1^2}\Big\{L_4\big[p_2^2(p_1^2-p_2^2-p_3^2)F_9-2p_2^2p_3^2F_{11}-2F_{13}\big]+2R\big[(p_1^2-p_2^2-p_3^2)F_{13}\big]\notag\\
&\hspace{0.5cm}-4p_1^2\big[p_2^2(p_1^2-p_2^2-p_3^2)F_9-2p_2^2p_3^2F_{11}-2F_{13}\big]\Big\}+\sdfrac{4d}{p_1^2}\big[p_2^2p_3^2\,p_1\cdot p_3 F_{11}+p_2^2F_{13}+p_1\cdot p_2 p_2^2 p_3\cdot p_2 F_9\big].
\end{align}
We are now going to use the results above in order to identify the link between the two transverse sections in the $F$-basis introduced by the perturbative expansion and the $A-$basis of the transverse traceless sector. Notice that the 13 form factors of the F-basis form a complete basis in $d$-dimensions, and have some nice properties, as we are going to emphasize below.

\subsection{Connection between the \texorpdfstring{$A-$}{} and the \texorpdfstring{$F-$}{} basis}
By a direct analysis of the previous primary and secondary constraint in the $F-$basis, using the equations given in Sections \ref{primsection} and \ref{secsection} for the 
$A_i$ form factors, we obtain the relations which define the mapping between the transverse traceless sectors in the two basis, which is given by
\begin{subequations}
\begin{align}
&A_1=4(F_7-F_3-F_5)-2p_2^2F_9-2p_3^2F_{10}\\
&A_2=2(p_1^2-p_2^2-p_3^2)(F_7-F_5-F_3)-4p_2^2p_3^2(F_6-F_8+F_4)-2F_{13}\\
&A_3=p_3^2(p_1^2-p_2^2-p_3^2)F_{10}-2p_2^2\,p_3^2 F_{12}-2F_{13}\\
&A_3(p_2\leftrightarrow p_3)=p_2^2(p_1^2-p_2^2-p_3^2)F_9-2p_2^2p_3^2F_{11}-2F_{13}\\
&A_4=(p_1^2-p_2^2-p_3^2)F_{13}.
\end{align} \label{mapping}
\end{subequations}
It is worth noticing that the form factor $A_3$ and its corresponding $A_3(p_2\leftrightarrow p_3)$ are well-defined since 
\begin{equation}
F_{10}(s;s_1,s_2,0)=F_9(s;s_2,s_1,0),\quad F_{12}(s;s_1,s_2,0)=F_{11}(s;s_2,s_1,0).
\end{equation}

Going back to the full perturbative amplitude we can re-express the entire correlator as
\begin{equation}
\braket{T^{\m_1\n_1}(p_1)\,J^{\m_2}(p_2)\,J^{\m_3}(p_3)}=\braket{t^{\m_1\n_1}(p_1)\,j^{\m_2}(p_2)\,j^{\m_3}(p_3)}+\braket{t^{\m_1\n_1}_{loc}(p_1)\,J^{\m_2}(p_2)\,J^{\m_3}(p_3)}
\end{equation}
where the semi-local term is expressed exactly as
\begin{align}
&\braket{t^{\m_1\n_1}_{loc}(p_1)\,J^{\m_2}(p_2)\,J^{\m_3}(p_3)}=\notag\\
&\qquad=\sdfrac{p_{1\b_1}}{p_1^2}\bigg[2p_1^{(\m_1}\d^{\n_1)}_{\a_1}-\sdfrac{p_{1\a_1}}{(d-1)}\left(\d^{\m_1\n_1}+(d-2)\sdfrac{p_1^{\m_1}p_1^{\n_1}}{p_1^2}\right)\bigg]\,\braket{T^{\a_1\b_1}(p_1)\,J^{\m_2}(p_2)\,J^{\m_3}(p_3)}
\label{loc}
\end{align}
and the transverse traceless part is reconstructed as 
\begin{align}
&\braket{{t^{\m_1\n_1}(p_1)\,j^{\m_2}(p_2)\,j^{\m_3}(p_3)}}_{pert}=\Pi^{\m_1\n_1}_{\a_1\b_1}(p_1)\pi^{\m_2}_{\a_2}(p_2)\pi^{\m_3}_{\a_3}(p_3)\times\notag\\
&\hspace{0.5cm}\times \bigg\{\ \big[4(F_7-F_3-F_5)-2p_2^2F_9-2p_3^2F_{10}\big]\,p_2^{\a_1}p_2^{\b_1}p_3^{\a_2}p_1^{\a_3}\notag\\
&\hspace{1cm}+ \big[2(p_1^2-p_2^2-p_3^2)(F_7-F_5-F_3)-4p_2^2p_3^2(F_6-F_8+F_4)-2F_{13}\big]\ \d^{\a_2\a_3} p_2^{\a_1}p_2^{\b_1}\notag\\
&\hspace{1cm} +\big[\,p_3^2(p_1^2-p_2^2-p_3^2)F_{10}-2p_2^2\,p_3^2 F_{12}-2F_{13}\big]\,\d^{\a_1\a_2}p_2^{\b_1}p_1^{\a_3}\notag\\
&\hspace{1cm} +\big[p_2^2(p_1^2-p_2^2-p_3^2)F_9-2p_2^2p_3^2F_{11}-2F_{13}\big]\,\d^{\a_1\a_3}p_2^{\b_1}p_3^{\a_2}+ \big[(p_1^2-p_2^2-p_3^2)F_{13}\big]\d^{\a_1\a_3}\d^{\a_2\b_1}\,\bigg\}.
\end{align}
Notice that neither $F_1$ nor $F_2$ will be part of the local contributions since they are both 
completely traceless. Therefore in the $F$-basis, the contributions appearing in \eqref{loc} will 
be combinations of $F_3\ldots F_{13}$ which are independent from the 4 combinations of the $F'$s identified by the mapping \eqref{mapping}.\\
Since we are still defining the correlator in $d-$dimensions, and it is conformal in this case, then its 
d-dimensional trace has to vanish. 
This condition brings in two additional constraints on the two form factors $F_1$ and $F_2$, which now enter into the analysis,  
\begin{align}
F_1&=\frac{(d-4)}{p_1^2(d-1)}\big[F_{13}-p_2^2\,F_3-p_3^2\,F_5-p_2\cdot p_3\, F_7\big]\\
F_2&=\sdfrac{(d-4)}{p_1^2(d-1)}\big[p_2^2\,F_4+p_3^2\,F_6+p_2\cdot p_3\,F_8\big],
\label{confeq}
\end{align}
which will be important for the renormalization procedure and the identification of the anomaly term. \\
We remark that the independent analysis of $A_4$ \cite{Bzowski:2017poo}, which has essentially the same structure 
as $F_{13}$, as one can immediately realize from \eqref{mapping}, shows that in a general conformal field theory the singularity of $F_{13}$ can only be of order $1/(d-4)$ 
and not any higher. We are now going to test such general analysis to the specific case of QED at one-loop.
\subsection{\texorpdfstring{$F_{13}$}{} in QED}
The expressions of the 13 $F_i$ form factors in $d-$ dimensions are shown in the appendix. The renormalized results in $d=4$ have been given in \cite{Armillis:2009pq}. Notice that we have expressed their expressions directly in terms of the two master integrals   
\bea
B_0(s,0,0)&=&\int \frac{d^d l}{i\pi^2}\frac{1}{l^2(l-p_1)^2}\equiv B_0(s)\nn
C_0 (s, s_1, s_2) &=&
 \frac{1}{i \pi^2} \int d^d l \, \frac{1}{(l^2 ) \, ((l -q )^2  ) \, ((l + p )^2  )} 
\eea
which will be useful for the discussion of the action of the conformal generators on each of them. 
Introducing the variables $\s\equiv s^2-2s(s_1+s_2)+(s_1-s_2)^2$ and $\g\equiv s-s_1-s_2$, $F_{13}$ takes the form 
 
\begin{align}
&F_{13, d}(s,s_1,s_2)=\frac{2 \pi ^2 e^2 s^2}{( d-2) ( d-1)  d \sigma ^2} \bigg\{-2 s_2 \big[( d^2-3 d+4) (s+s1)-4( d-1) s_1\big]\notag\\
&+( d^2-3 d+4) \big[(s-s_1)^2+s_2^2\big]\bigg\}B_0(s)-\sdfrac{\pi ^2 e^2 s_1}{( d-2) ( d-1)  d \sigma ^2 \,\g} \bigg\{ d^3 \sigma ^2- d^2 \big[3 s^4-2 s^3 (7 s_1+9 s_2)\notag\\
&+8 s^2 \left(3 s_1^2+3 s_1 s_2+4 s_2^2\right)-2 s (s_1-s_2)^2 (9 s_1+11 s_2)+5 (s_1-s_2)^4\big]+2  d \big[2 s^4-11 s^3 (s_1+s_2)\notag\\
&+s^2 \left(21 s_1^2+24 s_1 s_2+19 s_2^2\right)+s \left(-17 s_1^3+5 s_1^2 s_2+25 s_1 s_2^2-13 s_2^3\right)+(s_1-s_2)^3 (5 s_1-3 s_2)\big]\notag\\
&+8 s_1 \left(s^3-3 s^2 (s_1+s_2)+3 s \left(s_1^2-s_2^2\right)-(s_1-s_2)^3\right)\bigg\}B_0(s_1)-\sdfrac{\pi ^2 e^2 s_2}{( d-2) ( d-1)  d \sigma ^2\g} \bigg\{ d^3 \sigma ^2\notag\\
&- d^2 \big[3 s^4-2 s^3 (9 s_1+7 s_2)+8 s^2 \left(4 s_1^2+3 s_1 s_2+3 s_2^2\right)-2 s (s_1-s_2)^2 (11 s_1+9 s_2)+5 (s_1-s_2)^4\big]\notag\\
&+2  d \big[2 s^4-11 s^3 (s_1+s_2)+s^2 \left(19 s_1^2+24 s_1 s_2+21 s_2^2\right)+s \left(-13 s_1^3+25 s_1^2 s_2+5 s_1 s_2^2-17 s_2^3\right)\notag\\
&+(s_1-s_2)^3 (3 s_1-5 s_2)\big]+8 s_2 \big[s^3-3 s^2 (s_1+s_2)-3 s \left(s_1^2-s_2^2\right)+(s_1-s_2)^3\big]\bigg\}B_0(s_2)\notag\\
&+\sdfrac{4 \pi ^2 e^2 s^2 s_1 s_2 ( d \sigma +8 s_1 s_2)}{( d-2)  d \sigma ^2 \g}C_0(s,s_1,s_2).
\end{align}
which we will study in the $d\to 4$ limit. As discussed in \cite{Giannotti:2008cv} \cite{Armillis:2009pq}, the singularity of this form factor comes from the scalar form factor $\Pi(p^2)$ of the photon 2-point function $\braket{JJ}$. 
The singularity of this correlator will be at all orders of the form $1/\epsilon$ in a conformal theory and not higher. This is 
a crucial point in the proof which is clearly not satisfied in a non-conformal theory. In fact, the only available counterterm to regulate a conformal theory is given by 
\be
\frac{1}{\epsilon}\int d^4 x \sqrt{g} F_{\mu\nu}F^{\mu\nu}
\ee
which renormalizes the 2-point function $\langle JJ \rangle$ and henceforth $F_{13}$. Explicit computations in QED at one-loop, where the theory is conformal, show that 
\be
\label{f13}
F_{13}= G_0(p_1^2, p_2^2,p_3^3) -\frac{1}{2} \, [\Pi (p_2^2) +\Pi (p_3^2)]
\ee
We just recall that the structure of the two-point function of two conserved vector currents of scaling dimensions $\eta_1$ and $\eta_2$ is given by \cite{Coriano:2013jba}
\be
\label{TwoPointVector}
G_V^{\alpha \beta}(p) = \delta_{\eta_1 \eta_2}  \, c_{V 12}\, 
\frac{\pi^{d/2}}{4^{\eta_1 - d/2}} \frac{\Gamma(d/2 - \eta_1)}{\Gamma(\eta_1)}\,
\left( \eta^{\alpha \beta} -\frac{p^\alpha p^\beta}{p^2} \right)\
(p^2)^{\eta_1-d/2} \,,
\ee
with $c_{V12}$ being an arbitrary constant. It will be nonvanishing only if the two currents share the same dimensions, and it is characterized just by a single pole (to all orders) ${1/\epsilon}$ in dimensional regularization. The divergence can be regulated with $d \to d - 2 \epsilon$, and expanding the product $\Gamma(d/2-\eta)\,(p^2)^{\eta - d/2}$ in \ref{TwoPointVector} in a Laurent series around $d/2 - \eta = -n$ (integer) one can extract the single pole in $1/\epsilon$ in the form \cite{Coriano:2013jba}
\bea
\label{expansion}
\Gamma\left(d/2-\eta\right)\,(p^2)^{\eta-d/2} = \frac{(-1)^n}{n!} \left( - \frac{1}{\epsilon} + \psi(n+1)  + O(\epsilon) \right) (p^2)^{n + \epsilon} \,,
\eea
where $\psi(z)$ is the logarithmic derivative of the Gamma function, and $\epsilon$ takes into account the divergence of the two-point correlator for particular values of the scale dimension $\eta$ and of the space-time dimension $d$.
In the QED case, the renormalization involves only the master integrals $B_0(s_i)$, which gives ($d=4 -2 \epsilon$)
\be
\label{f13ren}
F_{13}=\frac{2 \pi^2 e^2}{3 \epsilon} + F_{13}^{fin},
\ee
implying that $F_1$ (from \eqref{confeq}) will be given by
\be
\label{limit}
F_1=\left(\sdfrac{2}{3}\sdfrac{\epsilon}{s} \right)\left(\frac{2 \pi e^2}{3\epsilon} + F_{13}^{fin} \right),
\ee
which in the $d\to 4$ limit gives 
\be
\lim_{d\to 4} F_1=-\frac{4}{9}\frac{\pi^2}{s},
\ee
showing the appearance of an anomaly pole in the single form factor which is responsible for the trace anomaly.\\
 It is quite obvious that the non-perturbative analysis of \cite{Bzowski:2017poo} in the $A-$basis and the perturbative ones in the $F-$basis are consistent. There is some additional  important information that we can extract in the latter basis if we go back to the two equations in \eqref{confeq}.
 
  1. From the finiteness of all the form factors, except for $F_{13}$ which is regulated with a $1/\epsilon$ divergence, it is obvious that in the limit of $d\to 4$, as a result of \eqref{limit},
  $F_1$ is nonvanishing and exhibits a $1/p_1^2\equiv 1/s$ behaviour. Therefore, the emergence of $F_1$ in $d=4$ as a form factor which accounts for the anomaly is a nice feature if this general analysis. It shows how to link an anomaly pole to the renormalization of a single form factor in the expansion of the correlator. 
  
  2. At the same time, it is possible to check explicitly from its $d-$dimensional expression shown in \eqref{F2} that the second form factor $F_2$ vanishes as $d\to 4$, proving that there will be one and only one tensor structure of nonzero trace.
  
  3. Although the results above are fully confirmed by the previous perturbative analysis, 
  they hold generically (non perturbatively) in the context of the conformal realizations of such correlator.\\ 
   A natural question to ask is what happens to the tensor structures $t_2,\ldots t_{13}$ as we move from $d$ to 4 dimensions. The answer is quite immediate.  We contract such 
 structures with the $d-$dimensional metric $g_{\mu\nu}(d)$ and perform the $d\to 4$ limit. One can easily check that $t_9,t_{10},t_{11}$ and $t_{12}$, remain traceless in any dimensions, 
while the remaining ones become traceless in this limit. For instance 
 \bea
g^{\mu\nu}(d)t_1^{\mu\nu\alpha\beta}&=&(d-4) k^2 u^{\alpha\beta}(p,q)\nn 
g^{\mu\nu}(d)t_2^{\mu\nu\alpha\beta}&=&(d-4) k^2 w^{\alpha\beta}(p,q),
\eea
and similarly for the others. Therefore, in the $d\to 4$ limit, the $F-$basis satisfies all the original constraints and the separation between traceless and trace-contributions which were described in Sect. \eqref{pchecks}.
\begin{figure}[t]
\centering
\vspace{-3.5cm}
\includegraphics[scale=0.4]{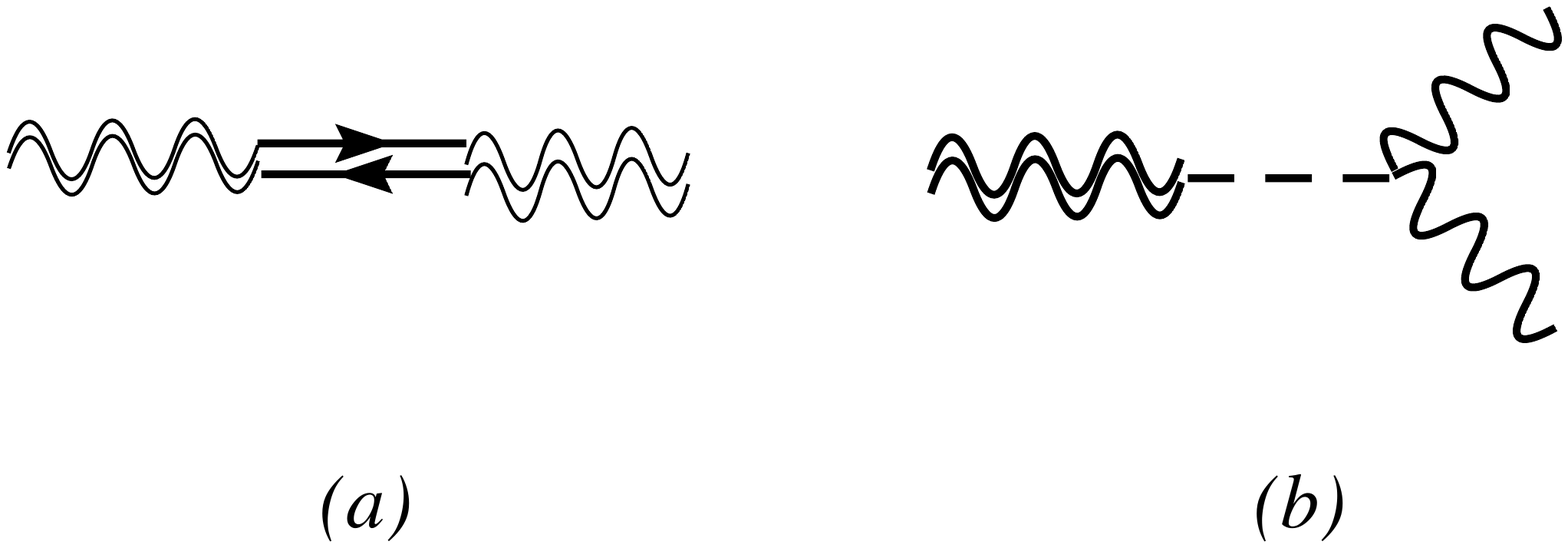}
\vspace{-3.3cm}
\caption{\small Fig. (a): Dispersive description of the singularity of the spectral density $\rho(s)$ as a spacetime process. Fig. (b): The exchange of a pole as the origin of the conformal anomaly in the TJJ viewed in perturbation theory.}
\label{collinear}
\end{figure}
\subsection{Implications}
We may summarize the result of this section by saying that the emergence of an anomaly pole in the $TJJ$ is not limited to perturbation theory but is a specific feature of the non-perturbative solution as well. The perturbative description offers a simple view of why this phenomenon takes place. In the dispersive representation of the unique form factor which is responsible for the appearance of the anomaly $(F_1)$, this phenomenon is related to the exchange of a collinear fermion/antifermion pair in the $s$ variable $(\rho(s)\sim \delta(s))$ (see Fig. \ref{collinear}). This configuration provides a contribution to the anomaly action of the form 
\be
 \label{pole}
\mathcal{S}_{pole}= - \frac{e^2}{ 36 \pi^2}\int d^4 x d^4 y \left(\square h(x) - \partial_\mu\partial_\nu h^{\mu\nu}(x)\right)  \square^{-1}_{x\, y} F_{\alpha\beta}(x)F^{\alpha\beta}(y)
\ee
In view of the equivalence between the perturbative and the nonperturbative solution for the $A_i$ (and henceforth for the $F$- basis), which will be discussed in section 13, it is obvious that this phenomenon is lifted from its perturbative origin and acquires a general meaning. We refer to \cite{Giannotti:2008cv,Coriano:2014gja} for a general perturbative analysis 
of the spectral densities for such type of vertices.

There is no doubt that this is a one-loop phenomenon in QED which is obviously violated at higher orders, since the theory, in this case, ceases to be conformal. However, as we are going to show in the next section, the one-loop expression in QED reproduces the {\em entire} non perturbative conformal BMS solution, and this explains why our proof should be considered a definitive prove of the fact that, at least for this correlator, the exchange of effective massless interactions is the key signature of the conformal anomaly. 

\section{Conformal Ward identities and the perturbative master integrals}
We now turn to illustrate the action of the conformal generators on the $d-$dimensional expressions of the $F_i$, showing that they are indeed solutions of the corresponding CWI's. We will first elaborate on the action of the conformal generators on the simple master integrals $B_0$, $C_0$. Such an action will be reformulated in terms of the external invariant of each master integral, starting from the original definitions of the special conformal $K^\kappa$ and dilatation $D$ operators. \\
 Understanding the way the conformal constraints work on perturbative realizations of conformal correlators is indeed important for various reasons. For instance, one can find, by a direct perturbative analysis simpler realizations of the general hypergeometric solutions of the CWI's for the $A_i$, while, at the same time, one can test the consistency of the general approach implemented in the solution of the conformal constraints which does not rely on an explicit Lagrangian but just on the data content of a given CFT. We are going to show that indeed such is the case and that the $d$-dimensional form factors given in the appendix satisfy all the conformal constraints corresponding to the dilatation and special conformal WI's.
 
 \subsection{The conformal constraints on the external invariants}
The action of the special conformal transformations $K^\kappa$ and of the dilatations $D$ on them, will require that $K_{ij}$, $L_N$, $R$ in \eqref{kij}, \eqref{Ldef} and \eqref{Rdef} be expressed in terms of the three invariants $s, s_1$ and $s_2$.  

Taking the four-momentum $\scalebox{0.8}{$p_1^\m$}$ as independent, we will be using the chain rules
\begin{align}
\sdfrac{\partial}{\partial p_2^\m}&=-\sdfrac{p_{1\m}}{p_1}\sdfrac{\partial}{\partial p_1}+\sdfrac{p_{2\m}}{p_2}\sdfrac{\partial}{\partial p_2}\\
\sdfrac{\partial}{\partial p_3^\m}&=-\sdfrac{p_{1\m}}{p_1}\sdfrac{\partial}{\partial p_1}+\sdfrac{p_{3\m}}{p_3}\sdfrac{\partial}{\partial p_3},
\end{align}
and by taking appropriate linear combinations of these relations we obtain the system of equations
\begin{equation}
\begin{pmatrix}
p_2^\m\sdfrac{\partial}{\partial p_2^\m}-p_2^\m\sdfrac{\partial}{\partial p_3^\m}\\[3ex]
p_3^\m\sdfrac{\partial}{\partial p_2^\m}-p_3^\m\sdfrac{\partial}{\partial p_3^\m}
\end{pmatrix}=
\begin{pmatrix}
&\\[-1ex]
p_2& -\sdfrac{p_2\cdot p_3}{p_3}\\[3ex]
\sdfrac{p_2\cdot p_3}{p_2}&-p_3\\[2ex]
\end{pmatrix} \begin{pmatrix}
\sdfrac{\partial}{\partial p_2}\\[3ex]
\sdfrac{\partial}{\partial p_3}
\end{pmatrix}\ .
\end{equation}

Solving the system above for the derivative of the magnitudes of the momenta we obtain the relations
\begin{subequations}
\begin{align}
\frac{\partial}{\partial p_2}&=\left(\frac{p_2\,p_3}{(p_2\cdot p_3)^2-p_2^2\,p_3^2}\right)\left[\frac{p_2\cdot p_3}{p_3}\left(p_3^\m\sdfrac{\partial}{\partial p_2^\m}-p_3^\m\sdfrac{\partial}{\partial p_3^\m}\right)-p_3\left(p_2^\m\sdfrac{\partial}{\partial p_2^\m}-p_2^\m\sdfrac{\partial}{\partial p_3^\m}\right)\right]\\[2ex]
\frac{\partial}{\partial p_3}&=\left(\frac{p_2\,p_3}{(p_2\cdot p_3)^2-p_2^2\,p_3^2}\right)\left[p_2\left(p_3^\m\sdfrac{\partial}{\partial p_2^\m}-p_3^\m\sdfrac{\partial}{\partial p_3^\m}\right)-\frac{p_2\cdot p_3}{p_2}\left(p_2^\m\sdfrac{\partial}{\partial p_2^\m}-p_2^\m\sdfrac{\partial}{\partial p_3^\m}\right)\right].
\end{align}\label{p2p3}
\end{subequations}
Afterwards, by taking other linear combinations we obtain
\begin{align}
\left\{\begin{matrix}
p_2^\m\sdfrac{\partial}{\partial p_2^\m}+p_3^\m\sdfrac{\partial}{\partial p_2^\m}=p_1\sdfrac{\partial}{\partial p_1}-\sdfrac{p_1\cdot p_2}{p_2}\sdfrac{\partial}{\partial p_2}\\[3ex]
p_3^\m\sdfrac{\partial}{\partial p_3^\m}+p_2^\m\sdfrac{\partial}{\partial p_3^\m}=p_1\sdfrac{\partial}{\partial p_1}-\sdfrac{p_1\cdot p_3}{p_3}\sdfrac{\partial}{\partial p_3}
\end{matrix}\right.\ .
\end{align}
which, combined together, give
\begin{align}
p_1\frac{\partial}{\partial p_1}=&\left(\frac{p_2\,p_3}{(p_2\cdot p_3)^2-p_2^2\,p_3^2}\right)\bigg[\frac{p_2\cdot p_3}{p_2\,p_3}\big((p_2\cdot p_3)+p_2^2\big)\,p_3^\m\sdfrac{\partial}{\partial p_3^\m}\notag\\[2ex]
&+\frac{p_2\cdot p_3}{p_2\,p_3}\big((p_2\cdot p_3)+p_3^2\big)\,p_2^\m\sdfrac{\partial}{\partial p_2^\m}-\frac{p_2}{p_3}\big((p_2\cdot p_3)+p_3^2\big)\,p_3^\m\sdfrac{\partial}{\partial p_2^\m}-\frac{p_3}{p2}\big((p_2\cdot p_3)+p_2^2\big)\,p_2^\m\sdfrac{\partial}{\partial p_3^\m}\bigg],\label{p1}
\end{align}

with the four-vector forms of the derivatives rearranged as
\begin{subequations}
\begin{align}
p_2^\m\frac{\partial}{\partial p_2^\m}\,B_0(s)&=\sdfrac{( d-4)}{2s}(s+s_1-s_2)\,B_0(s)\\ 
p_3^\m\frac{\partial}{\partial p_2^\m}\,B_0(s)&=\sdfrac{( d-4)}{2s}(s-s_1+s_2)\,B_0(s)\\
p_2^\m\frac{\partial}{\partial p_3^\m}\,B_0(s)&=\sdfrac{( d-4)}{2s}(s-s_1+s_2)\,B_0(s)\\
p_3^\m\frac{\partial}{\partial p_2^\m}\,B_0(s)&=\sdfrac{( d-4)}{2s}(s+s_1-s_2)\,B_0(s).
\end{align}
\end{subequations}
Inserting these expressions in \eqref{p2p3} and \eqref{p1} we obtain the relations
\begin{subequations}
\begin{align}
p_1\frac{\partial}{\partial p_1}B_0(s)=( d-4)\,B_0(s),\qquad\frac{\partial}{\partial p_2}B_0(s)=0,\qquad
\frac{\partial}{\partial p_3}B_0(s)=0.
\end{align}
\end{subequations}
The four-derivatives of $B_0(s_1)$ are computed in a similar way, obtaining
\begin{subequations}
	\begin{align}
	p_2^\m\frac{\partial}{\partial p_2^\m}\,B_0(s_1)&=( d-4)\,B_0(s_1)\\
	p_3^\m\frac{\partial}{\partial p_2^\m}\,B_0(s_1)&=\sdfrac{( d-4)}{2s_1}(s-s_1-s_2)\,B_0(s_1)\\
	p_2^\m\frac{\partial}{\partial p_3^\m}\,B_0(s_1)&=p_3^\m\frac{\partial}{\partial p_2^\m}\,B_0(s_1)=0,
	\end{align}
\end{subequations}
giving
\begin{subequations}
	\begin{align}
	p_1\frac{\partial}{\partial p_1}B_0(s_1)=0,\qquad\frac{\partial}{\partial p_2}B_0(s_1)=\sdfrac{( d-4)}{p_2}\,B_0(s_1),\qquad
	\frac{\partial}{\partial p_3}B_0(s_1)=0.
	\end{align}
\end{subequations}

Finally, for the scalar 2-point function $B_0(s_2)$ we obtain
\begin{subequations}
	\begin{align}
	p_2^\m\frac{\partial}{\partial p_2^\m}\,B_0(s_2)&=p_3^\m\frac{\partial}{\partial p_2^\m}\,B_0(s_2)=0\\
	p_2^\m\frac{\partial}{\partial p_3^\m}\,B_0(s_1)&=\sdfrac{( d-4)}{2s_2}(s-s_1-s_2)\,B_0(s_2)\\
	p_3^\m\frac{\partial}{\partial p_2^\m}\,B_0(s_1)&=( d-4)\,B_0(s_2),
	\end{align}
\end{subequations}
and
\begin{subequations}
	\begin{align}
	p_1\frac{\partial}{\partial p_1}B_0(s_2)=0,\qquad\frac{\partial}{\partial p_2}B_0(s_2,0,0)=0,\qquad
	\frac{\partial}{\partial p_3}B_0(s_2)=\sdfrac{( d-4)}{p_3}\,B_0(s_2).
	\end{align}
\end{subequations}

Finally, we consider the action of these operators on the scalar integral $C_0(s,s_1,s_2)$, obtaining the relations
\begin{subequations}
	\begin{align}
	p_2^\m\frac{\partial}{\partial p_2^\m}\,C_0(s,s_1,s_2)&=\sdfrac{( d-3)}{s}\big[B_0(s_1)-B_0(s_2)\big]+\sdfrac{( d-6)s+( d-4)(s_1-s_2)}{2s}C_0(s,s_1,s_2)\\[2ex]
	p_3^\m\frac{\partial}{\partial p_2^\m}\,C_0(s,s_1,s_2)&=\sdfrac{( d-3)}{s\,s_1}\big[(s_1-s)B_0(s_2)+sB_0(s)-s_1B_0(s_1)\big]\notag\\
	&\hspace{4cm}+\sdfrac{( d-4)(s-s_1)(s+s_1-s_2)}{2s\,s_1}C_0(s,s_1,s_2)\\[2ex]
	p_2^\m\frac{\partial}{\partial p_3^\m}\,C_0(s,s_1,s_2)&=\sdfrac{( d-3)}{s\,s_2}\big[(s_2-s)B_0(s_1)+sB_0(s)-s_2B_0(s_2)\big]\notag\\
	&\hspace{4cm}+\sdfrac{( d-4)(s-s_2)(s-s_1+s_2)}{2s\,s_2}C_0(s,s_1,s_2)\\[2ex]
	p_3^\m\frac{\partial}{\partial p_2^\m}\,C_0(s,s_1,s_2)&=\sdfrac{( d-3)}{s}\big[B_0(s_2)-B_0(s_1)\big]+\sdfrac{( d-6)s-( d-4)(s_1-s_2)}{2s}C_0(s,s_1,s_2).
	\end{align}
\end{subequations}
Using the relations given above, after some algebra, we obtain the expressions
\begin{subequations}
\begin{align}
\sdfrac{\partial}{\partial p_1}C_0&=\sdfrac{p_1}{\s\,s}\bigg\{ 2( d-3)\big[(s+s_1-s_2)\,B_0(s_1)+(s-s_1+s_2)\,B_0(s_2)-2s\,B_0(s)\big]\notag\\
&\qquad+\big[( d-4)(s_1-s_2)^2-( d-2)s^2+2s(s_1+s_2)\big]\,C_0(s,s_1,s_2)\bigg\}\\[2ex]
\sdfrac{\partial}{\partial p_2}C_0&=\sdfrac{p_2}{\s\,s_1}\bigg\{ 2( d-3)\big[(s+s_1-s_2)\,B_0(s)+(s_1+s_2-s)\,B_0(s_2)-2s_1\,B_0(s_1)\big]\notag\\
&\quad+\big[2s_2\big(s_1-( d-4)s\big)+(s-s_1)\big(( d-4)s+( d-2)s_1\big)+( d-4)s_2^2\big]C_0(s,s_1,s_2)\bigg\}\\
\sdfrac{\partial}{\partial p_3}C_0&=\sdfrac{p_3}{\s\,s_2}\bigg\{ 2( d-3)\big[(s-s_1+s_2)\,B_0(s)+(s_1+s_2-s)\,B_0(s_1)-2s_2\,B_0(s_2)\big]\notag\\
&\qquad+\big[( d-4)(s-s_1)^2-( d-2)s_2^2+2s_2(s+s_1)\big]C_0(s,s_1,s_2)\bigg\}.
\end{align}
\end{subequations}

At this point, one can write the operators $K$, $L_N$ and $R$ of the special CWI's in terms of the three invariants $s$, $s_1$ and $s_2$ as
\begin{equation}
\frac{\partial}{\partial s}=\frac{1}{2p_1}\frac{\partial}{\partial p_1},\qquad\frac{\partial}{\partial s_1}=\frac{1}{2p_2}\frac{\partial}{\partial p_2},\qquad \frac{\partial}{\partial s_2}=\frac{1}{2p_3}\frac{\partial}{\partial p_3},
\end{equation}
and from the explicit form of the $K_i$ operator in \eqref{kij} we obtain
\begin{equation}
K_i=\d_{i,j+1}\left[4s_j\sdfrac{\partial^2}{\partial s_j^2}+2( d+2-2\D_i)\,\sdfrac{\partial}{\partial s_j}\right],\quad i=1,2,3,
\end{equation}
 where we have set $s_0\equiv s$.\\
  In the same way, we find that the operators $L_N$ and $R$, from their defining expressions \eqref{Ldef} 
and \eqref{Rdef}, take the form

\begin{align}
L_N&=2(s+s_1-s_2)\,s\frac{\partial}{\partial s}+4s\,s_1\,\frac{\partial }{\partial s_1}+ \big[(2d-\D_1-2\D_2+N)\,s+(2\D_1-d)(s_2-s_1)\big]\\
R&=2\,s\,\frac{\partial}{\partial s}-(2\D_1-d),
\end{align}
while their symmetric versions are given by
\begin{align}
L'_N&=L_N\ \ \text{with $s\leftrightarrow s_1$ and $\D_1\leftrightarrow\D_2$} \\
R'&=R\ \ \text{with $s\mapsto s_1$ and $\D_1\mapsto\D_2$}.
\end{align}
We then obtain
\begin{subequations}
\begin{align}
K_{12}\ B_0(s)&=K_{13}\ B_0(s)=-\sdfrac{4( d-4)}{s}\,B_0(s,0,0),\qquad K_{23}\ B_0(s)=0\nn
K_{21}\ B_0(s_1)&=K_{23}\ B_0(s_1)=-\sdfrac{2( d-4)}{s_1}\,B_0(s_1),\qquad K_{13}\ B_0(s_1)=0\nn
K_{31}\ B_0(s_2)&=K_{32}\ B_0(s_2)=-\sdfrac{2( d-4)}{s_2}\,B_0(s_2),\qquad K_{12}\ B_0(s_2)=0,
\end{align}
\end{subequations}
and 
\begin{subequations}
	\begin{align}
	&K_{12}\ C_0(s,s_1,s_2)\notag\\
	&\hspace{0.2cm}=\sdfrac{4( d-3)}{s\,s_1\,\s}\bigg[-\big(s^2+(s-2s_1)(s_1-s_2)\big)\,B_0(s_2)+s(s+5s_1-s_2)\,B_0(s)-2s_1\,(2s+s_1-s_2)\,B_0(s_1)\bigg]\notag\\
	&\hspace{0.5cm}+\sdfrac{1}{s\,s_1\,\s}\big[4( d-3)(3s-3s_1-s_2)\,s\,s_1+2( d-4)(s-2s_1)\s\big]\,C_0(s,s_1,s_2)
	\\[2ex]
	&K_{23}\ C_0(s,s_1,s_2)\notag\\
	&\hspace{0.2cm}=\sdfrac{4( d-3)}{s_1\,s_2\,\s}\bigg[(s_1-s_2)(s-s_1-s_2)\,B_0(s)+s_1(s_1-s+2s_2)\,B_0(s_1)+s_2(s-3s_1-s_2)\,B_0(s_2)\bigg]\notag\\
	&\hspace{0.5cm}+\sdfrac{1}{s_1\,s_2\,\s}\big[2( d-4)(s_1-s_2)[(s-s_1)^2-2s\,s_2+s_2^2]+4( d-2)(s_1-s2)\,s_1\,s_2\big]\,C_0(s,s_1,s_2)\\[2ex]
	&K_{13}\ C_0(s,s_1,s_2)\nn
	&\hspace{0.2cm}=\sdfrac{4( d-3)}{s\,s_2\,\s}\bigg[-\big(s^2+(s-2s_2)(s_2-s_1)\big)\,B_0(s_1)+s(s-s_1+5s_2)\,B_0(s)-2s_2(2s-s_1+s_2)\,B_0(s_2)\bigg]\notag\nn
	&\hspace{0.5cm}+\sdfrac{1}{s\,s_2\,\s}\big[4( d-3)(3s-3s_2-s_1)\,s\,s_2+2( d-4)(s-2s_2)\s\big]\,C_0(s,s_1,s_2),
	\end{align}
\end{subequations}
while the action of the operators $L_N$ and $R$ on $B_0(s)$ is given by
\begin{subequations}
\begin{align}
L_N\,B_0(s)&=\big[(N-2)s-4s_1+4s_2\big]\,B_0(s)\nn
L'_N\,B_0(s)&=\big[(d+N-7)s_1+(d-2)(s_2-s)\big]\,B_0(s)\nn
R\,B_0(s)&=-4B_0(s)\nn
R'\,B_0(s)&=-(d-2)B_0(s),
\end{align}
\end{subequations}
and on $B_0(s_1)$ 
\begin{subequations}
\begin{align}
L_N\,B_0(s_1)&=\big[(d+N-6)s+d(s_2-s_1)\big]\,B_0(s_1)\nn
L'_N\,B_0(s_1)&=\big[(N-3)s_1-2s+2s_2\big]\,B_0(s_1)\nn
R\,B_0(s_1)&=-d\,B_0(s_1)\nn
R'\,B_0(s_1)&=-2\,B_0(s_1).
\end{align}
\end{subequations}
Similarly, for $B_0(s_2)$ we obtain
\begin{subequations}
\begin{align}
L_N\,B_0(s_2)&=\big[(N-d+2)s+d(s_2-s_1)\big]\,B_0(s_2)\nn
L'_N\,B_0(s_2)&=\big[(N-d+1)s_1+(d-2)(s_2-s)\big]\,B_0(s_2)\nn
R\,B_0(s_2)&=-d\,B_0(s_2)\nn
R'\,B_0(s_2)&=-(d-2)\,B_0(s_2),
\end{align}
\end{subequations}
and for $C_0(s,s_1,s_2)$
\begin{subequations}
\begin{align}
L_N\,C_0(s,s_1,s_2)&=2(d-3)\big[B_0(s_1)-B_0(s_2)\big]+\big[(N-4)s-4(s_1-s_2)\big]C_0(s,s_1,s_2)\nn
L'_N\,C_0(s,s_1,s_2)&=2(d-3)\big[B_0(s)-B_0(s_2)\big]+\big[(N-5)s_1-2(s-s_2)\big]C_0(s,s_1,s_2)\nn
R\,C_0(s,s_1,s_2)&=\sdfrac{2(d-3)}{\s}\big[(s+s_1-s_2)\,B_0(s_1)+(s-s_1+s_2)\,B_0(s_2)-2s\,B_0(s)\big]\notag\nn
&+\frac{2}{\s}\big[-(d-1)s^2+(d+1)s(s_1+s_2)-2(s_1-s_2)^2\big]C_0(s,s_1,s_2)\nn
R'\,C_0(s,s_1,s_2)&=\sdfrac{2(d-3)}{\s}\big[(s+s_1-s_2)\,B_0(s)+(-s+s_1+s_2)\,B_0(s_2)-2s_1\,B_0(s_1)\big]\notag\nn
&-\frac{2}{\s}\big[s(s_1-d\,s_1-2s_2)+(s_1-s_2)(d-2)s_1-s_2(s_1-s_2)+s^2\big]C_0(s,s_1,s_2) .
\end{align}
\end{subequations}
Using the explicit expressions of the form factors $F_i$ presented in the appendix, we have explicitly verified that they satisfy all the conformal constraints, once the action on the two master integrals is re-expressed according to the derivative rules that we have derived in this section in terms of the external invariants.

\section{Renormalization and Anomalous Ward Identities in QED }
In this final section we turn to the derivation of the anomalous CWI's for this correlator in QED. We recall that the singularity in the form factor $F_{13}$, which is the only one which is affected by the renormalization of the 
$F-$basis, can be removed by using the counterterm
\begin{equation}
\int d^4x\,\sqrt{-g} F_{\mu\nu}F^{\mu\nu}.
\end{equation}
After the renormalization procedure in the F-basis, the renormalized expressions of the $A_i$ form factors are obtained using the mapping \eqref{mapping}. One can then derive renormalized anomalous (dilatation, primary and secondary) WI's. Using the explicit expressions of the $A_i$ given in the appendix, and performing the renormalization, we obtain

\begin{align}
\left(\sum_{i}^{3}p_i\sdfrac{\partial}{\partial p_i}+2\right)\,A_1&=0=-\m\sdfrac{\partial}{\partial \m}A_1\\
\left(\sum_{i}^{3}p_i\sdfrac{\partial}{\partial p_i}\right)\,A_2^R&=\sdfrac{8\p^2\,e^2}{3}=-\m\sdfrac{\partial}{\partial \m}A_2^R\\
\left(\sum_{i}^{3}p_i\sdfrac{\partial}{\partial p_i}\right)\,A_3^R(p_2\leftrightarrow p_3)&=\sdfrac{8\p^2\,e^2}{3}=-\m\sdfrac{\partial}{\partial \m}A_3^R\\
\left(\sum_{i}^{3}p_i\sdfrac{\partial}{\partial p_i}-2\right)\,A_4^R&=-\sdfrac{4}{3}\p^2\,e^2(s-s_1-s_2)=-\m\sdfrac{\partial}{\partial \m}A_4^R.
\end{align}
for the dilatation WI's, while for the primary CWI's we obtain
\begin{equation}
\begin{split}
&K_{13}A_1=0\\
&K_{13}A^R_2=-2A_1\\
&K_{13}A^R_3=4A_1\\
&K_{13}A^R_3(p_2\leftrightarrow p_3)=0\\
&K_{13}A^R_4=2A^R_3(p_2\leftrightarrow p_3)-\sdfrac{16\,\pi^2e^2}{3}
\end{split}
\hspace{1.5cm}
\begin{split}
&K_{23}A_1=0\\
&K_{23}A^R_2=0\\
&K_{23}A^R_3=4A_1\\
&K_{23}A^R_3(p_2\leftrightarrow p_3)=-4A_1\\
&K_{23}A^R_4=-2A^R_3+2A^R_3(p_2\leftrightarrow p_3).
\end{split}
\end{equation}
In all the equations $e$ is the renormalized charge and can be traded for $\beta(e)$ by the relation $\beta(e)/e=e^2/(12 \pi^2)$. 
For the secondary CWI's we obtain
 \begin{equation}
 \begin{split}
&L_4 A_1+R A_3^R-R A_3^R(p_2\leftrightarrow p_3)=0\\[1.5ex]
&L_2\,A_2^R-s\,(A_3^R-A_3^R(p_2\leftrightarrow p_3))=\sdfrac{16}{9}\pi^2e^2\left[3s1\,B_0^R(s_1,0,0)-3s_2B_0^R(s_2,0,0)-s1+s2\right]+\sdfrac{24}{9}\pi^2e^2s\\[1.5ex]
&L_4\,A^R_3-2R\,A^R_4=\sdfrac{32}{9}\pi^2\,e^2s_2\left[1-3B_0^R(s_2,0,0)\right]+\sdfrac{48}{9}\pi^2\,e^2\,s\\[1.5ex]
&L_4\,A^R_3(p_2\leftrightarrow p_3)+2R\,A^R_4-4\,sA^R_3(p_2\leftrightarrow p_3)=\sdfrac{32}{9}\,\pi^2\,e^2s_1\left[3B_0^R(s_1,0,0)-1\right]\\[1.5ex]
&L'_3\,A_1^R-2R'A_2^R+2R'A_3^R=0\\[1.5ex]
&L'_1\,A_3^R(p_2\leftrightarrow p_3)+p_2^2(4A^R_2-2A^R_3)+2R'A^R_4=\sdfrac{16}{3}\pi^2\,e^2\,s_1,\\
 \end{split}
 \end{equation}
where we have used the relations 
\begin{equation}
\sdfrac{\partial}{\partial s_i}\,B_0^R(s_j,0,0)=-\sdfrac{\d_{ij}}{s_i}\  i=0,1,2
\end{equation}
where $s_0=s$ and 
\begin{align}
\sdfrac{\partial}{\partial s}\,C_0&= \sdfrac{1}{s\,\s}\,\big[s(s_1+s_2-s)C_0+B_{0,R}(s_1)(s+s_1-s_2)+B_{0,R}(s_2)(s-s_1+s_2)-2s\,B_{0,R}(s)\big]\\
\sdfrac{\partial}{\partial s_1}\,C_0&= \sdfrac{1}{s_1\,\s}\,\big[s_1(s+s_2-s_1)C_0+B_{0,R}(s)(s+s_1-s_2)+B_{0,R}(s_2)(s_1-s+s_2)-2s_1\,B_{0,R}(s_1)\big]\\
\sdfrac{\partial}{\partial s_2}\,C_0&= \sdfrac{1}{s_2\,\s}\,\big[s_2(s+s_1-s_2)C_0+B_{0,R}(s_1)(s_2+s_1-s)+B_{0,R}(s)(s-s_1+s_2)-2s_2\,B_{0,R}(s_2)\big].
\end{align}
We have defined $\s=s^2-2s(s_1+s_2)+(s_1-s_2)^2$,  $B_{0,R}(s_i)\equiv B_{0,R}(s_i,0,0)=2-\log\left(-\sdfrac{s_i}{\mu^2}\right)$ and, for simplicity, $C_0\equiv C_0(s,s_1,s_2)$. \\
In $d=4$ the operator $K_i$ and $L_N$ take the forms
\begin{equation}
K_1\equiv 4s\sdfrac{\partial^2}{\partial s^2}-4\sdfrac{\partial}{\partial s},\quad K_2\equiv 4s_1\sdfrac{\partial^2}{\partial s_1^2},\quad K_3\equiv 4s_2\sdfrac{\partial^2}{\partial s_2^2}
\end{equation}
\begin{align}
L_N&\equiv 2s\,(s+s_1-s_2)\sdfrac{\partial}{\partial s}+4s\,s_1\sdfrac{\partial}{\partial s_1}+\left[(N-2)s+4(s_2-s_1)\right],\qquad R\equiv2s\,\sdfrac{\partial}{\partial s}-4\\
L'_N&\equiv 2s_1\,(s+s_1-s_2)\sdfrac{\partial}{\partial s_1}+4s\,s_1\sdfrac{\partial}{\partial s}+\left[(N-3)s_1+2(s_2-s)\right],\qquad R'\equiv2s_1\,\sdfrac{\partial}{\partial s_1}-2.
\end{align}
\section{Comparing the $3K$ solutions of BMS with the perturbative result in QED}
The perturbative analysis presented above provides a significant check of the consistency of the BMS approach for 3-point functions. However, at the same time, it carries a lot of insight about the connection between CFT's realized by free field theories and the structure of the corresponding nonperturbative solutions. In order to clarify this point we proceed 
with a direct comparison between the expressions of the $A_i$ given in the appendix and the analogous results for the same correlator given in \cite{2014JHEP...03..111B}. The $A_i$'s given in the appendix have been obtained by the recomputed $F_i$'s.

 The final outcome of this analysis will be, by this direct check, that free field theories 
completely saturate the general nonperturbative solution, providing  drastic simplifications of the results presented in \cite{2014JHEP...03..111B}. As already mentioned, the latter have been presented in the form of parametric integrals of 3 Bessel functions, which obviously amount to linear combinations of Appell functions, originally introduced in studies of the AdS/CFT correspondence. \\
To make our treatment self-contained we need to provide some basic description of the structure 
of the solutions obtained by BMS in their analysis and then we will establish a direct link between these expressions and the simplified ones given in the appendix, corresponding to the perturbative $A_i$'s. 
The BMS solutions take the form 
\beqa
A_1 &=&\alpha_1 J_{4[000]}  \nn
A_2 &=& \alpha_1 J_{3[100]} +\alpha_3 J_{2[000]}\nn
A_3 &=&2 \alpha_1 J_{3[001]} +\alpha_3 J_{2[000]} \nn
A_4 &=&  2 \alpha_1 J_{2[011]} +\alpha_3 J_{1[010]} +\alpha_4 J_{0[000]} 
\label{nonper}
\eeqa
where the $J_{n[k_1 k_2 k_3]}$ are integrals corresponding to hypergeometric functions $F_4$ of 2 variables. In general  they are defined as 
\be
J_{n[k_1 k_2 k_3]}=I_{d/2 -1 + n[\beta_1\beta_2\beta_3]}
\label{tworel}
\ee
where 
\be
I_{\alpha[\sigma_1\sigma_2\sigma_3]}(p_1,p_2,p_3)=\int_0^\infty dx x^\alpha \prod_{j=1}^3 p_j^{\sigma_j}K_{\sigma_j}(p_j x)
\ee
with $\beta_i=\Delta_i -d/2 + k_i$. In our case $\Delta_1=d$ and $\Delta_2=\Delta_3=d-1$. 
The parametric integral is expressed in terms of products of modified Bessel functions $K_\nu(x)$ of second kind.  The explicit expressions of such integrals have been worked out in \cite{2014JHEP...03..111B}. All the $J$ integrals appearing in the solution correspond to master integrals of the form \cite{Coriano:2013jba}
\be
\label{davy}
J(\nu_1,\nu_2,\nu_3) = \int \frac{d^d l}{(2 \pi)^d} \frac{1}{(l^2)^{\nu_3} ((l+p_1)^2)^{\nu_2} ((l-p_2)^2)^{\nu_1}}\, ,
\ee
which can be directly connected to 3-point functions of scalar operators, of suitable scaling dimensions $\Delta_i$ by the relations 
\bea
&& \int \frac{d^d p_1}{(2\pi)^d} \frac{d^d p_2}{(2\pi)^d} \frac{d^d p_3}{(2\pi)^d} \, (2\pi)^d \delta^{(d)}(p_1 + p_2 + p_3) \, 
J(\nu_1,\nu_2,\nu_3) e^{- i p_1 \cdot x_1 - i p_2 \cdot x_2 - i p_3 \cdot x_3} \nn 
&& = \frac{1}{4^{\nu_1+\nu_2+\nu_3} \pi^{3 d/2}}  \frac{\Gamma(d/2 - \nu_1) \Gamma(d/2 - \nu_2) \Gamma(d/2 - \nu_3)}{\Gamma(\nu_1) 
\Gamma(\nu_2) \Gamma(\nu_3)}  \frac{1}{(x_{12}^2)^{d/2- \nu_3} (x_{23}^2)^{d/2- \nu_1} (x_{31}^2)^{d/2- \nu_2}}\,, \nn 
\label{oner}
\eea
with 
\bea
\label{etafromnu}
\Delta_1 = d - \nu_2 - \nu_3 \,, \qquad
\Delta_2 = d - \nu_1 - \nu_3 \,, \qquad
\Delta_3 = d - \nu_1 - \nu_2 \, 
\eea
\bea
\label{etafromnu}
\nu_1= \frac{1}{2} (d + \Delta_1 -\Delta_2-\Delta_3) \qquad
\nu_2 = \frac{1}{2}(d - \Delta_1 + \Delta_2 -\Delta_3)\,, \qquad
\nu_3 = \frac{1}{2}(d - \Delta_1 - \Delta_2 +\Delta_3)\,
\eea
An equivalent expression of the master integral $J(\nu_1,\nu_2,\nu_3)$ can be obtained combining the expressions above in the form
\beqa
J(\nu_1,\nu_2,\nu_3)
&=&\frac{\pi^{-d/2} 2^{4 - 3 d/2} }{\Gamma(\nu_1)\Gamma(\nu_2)\Gamma(\nu_3)\Gamma(d-\nu_1-\nu_2-\nu_3)} p_1^{d/2-\nu_2-\nu_3} p_2^{d/2-\nu_1-\nu_3}p_3^{d/2-\nu_1-\nu_2}\nn
&& \times \int_0^\infty d x x^{d/2-1}K_{d-\nu_1-\nu_3} (p_1 x) K_{d-\nu_2-\nu_3} (p_2 x) K_{d-\nu_1-\nu_2} (p_3 x) 
\eeqa
An alternative expression for this integral can be found in \cite{Coriano:2013jba}. 
Notice that if we plug the scaling dimensions of $T^{\mu\nu}$ and of the current $J^\rho$, for instance in $J_{0[000]}$, we encounter divergences which need to be regulated using a rather complex scheme which has been discussed in \cite{2014JHEP...03..111B}. This implies that the coefficients of the scalar 3-point functions contained in the $J$ integrals 
correspond to generalized master integrals with indices $\nu_i$ which are {\em real} numbers rather than just integers, as in the case of ordinary perturbation theory, for instance in QCD. This sets them apart from the standard integrals appearing in massless theories at higher orders. \\ 
Because of this, it is not obvious that the general 3K solution can be directly related to the simple master integrals which appear in the $A_i$, given in the appendix, containing the two master integrals $B_0$ and $C_0$. In fact, all the relations presented in \cite{2014JHEP...03..111B} which allow to connect various $J$ integrals are, therefore, unrenormalized and need to be expanded in a regulator in order to generate the final expressions for each form factor. This, unfortunately does not allow to recognize that such solutions can be drastically simplified. By the same token, the solutions that we have provided for the $A_i$ using the Fuchsian analysis of section (\ref{fuchs}) cannot be easily recognized for being related just to ordinary $B_0$ and $C_0$ master integrals. \\
We are now going to show that the same information provided by the $J$ integrals is entirely reproduced by the perturbative solution in a simplified way. This clearly implies that 
there are significant cancellations among the contributions coming from different $J$ integrals in the BMS solution, in whatever form they are written, which are far from being evident. The latter, obviously, remains essential in order to determine the minimal set of constants which characterize the general conformal solution and allow to establish a link between this and the perturbative result, but their significance probably stops here.   \\
In order to establish this link, we consider different field theories in various dimensions characterized by a generic 
number of degenerate massless fermions, say $n_\psi$, which, as we are going to show, will be taking the role of $\alpha_1$ in the final solution. Let's investigate this point.

As shown in \cite{2014JHEP...03..111B}, the secondary CWI's allow to express the 4 constants $\alpha_i$ in terms of $\alpha_1$ and the normalization of the 2-point function $C_J$. These relations, in general, involve a regulator, except for specific dimensions. In $d=3$ and $d=5$, for instance, the correlator is finite and the relation between the perturbative and the non perturbative expressions of the $A_i$ is transparent and can be worked out analytically. \\ 
We start from the two point functions,  since the $C_J$ presented \cite{2014JHEP...03..111B} and in free field theory (QED) must be matched. In \cite{2014JHEP...03..111B} the normalization of the $JJ$ correlator is defined by the relation
\begin{equation}
\braket{J^{\m}(p)\,J^{\n}(-p)}=C_J\,\pi^{\m\n}(p)\G\left(1-\frac{d}{2}+\frac{\epsilon}{2}\right)\,p^{d-2-\epsilon}\label{SkendCJ}
\end{equation}
while in our case, in conformal QED, we find in dimensional regularization
\begin{align}
\braket{\ J^{\m}(p)\,J^{\n}(-p)\ }&=\frac{2\p^\frac{D}{2}\,e^2\,(d-2)}{d-1}\,p^2\,\p^{\m\n}(p)\,\mathcal{B}_0(p^2,0,0)\notag\\
&=\frac{2\p^\frac{d}{2}\,e^2\,(d-2)\,\p^{\m\n}(p)}{(d-1)\,\G\left(d-2\right)}\ \left[\G\left(\frac{d}{2}-1\right)\right]^2\G\left(2-\frac{D}{2}\right)p^{d-2}\label{QEDCJ},
\end{align}
having used the explicit expression of the two point scalar integral \eqref{B0}.\\
In odd spacetime dimension $n$ ($d=n +\epsilon$), the limit $\epsilon\to0$ is finite and then the comparison between \eqref{SkendCJ} and \eqref{QEDCJ} gives the value for the normalization constant $C_J$ of the two point function $\braket{JJ}$ as
\begin{equation}
C_J=\frac{2\p^\frac{d}{2}\,e^2\,(d-2)}{(d-1)\,\G\left(d-2\right)}\ \frac{\left[\G\left(\frac{d}{2}-1\right)\right]^2\G\left(2-\frac{d}{2}\right)}{\G\left(1-\frac{d}{2}\right)}\label{CJ}.
\end{equation}
It is simple to verify that the values of $C_J$ in the $d=3$ and $d=5$ are those given by \eqref{CJ}, in fact
\begin{align}
C_J&\ \ \mathrel{\stackrel{\makebox[0pt]{\mbox{\normalfont\tiny $d=3$}}}{=}}\ \ \p^\frac{3}{2}e^2\, \left[\G\left(\frac{1}{2}\right)\right]^3\left[\G\left(-\frac{1}{2}\right)\right]^{-1}=-\frac{\pi^\frac{5}{2}\,e^2}{2}\\[2ex]
C_J&\ \ \mathrel{\stackrel{\makebox[0pt]{\mbox{\normalfont\tiny $d=5$}}}{=}}\ \ \frac{3\,\p^\frac{5}{2}e^2}{2\,\G(3)}\, \left[\G\left(\frac{3}{2}\right)\right]^2\,\G\left(-\frac{1}{2}\right)\,\left[\G\left(-\frac{3}{2}\right)\right]^{-1}=-\frac{9\,\pi^\frac{7}{2}\,e^2}{32}
\end{align}
With this information, we can proceed further, showing that 
the expressions of the $A_i$'s presented in section \ref{ffs} agree with those determined in the nonperturbative solution. 

\subsection{Explicit result in $d=3$ (QED and scalar QED)}

We consider the fermion and the scalar contribution combined together, selecting an arbitrary number of massless fermions and scalars running inside the loops, with constants $n_F$ and $n_S$ related to this condition. \\
In the particular case of $d=3$  there are simplifications both in the exact and the perturbative solutions. The scalar integral ${B}_0$ is given by
\begin{equation}
{B}_0(s,0,0)=\sdfrac{1}{i\p^\frac{d}{2}}\int\,d^d \ell\ \frac{l}{\ell^2(\ell-p_1)^2}=\frac{\p^\frac{d}{2}\ \left[\G\left(\frac{d}{2}-1\right)\right]^2\G\left(2-\frac{d}{2}\right)}{\p^\frac{d}{2}\,\G\left(d-2\right)(p_1^2)^{2-\frac{d}{2}}}\label{B0ex}
\end{equation}
for which in $d=3$ becomes
\begin{equation}
{B}_0(s,0,0)=\frac{\p^{3/2}}{\,p_1}\label{B0}
\end{equation}
where $p_1=|p_1|=\sqrt{p_1^2}$. The scalar 3-point function ${C}_0$ can be simplified using the star-triangle rule for which
\begin{equation}
\int\frac{d^{d}x}{[(x-x_1)^2]^{\a_1}\,[(x-x_2)^2]^{\a_2}\,[(x-x_3)^2]^{\a_3}}\ \  \mathrel{\stackrel{\makebox[0pt]{\mbox{\normalfont\tiny $\sum_i\a_i=d$}}}{=}}\ \  \frac{i\pi^{d/2}\n(\a_1)\n(\a_2)\n(\a_3)}{[(x_2-x_3)^2]^{\frac{d}{2}-\a_1}\,[(x_1-x_2)^2]^{\frac{d}{2}-\a_3}\,[(x_1-x_3)^2]^{\frac{d}{2}-\a_2}}\label{startriangle}
\end{equation}
where
\begin{equation}
\n(x)=\frac{\G\left(\frac{d}{2}-x\right)}{\G(x)},
\end{equation}
which holds only if the condition $\sum_i\a_i=d$ is satisfied. In the case $d=3$  \eqref{startriangle} is proportional to the three point scalar integral, and in particular
\begin{align}
{C}_0(p_1^2,p_2^2,p_3^2)&=\int\,\frac{d^d\ell}{i\p^\frac{d}{2}}\frac{1}{\ell^2(\ell-p_2)^2(\ell+p_3)^2}=\int\,\frac{d^dk}{i\p^\frac{d}{2}}\frac{1}{(k-p_1)^2(k+p_3)^2(k+p_3-p_2)^2}\notag\\[2ex]
&=\frac{\left[\G\left(\frac{d}{2}-1\right)\right]^3}{\,(p_1^2)^{\frac{d}{2}-1}(p_2^2)^{\frac{d}{2}-1}(p_3^2)^{\frac{d}{2}-1}}\ \mathrel{\stackrel{\makebox[0pt]{\mbox{\normalfont\tiny $d=3$}}}{=}}\ \frac{\p^{3/2}}{\,p_1\,p_2\,p_3}.\label{C0}
\end{align}

The explicit expression of the form factors in $d=3$ using the perturbative approach to one loop order, can be found by substituting the expression of the scalar integral, using \eqref{B0} and \eqref{C0}, and then considering the limit $d\to3$ for all the form factors. The scalar and the fermion cases contribute equally, modulo an overall constant, giving 
\begin{align}
A_{1,D=3}&=\left(\frac{\p^3\,e^2(8n_F+n_S)}{6}\right)\frac{2\big(4p_1+p_2+p_3\big)}{(p_1+p_2+p_3)^4}\\
A_{2,D=3}&=\left(\frac{\p^3\,e^2(8n_F+n_S)}{6}\right)\frac{2\,p_1^2}{(p_1+p_2+p_3)^3}-\left(\frac{\p^3\,e^2\,(8n_F-n_S)}{2}\right)\frac{(2p_1+p_2+p_3)}{(p_1+p_2+p_3)^2}\\
A_{3,D=3}&=\left(\frac{\pi ^3 e^2 (8 n_F+n_S)}{6} \right)\frac{\left(-2 p_1^2-3 p_1 p_2+3 p_1 p_3-p_2^2+p_3^2\right)}{(p_1+p_2+p_3)^3}-\left(\frac{\pi ^3 e^2 (8 n_F-n_S) }{2}\right)\frac{(2 p_1+p_2+p_3)}{ (p_1+p_2+p_3)^2}\\
A_{4,D=3}&=\left(\frac{\pi ^3 e^2 (8 n_F+n_S)}{6}\right)\frac{ (2 p_1+p_2+p_3) \left(p_1^2-(p_2+p_3)^2+4 p_2 p_3\right)}{2(p_1+p_2+p_3)^2}\notag\\
&\hspace{4cm}+\left(\frac{\pi ^3 e^2 (8 n_F-n_S) }{4}\right)\left(\frac{2p_1^2}{ (p_1+p_2+p_3)-p_2-p_3}\right)
\end{align}
which coincide with the solution given by BMS in the limit $d=3$, modulo the identification of the constants
\begin{equation}
\a_1=\left(\frac{\p^3\,e^2(8n_F+n_S)}{6}\right),\qquad c_J=-\left(\frac{\p^\frac{5}{2}\,e^2\,(8n_F-n_S)}{8}\right).
\end{equation}
\subsection{Explicit result in $d=5$ (QED and scalar QED)}

The case $d=5$ is slightly more involved. In fact, while the result of ${B}_0$ is still the same, the explicit form of the ${C}_0$ needs some manipulations. We start from the star-triangle relation \eqref{startriangle} 
\begin{align}
\int\frac{d^{d}x}{[(x-x_1)^2]^{2}\,[(x-x_2)^2]\,[(x-x_3)^2]^{2}}&\ \mathrel{\stackrel{\makebox[0pt]{\mbox{\normalfont\tiny $d=5$}}}{=}}\ \frac{i\pi^{d/2}\n(2)\n(1)\n(2)}{[(x_2-x_3)^2]^{\frac{d}{2}-2}\,[(x_1-x_2)^2]^{\frac{d}{2}-2}\,[(x_1-x_3)^2]^{\frac{d}{2}-1}}\\[2ex]
&=\frac{i\p^4}{2}\frac{1}{[(x_2-x_3)^2]^{1/2}\,[(x_1-x_2)^2]^{1/2}\,[(x_1-x_3)^2]^{3/2}}\label{eq}
\end{align}
and use an integration by parts to reduce the left-hand side. In particular, by setting $x_1\to p_1$, $x_2\to -p_3$, $x_3\to p_1-p_3$ we obtain
\begin{align}
\int\frac{d^{d}x}{[(x-x_1)^2]^{2}\,[(x-x_2)^2]\,[(x-x_3)^2]^{2}}&=\int\frac{d^{d}x}{[(x-p_1)^2]^{2}\,[(x+p_3)^2]\,[(x-p_1+p_3)^2]^{2}}\notag\\[2ex]
&=\int\frac{d^{d}\ell}{[(\ell)^2]^{2}\,(\ell-p_2)^2\,[(\ell+p_3)^2]^{2}}=\frac{i\p^4}{2\,p_1\,p_2\,p_3^3}
\end{align}
and 
\begin{align}
\int\frac{d^{d}\ell}{[(\ell)^2]^{2}\,(\ell-p_2)^2\,[(\ell+p_3)^2]^{2}}&=-\frac{i\p^\frac{d}{2}(d-4)\big[(d-6)(s-s_1)^2-(d-4)s_2^2+2s_2(s+s_1)\big]}{4s\,s_1\,s_2^2}\ {C}_0(s,s_1,s_2)\notag\\[2ex]
&\hspace{-5cm}-\frac{i\p^\frac{d}{2}\,(d-3)\big[(d-6)(s-s_1)+(d-4)s_2\big]}{2s\,s_1\,s_2^2}\,{B}_0(s,0,0)+\frac{i\p^\frac{d}{2}\,(d-3)\big[(d-6)(s-s_1)-(d-4)s_2\big]}{2s\,s_1\,s_2^2}\,{B}_0(s_1,0,0)\notag\\[2ex]
&+\frac{i\p^\frac{d}{2}\,(d-3)\big[(d-6)(s+s_1)+(d-4)s_2\big]}{2s\,s_1\,s_2^2}\,{B}_0(s_2,0,0)
\end{align}
where $s=p_1^2$, $s_1=p_2^2$ and $s_2=p_3^2$ as previously. Inserting this expression into Eq. \eqref{eq} and solving for ${C}_0$, in $d=5$ one finds
\begin{equation}
{C}_0(s,s_1,s_2)=\frac{\p^{3/2}}{p_1+p_2+p_3}.
\end{equation}
From \eqref{B0ex} the $B_0$ is calculated in $d=5$ as
\begin{equation}
{B}_0(s,0,0)=-\frac{\p^{3/2}}{4} p_1.
\end{equation}
Plugging these results into the general form of the $A_i$ given in appendix \ref{ffs} we find
\begin{align}
A_{1,d=5}&=\frac{\pi ^4 e^2 (8 n_F+n_S)}{120 (p_1+p_2+p_3)^5} \Big[4 p_1^4+20 p_1^3 (p_2+p_3)+4 p_1^2 \left(7 \big(p_2+p_3\big)^2+6p_2 p_3\right)\notag\\
&\hspace{2cm}+15 p_1 (p_2+p_3) \left(\big(p_2+p_3\big)^2+ p_2 p_3\right)+3 (p_2+p_3)^2 \left(\big(p_2+p_3\big)^2+ p_2 p_3\right)\Big]\\[2ex]
A_{2,d=5}&=\frac{\pi ^4 e^2 p_1^2 (8 n_F+n_S) }{120 (p_1+p_2+p_3)^4}\Big[(p_1+p_2+p_3)^3+(p_1 p_2+p_1 p_3+p_2 p_3) (p_1+p_2+p_3)+3 p_1 p_2 p_3\Big]\notag\\
&\quad+\frac{\pi ^4 e^2 (24 n_F-n_S)}{48 (p_1+p_2+p_3)^3} \Big[2 p_1^4+6 p_1^3 (p_2+p_3)+2 p_1^2 \left(5 \big(p_2+p_3\big)^2+p_2 p_3\right)\notag\\
&\hspace{2cm}+9 p_1 \left(p_2^3+2 p_2^2 p_3+2 p_2 p_3^2+p_3^3\right)+3 (p_2+p_3)^2 \left(\big(p_2+p_3\big)^2-p_2 p_3\right)\Big]
\end{align}
\begin{align}
A_{3,d=5}&=\frac{\pi ^4 e^2 (8 n_F+n_S)}{240 (p_1+p_2+p_3)^4} \Big[(-2 p_1^5-8 p_1^4 (p_2+p_3)-8 p_1^3 p_2 (2 p_2+3 p_3)\notag\\
&+p_1^2 \left(-19 p_2^3-40 p_2^2 p_3+24 p_2 p_3^2+15 p_3^3\right)-3 (p_2-p_3) (p_2+p_3) \left(p_2^2+3 p_2 p_3+p_3^2\right) (4 p_1+p_2+p_3)\Big]\notag\\
&+\frac{\pi ^4 e^2 (24 n_F-n_S)}{48 (p_1+p_2+p_3)^3} \Big[2 p_1^4+6 p_1^3 (p_2+p_3)+2 p_1^2 \left(5 p_2^2+6 p_2 p_3+5 p_3^2\right)\notag\\
&\hspace{2cm}+9 p_1 \left(p_2^3+2 p_2^2 p_3+2 p_2 p_3^2+p_3^3\right)+3 (p_2+p_3)^2 \left(p_2^2+p_2 p_3+p_3^2\right)\Big]
\\[2ex]
A_{4,d=5}&=\frac{\pi ^4 e^2 (8 n_F+n_S) }{480 (p_1+p_2+p_3)^3}\Big[2 p_1^6+6 p_1^5 (p_2+p_3)+4 p_1^4 \left(2 (p_2+p_3)^2-p_2 p_3\right)\notag\\
&\hspace{2cm}+p_1^2 \left((p_2+p_3)^2+p_2 p_3\right) \left(3 p_1 (p_2+p_3)-7 (p_2+p_3)^2+32 p_2 p_3\right)\notag\\
&\hspace{3cm}-3 (p_2+p_3) \left((p_2+p_3)^2-4 p_2 p_3\right) \left((p_2+p_3)^2+p_2 p_3\right) (3 p_1+p_2+p_3)\Big]\notag\\
&-\frac{\pi ^4 e^2 (24 n_F-n_S) }{96 (p_1+p_2+p_3)^2}\Big[2 p_1^5+4 p_1^4 (p_2+p_3)+4 p_1^3 \left(p_2^2+p_2 p_3+p_3^2\right)\notag\\
&\quad-p_1^2 (p_2+p_3) \left(p_2^2-5 p_2 p_3+p_3^2\right)-6 p_1 \left(p_2^4+p_2^3 p_3+p_2 p_3^3+p_3^4\right)-3 (p_2+p_3)^3 \left(p_2^2-p_2 p_3+p_3^2\right)\Big]
\end{align}
in agreement with the solution given by BMS in the limit $d=5$, with the identifications 
\begin{equation}
\a_1=\left(\frac{\pi ^4 e^2 (8 n_F+n_S)}{240} \right),\qquad c_J=-\left(\frac{3\pi ^\frac{7}{2} e^2 (24 n_F-n_S)}{128 }\right).
\end{equation}
This shows that, after fixing the normalization of the 2-point function, we are esentially left, both from the perturbative and the non perturbative side, with the same solution. 


\section{Comments} 
There are several obvious conclusions that we can draw from this comparative study of the $TJJ$ that we are going to briefly summarize.\\
The $A_i$ computed in perturbation 
theory satisfy the same anomalous conformal Ward identities as the nonperturbative ones, as shown by us in the previous sections for $d=4$.\\
They both satisfy homogeneous (non anomalous) CWI's in general $(d)$ dimensions. \\
In $d=3$ and $d=5$ the two solutions completely match, having consistently matched the normalization of the $JJ$ 2-point function in the two separate cases.\\
In $d=5$, where the corresponding field theory is nonrenormalizable, the perturbative computation still matches the nonperturbative one. This obviously occurs because the non-renormalizability of the theory does not play any role, being the matching between the two theories a purely one loop result (one loop saturation).\\ 
We conclude that, at least for the $TJJ$, free field theories in momentum space at one loop provide the same information derived from the non-perturbative solutions, and the two can be freely interchanged, being equivalent. The two analytical derivations are therefore expected to coincide for any dimension. Obviously, as already mentioned, this implies that there should be significant cancellations among the contributions of the 3K integrals or those given by us in \ref{finfin} in such a way that they can be expressed in terms of the elementary master integrals $B_0$ and $C_0$.\\
 The two expressions of $A_1$,  which is affected only by a single integration constant $\alpha_1$, in the two forms given in section \ref{ffs} and in Eq. (\ref{nonper}), show quite directly that $J_{4 [000]}$ can be expressed in terms of the elementary master integrals present in the perturbative expansion. Direct checks for the other $J$ integrals appearing in the expressions of $A_2, A_3...$ and so on, are harder to perform, since each of these form factors depends on at least two $J$ integrals, as evident from Eq. (\ref{nonper}).\\
 \subsection{Coordinate space analysis}
  We should mention that this result {\em is not unexpected}, since it had been shown in (\cite{Coriano:2012wp} section 3) that the solution for the $TJJ$ (but also for the TOO and $JJJ$) in coordinate space presented by Osborn and Petkou \cite{Osborn:1993cr} could be {\em completely reproduced } by the simple one loop diagrams of scalar QED and QED using fermion and scalar sectors. These correlators had been shown to be "completely integrable", in the words of \cite{Coriano:2012wp}, which meant that it was possible for such expressions to proceed with a direct Fourier transform to momentum space without the need of introducing an extra regulator ($\omega$) for the transform. 
  
  The fact that the nonperturbative solutions are mirrored by the perturbative ones, shows that result obtained 
at one loop in the perturbative description are automatically inherited by the nonperturbative one, being identical in the case of 3-point functions. The complex machinery of the general solutions, both in coordinate and in position space, 
can be completely reproduced by a simple one-loop analysis. Since  conformal symmetry for 3-point functions is "saturated" at one loop, this explains why radiative corrections in such correlators which preserve the conformal symmetry are necessarily bound to be proportional to the one-loop result, as in the case of the $AVV$ diagram, for instance.  

We are then entitled to come to immediate conclusions concerning the analysis presented in \cite{Armillis:2009pq} in regards to the emergence of anomaly poles in one-loop QED, which are not artifact of perturbation theory, but are  naturally part of the nonperturbative solution and are {\em protected by the conformal symmetry}. 

\section{Summary and Conclusions}
 We have  presented a general discussion of the transition from position to momentum space in the analysis of tensor correlators, clarifying some aspects of 
the realization of the conformal generators in this space. We have illustrated how to handle the derivative of the $\delta$ functions of momentum conservation, giving a direct derivation of the correct forms of the equations. We have also shown how to proceed with the treatment of the 
dependent momentum, when acting with the conformal generator on each of the independent momenta.  \\
Then we have moved toward a direct analysis of the $TJJ$ vertex, detailing some of the involved intermediate steps, technically demanding, used in the BMS reconstruction of such correlator.  

In a recent work \cite{Coriano:2018zdo} we have presented the physical motivations, derived from combined perturbative and non perturbative studies of the $TJJ$ vertex, why anomaly poles in anomaly vertices should be considered the key signature of the conformal anomaly. The goal of this work has been to fill in the intermediate steps of our previous 
analysis. The presence of such massless degrees of freedom generated in the presence of anomalies shows that they are not an artifact of perturbation theory.  

 We have studied a simple instantiation of 
such vertex in massless QED, and confronted it with the 
general approach for solving the conformal constraints in momentum space for scalars \cite{Armillis:2009pq,Bzowski:2013sza} and tensor correlators \cite{Bzowski:2013sza,Bzowski:2017poo}. 
Such solutions do not rely on any Lagrangian realization and are, therefore, very general. In this way it is possible to establish an important connection between perturbative and non-perturbative approaches, in the analysis of specific correlators, which brings to significant simplifications of the general result.

Even though massless QED is not a conformal theory, the violation of conformality is associated to the $\beta$-function of the running coupling constant  and, exactly for this reason, specific correlation functions containing one or more insertions of stress-energy tensors play, in this context, an important role, due to the trace anomaly. \\
There has been considerable interest in the analysis of the breaking of the conformal symmetry in realistic non-conformal theories such as QED and QCD, and in their manifestations through higher perturbative orders \cite{Kataev:1996ce,Broadhurst:1993ru}. In this respect, the analysis of exact solutions, at least for 3-point functions, may shed light on the manifestation of the conformal breaking, which is obviously associated to a non-zero $\beta$ function and, according to our analysis and to the analysis of \cite{Giannotti:2008cv,2009PhLB..682..322A,Armillis:2009pq}, is accompanied by the appearance of an anomaly pole in the effective action.\\
We have shown that the one loop result in perturbation theory saturates the nonperturbative solutions. This implies that the main master integrals necessary to expand the nonperturbative result should turn out to be just $B_0$ and $C_0$, in agreement with former analysis in coordinate space. These analysis share the goal of establishing the form of the anomaly action and address its nonlocality, given their role in describing the breaking of conformal symmetry. It is not surprising, therefore, 
that such massless interactions have received attention in the investigation of the role of the chiral and of the conformal anomalies in topological insulators and in Weyl semi-metals, as recently emphasized in \cite{Chernodub:2017jcp,Rinkel:2016dxo}. 
The study of such materials provides a direct application of the properties of the vertices that we have been interested in.
Obviously, the parallel study of perturbative and non-perturbative methods stops at the level of 3-point functions, since higher point functions need to be bootstrapped. However, the connection found in our work clearly indicate that the possibility of studying a skeleton expansion of realistic field theories, with conformal vertices arrested to one-loop order,  
could provide a complementary way to investigate the bootstrap program for higher point functions is CFT's.\\
One obvious question, emerging from this analysis, is if this result can be generalized to more complex correlators, such as the TTT, where more integration constants are present and for which a coordinate space analysis of such correlator has not been related to the free field theory result. We plan to address this issue in related work.
\newpage

\centerline{ \bf Acknowledgements} 
We are particularly grateful to Paul McFadden for exchanges and clarifications on the BMS reconstruction method. We thank  Emil Mottola, Kostas Skenderis, Mirko Serino, Luigi Delle Rose, Fiorenzo Bastianelli, Marco Guzzi, Olindo Corradini and Andrei Kataev for discussions. C.C. thanks the theory group at 
ETH, and in particular Vittorio Del Duca and Charalampos Anastasiou, Maxim Chernodub for discussions and hospitality at ETH and LMTP Tours. Finally, we thank Pietro Colangelo for discussions and comments on the manuscript.
This work is partially supported by INFN under the Iniziativa Specifica QFT-HEP.  
\vspace{2cm}
\appendix
\section{ Form factors contributions} 
\label{ffs}

\subsection{Form Factors \texorpdfstring{$A_i$}{} in \texorpdfstring{$d$}{} dimensions}

We present the expressions of the form factors $A_i$ in the massless limit in $D$- dim. These relations will be functions of the three invariants $s$, $s_1$ and $s_2$ and we have defined $\s=s^2-2s(s_1+s_2)+(s_1-s_2)^2$ and $\g=s-s_1-s_2$
{\small
\begin{align}
&A_{1,d}(s,s_1,s_2)=\sdfrac{16\pi^2e^2s^2}{(d-2)(d-1)d\s^4}\bigg\{2\g\big[-2 s_2 (s+2 s_1)+(s-s_1)^2+s_2^2\big] \left(s^2+s\,\g-(s_1-s_2)^2\right)\notag\\
&-2d^3 s_1 s_2\g\big[(s_1-s_2)^2-s^2\big]+4d^2 s_1 s_2 \big[s^2 (s_1+s_2)-(\g+s) (s_1-s_2)^2-4s\,s_1\,s_2\big]\notag\\
&+s_2d \left(5 s^2+12 s s_1-5 s_1^2\right) (s-s_1)^2-2 s_2^2 \left(5 s^3+12 s^2 s_1+17 s s_1^2-2 s_1^3\right)-5 s_2^4 (s+s_1)-(s-s_1)^5+s_2^5\notag\\
&+2 s_2^3 (5 s+s_1) (s+2s_1)\bigg\}B_0(s)-\sdfrac{4\pi^2e^2\,s_1}{(d-2)(d-1)d\,\s^4}\bigg\{8\,d^3s^2\,s_2(s+s_1-s_2)\g^2+2d \big[s^6-2 s^5 (s_1+3 s_2)\notag\\
&-s^4 (s_1-3 s_2) (5 s_1+3 s_2)+4 s^3 s_1 \left(5 s_1^2+2 s_1 s_2-11 s_2^2\right)-s^2 (s_1-s_2) \left(25 s_1^3+37 s_1^2 s_2+59 s_1 s_2^2-9 s_2^3\right)\notag\\
&+2 s (s_1-s_2)^3 \left(7 s_1^2+12 s_1 s_2-3 s_2^2\right)-(s_1-s_2)^5 (3 s_1-s_2)\big]+d^2\big[s^6-2 s^5 (3 s_1+5 s_2)+(s_1-s_2)^6\notag\\
&+s^4 \left(15 s_1^2+50 s_1 s_2+31 s_2^2\right)-4 s^3 \left(5 s_1^3+17 s_1^2 s_2-9 s_1 s_2^2+11 s_2^3\right)+s^2 (s_1-s_2) \big(15 s_1^3+35 s_1^2 s_2+77 s_1 s_2^2\notag\\
&-31 s_2^3\big)-2 s (s_1-s_2)^4 (3 s_1+5 s_2)\big]+8s_1 (\g+2s_2) \left(2 s^2-s (s_1+s_2)-(s_1-s_2)^2\right) \left(\s-2 s s_2\right)\bigg\}B_0(s_1)\notag\\
&-\sdfrac{4 \pi ^2 e^2 s_2}{(d-2) (d-1) d \sigma ^4} \bigg\{8 d^3 s^2 s_1 (s-s_1+s_2) \g^2+d^2 \big[s^6-2 s^5 (5 s_1+3 s_2)+s^4 \left(31 s_1^2+50 s_1 s_2+15 s_2^2\right)\notag\\
&-4 s^3 \big(11 s_1^3-9 s_1^2 s_2+17 s_1 s_2^2+5 s_2^3\big)+s^2 \big(31 s_1^4-108 s_1^3 s_2+42 s_1^2 s_2^2+20 s_1 s_2^3+15 s_2^4\big)-2 s (s_1-s_2)^4 (5 s_1\notag\\
&+3 s_2)+(s_1-s_2)^6\big]+2 d \big[s^6-2 s^5 (3 s_1+s_2)+s^4 \left(9 s_1^2+12 s_1 s_2-5 s_2^2\right)+4 s^3 s_2 \left(-11 s_1^2+2 s_1 s_2+5 s_2^2\right)\notag\\
&-s^2 \big(9 s_1^4-68 s_1^3 s_2+22 s_1^2 s_2^2+12 s_1 s_2^3+25 s_2^4\big)+2 s (s_1-s_2)^3 \left(3 s_1^2-12 s_1 s_2-7 s_2^2\right)-(s_1-3 s_2) (s_1-s_2)^5\big]\notag\\
&+8 s_2 \big[2 s^5-7 s^4 (s_1+s_2)+s^3 \left(-4 s_1^2+8 s_1 s_2+8 s_2^2\right)+s^2 \left(8 s_1^3-6 s_1^2 s_2-2 s_2^3\right)+2 s (s_1-s_2)^3 (s_1+s_2)\notag\\
&-(s_1-s_2)^5\big]\bigg\}B_0(s_2)+\frac{16 \pi ^2 e^2 s^2 s_1 s_2}{(d-2) d \sigma ^4} \bigg\{d^2 \left(s^2-(s_1-s_2)^2\right) \g^2+2 d \big[s^3 (s_1+s_2)+s^2 \left(-3 s_1^2+8 s_1 s_2-3 s_2^2\right)\notag\\
&+s \left(3 s_1^3-7 s_1^2 s_2-7 s_1 s_2^2+3 s_2^3\right)-(s_1-s_2)^2 \left(s_1^2+4 s_1 s_2+s_2^2\right)\big]-8 s_1 s_2 \left(\s-3\g\,s\right)\bigg\}C_0(s,s_1,s_2)
\end{align}
\begin{align}
&A_{2,d}(s,s_1,s_2)=-\sdfrac{4\pi^2e^2\,s^2}{(d-2)(d-1)d\s^3}\bigg\{-d\big[-2 s_2 (s+3 s_1)+(s-s_1)^2+s_2^2\big] \s+4d^2\s\,s_1\,s_2\notag\\
&+d^3\big[-4 s_2 \left(s^2+4 s s_1-3 s_1^2\right) (s-s_1)-4 s_2^3 (s+3 s_1)+2 s_2^2 (s+s_1) (3 s+11 s_1)+(s-s_1)^4+s_2^4\big]\bigg\}B_0(s)\notag\\
&+\sdfrac{2\pi^2e^2\,s_1}{(d-2)(d-1)d\s^3}\bigg\{-2 d\sigma  \left(s^3-12 s^2 s_2-3 s (s_1-s_2) (s_1+3 s_2)+2 (s_1-s_2)^3\right)+2d^2\big[s^4 (6 s_1-8 s_2)\notag\\
&-s^5+2 s^3 \left(-7 s_1^2+s_1 s_2+14 s_2^2\right)+2 s^2 (s_1-s_2) \left(8 s_1^2+9 s_1 s_2+15 s_2^2\right)-s (s_1-s_2)^3 (9 s_1+13 s_2)+2 (s_1-s_2)^5\big]\notag\\
&-8s\,s_1(s-s_1+s_2)(\s-2s\,s_2)+d^3\sigma  \big(s^3-3 s^2 s_1+3 s^2 s_2+3 s s_1^2+2 s s_1 s_2-5 s s_2^2-s_1^3+3 s_1^2 s_2-3 s_1 s_2^2\notag\\
&+s_2^3\big)\bigg\}\,B_0(s_1)+\sdfrac{2\pi^2e^2s_2}{(d-2)(d-1)d\,\s^3}\bigg\{d^3\sigma  \big[s^3+3 s^2 (s_1-s_2)-s (s_1-s_2) (5 s_1+3 s_2)+(s_1-s_2)^3\big]-2d^2 \big[s^5\notag\\
&+s^4 (8 s_1-6 s_2)-2 s^3 \left(14 s_1^2+s_1 s_2-7 s_2^2\right)+2 s^2 (s_1-s_2) \left(15 s_1^2+9 s_1 s_2+8 s_2^2\right)-s (s_1-s_2)^3 (13 s_1+9 s_2)+\notag\\
&2 (s_1-s_2)^5\big]-2 d\sigma\big[s^3-12 s^2 s_1+3 s (s_1-s_2) (3 s_1+s_2)-2 (s_1-s_2)^3\big]-8s\,s_2(s+s_1-s_2)(\s-2s\,s_1)\bigg\}B_0(s_2)\notag
\end{align}
\begin{align}
&+\sdfrac{8 \pi ^2 e^2 s^2 }{(d-2) d \sigma ^3}s_1 s_2 \bigg\{-d^2 \sigma\,\g+d \left(2 s\,\g+\s\right) \g+8 s s_1 s_2\bigg\}C_0(s,s_1,s_2)\\[3.5ex]
&A_{3,d}(s,s_1,s_2)=-\frac{4 \pi ^2 e^2 s^2}{(d-2) (d-1) d \sigma ^3} \bigg\{4\g (s-s_1+s_2) \left(\s-2s_2\,s_1\right)-3d\s \left(-2 s_2 (s+3 s_1)+(s-s_1)^2+s_2^2\right)\notag\\
&+d^2\big[2 s_2^2 \left(3 s^2+18 s s_1+19 s_1^2\right)-4 s_2^3 (s+7 s_1)-4 s_2 (s+3 s_1) (s-s_1)^2+(s-s_1)^4+s_2^4\big]+4 d\s\,s_1 s_2 \bigg\}B_0(s)\notag\\
&+\sdfrac{2 \pi ^2 e^2 s_1}{(d-2) (d-1) d \sigma ^3} \bigg\{d^3 \sigma  \big[s^3+3 s^2 (s_2-s_1)+s (s_1-s_2) (3 s_1+5 s_2)-(s_1-s_2)^3\big]+d^2 \big[s^4 (17 s_1+7 s_2)\notag\\
&-3 s^5-2 s^3 \left(19 s_1^2+14 s_1 s_2-9 s_2^2\right)+2 s^2 \left(21 s_1^3+5 s_1^2 s_2+31 s_1 s_2^2-25 s_2^3\right)-s (s_1-s_2)^3 (23 s_1+33 s_2)\notag\\
&+5 (s_1-s_2)^5\big]+2 d \sigma  \big[2 s^3-3 s^2 (3 s_1+s_2)+2 s (2 s_1-s_2) (3 s_1+5 s_2)-(5 s_1-3 s_2) (s_1-s_2)^2\big]\notag\\
&+8 s_1 \s(s-s_1+s_2)^2\bigg\}B_0(s_1)+\sdfrac{2 \pi ^2 e^2 s_2}{(d-2) (d-1) d \sigma ^3} \bigg\{d^3 \sigma  \big[s^3+3 s^2 (s_1-s_2)+s \left(-5 s_1^2+2 s_1 s_2+3 s_2^2\right)\notag\\
&+(s_1-s_2)^3\big]-d^2 \big[s^5+7 s^4 (s_1-s_2)-6 s^3 \left(5 s_1^2+6 s_1 s_2-3 s_2^2\right)+2 s^2 \left(19 s_1^3+7 s_1^2 s_2+17 s_1 s_2^2-11 s_2^3\right)\notag\\
&-s (s_1-s_2)^3 (19 s_1+13 s_2)+3 (s_1-s_2)^5\big]+2 d \sigma  \big[s^2 (3 s_1+s_2)-2 s \left(2 s_1^2+5 s_1 s_2+s_2^2\right)+(s_1-s_2)^2 (s_1+s_2)\big]\notag\\
&-8 s_2 \big[-s^4+2 s^3 (2 s_1+s_2)-2 s (s_1-s_2)^2 (2 s_1+s_2)+(s_1-s_2)^4\big]\bigg\}B_0(s_2)-\sdfrac{8 \pi ^2 e^2 s^2 s_1 s_2}{(d-2) d \sigma ^3} \bigg\{d^2 \big[\g\s\notag\\
&+2(s_1+s_2)(s_1-s_2)^2\big]-d \big[s^3-s^2 (3 s_1+7 s_2)+s \left(3 s_1^2+2 s_1 s_2+11 s_2^2\right)-s_1^3+5 s_1^2 s_2+s_1 s_2^2-5 s_2^3\big]\notag\\
&+8 s_1 s_2 (s-s_1+s_2)\bigg\}C_0(s,s_1,s_2)
\end{align}

\begin{align}
&A_{4,d}(s,s_1,s_2)=\sdfrac{2 \pi ^2 e^2 s^2 \g }{(d-2) (d-1) d \sigma ^2}\bigg\{d^2\s +4(\s-2s_1s_2)-d\big[3\s-8s_1s_2\big]\bigg\}B_0(s)\notag\\
&-\sdfrac{\pi ^2 e^2 s_1}{(d-2) (d-1) d \sigma ^2} \bigg\{d^3\s^2+8 s_1(s-s_1+s_2) (\s-2s\,s_2)-d^2\s\left(3 s^2-4 s (2 s_1+3 s_2)+5 (s_1-s_2)^2\right)\notag\\
&+2d \big[2 s^4-11 s^3 (s_1+s_2)+s^2 \left(21 s_1^2+24 s_1 s_2+19 s_2^2\right)-s (s_1-s_2) \left(17 s_1^2+12 s_1 s_2-13 s_2^2\right)\notag\\
&+(s_1-s_2)^3 (5 s_1-3 s_2)\big]\bigg\}B_0(s_1,0,0)-\sdfrac{\pi ^2 e^2 s_2 }{(d-2) (d-1) d \sigma ^2}\bigg\{d^3 \sigma ^2+d^2 \sigma  \big[-3 s^2+4 s (3 s_1+2 s_2)\notag\\
&-5 (s_1-s_2)^2\big]+2 d \big[2 s^4-11 s^3 (s_1+s_2)+s^2 \left(19 s_1^2+24 s_1 s_2+21 s_2^2\right)+s \left(-13 s_1^3+25 s_1^2 s_2+5 s_1 s_2^2-17 s_2^3\right)\notag\\
&+(s_1-s_2)^3 (3 s_1-5 s_2)\big]+8 s_2 \left(s^3-3 s^2 (s_1+s_2)-3 s \left(s_1^2-s_2^2\right)+(s_1-s_2)^3\right)\bigg\}B_0(s_2,0,0)\notag\\
&+\sdfrac{4 \pi ^2 e^2 s^2 s_1 s_2}{(d-2) d \sigma ^2} (d \sigma +8 s_1 s_2)C_0(s,s_1,s_2)
\end{align}
}%

\subsection{Form Factors \texorpdfstring{$F_i$}{} in d-dim}

We present here the expressions of the invariant amplitudes in the massless limit in d-dimensions, previously given in $d=4$. These expressions will be functions of the three invariants $s=p_1^2$, $s_2=p_1^2$ and~$s_2=p_3^2$ and are explicitly given as
{\small
\begin{align}
&F_{1,d}(s,s_1,s_2,0)=\sdfrac{2\pi^2\,e^2(d-4)}{3(d-2)(d-1)d\,\s^3}\bigg\{ 2s_2^3\big[(8-9d)s^2+2s\,s_1d(3d-8)+4s\,s_1+(2d-3)d\,s_1^2\big]\notag\\
&-s_2(s-s_1)^2\big[(13d-16)s^2-2(d-4)(2d-1)s\,s_1+d(4d-7)s_1^2\big]+2s_2^2\big[(11d-12)s^3+3(5-2d)ds^2\,s_1\notag\\
&+3(3-4d)d\,s\,s_1^2+12s\,s_1^2+d(2d-3)s_1^3\big]+s_2^4\big[(7d-4)s+(7-4d)ds_1\big]+(s-s_1)^4\big[(3d-4)s-d\,s_1\big]\notag\\
&-d\,s_2^5\bigg\}\,B_0(s)+\sdfrac{2\pi^2\,e^2(d-4)s_1}{3(d-2)(d-1)d\,s\,\s^3}\bigg\{ 2s_2^3\big[-d^2(s^2+10s\,s_1+5s_1^2)+d(s+s_1)(s+9s_1)+4s_1(2s+s_1)\big]\notag\\
&-2ds_2^2\big[(d-1)s^3+(13-7d)s^2\,s_1+(9-5d)s\,s_1^2+(11-5d)s_1^3\big]+s_2^4\big[3(d-1)d\,s+d\,s_1(5d-7)-4s_1\big]\notag\\
&+(d-1)s_2(s-s_1)^2\big[d(s-s_1)(3s+5s_1)+8s_1(2s+s_1)\big]-(s-s_1)^4\big[(d-1)d\,s-s_1(d-3)-4s_1\big]\notag\\
&-(d-1)d\,s_2^5\bigg\}\,B_0(s_1)+\sdfrac{2\pi^2\,e^2(d-4)s_2}{3(d-2)(d-1)d\,s\,\s^3}\bigg\{ d^2\big[2s_2^3(5s^2+6s\,s_1+5s_1^2)-2s_2^2(5s^3+3s^2\,s_1-5s\,s_1^2+5s_1^3)\notag\\
&+s_2(s-s_1)(5s^3+s^2\,s_1+15s\,s_1^2-5s_1^3)-5s_2^4(s+s_1)-(s-s_1)^4(s+s_1)+s_2^5\big]\notag\\
&+d\s\big[7s_2^2(s+s_1)-5s_2(s-s_1)^2+(s-s_1)^2(s+s_1)-3s_2^3\big]+4s_2\g(s+s_1-s_2)\big[\s-2s\,s_1\big]\bigg\}\,B_0(s_2)\notag\\
&-\sdfrac{4\pi^2\,e^2(d-4)s_1\,s_2\,\g}{3(d-2)d\,\s^3}\bigg\{d\big[-s^3+3s^2(s_1+s_2)-(s_1-s_2)^2(\g+2s)\big]+8s\,s_1\,s_2\bigg\}\,C_0(s,s_1,s_2)
\end{align}

\begin{align}
\label{F2}
&F_{2,d}(s,s_1,s_2)=\sdfrac{4\pi^2\,e^2(d-4)}{3(d-2)(d-1)d\,\s^3}\bigg\{2s^2\big[2d^2\,s_1\,s_2-3(d-2)s_1^2-3(d-2)s_2^2\big]-(d-4)s^4\notag\\
&+4(d-3)s^3(s_1+s_2)+4(d-1)s(s_1+s_2)\big[s_1^2-4s_1\,s_2+s_2^2\big]-d(s_1-s_2)^2\big[2(2d-3)s_1\,s_2+s_1^2+s_2^2\big]\bigg\}\,B_0(s)\notag\\
&-\sdfrac{4\pi^2\,e^2(d-4)s_1}{3(d-2)(d-1)d\,s\,\s^3}\bigg\{ 4d^2\,s\,s_2\big[(s-s_2)^2-s_1^2\big]+d\,\s(\s -4s\,s_2)+4s_1(s-s_1+s_2)(\s-2s_2\,s)\bigg\}B_0(s_1,0,0)\notag\\
&-\sdfrac{4\pi^2\,e^2(d-4)s_2}{3(d-2)(d-1)d\,s\,\s^3}\bigg\{4d^2s\,s_1\big[(s-s_1)^2-s_2^2\big]+d\s(\s-4s\,s_1)+4s_2(s+s_1-s_2)(\s-2s\,s_1)\bigg\}B_0(s_2,0,0)\notag\\
&+\sdfrac{8\pi^2(d-4)e^2\,s_1\,s_2}{3(d-2)d\s^3}\bigg\{8s\,s_1\,s_2+d\,\g\,\big[s^2-(s_1-s_2)^2\big]\bigg\}\,C_0(s,s_1,s_2)
\end{align}
}%
{\small
\begin{align}
&F_{3,d}(s,s_1,s_2)=F_{5,d}(s,s_2,s_1)=\sdfrac{\pi^2\,e^2}{3(d-2)(d-1)d\s^4}\bigg\{s_2^5\big[-3(d(5d-21)+8)s^2+4(d(11-d(4d+9))\notag\\
&+8)s\,s_1+3(d-1)d(4d-7)s_1^2\big]-s_2(s-s_1)^4\big[\big((d-1)d+24\big)s^2+8\big(d((d-2)d+8)-4\big)s\,s_1\notag\\
&+(d-1)d(4d-9)s_1^2\big]+s_2^4\big[2\big(5d(2d-11)+84\big)s^3+\big(d(d(32d-35)-121)+64\big)s^2\,s_1+2\big(d(3d(4d\notag\\
&-17)+74)+4\big)s\,s_1^2-(d-1)d(8d-13)s_1^3\big]+s_2^2(s-s_1)^2\big[6\big((d-6)d+28\big)s^3+\big(d(d(4d-93)+37)\notag\\
&+64\big)s^2\,s_1+2\big((149-84d)d+4\big)s\,s_1^2+3(d-1)d(4d-7)s_1^3\big]+8\big(d(d(d+33)-50)-8\big)s\,s_1^3\notag\\
&-(d-1)d(8d-13)s_1^4\big]+s_2^3\big[\big(5(19-3d)d-272\big)s^4-8d\big((d-18)d-7\big)s^3\,s_1+2\big(d(67-d(28d\notag\\
&+111))+168\big)s^2\,s_1^2+s_2^6\big[2\big(d(3d-8)-4\big)s-(d-1)d(4d-9)s_1\big]+(s-s_1)^6\big[2(d-4)s-(d-1)d\,s_1\big]\notag\\
&-(d-1)d\,s_2^7\bigg\}\,B_0(s,0,0)+\sdfrac{\pi ^2 e^2 s_1 }{6 (d-2) (d-1) d s \sigma ^4} \bigg\{d^3 \big[s_2 ^5 \left(21 s^2+46 s s_1 +21 s_1 ^2\right)+s_2  \left(7 s^2+10 s s_1 +7 s_1 ^2\right)\notag\\
&\times(s-s_1 )^4+s_2 ^4 \left(19 s^3-105 s^2 s_1 -39 s s_1 ^2-35 s_1 ^3\right)-s_2 ^2 \left(3 s^3-17 s^2 s_1 +25 s s_1 ^2+21 s_1 ^3\right) (s-s_1 )^2+(s-s_1 )^7\notag\\
&+s_2 ^3 \left(-29 s^4+68 s^3 s_1 +50 s^2 s_1 ^2+4 s s_1 ^3+35 s_1 ^4\right)-s_2 ^6 (17 s+7 s_1 )+s_2 ^7\big]+d^2 \big[-3 s_2 ^5 \left(47 s^2+66 s s_1 +39 s_1 ^2\right)\notag\\
&+s_2  (s-s_1 )^4 \left(25 s^2-34 s s_1 -47 s_1 ^2\right)+s_2 ^4 \left(169 s^3-85 s^2 s_1 +287 s s_1 ^2+205 s_1 ^3\right)-3 s_2 ^2 (s-s_1 )^2 \big(7 s^3-107 s^2 s_1\notag\\
& -47 s s_1 ^2-45 s_1 ^3\big)-s_2 ^3 \left(71 s^4+20 s^3 s_1 -38 s^2 s_1 ^2+116 s s_1 ^3+215 s_1 ^4\right)+s_2 ^6 (49 s+37 s_1 )-(5 s-7 s_1 ) (s-s_1 )^6\notag\\
&-5 s_2 ^7\big]+2 d \big[2 s_2 ^5 \left(51 s^2+32 s s_1 +29 s_1 ^2\right)-2 s_2  (s-s_1 )^4 \left(11 s^2-24 s s_1 -23 s_1 ^2\right)-s_2 ^4 \big(169 s^3-2 s^2 s_1 +65 s s_1 ^2\notag\\
&+120 s_1 ^3\big)-s_2 ^2 (s-s_1 )^2 \left(3 s^3+82 s^2 s_1 +123 s s_1 ^2+112 s_1 ^3\right)+2 s_2 ^3 \left(55 s^4-36 s^3 s_1 -34 s^2 s_1 ^2+4 s s_1 ^3+75 s_1 ^4\right)\notag\\
&-s_2 ^6 (25 s+16 s_1 )+(s-s_1 )^6 (5 s-8 s_1 )+2 s_2 ^7\big]+16 s_1  \left(s^2+s (7 s_2 -2 s_1 )+(s_1 -s_2 )^2\right) \left(\s-2 s\,s_2\right)\g\notag\\
&\times(s-s_1 +s_2 )\bigg\}\,B_0(s_1,0,0)+\sdfrac{\pi ^2 e^2 s_2}{6 (d-2) (d-1) d s \sigma ^4} \bigg\{d^3 \big[s^7+7 s^6 (s_1-s_2)-3 s^5 (s_1-s_2) (17 s_1+7 s_2)\notag\\
&+s^4 \left(115 s_1^3-73 s_1^2 s_2-39 s_1 s_2^2-35 s_2^3\right)+s^3 \left(-125 s_1^4+116 s_1^3 s_2+82 s_1^2 s_2^2-44 s_1 s_2^3+35 s_2^4\right)+s^2 (s_1-s_2)\notag\\
&\times \left(69 s_1^4-36 s_1^3 s_2-82 s_1^2 s_2^2-36 s_1 s_2^3+21 s_2^4\right)-s (s_1-s_2)^3 \left(17 s_1^3+5 s_1^2 s_2+3 s_1 s_2^2+7 s_2^3\right)+(s_1-s_2)^7\big]\notag\\
&+d^2 \big[s^7+s^6 (s_2-35 s_1)+s^5 \left(165 s_1^2+166 s_1 s_2-27 s_2^2\right)-s^4 \left(335 s_1^3+381 s_1^2 s_2+169 s_1 s_2^2-85 s_2^3\right)\notag\\
&+s^3 \left(355 s_1^4+100 s_1^3 s_2+498 s_1^2 s_2^2+4 s_1 s_2^3-125 s_2^4\right)-s^2 (s_1-s_2) \left(201 s_1^4-110 s_1^3 s_2+452 s_1^2 s_2^2+62 s_1 s_2^3+99 s_2^4\right)\notag\\
&+s (s_1-s_2)^3 \left(55 s_1^3-69 s_1^2 s_2+5 s_1 s_2^2+41 s_2^3\right)-(5 s_1-7 s_2) (s_1-s_2)^6\big]+2 d \big[-s^7+2 s^6 (7 s_1+11 s_2)\notag\\
&-s^5 \left(57 s_1^2+176 s_1 s_2+87 s_2^2\right)+2 s^4 \left(55 s_1^3+190 s_1^2 s_2+221 s_1 s_2^2+70 s_2^3\right)-s^3 \big(115 s_1^4+288 s_1^3 s_2+502 s_1^2 s_2^2\notag\\
&+312 s_1 s_2^3+95 s_2^4\big)+2 s^2 (s_1-s_2) \left(33 s_1^4+40 s_1^3 s_2+162 s_1^2 s_2^2+8 s_1 s_2^3-3 s_2^4\right)-s (s_1-s_2)^3 \big(19 s_1^3-7 s_1^2 s_2\notag\\
&+77 s_1 s_2^2+23 s_2^3\big)+2 (s_1-s_2)^5 \left(s_1^2-3 s_1 s_2+4 s_2^2\right)\big]+16 s_2 \left(s^2+s (7 s_2-2 s_1)+(s_1-s_2)^2\right) \left(\s-2s\,s_1\right) \g\,\notag\\
&\times(s+s_1-s_2)\bigg\}\,B_0(s_2,0,0)+\frac{2 \pi ^2 e^2 s_1 s_2}{3 (d-2) d \sigma ^4} \bigg\{\big[s_2^4 \left(-12 (d-1) d s^2+((d+55) d+16) s s_1-(d-1) d s_1^2\right)\notag\\
&+s_2 (s-s_1)^3 \left(3 (d-7) d s^2-((d+37) d+16) s s_1+2 (d-1) d s_1^2\right)+2 s_2^3 \big((5 d-23) d s^3+6 ((d+3) d+8) \notag\\
&\times s^2 s_1-(d+8) (5 d+1) s s_1^2+2 (d-1) d s_1^3\big)-s_2^2 (s-s_1) \big(3 (d-17) d s^3+((13 d+59) d+96) s^2 s_1\notag\\
&+((d-9) d-16) s s_1^2-(d-1) d s_1^3\big)+d s_2^5 (3 (d+1) s-2 (d-1) s_1)-(d-1) d (s-s_1)^5 (2 s+s_1)\notag\\
&+(d-1) d s_2^6\bigg\}\,C_0(s,s_1,s_2,0,0,0)
\end{align}
\begin{align}
&F_{4,d}(s,s_1,s_2)=F_{6,d}(s,s_2,s_1)=\sdfrac{\pi ^2 e^2}{3 (d-2) (d-1) d \sigma ^4 s_1} \bigg\{2 s^4 \big[-\left((6 d^2+45 d-31) d+136\right) s_1 s_2^2\notag\\
&+5 (d (3 (2 d-7) d+35)-8) s_1^3+4 (d ((d-9) d+9)-34) s_1^2 s_2-30 ((d-3) d+4) s_2^3\big]\notag\\
&+2 s^2 \big[2 \left(-39 d^2+d+44\right) s_1^2 s_2^3+((3 (5 d-18) d+77) d+28) s_1^5-((3 (8 d-69) d+205) d+92) s_1^4 s_2\notag\\
&+2 (((3 d-64) d+19) d+60) s_1^3 s_2^2+((3 d (d+10)-67) d+28) s_1 s_2^4-3 (3 (d-3) d+4) s_2^5\big]\notag\\
&+3 ((d-3) d+8) s^7+2 s^6 ((d ((3 d-16) d+35)-52) s_1-3 (3 (d-3) d+20) s_2)\notag\\
&+s^5 \big[((3 (39-10 d) d-217) d+160) s_1^2+2 ((-3 (d-17) d-86) d+152) s_1 s_2+15 (3 (d-3) d+16) s_2^2\big]\notag\\
&+s^3 \big[-5 (((12 d-41) d+63) d+8) s_1^4+4 (((5 d-61) d+74) d+48) s_1^3 s_2+2 (d (11 (2 d+5) d+61)\notag\\
&-72) s_1^2 s_2^2+4 (((3 d-5) d+36) d+8) s_1 s_2^3+15 (3 (d-3) d+8) s_2^4\big]-s (s_1-s_2)^2 \big[((3 (2 d-9) d+35) d\notag\\
&+16) s_1^4-2 (((11 d-78) d+55) d+24) s_1^3 s_2+2 (d (d (13 d+27)-34)-24) s_1^2 s_2^2+2 ((3 d (d+2)-5) d\notag\\
&+8) s_1 s_2^3-3 (d-3) d s_2^4\big]-2 (d-1) d s_1 (s_1-s_2)^4 \big[2 (2 d-3) s_1 s_2+s_1^2+s_2^2\big]\bigg\}B_0(s,0,0)\notag\\
&+\sdfrac{\pi ^2 e^2}{3 (d-2) (d-1) d s \sigma ^4} \bigg\{2 d^3 s \big[s_2^4 \left(9 s^2-5 s s_1+8 s_1^2\right)-2 s s_2^3 \left(3 s^2-12 s s_1+s_1^2\right)+s_2 \left(9 s^2-s s_1-4 s_1^2\right)\notag\\
&\times (s-s_1)^3-2 s_2^2 \left(3 s^4+s^3 s_1+s^2 s_1^2-9 s s_1^3+4 s_1^4\right)-s_2^5 (3 s+4 s_1)-3 s (s-s_1)^5\big]\notag\\
&+d^2 \big[12 s^7-s^6 (59 s_1+48 s_2)+2 s^5 \left(57 s_1^2+23 s_1 s_2+36 s_2^2\right)+s^4 \left(-105 s_1^3+162 s_1^2 s_2+23 s_1 s_2^2-48 s_2^3\right)\notag\\
&+4 s^3 \left(10 s_1^4-69 s_1^3 s_2-45 s_1^2 s_2^2+13 s_1 s_2^3+3 s_2^4\right)+s^2 s_1 (s_1-s_2) \left(3 s_1^3+119 s_1^2 s_2+169 s_1 s_2^2+61 s_2^3\right)\notag\\
&-2 s s_1 (s_1-s_2)^3 \left(3 s_1^2+6 s_1 s_2-s_2^2\right)+s_1 (s_1-s_2)^6\big]+d \big[-9 s^7+16 s^6 (2 s_1+3 s_2)-s^5 \big(17 s_1^2+62 s_1 s_2\notag\\
&+105 s_2^2\big)-4 s^4 (s_1-s_2) \left(20 s_1^2+53 s_1 s_2+30 s_2^2\right)+s^3 \left(165 s_1^4+244 s_1^3 s_2+14 s_1^2 s_2^2-124 s_1 s_2^3-75 s_2^4\right)\notag\\
&-8 s^2 (s_1-s_2) \left(17 s_1^4+22 s_1^3 s_2+8 s_1^2 s_2^2+10 s_1 s_2^3+3 s_2^4\right)+s (s_1-s_2)^3 \left(53 s_1^3+57 s_1^2 s_2-s_1 s_2^2+3 s_2^3\right)\notag\\
&-4 s_1 (s_1-s_2)^5 (2 s_1-s_2)\big]-8 s_1 (s-s_1+s_2) \left(s^2-2 s (s_1+2 s_2)+(s_1-s_2)^2\right) \big[3 s^3-2 s^2 (2 s_1+3 s_2)\notag\\
&+s \left(-s_1^2+8 s_1 s_2+3 s_2^2\right)+2 s_1 (s_1-s_2)^2\big]\bigg\}\,B_0(s_1,0,0)+\sdfrac{\pi ^2 e^2}{6 (d-2) (d-1) d s \sigma ^4 s_1} \bigg\{d^3 s \big[s_2^5 \big(63 s^2+90 s s_1\notag\\
&+79 s_1^2\big)+21 s_2^3 (s-s_1)^2 \left(5 s^2+6 s s_1+5 s_1^2\right)+s_2 (s-s_1)^4 \left(21 s^2-18 s s_1+5 s_1^2\right)-s_2^4 \big(105 s^3+87 s^2 s_1\notag\\
&+23 s s_1^2+137 s_1^3\big)-s_2^2 (s-s_1)^2 \left(63 s^3-63 s^2 s_1+s s_1^2+31 s_1^3\right)-21 s_2^6 (s+s_1)-3 (s-s_1)^6 (s+s_1)+3 s_2^7\big]\notag\\
&+d^2 \big[-3 s_2^5 \left(93 s^3+100 s^2 s_1+89 s s_1^2-10 s_1^3\right)-s_2 (s-s_1)^4 \left(69 s^3+28 s^2 s_1+65 s s_1^2-2 s_1^3\right)\notag\\
&+s_2^4 \left(435 s^4+245 s^3 s_1+517 s^2 s_1^2+475 s s_1^3-40 s_1^4\right)+3 s_2^2 (s-s_1)^2 \left(75 s^4+s^3 s_1-119 s^2 s_1^2+79 s s_1^3-4 s_1^4\right)\notag\\
&+s_2^3 \left(-405 s^5+210 s^4 s_1+186 s^3 s_1^2+328 s^2 s_1^3-477 s s_1^4+30 s_1^5\right)+s_2^7 (2 s_1-15 s)+3 s_2^6 (s+s_1) (33 s-4 s_1)\notag\\
&+9 s (s-s_1)^6 (s+s_1)\big]+2 d \big[s_2^6 \left(-3 s^2+17 s s_1+44 s_1^2\right)-s_2^2 (s-s_1)^2 \left(5 s^2+23 s s_1-4 s_1^2\right) \left(9 s^2-8 s s_1+7 s_1^2\right)\notag\\
&+2 s_2^5 \left(9 s^3+52 s^2 s_1+39 s s_1^2-50 s_1^3\right)+2 s_2 (s-s_1)^4 \left(9 s^3+16 s^2 s_1+15 s s_1^2-2 s_1^3\right)\notag\\
&-s_2^4 \left(45 s^4+339 s^3 s_1+259 s^2 s_1^2+373 s s_1^3-120 s_1^4\right)+4 s_2^3 \left(15 s^5+82 s^4 s_1-80 s^3 s_1^2-86 s^2 s_1^3+121 s s_1^4-20 s_1^5\right)\notag\\
&-3 s (s-s_1)^6 (s+s_1)-8 s_1 s_2^7\big]-16 s_2^2 (s+s_1-s_2) \left(\s -2s\,s_1\right) \big[3 s^3-2 s^2 (2 s_1+3 s_2)+s \left(-s_1^2+8 s_1 s_2+3 s_2^2\right)\notag\\
&+2 s_1 (s_1-s_2)^2\big]\bigg\}\,B_0(s_2,0,0)\notag+\sdfrac{\pi ^2 e^2}{3 (d-2) d \sigma ^4} \bigg\{6 d s_2^5 \big[-(d-1) s^2+(d+11) s s_1-2 (d-1) s_1^2\big]\notag\\
&-2 d s_2 (s-s_1)^4 \left(3 (d-5) s^2+(d-25) s s_1-2 (d-1) s_1^2\right)-s_2^4 \big[3 (d+10) d s^3+2 (5 d (d+2)-48) s^2 s_1\notag\\
&+(d (11 d+38)-64) s s_1^2-8 (d-1) d s_1^3\big]-s_2^2 (s-s_1)^2 \big[3 (d+20) d s^3-6 ((3 d+2) d+16) s^2 s_1-((13 d\notag\\
&+4) d+64) s s_1^2+12 (d-1) d s_1^3\big]+4 s_2^3 \big[3 (d+5) d s^4-(7 (d+5) d+48) s^3 s_1+((7 d+55) d+64)s^2 s_1^2\notag\\
&-((5 d+17) d+32) s s_1^3+2 (d-1) d s_1^4\big]+d s_2^6 (3 d s+4 (d-1) s_1)+3 (d-2) d s (s-s_1)^6\bigg\}C_0(s,s_1,s_2)
\end{align}
\begin{align}
&F_{7,d}(s,s_1,s_2)=-\sdfrac{2 \pi ^2 e^2}{3 (d-2) (d-1) d \sigma ^4\g} \bigg\{4 (d-4) s^8-2 (d (d+10)-44) s^7 (s_1+s_2)\notag\\
&+s^6 \left((d (11 d+37)-192) s_1^2+2 (d ((4 d-5) d+5)-136) s_1 s_2+(d (11 d+37)-192) s_2^2\right)\notag\\
&-2 s^5 (s_1+s_2) \left((d (12 d+13)-100) s_1^2+2 (d (4 d (3 d-8)-23)-8) s_1 s_2+(d (12 d+13)-100) s_2^2\right)\notag\\
&+s^4 \big[5 (d (5 d-1)-16) s_1^4+4 ((3 d-7) (9 d-1) d+8) s_1^3 s_2+2 (d ((20 d-177) d+57)-128) s_1^2 s_2^2\notag\\
&+4 ((3 d-7) (9 d-1) d+8) s_1 s_2^3+5 (d (5 d-1)-16) s_2^4\big]-2 s^3 (s_1+s_2) \big[((5 d-8) d+12) s_1^4\notag\\
&+4 (((14 d-37) d+36) d+8) s_1^3 s_2-2 (((44 d-47) d+20) d+172) s_1^2 s_2^2+4 (((14 d-37) d+36) d\notag\\
&+8) s_1 s_2^3+((5 d-8) d+12) s_2^4\big]-s^2 \big[(d (3 d+5)-32) s_1^6+2 (((61-24 d) d-129) d+8) s_1^5 s_2\notag\\
&+(((48 d-355) d+363) d+448) s_1^4 s_2^2+4 (d (147 d+73)-376) s_1^3 s_2^3+(((48 d-355) d+363) d+448) s_1^2 s_2^4\notag\\
&+2 (((61-24 d) d-129) d+8) s_1 s_2^5+(d (3 d+5)-32) s_2^6\big]+2 s (s_1-s_2)^2 (s_1+s_2) \big[(d (2 d-1)-4) s_1^4\notag\\
&+8 (1-4 d) s_1^3 s_2+2 ((69-2 d (2 d+25)) d+12) s_1^2 s_2^2+8 (1-4 d) s_1 s_2^3+(d (2 d-1)-4) s_2^4\big]\notag\\
&-(d-1) d (s_1-s_2)^4 (s_1+s_2)^2 \left(2 (2 d-3) s_1 s_2+s_1^2+s_2^2\right)\bigg\}B_0(s,0,0)-\sdfrac{e^2 \pi ^2 s_1}{3 (d-2) (d-1) d s \g^2 \sigma ^4} \bigg\{\big[-s_2^9\notag\\
&+(28 s+5 s_1) s_2^8-2 \left(79 s^2+45 s_1 s+4 s_1^2\right) s_2^7+2 s \left(209 s^2+214 s_1 s+101 s_1^2\right) s_2^6-2 \big(316 s^4+427 s_1 s^3+195 s_1^2 s^2\notag\\
&+189 s_1^3 s-7 s_1^4\big) s_2^5+2 \left(293 s^5+373 s_1 s^4-46 s_1^2 s^3+130 s_1^3 s^2+217 s_1^4 s-7 s_1^5\right) s_2^4-2 s (s-s_1) \big(169 s^4+236 s_1 s^3\notag\\
&-2 s_1^2 s^2-68 s_1^3 s-143 s_1^4\big) s_2^3+2 (s-s_1)^3 \left(59 s^4+75 s_1 s^3+33 s_1^2 s^2-67 s_1^3 s-4 s_1^4\right) s_2^2-(s-s_1)^5 \big(23 s^3-3 s_1 s^2\notag\\
&-39 s_1^2 s-5 s_1^3\big) s_2+(s-s_1)^8 (2 s+s_1)\big] d^3+\big[5 s_2^9-(98 s+27 s_1) s_2^8+2 \left(278 s^2+229 s_1 s+24 s_1^2\right) s_2^7-2 \big(784 s^3\notag\\
&+875 s_1 s^2+533 s_1^2 s+4 s_1^3\big) s_2^6+2 \left(1295 s^4+1325 s_1 s^3+1790 s_1^2 s^2+861 s_1^3 s-39 s_1^4\right) s_2^5-2 \big(1330 s^5+618 s_1 s^4\notag\\
&+1515 s_1^2 s^3+743 s_1^3 s^2+1023 s_1^4 s-45 s_1^5\big) s_2^4+2 (s-s_1) \left(854 s^5+317 s_1 s^4+185 s_1^2 s^3+19 s_1^3 s^2-803 s_1^4 s+4 s_1^5\right) s_2^3\notag\\
&-2 (s-s_1)^3 \left(328 s^4+221 s_1 s^3-358 s_1^2 s^2-407 s_1^3 s-24 s_1^4\right) s_2^2+(s-s_1)^5 \left(133 s^3+39 s_1 s^2-211 s_1^2 s-33 s_1^3\right) s_2\notag\\
&-(5 s-7 s_1) (s-s_1)^7 (2 s+s_1)\big] d^2+2 \big[-2 s_2^9+(41 s+12 s_1) s_2^8-\left(247 s^2+198 s_1 s+28 s_1^2\right) s_2^7+\big(743 s^3\notag\\
&+755 s_1 s^2+706 s_1^2 s+20 s_1^3\big) s_2^6-\left(1315 s^4+1048 s_1 s^3+2423 s_1^2 s^2+1470 s_1^3 s-32 s_1^4\right) s_2^5+\big(1457 s^5+101 s_1 s^4\notag\\
&+2431 s_1^2 s^3+2011 s_1^3 s^2+1748 s_1^4 s-68 s_1^5\big) s_2^4-(s-s_1) \big(1021 s^5-213 s_1 s^4+517 s_1^2 s^3-819 s_1^3 s^2-1302 s_1^4 s\notag\\
&+28 s_1^5\big) s_2^3+(s-s_1)^3 \left(437 s^4-20 s_1 s^3-595 s_1^2 s^2-706 s_1^3 s-28 s_1^4\right) s_2^2-(s-s_1)^5 \big(103 s^3-41 s_1 s^2-224 s_1^2 s\notag\\
&-30 s_1^3\big) s_2+(s-s_1)^7 (2 s+s_1) (5 s-8 s_1)\big] d+16 s_1 \left(s^2-2 (s_1+2 s_2) s+(s_1-s_2)^2\right) (s-s_1+s_2) \big[2 s^5\notag\\
&-7 (s_1+s_2) s^4+2 \left(4 s_1^2+7 s_2 s_1+4 s_2^2\right) s^3-2 (s_1+s_2)^3 s^2-2 \left(s_1^4+s_2 s_1^3-24 s_2^2 s_1^2+s_2^3 s_1+s_2^4\right) s\notag\\
&+(s_1-s_2)^2 (s_1+s_2)^3\big]\bigg\}B_0(s_1)-\sdfrac{e^2 \pi ^2 s_2}{3 (d-2) (d-1) d s \g^2 \sigma ^4} \bigg\{\big[s_2^9-(6 s+5 s_1) s_2^8+2 \left(6 s^2-7 s_1 s+4 s_1^2\right) s_2^7\notag\\
&+2 s s_1 (71 s+55 s_1) s_2^6-2 \left(21 s^4+151 s_1 s^3+222 s_1^2 s^2+143 s_1^3 s+7 s_1^4\right) s_2^5+2 \big(42 s^5+110 s_1 s^4+221 s_1^2 s^3+75 s_1^3 s^2\notag\\
&+217 s_1^4 s+7 s_1^5\big) s_2^4+2 s \left(-42 s^5+35 s_1 s^4+66 s_1^3 s^2+130 s_1^4 s-189 s_1^5\right) s_2^3+2 (s-s_1)^2 \big(24 s^5-55 s_1 s^4-149 s_1^2 s^3	\notag\\
&-5 s_1^3 s^2+93 s_1^4 s-4 s_1^5\big) s_2^2-(s-s_1)^4 \left(15 s^4-58 s_1 s^3-118 s_1^2 s^2+70 s_1^3 s-5 s_1^4\right) s_2+(s-s_1)^6 \big(2 s^3-11 s_1 s^2\notag\\
&+22 s_1^2 s-s_1^3\big)\big] d^3+\big[-7 s_2^9+(40 s+33 s_1) s_2^8-2 \left(37 s^2-23 s_1 s+24 s_1^2\right) s_2^7-2 \big(7 s^3+382 s_1 s^2+335 s_1^2 s\notag\\
&+4 s_1^3\big) s_2^6+2 \left(140 s^4+921 s_1 s^3+791 s_1^2 s^2+807 s_1^3 s+45 s_1^4\right) s_2^5-2 \big(259 s^5+835 s_1 s^4-98 s_1^2 s^3+822 s_1^3 s^2\notag\\
&+1023 s_1^4 s+39 s_1^5\big) s_2^4+2 \left(245 s^6+41 s_1 s^5-1002 s_1^2 s^4-166 s_1^3 s^3-743 s_1^4 s^2+861 s_1^5 s-4 s_1^6\right) s_2^3-2 (s-s_1)^2\notag\\
&\times \left(133 s^5-196 s_1 s^4-562 s_1^2 s^3-796 s_1^3 s^2+485 s_1^4 s-24 s_1^5\right) s_2^2+(s-s_1)^4 \big(79 s^4-310 s_1 s^3-188 s_1^2 s^2+350 s_1^3 s\notag\\
&-27 s_1^4\big) s_2-(s-s_1)^7 \left(10 s^2-63 s_1 s+5 s_1^2\right)\big] d^2+2 \big[8 s_2^9-15 (3 s+2 s_1) s_2^8+\left(81 s^2-74 s_1 s+28 s_1^2\right) s_2^7+\big(21 s^3\notag\\
&+779 s_1 s^2+622 s_1^2 s+28 s_1^3\big) s_2^6-\left(315 s^4+1632 s_1 s^3+1439 s_1^2 s^2+1330 s_1^3 s+68 s_1^4\right) s_2^5+\big(567 s^5+1165 s_1 s^4\notag\\
&+325 s_1^2 s^3+483 s_1^3 s^2+1748 s_1^4 s+32 s_1^5\big) s_2^4+\big(-525 s^6+350 s_1 s^5+582 s_1^2 s^4+1336 s_1^3 s^3+2011 s_1^4 s^2-1470 s_1^5 s\notag\\
&+20 s_1^6\big) s_2^3+(s-s_1)^2 \left(279 s^5-453 s_1 s^4-409 s_1^2 s^3-1095 s_1^3 s^2+650 s_1^4 s-28 s_1^5\right) s_2^2\notag
\end{align}
\begin{align}
&-(s-s_1)^4 \left(81 s^4-232 s_1 s^3-83 s_1^2 s^2+150 s_1^3 s-12 s_1^4\right) s_2+(s-s_1)^6 \left(10 s^3-43 s_1 s^2+29 s_1^2 s-2 s_1^3\right)\big] d\notag\\
&+16 \left(\s-2s\,s_1\right) (s+s_1-s_2) s_2 \big[2 s^5-7 (s_1+s_2) s^4+2 \left(4 s_1^2+7 s_2 s_1+4 s_2^2\right) s^3-2 (s_1+s_2)^3 s^2\notag\\
&-2 \left(s_1^4+s_2 s_1^3-24 s_2^2 s_1^2+s_2^3 s_1+s_2^4\right) s+(s_1-s_2)^2 (s_1+s_2)^3\big]\bigg\}B_0(s_2,0,0)+\sdfrac{4 \pi ^2 e^2 s_1 s_2}{3 (d-2) d \sigma ^4 \g^2} \bigg\{s_2^6 \big[d (17\notag\\
&-11 d) s^2+2 ((3 d+19) d+8) s s_1-4 (d-1) d s_1^2\big]+2 s s_2^5 \big(2 (13 d-22) d s^2-((11 d+69) d+16) s s_1\notag\\
&+(5 d+1) (d+8) s_1^2\big)+2 s s_2 (s-s_1)^4 \left(2 (4 d-13) d s^2+((d+7) d+16) s s_1+((3 d+19) d+8) s_1^2\right)\notag\\
&-d (s-s_1)^6 \big[2 (d-4) s^2-4 (d-1) s s_1-(d-1) s_1^2\big]+s_2^4 \big[5 d (37-19 d) s^4+4 (d (11 d+30)-8) s^3 s_1\notag\\
&+(d (11 d+63)-32) s^2 s_1^2-2 ((7 d+61) d+16) s s_1^3+6 (d-1) d s_1^4\big]+2 s s_2^3 \big[(47 d-107) d s^4+2 (d (35\notag\\
&-19 d)+32) s^3 s_1-24 ((d+9) d+2) s^2 s_1^2+2 ((11 d+61) d+192) s s_1^3-((7 d+61) d+16) s_1^4\big]-s_2^2 (s-s_1)^2\notag\\
&\times \big[(53 d-143) d s^4+4 ((3 d+11) d+28) s^3 s_1-d (19 d+239) s^2 s_1^2-2 ((d+45) d+8) s s_1^3+4 (d-1) d s_1^4\big]\notag\\
&-2 (d-1) d s s_2^7+(d-1) d s_2^8\bigg\}C_0(s,s_1,s_2)
\end{align}
}%
{\small
\begin{align}
&F_{8,d}(s,s_1,s_2)=-\sdfrac{4 \pi ^2 e^2}{3 (d-2) (d-1) d \sigma ^4} \bigg\{\left(d^2+d+16\right) s^6+(d (3 (d-4) d+11)-56) s^5 (s_1+s_2)\notag\\
&+s^4 \big[((3 (13-4 d) d-55) d+64) s_1^2+2 ((d (4 d+21)-21) d+56) s_1 s_2+((3 (13-4 d) d-55) d+64) s_2^2\big]\notag\\
&+2 s^3 (s_1+s_2) \big[(d ((9 d-28) d+45)-8) s_1^2-4 (((5 d-8) d+14) d+10) s_1 s_2+(d ((9 d-28) d+45)\notag\\
&-8) s_2^2\big]+s^2 \big[-((3 (4 d-13) d+65) d+16) s_1^4+4 ((3 (d-9) d+34) d+20) s_1^3 s_2+2 (d (85 d-7)-144) s_1^2 s_2^2\notag\\
&+4 ((3 (d-9) d+34) d+20) s_1 s_2^3-((3 (4 d-13) d+65) d+16) s_2^4\big]+s (s_1-s_2)^2 (s_1+s_2) \big[((3 (d-4) d\notag\\
&+19) d+8) s_1^2-2 (((d-36) d+37) d+16) s_1 s_2+((3 (d-4) d+19) d+8) s_2^2\big]+(d-1) d (s_1-s_2)^4 \notag\\
&\times\left(2 (2 d-3) s_1 s_2+s_1^2+s_2^2\right)\bigg\}B_0(s)+\sdfrac{\pi ^2 e^2}{3 (d-2) (d-1) d s \sigma ^4 (s-s_1-s_2)} \bigg\{s \big[3 s_2^7-5 (3 s+s_1) s_2^6+\big(27 s^2\notag\\
&+82 s_1 s+47 s_1^2\big) s_2^5-\left(15 s^3+287 s_1 s^2+105 s_1^2 s+137 s_1^3\right) s_2^4+\left(-15 s^4+428 s_1 s^3-66 s_1^2 s^2+92 s_1^3 s+137 s_1^4\right) s_2^3\notag\\
&+(s-s_1) \left(27 s^4-280 s_1 s^3-18 s_1^2 s^2+32 s_1^3 s+47 s_1^4\right) s_2^2-(s-s_1)^3 (s+s_1) \left(15 s^2-68 s_1 s+5 s_1^2\right) s_2\notag\\
&+3 (s-s_1)^5 (s+s_1)^2\big] d^3+\big[-9 s^8+(53 s_1+45 s_2) s^7-\left(155 s_1^2+264 s_2 s_1+81 s_2^2\right) s^6+\big(297 s_1^3+691 s_2 s_1^2\notag\\
&+599 s_2^2 s_1+45 s_2^3\big) s^5-\left(385 s_1^4+822 s_2 s_1^3+196 s_2^2 s_1^2+818 s_2^3 s_1-45 s_2^4\right) s^4+\big(319 s_1^5+275 s_2 s_1^4-2 s_2^2 s_1^3\notag\\
&-1346 s_2^3 s_1^2+707 s_2^4 s_1-81 s_2^5\big) s^3-(s_1-s_2) \left(153 s_1^5-59 s_2 s_1^4+590 s_2^2 s_1^3+990 s_2^3 s_1^2-311 s_2^4 s_1+45 s_2^5\right) s^2\notag\\
&+(s_1-s_2)^3 \left(35 s_1^4-42 s_2 s_1^3+116 s_2^2 s_1^2-54 s_2^3 s_1+9 s_2^4\right) s-2 s_1 (s_1-s_2)^6 (s_1+s_2)\big] d^2+2 \big[3 s^8-(13 s_1\notag\\
&+15 s_2) s^7+\left(53 s_1^2+32 s_2 s_1+27 s_2^2\right) s^6-\left(183 s_1^3+77 s_2 s_1^2+29 s_2^2 s_1+15 s_2^3\right) s^5+\big(365 s_1^4+108 s_2 s_1^3\notag\\
&-390 s_2^2 s_1^2+60 s_2^3 s_1-15 s_2^4\big) s^4+\left(-403 s_1^5+79 s_2 s_1^4+490 s_2^2 s_1^3+1030 s_2^3 s_1^2-135 s_2^4 s_1+27 s_2^5\right) s^3\notag\\
&+(s_1-s_2) \left(243 s_1^5-37 s_2 s_1^4+94 s_2^2 s_1^3+758 s_2^3 s_1^2-113 s_2^4 s_1+15 s_2^5\right) s^2-(s_1-s_2)^3 \big(73 s_1^4+30 s_2 s_1^3\notag\\
&+156 s_2^2 s_1^2-38 s_2^3 s_1+3 s_2^4\big) s+4 s_1 (s_1-s_2)^5 (2 s_1-s_2) (s_1+s_2)\big] d+32 s_1^2 \left(s^2-2 (s_1+2 s_2) s+(s_1-s_2)^2\right) \notag\\
&\times(s-s_1+s_2) \big[2 s^3-3 (s_1+s_2) s^2+10 s_1 s_2 s+(s_1-s_2)^2 (s_1+s_2)\big]\bigg\}B_0(s_1)\notag\\
&+\sdfrac{\pi ^2 e^2}{3 (d-2) (d-1) d s \sigma ^4 (s-s_1-s_2)} \bigg\{s \big[-3 s_2^7+(9 s+5 s_1) s_2^6-\big(3 s^2+78 s_1 s+47 s_1^2\big) s_2^5+\big(-15 s^3\notag\\
&+151 s_1 s^2+15 s_1^2 s+137 s_1^3\big) s_2^4+(s-s_1) \left(15 s^3-5 s_1 s^2+45 s_1^2 s+137 s_1^3\right) s_2^3+(s-s_1)^2 \big(3 s^3-135 s_1 s^2\notag\\
&-11 s_1^2 s+47 s_1^3\big) s_2^2-(s-s_1)^4 \left(9 s^2-62 s_1 s+5 s_1^2\right) s_2+3 (s-s_1)^6 (s+s_1)\big] d^3+\big[-2 s_2^8+5 (7 s+2 s_1) s_2^7\notag
\end{align}
\begin{align}
&-3 \left(51 s^2+49 s_1 s+6 s_1^2\right) s_2^6+\left(319 s^3+212 s_1 s^2+347 s_1^2 s+10 s_1^3\right) s_2^5+\big(-385 s^4+275 s_1 s^3-649 s_1^2 s^2\notag\\
&-563 s_1^3 s+10 s_1^4\big) s_2^4+\left(297 s^5-822 s_1 s^4-2 s_1^2 s^3-400 s_1^3 s^2+561 s_1^4 s-18 s_1^5\right) s_2^3-(s-s_1)^2 \big(155 s^4-381 s_1 s^3\notag\\
&-721 s_1^2 s^2+285 s_1^3 s-10 s_1^4\big) s_2^2+(s-s_1)^4 \left(53 s^3-52 s_1 s^2+73 s_1^2 s-2 s_1^3\right) s_2-9 s (s-s_1)^6 (s+s_1)\big] d^2\notag\\
&+2 \big[8 s_2^8-(73 s+36 s_1) s_2^7+\left(243 s^2+189 s_1 s+56 s_1^2\right) s_2^6-\left(403 s^3+280 s_1 s^2+285 s_1^2 s+20 s_1^3\right) s_2^5+\big(365 s^4\notag\\
&+79 s_1 s^3+131 s_1^2 s^2+489 s_1^3 s-40 s_1^4\big) s_2^4+\left(-183 s^5+108 s_1 s^4+490 s_1^2 s^3+664 s_1^3 s^2-555 s_1^4 s+52 s_1^5\right) s_2^3\notag\\
&+(s-s_1)^2 \left(53 s^4+29 s_1 s^3-385 s_1^2 s^2+231 s_1^3 s-24 s_1^4\right) s_2^2-(s-s_1)^4 \left(13 s^3+20 s_1 s^2+31 s_1^2 s-4 s_1^3\right) s_2\notag\\
&+3 s (s-s_1)^6 (s+s_1)\big] d+32 \left(\s-2s\,s_1\right) (s+s_1-s_2) s_2^2 \big[2 s^3-3 (s_1+s_2) s^2+10 s_1 s_2 s+(s_1-s_2)^2\notag\\
&\times (s_1+s_2)\big]\bigg\}B_0(s_2,0,0)-\sdfrac{2 \pi ^2 e^2}{3 (d-2) d \sigma ^4 (s-s_1-s_2)} \bigg\{d^2 \big[3 s^5 (s_1+s_2)-4 s^4 \left(3 s_1^2-2 s_1 s_2+3 s_2^2\right)\notag\\
&+2 s^3 (s_1+s_2) \left(9 s_1^2-20 s_1 s_2+9 s_2^2\right)-12 s^2 (s_1-s_2)^2 \left(s_1^2+s_1 s_2+s_2^2\right)+s (s_1-s_2)^2 (s_1+s_2) \big(3 s_1^2-2 s_1 s_2\notag\\
&+3 s_2^2\big)+4 s_1 s_2 (s_1-s_2)^4\big] (-s+s_1+s_2)^2-2 d \big[3 s^7 (s_1+s_2)-2 s^6 \left(9 s_1^2+13 s_1 s_2+9 s_2^2\right)+s^5 (s_1+s_2)\notag\\
&\times \left(45 s_1^2+8 s_1 s_2+45 s_2^2\right)-2 s^4 \left(30 s_1^4+5 s_1^3 s_2+74 s_1^2 s_2^2+5 s_1 s_2^3+30 s_2^4\right)+s^3 (s_1+s_2) \notag\\
&\times\left(45 s_1^4-120 s_1^3 s_2+286 s_1^2 s_2^2-120 s_1 s_2^3+45 s_2^4\right)-2 s^2 \left(9 s_1^6-41 s_1^5 s_2+7 s_1^4 s_2^2+82 s_1^3 s_2^3+7 s_1^2 s_2^4-41 s_1 s_2^5+9 s_2^6\right)\notag\\
&+s (s_1-s_2)^2 (s_1+s_2)^3 \left(3 s_1^2-32 s_1 s_2+3 s_2^2\right)+2 s_1 s_2 (s_1-s_2)^4 (s_1+s_2)^2\big]+64 s s_1^2 s_2^2 \big(2 s^3-3 s^2 (s_1+s_2)\notag\\
&+10 s s_1 s_2+(s_1-s_2)^2 (s_1+s_2)\big)\bigg\}C_0(s,s_1,s_2)
\end{align}
\begin{align}
&F_{9,d}(s,s_1,s_2)=F_{10,d}(s,s_2,s_1)=\frac{4 \pi ^2 e^2 s}{(d-2) (d-1) d \sigma ^3 \g} \bigg\{-2 s_2 (s-s_1)^2 \big[\left(d^3-5 d+8\right) s-d ((d-8) d\notag\\
&+11) s_1\big]+2 s_2^2 \big[(d+4) ((d-3) d+4) s^2+2 (d+2) ((d-7) d+8) s s_1-(d-4) d (3 d-5) s_1^2\big]+2 s_2^3 \big[(d \notag\\
&\times((d-4) d+7)-8) s+(3 (d-4) d+13) d s_1\big]+(d-3) d (s-s_1)^4+((5-2 d) d-5) d s_2^4\bigg\}B_0(s)\notag\\
&-\frac{2 \pi ^2 e^2}{(d-2) (d-1) d \sigma ^3 \g^2} \bigg\{d^3 \sigma  \big[s^4-4 s^3 (s_1+2 s_2)+2 s^2 \left(3 s_1^2+6 s_1 s_2+7 s_2^2\right)-4 s \left(s_1^3+s_1 s_2^2+2 s_2^3\right)\notag\\
&+(s_1-s_2)^4\big]+d^2 \big[-s_2^4 \left(123 s^2+112 s s_1+69 s_1^2\right)+2 s_2 (s-s_1)^3 \left(21 s^2-19 s s_1-18 s_1^2\right)+4 s_2^3 \big(43 s^3+10 s^2 s_1\notag\\
&+11 s s_1^2+24 s_1^3\big)-s_2^2 (s-s_1) \left(123 s^3-53 s^2 s_1-23 s s_1^2-79 s_1^3\right)+14 s_2^5 (3 s+2 s_1)-(5 s-7 s_1) (s-s_1)^5-5 s_2^6\big]\notag\\
&+2 d \sigma  \big[s_2^2 \left(26 s^2+19 s s_1+23 s_1^2\right)+s_2 \left(-16 s^3+33 s^2 s_1+2 s s_1^2-19 s_1^3\right)-s_2^3 (16 s+13 s_1)+3 (s-s_1)^3 (s-2 s_1)\notag\\
&+3 s_2^4\big]-32 s_1^2 s_2 (s-s_1+s_2) \left(s^2-2 s (s_1+2 s_2)+(s_1-s_2)^2\right)\bigg\}B_0(s_1)+\frac{4 \pi ^2 e^2 s_2}{(d-2) (d-1) d \sigma ^3 \g^2} \notag\\
&\times\bigg\{2 (d-1) s_2^3 \big[((7 d-19) d+24) s^2+4 (d-4) d s s_1+((5 d-21) d+24) s_1^2\big]+d s_2 (s-s_1)^2\notag\\
&\times \big[((d-6) d+13) s^2-2 ((3 d-22) d+27) s s_1+((5 d-22) d+25) s_1^2\big]-2 s_2^2 \big[(d ((5 d-18) d+29)-8) s^3\notag\\
&-(5 d-3) ((d-5) d+8) s^2 s_1+((-5 (d-6) d-57) d+24) s s_1^2+(d ((5 d-24) d+35)-8) s_1^3\big]\notag\\
&+s_2^4 \big[(((30-7 d) d-55) d+48) s-(d-3) ((5 d-13) d+16) s_1\big]+(d-1) d (s-s_1)^4 \big[(d-1) s\notag\\
&-(d-3) s_1\big]+(d ((d-6) d+13)-16) s_2^5\bigg\}B_0(s_2)\notag-\frac{4 \pi ^2 e^2 s s_2}{(d-2) d \sigma ^3 \g^2} \bigg\{d^2 \sigma  (s-s_1+s_2) \g^2+2 d \big[-2 s_2^2\notag\\
&\times \left(s^3+5 s^2 s_1-s s_1^2+3 s_1^3\right)+s_2^4 (3 s+4 s_1)+2 s s_2^3 (s_1-s)+s_2 (s-s_1)^2 (s+s_1) (3 s+s_1)-(s-2 s_1) (s-s_1)^4\notag\\
&-s_2^5\big]+64 s s_1^2 s_2^2\big\}C_0(s,s_1,s_2)
\end{align}
\begin{align}
&F_{11,d}(s,s_1,s_2)=F_{12,d}(s,s_2,s_1)=\frac{2 \pi ^2 e^2 s }{(d-2) (d-1) d \sigma ^3 s_2}\bigg\{2 s_2 (s-s_1) \big[4 \left(d^2+d-3\right) s s_1\notag\\
&+(d-3) ((d-1) d+4) s^2-d ((d-8) d+11) s_1^2\big]-2 s_2^2 \big[3 (d-2) ((d-1) d+2) s^2+2 (d-1) ((d-2) d\notag\\
&+6) s s_1+(3 d-5) (d-4) d s_1^2\big]+2 s_2^3 \big[(d-1) ((3 d-5) d+4) s+(3 (d-4) d+13) d s_1\big]+(s-s_1)^3\notag\\
&\times \big[((d-3) d+8) s-(d-3) d s_1\big]+((5-2 d) d-5) d s_2^4\bigg\}B_0(s,0,0)-\sdfrac{\pi ^2 e^2}{(d-2) (d-1) d \sigma ^3 s_2} \bigg\{d^3 \sigma \notag\\
&\times \big[s_2 \left(-s^2+6 s s_1+3 s_1^2\right)-s_2^2 (s+3 s_1)+(s-s_1)^3+s_2^3\big]+d^2 \big[-2 s_2^3 \left(3 s^2+10 s s_1+19 s_1^2\right)-2 s_2^2 (s-3 s_1)\notag\\
&\times \left(3 s^2+6 s s_1+7 s_1^2\right)+s_2 (s-s_1) \left(9 s^3-11 s^2 s_1+11 s s_1^2+23 s_1^3\right)\notag\\
&+s_2^4 (9 s+17 s_1)-(3 s-5 s_1) (s-s_1)^4-3 s_2^5\big]+2 d \sigma \big[s^3-s^2 (2 s_1+s_2)+s \left(s_1^2-8 s_1 s_2-s_2^2\right)+s_2 (s_1-s_2)^2\big]\notag\\
&+16 s_1^2 \left(s^2-2 s (s_1+2 s_2)+(s_1-s_2)^2\right) (s-s_1+s_2)\bigg\}B_0(s_1)-\frac{2 \pi ^2 e^2}{(d-2) (d-1) d \sigma ^3} \bigg\{2 d^3 s \sigma  (s-s_1-s_2)\notag\\
&+d^2 \big[-5 s^4+8 s^3 (3 s_1+2 s_2)-2 s^2 \left(17 s_1^2+6 s_1 s_2+9 s_2^2\right)+8 s (s_1-s_2) (2 s_1-s_2) (s_1+s_2)-(s_1-s_2)^4\big]\notag\\
&+d \sigma  \left(5 s^2-8 s (2 s_1+s_2)+3 (s_1-s_2)^2\right)+8 s_2 \left(s^2-2 s (2 s_1+s_2)+(s_1-s_2)^2\right) (s+s_1-s_2)\bigg\}B_0(s_2)\notag\\
&+\frac{2 \pi ^2 e^2 s}{(d-2) d \sigma ^3} \bigg\{2 d s_2^2 \left(3 (d-2) s^2+(d-5) s s_1-5 s_1^2\right)+d (s-s_1)^2 \left((d-2) s^2-(d-4) s_1^2+6 s s_1\right)\notag\\
&+2 s_2 \left(-2 (d-2) d s^3+(d-5) d s^2 s_1+2 (d+8) s s_1^2+(d-1) d s_1^3\right)-2 d s_2^3 (2 (d-2) s+(d-5) s_1)\notag\\
&+(d-2) d s_2^4\bigg\}C_0(s,s_1,s_2)
\end{align}
\begin{align}
&F_{13,d}(s,s_1,s_2)=\frac{2 \pi ^2 e^2 s^2}{(d-2) (d-1) d \sigma ^2} \bigg\{-2 s_2 \big[((d-3) d+4) s+\big[(d-7) d+8\big] s_1\big]+((d-3) d+4)\notag\\
&\times(s-s_1)^2+((d-3) d+4) s_2^2\bigg\}B_0(s)-\sdfrac{\pi ^2 e^2 s_1}{(d-2) (d-1) d \sigma ^2 (s-s_1-s_2)} \bigg\{d^3 \sigma ^2-d^2 \big[3 s^4-2 s^3 (7 s_1+9 s_2)\notag\\
&+8 s^2\left(3 s_1^2+3 s_1 s_2+4 s_2^2\right)-2 s (s_1-s_2)^2 (9 s_1+11 s_2)+5 (s_1-s_2)^4\big]+2 d \big[2 s^4-11 s^3 (s_1+s_2)+s^2 \big(21 s_1^2\notag\\
&+24 s_1 s_2+19 s_2^2\big)+s \left(-17 s_1^3+5 s_1^2 s_2+25 s_1 s_2^2-13 s_2^3\right)+(s_1-s_2)^3 (5 s_1-3 s_2)\big]+8 s_1 \big(s^3-3 s^2 (s_1+s_2)\notag\\
&+3 s \left(s_1^2-s_2^2\right)-(s_1-s_2)^3\big)\bigg\}B_0(s_1)-\sdfrac{\pi ^2 e^2 s_2}{(d-2) (d-1) d \sigma ^2\g} \bigg\{d^3 \sigma ^2-d^2 \big[3 s^4-2 s^3 (9 s_1+7 s_2)+8 s^2 \big(4 s_1^2\notag\\
&+3 s_1 s_2+3 s_2^2\big)-2 s (s_1-s_2)^2 (11 s_1+9 s_2)+5 (s_1-s_2)^4\big]+2 d \big[2 s^4-11 s^3 (s_1+s_2)+s^2 \big(19 s_1^2+24 s_1 s_2\notag\\
&+21 s_2^2\big)+s \left(-13 s_1^3+25 s_1^2 s_2+5 s_1 s_2^2-17 s_2^3\right)+(s_1-s_2)^3 (3 s_1-5 s_2)\big]+8 s_2 \big[s^3-3 s^2 (s_1+s_2)-3 s \left(s_1^2-s_2^2\right)\notag\\
&+(s_1-s_2)^3\big]\bigg\}B_0(s_2)+\sdfrac{4 \pi ^2 e^2 s^2 s_1 s_2 (d \sigma +8 s_1 s_2)}{(d-2) d \sigma ^2 \g}C_0(s,s_1,s_2)
\end{align}
}%



\end{document}